\documentclass[11pt, twoside,openright]{report}

\usepackage{amsmath,amssymb,array}
\usepackage{amsthm}
\usepackage{graphicx}
\usepackage{pgf,pgfarrows,pgfautomata,pgfheaps,pgfnodes,pgfshade}
\usepackage{myheaders}
\usepackage{fancyhdr}
\usepackage{titletoc} % replacement for titles, headers, contents see http://www.tex-tipografia.com/archive/titlesec.pdf
\usepackage{mathrsfs}
\usepackage{pdfpages}
\newcommand{\beqs}{\begin{eqnarray}}
\newcommand{\eeqs}{\end{eqnarray}}

\newcommand{\beq}{\begin{equation}}
\newcommand{\eeq}{\end{equation}}
\newcommand{\tr}{\mathrm{tr\,}}
\newcommand{\Q}{{\cal Q}}

\newtheorem{prop}{Proposition}%[section]
\newtheorem{lemma}[prop]{Lemma}

\definecolor{FD}{rgb}{0.65,0.0,0}
\definecolor{geo}{rgb}{0,0.0,0.65}
\definecolor{GC}{rgb}{0,0.0,0.65}

  \newcommand{\rom}[1]{\mathrm{#1}}

  \newcommand{\be}{\begin{equation}}
  \newcommand{\ee}{\end{equation}}
  \newcommand{\bea}{\begin{eqnarray}}
  \newcommand{\eea}{\end{eqnarray}}
  
    \newcommand{\eps}{\epsilon}
  
  \newcommand{\p}{\partial}
  
  \newcommand{\s}{\sigma}
  
  \newcommand{\nn}{\nonumber}
\newcommand{\curl}{\mbox{curl}\,}
\newcommand{\sech}{\,\text{sech}\,}

  \newcommand{\n}{\nabla }

\def\cH{\mathcal{H}}

\def\cR{\mathcal{R}}

\newcommand{\DD}{{\mathcal{D}}}
\newcommand{\D}{{\nabla}}

\newcommand{\g}{\hspace{1pt}\mbox{}^3\hspace{-2pt}g}

\renewcommand{\th}{\theta}

\titlecontents{chapter}% le niveau que tu veux modifier
[0pc] % un peu d'espace vertical avant.
{}% place dans la page
{\\*\sffamily{ \slshape Chapter\ \thecontentslabel} - \large  \slshape  \sffamily     }% mise en forme du titre.
{\large \sffamily \slshape } % mise en forme du titre quand il n'y a pas de label.
{ \hfill \bfseries \contentspage}% l'espace est complŽtŽ par des points, et on indique le numŽro de la page
[\addvspace{.5pc}]% un peu d'espace vertical aprs

\makeatletter
\def\cleardoublepage{\clearpage\if@twoside \ifodd\c@page\else
  \hbox{}
  \vspace*{\fill}
  \begin{center}

  \end{center}
  \vspace{\fill}
  \thispagestyle{empty}
  \newpage
  \if@twocolumn\hbox{}\newpage\fi\fi\fi}
\makeatother
\begin{document}

%%%%%%%%%%%%%%%%%%%%%%%%%%%%%%% COVERPAGE %%%%%%%%%%%%%%%%%%%%%%%%%%%%%%

\begin{titlepage}

\begin{center}
\begin{tabular}{lc}
\parbox[c]{3cm}{\includegraphics[width=2.5cm]{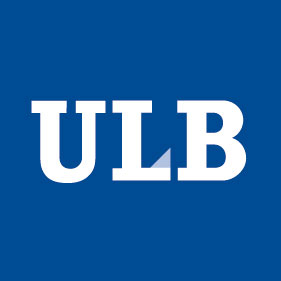}} &
\begin{minipage}[c]{0.6\linewidth}
  \begin{center}
   \textsc{Universit\'e Libre de Bruxelles}\\
   \bigskip
   \textsf{Facult\'{e} des Sciences}\\
   \bigskip
   \textsf{Physique Math\'ematique des Interactions Fondamentales} \\
  \end{center}
 \end{minipage}
\end{tabular}
\end{center}

%\begin{center}
%\textsc{Universit\'e Libre de Bruxelles}\\
%\medskip
%\textsf{Facult\'{e} des Sciences}\\
%\medskip
%\textsf{Physique Th\'{e}orique et Math\'ematique} \\
%\end{center}

%
%\begin{figure}
%\centerline{\includegraphics[width=2cm]{logoulb}}
%\end{figure}
%

\vfill

\begin{center}
\Huge{ \textit{Electric and magnetic aspects of gravitational theories}}
%$\mathZ{Of}$ $\mathcal N=1$
%$\mathZ{supersymmetric}$ $\mathZ{gauge}$}\\
%\medskip
%\Huge{$\mathZ{theories}$ $\mathZ{and}$ $\mathZ{localization}$}
\end{center}

\vfill

\begin{center}
{\large \bfseries{Fran\c cois \textsc{Dehouck}}}
\end{center}

\vfill

\begin{center}\textit{Th\`ese pr\'esent\'ee en vue de l'obtention du
titre de Docteur en Sciences.}\end{center}

\vfill

\begin{tabular}{rl}
\text{Directeur de th\`ese:} & \textsc{Riccardo Argurio}  
\end{tabular}

\begin{tabular}{rl}
\text{Membres du Jury:} & \textsc{Riccardo Argurio}  \\
&  \textsc{Glenn Barnich} \\
& \textsc{Marc Henneaux}   \\
& \textsc{Kostas Skenderis}  \\
& \textsc{Philippe Spindel}  \\
& \textsc{Antoine Van Proeyen}  
\end{tabular}

\hfill Ann\'ee acad\'emique 2010-2011

\newpage

\thispagestyle{empty}
\mbox{}

\newpage

\end{titlepage}

\newpage

%%%%% Reset pagenumbering to arabic style
\pagenumbering{roman} 
\section*{Remerciements}

{\small{
Comme tout physicien qui se respecte, je ne pourrais me permettre de remercier qui que ce soit sans indiquer au pr\'ealable mes conventions. C'est dans cet esprit que je tiens \`a preciser mon choix arbitraire de vous remercier en onze dimensions \`a l'aide d'une m\'etrique essentiellement positive (+,+,+,+,...). Vous remarquerez peut-\^etre  que certaines conventions non-sp\'ecifi\'ees, s\^urement consid\'er\'ees comme triviales par l'auteur pour le public auquel ceci s'adresse, peuvent substantiellement varier au cours du temps et de l'espace. Je ne chercherai jamais \`a m'excuser de telles actions volontaires mais seulement vous souhaiter la bienvenue dans le monde de la physique. Merci d\'ej\`a \`a Mike, Yves, Antonin, Josef et Amitabh pour avoir relu certaines parties de cette th\`ese et Riccardo pour l'avoir parcourue de bout en bout. Un tout grand merci \'egalement a mon jury de th\`ese pour toutes leurs remarques, questions, explications et interrogations.\\

La premi\`ere dimension, et s\^urement la plus importante, est d\'edi\'ee \`a mon promoteur Riccardo Argurio. Je ne saurais trop te remercier pour tout le temps que tu m'as consacr\'e et c'est selon toute logique que cette premi\`ere dimension, celle qu'on associe au temps, se devait de t'\^etre consacr\'ee. Tout a commenc\'e une fin d'apr\`es-midi de d\'ecembre 2006 o\`u tu m'as cont\'e, avec beaucoup d'engouement, la beaut\'e de la brisure dynamique de supersymm\'etrie. Ceci me plongea alors, sans r\'esistance, dans le doux refuge (en tout cas pour ceux qui ne savent pas r\'eellement ce que cela implique) du doctorat en septembre 2007 sous ta direction. En toute honn\^etet\'e,  je dois avouer maintenant que je n'ai \`a l'\'epoque strictement rien compris \`a cette brisure ! S\^urement encore trop ing\'enieur dans ma t\^ete ! D'ailleurs, \'evitez \`a l'avenir de me demander ce que c'est pr\'ecis\'ement, ca n'a jamais \'et\'e mon sujet de th\`ese. Pour revenir au propos, je voudrais te remercier pour ton enthousiasme tout au long de ce coaching de 4 ans, pour m'avoir fait rire, pour m'avoir soutenu, pour m'avoir appris la physique et le m\'etier de chercheur, et pour m'avoir guid\'e sur des routes difficiles, mais \^o combien passionnantes, de notre physique contemporaine. Je voudrais \'egalement te remercier pour m'avoir fait taire quand l'insouciance me faisait commuter des variables grassmaniennes ou remettre en doute le processus de s\'election des post-doctorants, mais aussi, pour avoir toujours cru en moi  et m'avoir fait confiance tout au long de ce parcours initiatique.\\

Ma deuxi\`eme dimension sera consacr\'ee \`a Laurent Houart. M\^eme s'il nous a quitt\'e ce 24 f\'evrier, je sais qu'il restera toujours au fond de moi quelques points noirs (ou blancs) reli\'es entre eux avec un petit +++ juste \`a c\^ot\'e, mais aussi quelques vannes digne des plus grands,... Finalement, quand j'y pense, je me rends compte que tu n'as jamais \'et\'e tr\`es loin tout au long de mon parcours de physicien. La premi\`ere fois que je t'ai rencontr\'e \`a la fin de mes \'etudes en Belgique, j'ai eu peur en serrant la main de ce personnage qu'on me pr\'esentait comme ``le professeur de th\'eorie des cordes". Ensuite, au Chili, c'est suite \`a un de tes fameux discours trilingues que je suis devenu un v\'eritable fan en co-cr\'eant le groupe des ``Dynkin boys" (par souci de discr\'etion, je tairai le nom de l'unique autre membre des Dynkin boys). C est aussi \`a ce moment l\`a que j'ai d\'ecid\'e qu'un jour j'\'ecrirais de la physique avec toi. Et voil\`a, comme en t\'emoigne notre ami Spires, maintenant c'est fait ! Un jour tu \'etais l\`a et le lendemain tu avais disparu pour toujours. Tu nous manqueras Laurent, tu me manqueras... Ca peut para\^itre d\'ebile mais merci \`a toi d'avoir \'et\'e toi !\\

Ma troisi\`eme dimension sera pour Marc Henneaux, notre chef \`a tous, pour avoir rendu cette aventure possible \`a l'aide d'un petit mail de seulement quelques lignes, envoy\'e \`a 13000 km de l\`a, qui aura fix\'e ces quatres ann\'ees v\'ecues ici. Ce fut un honneur pour moi d'avoir fait partie de ton groupe.\\

Ma quatri\`eme dimension est d\'edi\'ee \`a ce physicien hors pair, et hors normes, qui a pass\'e quatre longues ann\'ees \`a mes c\^ot\'es. Et quand je dis ``c\^ot\'e", je ne veux pas seulement dire sur une chaise d\'elabr\'ee \`a un m\`etre et beaucoup de poussi\`eres de moi bien s\^ur. Mes longues discussions avec C\'edric Troessaert \`a propos de la physique, et d'une vari\'et\'e (souvent inavouable) d'autres choses de la vie, m'ont permis de voir la vie de physicien autrement, mais aussi et surtout, d'approcher, d'attaquer, voire souvent de r\'esoudre des probl\`emes autrement... N'\'etait-ce pas R. Feynman qui disait "un physicien sait r\'esoudre un probl\`eme quand il conna\^it au moins sept mani\`eres diff\'erentes de le faire". Je voudrais te remercier C\'edric pour m'avoir (quasiment tout le temps) trouv\'e au moins une solution \`a des probl\`emes qui n'en poss\'edaient alors pas du tout \`a mes yeux.\\

Ma cinqui\`eme dimension, la plus \'etrange pour quelqu'un qui se serait arr\^et\'e au concept d'espace-temps quadri-dimensionnel mais tellement adapt\'ee au personnage, revient \`a Josef Lindman Hornlund, alias Tefatsexpert Ufo Lindemn... for ever. Cet extraordinaire physicien, tout droit sorti des contr\'ees froides du nord de l'Europe, a \'et\'e d'une grande importance pour moi tout au long de ce programme. Je suis heureux et fier de t'avoir connu Josef m\^eme si tu ne manges pas et ne t'habilles pas comme nous, et que tu \'ecoutes de la musique  \'etrange de 1981. Je te souhaite tout le bonheur du monde pour ta fin de parcours \`a Bruxelles et ta prochaine vie de retour en Su\`ede... ou ailleurs.\\

La sixi\`eme sera attribu\'ee \`a quelques (ex-)\'etudiants doctorants particuliers, tous aussi incroyables les uns que les autres, autant de l'ULB, que de la VUB et de la KUL. Pour commencer du c\^ot\'e wallon des choses, fiert\'e oblige (pas vrai Wieland ?), je voudrais remercier l'instanton du $\Omega$ background Vincent Wens pour ses nombreux conseils judicieux, ses d\'elires princetonniens, et cette d\'emonstration de "comment br\^uler plusieurs feux rouges \`a la chaine" qui restera grav\'ee \`a jamais dans ma m\'emoire. Merci aussi \`a Nassiba-Kac Tabti-Moody et Ella Jamsin pour m'avoir montr\'e que les filles dans la physique, ce n'est pas juste une l\'egende... mais une r\'ealit\'e \^o combien r´\'eelle, m\^eme si parfois trop \'eph\'em\`ere. Merci \`a vous deux pour tous ces d\'elires partag\'es. Merci \'egalement \`a ce bon vieux fractionnaire de Cyril Closset \`a qui j'ai parfois, je l'avoue, men\'e la vie dure. Merci \`a toi pour toutes ces histoires de physique, et puis aussi de physique (je te charierai jusqu'au bout... qu'est-ce que tu crois toi !). Du c\^ot\'e flamand, merci \`a Bert Vercnocke, ``mon ami", et Jan de Rydt pour nos escapades sur les routes endiabl\'ees des supergravit\'ees jaug\'ees. Merci aussi \`a Walter Van Herck pour nos longues discussions, lors de conf\'erences \`a l'\'etranger, et en particulier celle sur le statut des ing\'enieurs en physique th\'eorique, un bijou ! Du c\^ot\'e flaminguant (ceci est \`a consid\'erer sur un ton ironique bien s\^ur), merci \`a Wieland Staessens pour ce duo de choc lors de l'organisation de la sixi\`eme \'edition de la Modave Summer School et aussi pour toutes ces (parfois longues) histoires sur les myst\`eres de la physique contemporaine. C'est bien toi qui m'a expliqu\'e le premier ce qu'\'etait une brane et pourquoi ces maudites cordes ouvertes \'etaient accroch\'ees dessus ! Allez, dank u wel meneertje... ou meneereke ?\\

Dans la septi\`eme dimension, I would like to thank two particular post-docs that I have had great pleasure to work with. It has been very rewarding for me to work with Amitabh Virmani, a person definitely full of ideas who came to my office to give me real interesting work. I am valuable to you for all the advices you gave me, the answers you brought me, the long discussions we had in your office, the non-singular laughs and the really crazy laughs during our moments of doubts, or should I say of fun, all along this NUT momenta of ours. Next to him, there is also Geoffrey Comp\`ere, another married man. I dont know if I should switch again conventions but I guess I won't because this can get dangerous as you do not remember in the end with what conventions you started with, or what is true on-shell or off-shell, right Geoffrey ? So, I would like to thank you for your incredible calm during my incomprehensible fights about signs, factors of two,..., for your great ideas and following required explanations all along these two projects we have been working on, and eventually for your support and gestures of sympathy during the though moments I have gone through with the post-doc procedure. Wish all the best to you guys, professors of the future !\\

La huiti\`eme dimension sera d\'edi\'ee \`a tous ces autres physiciens que j'ai eu le plaisir de c\^otoyer durant ces quatre longues ann\'ees et qui ont tous, que ce soit lors d'une br\`eve discussion ou lors de longues \'echappades parfois m\^eme alcoolis\'ees, contribu\'e d'une facon ou d'une autre \`a mon bien \^etre en tant que physicien de type humanoide et \`a ce que cette th\`ese soit ce qu'elle est. Je remercierai dans un ordre tout \`a fait al\'eatoire, et pourtant \^o combien gr\^acieux:  Alice Bernamonti, Sophie et St\'ephane Detournay et leur visa, Jarah Evslin et son appart, Frank Ferrari, Davide Forcella, Bram Gaasbeek, Federico Galli, Sung-Soo Kim, Axel Kleinschmidt, Semyon Klevtsov, Chetan Krishnan, Pierre-Henry Lambert, Gustavo Lucena G\'omez, Carlo Maccafferi, Alberto Mariotti, Micha Moskovic, Johannes Oberreuter, Jakob Palmkvist, Daniel Persson, Diego Redigolo, Antonin Rovai,  Cl\'ement Ruef,... et beaucoup d'autres encore s\^urement que vous me pardonnerez d'avoir oubli\'e... I can not have a word for each of you...but I guess you can picture it yourself, right ? \\

Ma neuvi\`eme dimension sera d\'edi\'ee \`a ma famille. Tout d'abord, je voudrais remercier mes parents qui m'ont permis cette escapade au Chili, mon passage initiatique et obligatoire dans la physique, et qui m'ont toujours soutenu dans mes aventures. Egalement, je remercierai  mon fr\`ere pour m'avoir donn\'e go\^ut aux sciences, pour m'avoir expliqu\'e beaucoup de choses alors que j'\'etais plus jeune et pour avoir, plus r\'ecemment, essay\'e d'\'ecouter d'une oreille avertie mes d\'elires de physicien (peut-\^etre tu faisais semblant, mais j'aurais du mal \`a t'en vouloir au vu de ma connaissance exemplaire du prion). Merci \`a ma soeur C\'eline pour un million de choses et surtout pour m'avoir montr\'e la seule chose peut-\^etre importante dans la vie:  "il existe une autre vie apr\`es le doctorat !". Merci aussi \`a Mike pour m'avoir mis 6-0 quand je finissais par penser que j'\'etais plus malin que tout le monde. Merci \'egalement \`a mon cousin Dirk pour nos nombreuses discussions physico-philosophiques et la splendide (pas vrai ?) cover de cette th\`ese. Finalement, merci \`a Caroline d'\^etre et d'avoir toujours \'et\'e une t\'emoin si attentive de mes tribulations.\\

La dixi\`eme dimension sera d\'edi\'ee \`a tous ces amis et toutes ces amies, dont beaucoup n'aiment pas la physique et c'est tr\`es bien comme ca, indispensables \`a la sant\'e mentale d'un chercheur. A Sarah et Magali (pas pr\^etes \`a oublier ces deux heures de questionnements sur la physique je pr\'esume, ou c'est d\'ej\`a fait ? ), Firenze, Cadichou, AM, Fred et Sophie, Iron Pat, Frans de Cotonou et AC, le Tchim et sa m\'elodie, Phil et sa technique de triche \`a Carcassonne, C\'eline et son aide logistique contre les techniques frauduleuses de la police Schaerbeekoise, Hieu pour son accueil \`a Bangkok et son raid manqu\'e \`a mon mariage, Dr. Krizz et ses c\^ables, MG et les souvenirs de nos discussions endiabl\'ees, \`a tous les animateurs de la 48\`eme Unit\'e scoute pour leur d\'evouement (m\^eme dans sa th\`ese, il cherche la guerre lui...), à la psycho team et leurs amoureux Marie, Val, Crucifix, Nath, Cath-Laurent et la pluie, Alicia et Seb (au moment o\`u j'\'ecris ces lignes, l'info est confirm\'ee), Alexia et grand aigle bleu, Ju et Lau, No et Henry. Et puis merci aussi \`a tous les autres que j'ai oubli\'e de citer ou que je ne citerai pas en pensant qu'ils ne liront s\^urement jamais ces lignes...\\

La onzi\`eme, parce que tout le monde sait qu'il n'y a pas de douzi\`eme dimension sauf en $F$-theory, sera d\'edi\'ee \`a celle avec qui j'ai partag\'e mon spin 2 tous les jours qu'Einstein nous a donn\'e. Le 23 octobre 2010, on a d\'ecid\'e de faire un pas de plus dans la vie. M\^eme si le spin est rest\'e le m\^eme (heureusement en fait parce que les interactions entre particules de spin plus grand que 2, c'est gal\`ere), je suis fier et heureux de l'avoir fait avec toi et de te savoir toujours \`a mes c\^ot\'es. Voil\`a, on en a v\'ecu des aventures,... M\^eme si celles de la physique ont un go\^ut de fin, je garde en t\^ete le fait que tu adores d\'efinir la supersymm\'etrie. Je t'avais d\'ej\`a dit que j'y serais pas arriv\'e sans toi ? Ah bon...

}}

\newpage

\thispagestyle{empty}
\mbox{}

\newpage

\cleardoublepage

%%%%%%%%%%%%%%%%%%%%%%%%%%%%%%%%%%%%%%%%%%%%%%%%%%%%%%%%%%%%%%%%%

\chapter*{A first step into theoretical physics}\label{chap:intro}
\addcontentsline{toc}{chapter}{\textit{A first step into theoretical physics}}

Theoretical physics is a branch of physics whose ultimate goal is to formulate a (hopefully) unified theory that would be able to describe all phenomena that surround us, all laws of Mother Nature. One could picture it as the search for the DNA code of Nature.  Our most serious candidate for this unified theory, the ``theory of everything", is known as String Theory. As of today, the ideas developed in theoretical physics may rather sound like mathematical curiosities disconnected from reality. However, one should be aware that a certain amount of mathematical abstraction has always been required to construct theories that describe phenomenas of Nature. This awkward feeling one may experience with respect to contemporary theoretical physics is certainly strongly correlated to the absence of experimental verifications of the theories developed and studied by the actual community of theoretical physicists. 

Once upon a time, physicists used to formulate theories to describe phenomena of Nature they could experience.  Isaac Newton's theory of classical mechanics and law of universal gravitation were formulated in 1687 relying upon experiments. However, we know today that Newton's theory of classical mechanics only describes systems of particles at sizes and velocities we can experience in daily life.  An experimentally tested theory is a valid theory if it gives a good description of some phenomena ``within the limits of accuracy of the experiments one can design to check its validity".

The first hint that Newton's classical mechanics is only useful to describe systems in specific regimes appeared through the formulation of the laws of electromagnetism, describing electric, magnetic and optical phenomena, as given by James Clerck Maxwell in 1861. From one point of view, the theory of electromagnetism is clearly different from classical mechanics as it is a field theory where the variables can depend on the time coordinate but also on the coordinates of space. From another point of view, electromagnetism is also a theory whose formulation was dictated by experiments. As such, if the theories of both Maxwell and Newton were describing physical systems, they should rely on the same underlying physics.  The discovery of electromagnetic plane waves, as solutions of Maxwell's equations, and the Michelson-Morley experiment led to conclude that light was an electromagnetic wave that traveled at the same velocity in any Galilean reference frame. This was obviously in contradiction with the underlying concepts of Newton's classical mechanics. This situation is maybe one of the most illustrative examples where  inconsistency between theories relying on experiments was dealt with through theoretical considerations. 

The resolution of this contradiction was brought in by Einstein in 1905 in his theory of special relativity. Einstein postulated that one can not go faster than light and that physics should be equivalent in every inertial reference frame. To agree with this, one has to generalize the notion of Galilean reference frames. Relativistic classical mechanics is a generalization of Newton's classical mechanics that takes into account effects predicted by special relativity. The beauty of special relativity is that Maxwell's field theory of electromagnetism was already cast in a form that agrees with Einstein's theory. As such, Maxwell's theory is a relativistic field theory. 

 From experimental facts, it was already quite clear at the time that Newton's theory of classical mechanics was also not valid to describe very small sized objects such as subatomic particles. The description of such systems was explained  through the theory of quantas at the beginning of the XX$^{th}$ century. The quantum theory of electromagnetism, a relativistic quantum field theory, required the matching of the quantum theory of particles with special relativity. The formulation of such a theory led to the construction of relativistic quantum field theories which provide a unified framework to describe field-like objects and particle-like objects. 

Along the XX$^{th}$ century, two new interactions known as the weak and strong interactions were uncovered. The range of these interactions is very small. This makes them purely quantum-like in Nature. Just like electromagnetism, these interactions are also described by specific quantum field theories.  The unification of the electromagnetic, weak and strong interactions is formulated by the Standard Model of particles. In this model, the interactions mediate the dynamics of the elementary parts of matter, called elementary particles. These elementary particles are understood as sources of the fields.

In this whole discussion, we have left aside Newton's law of universal gravitation. Along with the discovery of special relativity, it seemed logical that one should modify such a law in a way that takes into account relativistic effects.  Instead of dealing with such a reformulation, Einstein introduced a totally new concept of space and time based on the principle of equivalence, which roughly states that all observers fall the same way in a gravitational field. Its main idea was to translate the free-fall trajectory followed by an object submitted to a gravitational field into a trajectory in a curved spacetime background. The gravitational field is also understood as the curving of spacetime, while energy acts as its source. This theory of General Relativity was formulated by Einstein in 1915. 

The experiments one can design today allow us to appreciate the beauty of the quantum field theories of the electromagnetic, weak and strong interactions but also of the theory of General Relativity describing the gravitational interaction. 
These theories could be thought, at our time, just like Newton's law at the time it was formulated;  they are valid theories in the sense that they have survived all experimental tests one was able to design to check their validity.  

However, there is a main difference. Indeed,  for example, the theory of general relativity has predicted its own domain of validity  through the existence of black hole singularities. For describing such entities, one needs an understanding of the gravitational interaction at the quantum level. Also, the formulated quantum field theories are not well understood in non-perturbative regimes. As such, and in the ``absence" of experimental facts, new insight is required.

In the last thirty years or so, theoretical physics has become more and more a branch of physics in itself which tries to approach the formulation of a quantum theory of gravitation, and the description of non-perturbative phenomena, through the study of symmetries, dualities and correspondences between various theories. The most famous theory that has come out of these analyses is known as String Theory. In this theory, one considers elementary particles and interactions as different vibrations of extremely small unidimensional strings. The main success of string theory is that it provides a quantum theory that contains gravity, i.e. the spin 2 particle is also understood as a mode of the vibrating string. 

To deal with the many puzzling phenomena Nature wants to reveal us, one needs to be equipped with appropriate mathematical tools. The work presented in this thesis deals with aspects of the theory of General Relativity and its symmetries. I would like to consider this work as a piece of the puzzle.

%%%%%%%%%%%%%%%%%%%%%%%%%%%%%%%%%%%%%%%%%%%%%%%%%%%%%%%%%%%%%%%%%%%%

\cleardoublepage

\chapter*{About this thesis}\label{aboutthisthesis}
\addcontentsline{toc}{chapter}{\textit{About this thesis}}

The research I conducted during these last four years, which is reported in this thesis, was motivated by the application of the electromagnetic duality idea to general relativity. It has been shown, in 2004, that such a duality symmetry exists at the level of the linearized action of general relativity \cite{Henneaux:2004jw}. It is known as gravitational duality. The hope that gravitational duality could be a symmetry of the non-linear theory finds its origin in the existence of a solution of the non-linear Einstein's equations, known as the Lorentzian Taub-NUT metric, which seems to describe a gravitational monopole. This solution presents several aspects of an \textit{ill-behaved} solution of Einstein's equations, as described by the usual notions and tools introduced by general relativity,  and is often  \textit{rejected on physical grounds}. However, if general relativity predicts its existence, I believe that this may not be the correct attitude. Indeed, I think one should rather try to explain how it can be described or, at least, try to formulate the appropriate framework where one could deal with such solutions. This thesis addresses the gravitational duality symmetry in the linearized theory and highlights, from several different perspectives, the problems underlying a complete understanding of gravitational duality, and the description of dyonic solutions, in the non-linear theory. 

During my work on gravitational duality, I got interested in the study of charges associated to asymptotically flat spacetimes at spatial infinity in general relativity or supersymmetric extensions of it. The usual ``Noether" charges one defines in general relativity are referred to as ``electric" charges. The topological ones, which we were able to define at the linearized level through gravitational duality, are referred to as ``magnetic" charges. It is using a specific framework, known as the Beig-Schmidt formalism, that I started wondering about a possible formulation of topological charges at the non-linear level. However, I realized that, even nowadays, some subtleties in the definitions of ``electric" conserved charges at spatial infinity or in hypotheses underlying validity of variational principles are not completely settled. In the last two years, I have mainly focused on trying to clarify these issues. The history will tell us if this was of any help in understanding ``magnetic" charges at the full non-linear level. 

This thesis describes the work that was presented in the following publications, given in chronological order,
\begin{itemize}
\item[1.] \textit{``Supersymmetry and Gravitational duality"}\\
R. Argurio, F. Dehouck, L. Houart \\
arXiv:0810.4999v3 [hep-th] \textbf{Phys. Rev. D79:
125001, 2009.}
\item[2.] \textit{``Boosting Taub-NUT to a BPS NUT-wave"} \\
R. Argurio, F. Dehouck, L. Houart\\
arXiv:0811.0538v1 [hep-th]
\textbf{JHEP 0901:045,2009.}
\item[3.] \textit{``Why not a di-NUT ? or Gravitational duality and rotating
solutions"} \\
R. Argurio, F. Dehouck \\
arXiv:0909.0542 [hep-th]  \textbf{Phys. Rev. D81:064010,2010.}
\item[4.] \textit{``Gravitational duality in General Relativity and Supergravity theories}\\
F.Dehouck\\
arXiv:1101.4020 [hep-th] \textbf{Nucl.Phys.Proc.Suppl.216:223-224,2011.}
\item[5.] \textit{``On Asymptotic Flatness and Lorentz Charges"}\\
G. Comp\`ere, F. Dehouck, A. Virmani\\
arXiv:1103.4078 [gr-qc] \textbf{Class.Quant.Grav.28:145007,2011.}
\item[6.] \textit{``Relaxing The Parity Conditions of Asymptotically Flat Gravity"}\\
G. Comp\`ere, F. Dehouck\\
arXiv:1106.4045v2 [hep-th] \textbf{Class.Quant.Grav.28:245016,2011.  }
\end{itemize}

We have chosen to split this thesis into three parts, in a way that seemed more appropriate for a presentation of conserved charges in gravity theories, as we now detail.

\subsubsection{Part I - Electric side: Asymptotic flatness and Poincar\'e charges }

The first part deals with the definition of ``electric" charges for asymptotically flat spacetimes at spatial infinity. As we have not been able to make sense of ``magnetic"  charges in the non-linear context, we will not have much to say about them in this first part. The original work presented here is contained in the two more recent publications listed here above. 

Conserved charges for gauge field theories can be constructed from the consideration of asymptotic gauge transformations, which act as ``global" transformations at large distances. As a consequence, the study of such charges must proceed through the description of the asymptotic properties of the fields. In more technical terms, we deal with specific boundary conditions which specify a particular class of solutions that \textit{behave asymptotically in the same way}. Given a set of such boundary conditions, one can study the asymptotic symmetries and construct the ``Noether" charges generated by these symmetries, in terms of surface integrals. 

General Relativity is a non-linear theory of space and time that is invariant under diffeomorphisms, i.e. under local reparametrizations of coordinates. It is thus also possible to define conserved charges as we have just explained. However, this theory presents two major difficulties. The first problem is related to the fact that the field in question is the metric. The background field is now also the dynamical field. For constructing charges associated to asymptotic diffeomorphisms, one has to describe first in what sense asymptotic properties of the metric should be understood.  Secondly, because it is a non-linear theory, conserved charges associated to specific asymptotic symmetries might turn out to present non-linearities in the asymptotic fields. The analysis is thus more complicated than for linear gauge field theories, such as electromagnetism. 

The study of conserved charges in general relativity was initiated by considering a class of spacetimes that approach Minkowski spacetime, in two different regimes known as null infinity and spatial infinity. These solutions are referred to as asymptotically flat spacetimes.  We only restrict in this first part to considerations at spatial infinity. The study of asymptotically flat spacetimes at spatial infinity has a long history. Nevertheless, the topic has  constantly been evolving through the years,  see e.g. \cite{Dirac:1958jc,Deser:1960zzc,Arnowitt:1961zz,Arnowitt:1962hi, Regge:1974zd,Geroch:1977jn,Ashtekar:1978zz} for a relevant sample of classic works before the eighties, \cite{Witten:1981mf,Abbott:1981ff,Beig:1987aa,Ashtekar:1991vb,Wald:1993nt,Iyer:1994ys,Wald:1999wa,Szabados:2003yn,Mann:2005yr,Barnich:2001jy,Barnich:2007bf} for a sample of works in the last thirty years. The traditional approach presented in the literature can be summed up as follows.  The set of boundary conditions are fixed so that they define a set of physically interesting spacetimes, such as the Schwarzschild or Kerr black holes, and so that charges can be made finite and conserved. These works have all found that the non-trivial asymptotic symmetries restrict to the isometries of the Minkowski metric. As such, one obtains a description of asymptotically flat spacetimes at spatial infinity in terms of Poincar\'e  charges. One important feature of all these constructions is that the set of conserved charges are linear in the fields. 

The specification of the asymptotic symmetries, through the choice of specific boundary conditions, is however a quite difficult task. In this first part, we would like to emphasize the importance of the study of the equations of motion to see what restrictions should be imposed on the asymptotic fields. Also, we will discuss the problem of regulation of infinities which seems to have been, up to now, the main guideline in the choice of boundary conditions. Through the study of the equations of motion and the definition of a good variational principle, we achieve a description of a class of asymptotically flat metrics that is more general than previously considered in the literature. Our conserved and finite charges represent a larger asymptotic symmetry group than the Poincar\'e group. Also, we find that the Lorentz charges may present non-linearities in the asymptotic fields.  \\

\textbf{Chapter 1:} This first chapter is intended as a broad review of conserved charges associated to global and gauge symmetries of an action as described by Noether's theorem.  We mainly focus on the construction of ``global" charges for general relativity. For example, we review the construction of Abbott and Deser \cite{Abbott:1981ff} who defined charges associated to isometries of a background metric. In the asymptotically flat regime, we present a review of the methods used to describe the asymptotic properties of Minkowski spacetime. At spatial infinity, we present the work of Regge and Teitelboim \cite{Regge:1974zd} who first obtained, from the Hamiltonian action, Poincar\'e surface charges as generators of asymptotic symmetries. \\

\textbf{Chapter 2:} We review the Beig-Schmidt formalism to describe asymptotically flat spacetimes at spatial infinity. We propose an extension of their definition of asymptotic flatness by considering an extended class of metrics. We study the generic construction of independent, conserved, finite, and non-trivial charges one can built for such spacetimes through the study of the equations of motion in the asymptotic expansion. 

Although we recover the standard results for the ``usual" boundary conditions, we stress that the equations of motion do impose less stringent restrictions. \\
 
\textbf{Chapter 3:} This chapter elaborates on the considerations presented in chapter 2 while making use of covariant methods to construct charges associated to asymptotic symmetries. Our first result is a clear understanding of the equivalence between counter-term charges constructed from the stress-energy tensor of Mann and Marolf \cite{Mann:2005yr} and the construction of Ashtekar and Hansen \cite{Ashtekar:1978zz}. From the study of the symplectic structure for our class of spacetimes, we show that the Einstein-Hilbert variational principle is ill-defined when specific parity conditions are not imposed. We propose a regulation of the phase space for a class of spacetimes, where we do not impose these parity conditions, through a fixation of the ambiguity in the off-shell Einstein-Hilbert action and the symplectic structure obtained from this action. This analysis generalizes the constructions that are present in the literature and provides a new way of looking at spatial infinity in the asymptotically flat regime. \\

\subsubsection{Part II: Magnetic theory through duality}

Gravitational duality is a symmetry of the linearized Hamiltonian action of General Relativity. If we can define ``electric" charges, one should be able to define ``magnetic" charges to characterize the solutions obtained through duality transformations. It is in this sense that the linearized Taub-NUT solution was first understood as a gravitational dyon, see for example \cite{ramaswamy,ashtekar,Mueller:1985ij}. 

The second part of this thesis deals with gravitational duality in the linearized theory. We use it as a playground to construct topological charges and study the sources of dual solutions obtained through duality rotations. \\

\textbf{Chapter 4:} This chapter is a review of the electromagnetic duality as a classical symmetry of Maxwell's equations. We also briefly comment on the great successes of this theoretical construction, such as the explanation of the quantization of the electric charge. \\

\textbf{Chapter 5:} In here, we review the gravitational duality as a symmetry of the equations of motion. We propose a definition of ten ``magnetic" charges at spatial infinity, these are referred to as the dual Poincar\'e charges. We use this construction as a playground to study the dual solutions of some ``electric" solutions such as the Schwarzschild and Kerr black holes, and the shock pp-waves.  In the presence of topological contributions, we point out the difficulty, already at the linearized level, of a definition of Lorentz charges in terms of surface integrals. \\

\subsubsection{Part III: Gravitational duality and Supersymmetry}

Supersymmetry has been one of the major ingredients in providing evidence for
dualities in the realm of string theories and M-theory. In particular, there
is a very tight relation between U-duality \cite{Hull:1994ys}, the most
general duality encompassing electric-magnetic duality, S-duality and
T-duality, and the existence of BPS bounds following from the most general
maximally extended supersymmetry algebra. This relation follows from the fact
that states (or supergravity solutions) which preserve some supersymmetries
also saturate a BPS bound which takes the form
$M=|Z|$, where $Z$ is a U-duality invariant combination of all the possible charges
arising in the specific theory one is considering. These charges, which
correspond to possibly extended charged objects, arise in the supersymmetry
algebra as central extensions \cite{Azcarraga:1989fk,Townsend:1995gp}, and this is the reason why they enter in the BPS bound.

It is however striking that U-duality acts only on the right hand side of the BPS bound,
while it leaves the left hand side, $M$, invariant. It is natural to ask
whether there are more general duality transformations that also act on $M$.
It is because of these considerations that we believe gravitational duality,  which maps the mass $M$ to a magnetic mass $N$, may play an important role, see also 
\cite{Henneaux:2004jw,Deser:2005sz,Bunster:2006rt,Cnockaert:2006gm,
Leigh:2007wf,Bergshoeff:2008vc,Alvarez:1997yp,Hull:1997kt}
 
 It is the purpose of this part III to study the ``magnetic" solutions as supersymmetric solutions of supergravity theories and their ``topological" contributions to the BPS bounds and the supersymmetry algebras. \\

\textbf{Chapter 6:} To deal with solutions of supergravity theories, we start by a review of several aspects relevant to the original work presented in the next chapter. We review how supergravities are local supersymmetric theories and present the $\mathcal{N}=1$ and $\mathcal{N}=2$ supergravities. We then elaborate on the construction of bosonic solutions to these theories. We then comment on a specific method to solve for the Kiling spinors, parameters of supersymmetry transformations.\\

\textbf{Chapter 7: } In this last chapter, we establish the supersymmetry properties of the Lorentzian charged Taub-NUT solution in $\mathcal{N}=2$ supergravity and review the appearance of the NUT charge in the BPS bound. We also recover the expressions for the dual momenta established in Part II by considering a complexified Witten-Nester two-form. This construction also illustrates how the NUT charge copes with the $\mathcal{N}=2$ supersymmetry algebra.  We end up by discussing these considerations in $\mathcal{N}=1$ supergravity through the study of pp-waves solutions. 

%%%%%%%%%%%%%%%%%%%%%%%%%%%%%%%%%%%%%%%%%%%%%%%%%%%%%%%%%%%%%%%%%%%%
\cleardoublepage

\tableofcontents

\cleardoublepage

%%%%% Reset pagenumbering to arabic style
\pagenumbering{arabic}

\part{ Electric side: Asymptotic flatness and Poincar\'e charges}

\renewcommand{\thesection}{\arabic{chapter}.\arabic{section}}

%%%%%%%%%%%%%%%%%%%%%%%%%%%%%%%%%%%%%%%%%%%%%%%%%%%%%%%%%%%%%%%%%%%%
\chapter{General relativity and conserved charges}\label{chap:lagvsham}

In this chapter, our main concern will be to review the fact that, for general relativity, conserved charges can be expressed as surface integrals and are associated to asymptotic symmetries which are to be understood as asymptotic diffeomorphisms that preserve the form of a given class of metrics at infinity.

To arrive at such statements, we first review, in section \ref{lagvshamm},  the Lagrangian and Hamiltonian reformulations of Newton's classical mechanics. This allows us to state Noether's theorem which shows that to any differentiable symmetry of an action, one can associate a conserved charge, known as the Noether charge. The study of the conserved charges of a system can thus be engineered from the study of the symmetries of the action. In Hamiltonian formalism, we see that the Noether charges of the system are also the generators of the symmetries of the action. To describe gauge systems, we briefly review the Hamiltonian formulation of constrained systems. We obtain that gauge symmetries are generated by first class constraints and are thus vanishing on-shell. 
In section \ref{einsteinn}, we present the Lagrangian and Hamiltonian formulations of General Relativity. We also review how Einstein's equations can be brought into a system with a well-posed initial value formulation. 
In section \ref{conservedy}, we apply Noether's theorem to gauge fields theories and show that conserved charges should be associated to asymptotic symmetries. For general relativity, we review the construction of Abbott and Deser \cite{Abbott:1981ff} who constructed conserved charges associated to isometries of a background metric. 
The section \ref{pathy} is devoted to the study of the asymptotic region of Minkowski spacetime, the spacetime of special relativity. We discuss the presence of two separated regions at infinity known as null and spatial infinity. Focusing on spatial infinity, we review two different ways the information reaching this region can be described by means of a limiting procedure of information contained on three-dimensional hypersurfaces.
The seminal work of Regge and Teitelboim  \cite{Regge:1974zd} which established the role of surface integrals as generators of asymptotic symmetries at spatial infinity is reviewed in section \ref{sec:RT}. 
Eventually, we end up in section \ref{sec:summaryyy} by a brief summary and a discussion about several aspects concerning the determination of asymptotic symmetries. This discussion is also intended as a motivation for the original work presented in the next two chapters.

\setcounter{equation}{0}
\section{Gauge symmetries and associated charges}
\label{lagvshamm}

Gauge theories are theories that are invariant under local transformations of the variables. In more formal language, one says that gauge transformations map an allowed state, described by the observables (in the case of electromagnetism by the values of the electric and magnetic fields,...), to another allowed \textit{equivalent} state. There is thus some redundancy in the description of the physical variables. The fact that gauge transformations are maps between \textit{equivalent} states is to be understood in constrast with global symmetries which are symmetries of the theory but \textit{do change} the state of the system.

In this section, we review basic facts about the Lagrangian and Hamiltonian formalisms of classical mechanics. We stress out how Noether's theorem is expressed in those two languages. In the Hamiltonian formalism, we see that the Noether charge, associated to a symmetry of the Hamiltonian action, is also the generator of that symmetry. We then see how gauge systems are constrained Hamiltonian systems and briefly review the study of such systems. The main result of this section is the expression of the generator of a gauge symmetry, i.e. the conserved charge. We see that the generating function can be expressed in the basis of (first class) constraints. It is thus zero on-shell. In the next section, we will comment on how these considerations are generalized to field theories and how one can make sense of non-trivial conserved charges associated to asymptotic gauge symmetries.

We should stress that the considerations in this section are largely borrowed from the book of M. Henneaux and C. Teitelboim \cite{Henneaux:1992ig} and also from unpublished notes prepared for lectures I gave in september 2010, in collaboration with C. Troessaert, at the sixth Modave Summer School, Modave, Belgium.

\subsection{Lagrangian and Hamiltonian formalisms}

The Lagrangian and Hamiltonian formalisms are reformulations of the theory of classical mechanics of Newton.
The basic tool of the Lagrangian formalism is to use, instead of the usual coordinates used in Newton's classical mechanics (and leading to vectorial equations), generalized coordinates $q_i$. When some of Newton's equations may be redundant and way more complicated to solve, the use of these independent coordinates describe the real degrees of freedom of the system and simplify greatly its study.
The Euler-Lagrange equations (1788), desribing a system with $N$ degrees of freedom, were obtained by re-expressing d'Alembert's principle of virtual works using  variational calculus. The result is
\beqs
\frac{d}{dt} \: \frac{\partial L}{\partial \dot{q}_i}-\frac{\partial L}{\partial q_i}=0,
\eeqs
where $i=1...N$.  Newton's vectorial equations are reduced to a system of  $N$ second order differential equations.
Here,  $L$ is the Lagrangian
\beqs
L=L(q_i, \dot{q}_i)=T-V,
\eeqs
where $T$ and $V$ are respectively the kinetic and potential energy of the system.

During his work on reformulating Lagrange classical mechanics, as we review below, Hamilton noticed that Euler-Lagrange equations can actually be obtained from an action principle.
This is known as Hamilton's principle and states that starting from the action
\beqs
S= \int L dt \; ,
\eeqs
 and demanding that the variation of this action is zero, or asking stationarity of this action, $\delta S=0 $ we recover Euler-Lagrange equations.  Derivation of equations of motion by means of a variational principle is by now a central concept in physics. We say that an action possesses a good variational principle if the variation of the action is zero upon imposing the equations of motion.

What Emmy Noether proved in 1918, is that to every symmetry of the action, one can associate a conserved charge. By invariance of the action we mean that under a  transformation of the generalized coordinates
\beqs
q_i \rightarrow \tilde{q}_i(q,s)= \tilde{q}_i(q,0)+\frac{d\tilde{q}_i}{ds}\biggr |_{s=0} s+O(s^2)\; ,
\eeqs
where
$\tilde{q}_i(q_j,0)=q_{i}(0)$,
the action is such that $\delta S=0$. Alternatively, this means that the variation of the Lagrangian is equal to a total derivative
\beqs
\delta L=\frac{dL}{ds} \biggr |_{s=0}=\frac{df}{dt}\; .
\eeqs
To recover Noether's result, we see that from considering a generic variation of $\delta L$, we can write 
\beqs\label{blih}
\frac{dL}{ds}=\frac{\partial L}{\partial q^i} \frac{dq^i}{ds}+\frac{\partial L}{\partial \dot{q}^i} \frac{d\dot{q}^i}{ds}=\frac{df}{dt} \; ,
\eeqs
and that, upon using the Euler-Lagrange equations of motion, we have
\beqs
\frac{d}{dt} \biggr [\frac{\partial L}{\partial \dot{q}^i} \frac{dq^i}{ds}-f \biggl ]=0\; .
\eeqs
A constant of the motion, or a conserved charge $Q$, is a function such that $dQ/dt=0$. We have thus proven that when a general transformation is a symmetry of the action, one can define an associated conserved charge which is
\beqs
Q\equiv \frac{\partial L}{\partial \dot{q}^i} \frac{dq^i}{ds}-f\; .
\eeqs
As an example, let us suppose we want to deal with a system that is invariant under translations of time. We set the parameter $s=t$ and from \eqref{blih} we have $f=L$ in the above demonstration. The conserved charge is just
\beqs
\frac{dH}{dt}=0 , \qquad H\equiv\frac{\partial L}{\partial \dot{q}^i} \frac{dq^i}{dt}-L=p_i \dot{q}^i-L.
\eeqs
We will see in the following that $H$ is known as the Hamiltonian. It represents the energy of the system. Through Noether's theorem, we see that the conserved energy of a system is associated to the invariance of this system under time translations.

Rowan Hamilton (1805-1865) noticed that the Lagrangian formalism may be confusing because the generalized velocities $\dot{q}_k \equiv dq_k/dt$ seem at first sight to depend on the generalized coordinates. However, given the $2 N$ initial coordinates $q$ and $\dot{q}$, we would like to specify, using the Euler-Lagrange equations\footnote{Note that the initial conditions are supposed to be completely independent here. This is not always the case as there can be relations between them, that we call constraints. Constrained systems will be discussed in the following.}, the future of the system at any time $t$. If this can be done, we say that the system has a good initial value formulation. We thus see that the $\dot{q}_i (t)$ are depending on the $N$ initial conditions $\dot{q}_i (0)$ of the generalized velocities but not on the generalized coordinates $q_i$. Hamilton's reformulation of the Lagrangian formalism resides in a change of variables to get rid of this potential confusion and set the $2N$ coordinates on the same footing. He introduced a set of $2 N$ independent coordinates $q_i$ and $p_i$ where $p_i$ are the conjugate momenta defined as
\beqs
p_i\equiv\frac{\partial L}{ \partial \dot{q}_i}\; .
\eeqs
We will refer to these coordinates as the canonical coordinates. To formulate the theory in terms of these canonical variables, Hamilton defines the function $H$, called the canonical Hamiltonian
\beqs\label{canoH}
H_c\equiv \: p_i  \: \dot{q}^i -L \; ,
\eeqs
where $L$ is the Lagrangian. The easiest way to see that $H$ is a function of $q$ and $p$ is to realize that the Hamiltonian is the Legendre transform of the Lagrangian which maps the space ($q,\dot{q}$) to the so-called "phase space" ($q,p$) and vice-versa.

The Legendre transform  of a function $f$ is the function $ \tilde{f} $ defined by
\beqs
\tilde{f}(q,p,y) \equiv \max_y \:  [p y-f(q,y)]\; ,
\eeqs
and  depends a priori on the three variables $q,p,y$. However, one easily sees that the right hand side of this last equation is maximized when
\beqs
\frac{d}{dy}(py-f(y))=p-\frac{df}{dy}=0 \rightarrow p=\frac{df}{dy}\; ,
\eeqs
which means that if we invert this last relation to obtain $y$ as a function of $p$, the Legendre transform only depends on $q$ and $p$ and is defined by
\beqs
\tilde{f} (q,p)= p y(p)-f(q,y(p)).
\eeqs

Said that, and given the definition \eqref{canoH}, one easily verifies that the Hamiltonian $H=H(q,p)$ is indeed the Legendre transform of the Lagrangian $L(q,\dot{q})$ for the variable $y=\dot{q}$. Doing this transformation, we have replaced the coordinate $\dot{q}$ by the coordinate $p$.  \\

The Lagrangian formalism starts from  the definition of a Lagrangian, the conjugate momenta and the Euler-Lagrange equations
\beqs
&& L=L(q_i,\dot{q}_i) \; , \qquad
p^i=\frac{\partial L}{\partial \dot{q}_i} \; ,\nonumber \\
&& \frac{d}{dt} \: \frac{\partial L}{\partial \dot{q}_i}-\frac{\partial L}{\partial q_i}=0 \; .
\eeqs
The properties of the Legendre transform allow to recover, from the Hamiltonian, all the information about the Lagrangian through the inverse relations that express the velocities in terms of the coordinates and the momenta. In Hamiltonian formalism, we have the Hamiltonian and Hamilton's equations
\beqs
&& H(q_i,p_i)=p_i \dot{q}^i-L \; , \qquad \frac{\partial H}{\partial q^i}=-\frac{\partial L}{\partial q^i} \; , \qquad \dot{q}^i=\frac{\partial{H}}{\partial p^i} \; , \label{fff} \\
&& \dot{p}^i= -\frac{\partial H}{\partial q^i} \; , \label{relhamilto}
\eeqs
where the last two equations in \eqref{fff}  come from the following two variations\footnote{The first equation is the formal variation of a quantity $H$ while the second one is the variation of $H=p_i\dot{q}^i-L$.}
\beqs\label{deltaH}
\delta H=\frac{\partial H}{\partial q^n} \delta q^n + \frac{\partial H}{\partial p^n} \delta p^n \; , \qquad
\delta H= \dot{q}_n \delta p^n -\frac{\partial L}{ \partial q^n} \delta q^n \; .
\eeqs
The relation (\ref{relhamilto}) is a rewriting of Lagrange's equations
\beqs
0= \frac{d}{dt} \: \frac{\partial L}{\partial \dot{q}_n}-\frac{\partial L}{\partial q_n}=\dot{p}^n-\frac{\partial L}{\partial q_n}
= \dot{p}^n + \frac{\partial H}{\partial q^n}\; .
\eeqs

One easy way to recover Hamilton's equations is  from the variational principle
\beqs
0=\delta S=\delta \int L dt &=&\delta \int (p^n \dot{q}_n-H) dt \nonumber \\
&=& \int \biggr [ (\dot{q}^n -\frac{\partial H}{\partial p^n})\delta p_n-( \dot{p}^n+ \frac{\partial H}{\partial q^n})\delta q_n \biggl ] + [p^n \delta q_n]_1^2\; .
\eeqs
Note that we also need to set $\delta q_i(t_1)=\delta q_i(t_2)=0$. Here, the action $S$ is called the Hamiltonian action. 

An important object one can define on the phase space ($q,p$) is the Poisson bracket
\beqs
[F,G]=\frac{\partial F}{\partial q^n}\frac{\partial G}{\partial p_n}-\frac{\partial F}{\partial p_n}\frac{\partial G}{\partial q^n},
\eeqs
where $F$ and $G$ are functions of the canonical variables $p$ and $q$. Using this bracket, the equations of motion take the simpler and more compact form
\beqs\label{32}
\dot{F}=[F,H]\; ,
\eeqs
whith $F=p$ or $F=q$, or more generally $F=F(q,p)$. We see that the Poisson bracket defines the evolution of the dynamical variables $F(q,p)$.

An important property of this bracket is that for every function $G(q,p,t)$ on phase space, we have
\beqs\label{33}
\frac{dG}{dt}= \frac{\partial G}{\partial q_i} \dot{q}_{i}+ \frac{\partial G}{\partial p_i} \dot{p}^{i}+\frac{\partial G}{\partial t}= [G,H]+\frac{\partial G}{\partial t} \; ,
\eeqs
where in the second equality we made use of Hamilton's equations.
As we already said, a function $G(q,p,t)$ is a constant of motion if it fulfills $dG/dt=0$. The Poisson bracket of two constant of motions will thus be another constant of motion. Indeed, given $G_1$ and $G_2$, two constants of motion, we have
\beqs
[[G_1,G_2],H] &=& [G_1,[G_2,H]]+[[G_1,H],G_2] = [G_1,-\frac{\partial G_2}{\partial t}] +[-\frac{\partial G_1}{\partial t},G_2] \nonumber \\
                          &=& -\frac{\partial }{\partial t} [G_1,G_2] \; ,
\eeqs
where we used the Jacobi identity, which the Poisson bracket satisfies, and \eqref{33}. This result states nothing else than
\beqs
\frac{dG_1}{dt}=0, \: \: \frac{dG_2}{dt}=0, \: \:  G_{3}\equiv [G_1,G_2] \qquad \rightarrow \qquad  \frac{dG_3}{dt}=0\; .
\eeqs
In other words, we have shown that the constants of motion form a closed algebra under the Poisson bracket. Note eventually that if a function $G$ does not depend explicitly on time, the condition for $G$ to be a constant of motion reduces  to
\beqs
[H,G]=0 \leftrightarrow \dot{G}=\frac{dG}{dt}=0 \; ,
\eeqs
meaning that it must Poisson commute with $H$. This also implies that if the Hamiltonian does not depend on time, it is automatically a constant of motion.

Let us now revisit Noether's theorem in the Hamiltonian formalism. Here, one could try to generalize the change of coordinates that left the Lagrangian action invariant by defining a general transformation of the form
\beqs
\delta q_i=Q_i (q,p,t) \; , \qquad \delta p_i=P_i (q,p,t) \; ,
\eeqs
that would leave the Hamiltonian action invariant. As previously said, we look for\footnote{Actually, one could also take a function $f$ that depends on $\dot{q}$ and $\dot{p}$ as we are off-shell.  However, one can show that redefinitions, involving trivial symmetries of the equations of motion, permit to get rid of this dependence.}
\beqs
\delta L=\frac{df(q,p,t)}{dt} \; .
\eeqs
Now, by computing $\delta L=\delta(p_i\dot{q}^i-H)$, we get
\beqs
\delta L=P_i \dot{q}^i + p^i \frac{d}{dt} Q_i -\frac{\partial H}{\partial p_i} P_i -\frac{\partial H}{\partial q_i} Q_i = \frac{df}{dt} \; .
\eeqs
By further expanding the total time derivatives, one obtains
\beqs\label{cantransetcstofmotion}
[P_i+p_j \frac{\partial Q^j}{\partial q^i}-\frac{\partial f}{\partial q^i}]\dot{q}^i + [ p_j \frac{\partial Q^j}{\partial p^i}-\frac{\partial f}{\partial p^i}] \dot{p}^i + [p^i \frac{\partial Q_i}{\partial t}-\frac{\partial f}{\partial t}-\frac{\partial H}{\partial p_i} P_i -\frac{\partial H}{\partial q_i} Q_i  ]=0\; .
\eeqs
Off-shell, the $\dot{q}$'s and the $\dot{p}$'s are independent of the $q$'s  and $p$'s. One sees that the equation actually  decouples into a set of three equations, the first two terms can easily be seen to rewrite as
\beqs
P_i&=&- \frac{\partial}{\partial q_i} (p_j Q^j-f) \; ,\nonumber \\
Q_i&=& \frac{\partial}{\partial p_i} (p_j Q^j-f) \; .
\eeqs
Defining $G\equiv p_j Q^j-f$ and using the definition of the Poisson bracket, we can write
\beqs\label{cantrans}
\delta q_i=Q_i= [q_i,G] \; , \qquad \delta p_i= P_i =[p_i,G] \; .
\eeqs
If we now plug this last result into the last term of (\ref{cantransetcstofmotion}), we obtain
\beqs
0&=& \frac{ \partial (p^i Q_i-f)}{\partial t}-\frac{\partial H}{\partial p_i} P_i -\frac{\partial H}{\partial q_i} Q_i
 =\frac{\partial G}{\partial t} + \frac{\partial H}{\partial p_i}\frac{\partial G}{\partial q_i} -\frac{\partial H}{\partial q_i}\frac{\partial G}{\partial p_i} \nonumber \\
 &=& \frac{\partial G}{\partial t}  + [G,H] \; ,
\eeqs
which is the same as saying that $G$ is a constant of motion.

From \eqref{cantrans}, we see that symmetries of the action are generated by a function $G=p_j Q^j-f$. We refer to them as the generators of the symmetries or the generating functions.
Note that transformations that are generated by a function $G$ are called canonical transformations as they leave the canonical Poisson bracket invariant\footnote{Canonical transformations will leave the equations of motion invariant if we also have $\delta H=0$. In this case, they are generated by a symmetry of the action which is a constant of motion. Note that equations of motion could a priori also possess symmetries that are not canonical. These will however not have Noether charges associated to it.}. 
What Noether's theorem tells us is that to each symmetry of the action, one associates a conserved charge, or constant of the motion, which is also $G$.  \\

\textit{
The role of $G$ is thus twofold: It is the generator of a symmetry of the action and also the conserved charge associated to this symmetry.}\\

In this thesis, we will always be concerned with symmetries of the action and their associated charges. For a specific symmetry, we will thus no longer make any distinction between its generator or its associated conserved charge.

Let us now see how one can describe gauge systems as Hamiltonian systems where some additional constraints are imposed.

\subsection{Gauge systems as constrained Hamiltonian systems}

As we already emphasized, gauge theories are theories that are invariant under local transformations.
In Lagrange's formalism, one sees that the presence of a gauge symmetry implies that the evolution of the dynamical quantities may allow for arbitrary functions of time. A gauge system is a system where the values of the generalized coordinates and velocities are given but where some transformations do not change the values of the accelerations which describe the physical quantities.  In a sense, there are less degrees of freedom than the apparent ones, i.e. the $q_i$.
Mathematically, gauge systems can be recognized as follows. If one starts from Lagrange's equations
\beqs
\frac{d}{dt} \: \frac{\partial L}{\partial \dot{q}^i}-\frac{\partial L}{\partial q^i}=0 \; ,
\eeqs
 and plugs in
 \beqs
 \frac{d}{dt}=\frac{\partial q^{j}}{\partial t} \: \frac{\partial}{\partial q^{j}}+\frac{\partial \dot{q}^{j}}  {\partial t}\:\frac{\partial }{\partial \dot{q}^{j}} \; ,
 \eeqs
 the Euler-Lagrange equations write
 \beqs
 \ddot{q}^{j} \: \frac{\partial^2 L}{\partial \dot{q}^{j} \partial \dot{q}^i}=\frac{\partial L}{\partial q^i}-\dot{q}^{j} \: \frac{\partial^2 L}{\partial q^{j} \partial \dot{q}^i} \; .
 \eeqs
 From this equation, one sees that the accelerations $ \ddot{q}^{j}$ at a given time are uniquely determined by the velocities and positions at that time if
 \beqs
 \text{det}(\frac{\partial^2 L}{\partial \dot{q}^{j} \partial \dot{q}^i}) \neq 0\; ,
 \eeqs
 meaning that this matrix can be inverted.  Gauge systems will precisely fall into the class of systems where this matrix cannot be inverted.

To discuss gauge systems, we consider systems for which the matrix
\beqs
A \equiv \frac{\partial^2 L}{\partial \dot{q}^{j} \partial \dot{q}^i}\; ,
\eeqs
is non-invertible, or not of maximal rank.  We will fix the rank, assumed to be constant throughout ($q$,$\dot{q}$)-space, of the matrix $A$ to be $2N-M$. By using the definition of the conjugate momenta, we see that
\beqs
 \text{det}(\frac{\partial^2 L}{\partial \dot{q}^{i'} \partial \dot{q}^i})=0 \rightarrow  \text{det}(\mathcal{J}_{ij})\equiv  \text{det}(\partial p_i/\partial \dot{q}^j)=0\; ,
\eeqs
meaning that the determinant of the Jacobian $\mathcal{J}_{ij}$ of the transformation from the $p_i$ to the $\dot{q}_i$  is zero. Alternatively said, this means that the Legendre transform is not invertible or equivalently that, from $p=\partial L/\partial \dot{q}$, one can not determine uniquely the velocities as functions of the canonical variables $p,q$. One immediate consequence is that  the conjugate momenta are not all independent but  there rather exists some relations between the canonical coordinates that we call "primary constraints"
\beqs\label{primary}
 \phi_m(q,p)=0\; ,
\eeqs
where we let $m=1...M$ in agreement with the rank of $A$.
The name "primary constraint" comes from the fact that the equations of motion are not used and  imply thus no restrictions on the $q^i$ or $\dot{q}^i$. Here and in the following, we will always assume that the constraints are all independent. In full generality, one could also imagine a (reducible) system of constraints where only some of them are independent. We refer the reader to \cite{Henneaux:1992ig} for a study of reducible systems.

An Hamiltonian system in the presence of a general set of constraints is called a constrained Hamiltonian system. Although we have understood gauge systems as constrained Hamiltonian systems, one should pay attention to the fact that the latter system is a more general one as not all constrained Hamiltonian systems can be obtained from a gauge principle.

The rest of this section aims at reviewing the theory of constrained Hamiltonian systems. Especially, we point out how one can recover the invertibility of the Legendre transform, the classification of constraints into primary and secondary constraints and then into first and second class constraints, and how one can ``solve" for the constraints. As previously mentioned, our aim is to show that first class constraints generate gauge symmetries. Indeed, reviewing Noether's theorem, we see that global symmetries are associated to first class functions which are constant of the motion while gauge symmetries are associated to first class functions which can be decomposed into the basis of first class constraints. As a consequence of this last result, the gauge symmetries are thus associated to trivial charges.

\subsubsection{Recovering invertibility of the Legendre transform}

To deal with constrained Hamiltonian systems, defined from a Lagrangian, one must see at what cost the invertibility of the Legendre transform can be recovered in the presence of primary constraints.

The first thing to notice is that the properties of the Legendre transform imply that $H=p_n \dot{q}^n -L$ is a function of the canonical variables because
\beqs
\delta H=\dot{q}^n \delta p_n-\delta q^n \frac{\partial L}{\partial q^n} \; .
\eeqs
However, in the presence of constraints, it is only a uniquely defined function of the canonical variables on the primary constraint surface, the ($2N-M$)-dimensional submanifold defined by $\phi_m=0$.

 If we want to restore the invertibility of the Legendre transform, we need to introduce $M$ extra variables, in agreement with the fact that the Legendre transformation can only be well defined between spaces of the same dimensionality. To achieve this, we will impose that the primary constraint surface is smoothly embedded in phase space and that it satisfies some regularity conditions that, roughly said, allows us to use the primary constraints as a local set of coordinates in the vicinity of the constraint surface.
Under these assumptions, one can check the following two theorems\footnote{Proofs of these theorems can be found in \cite{Henneaux:1992ig}.}

\textbf{Theorem 1}: If a (smooth) function $G$ on phase space vanishes on the primary constraint surface, then (locally) $G=g^m \phi_m$.\\

\textbf{Theorem 2:} If $\lambda_n \delta q^n +\mu^n \delta p_n=0$ for arbitrary variations $\delta q^n$, $\delta p_n$, then
\beqs
\lambda_n &=& u^m \frac{\partial \phi_m}{\partial q^n}\; , \nonumber \\
\mu^n &=& u^m \frac{\partial \phi_m}{\partial p_n}\; ,
\eeqs
for some $u^m$. These equalities are true on the primary constraint surface.\\

With the help of the second theorem, it is easy to see that by using (\ref{deltaH})
\beqs
(\frac{\partial H}{\partial q^n}+\frac{\partial L}{ \partial q^n}) \delta q^n + (\frac{\partial H}{\partial p^n} - \dot{q}_n) \delta p^n=0\; ,
\eeqs
we obtain the defining relations of the inverse Legendre transformation
 \beqs
 q^n=q^n\; , \qquad \dot{q}^n = \frac{\partial H}{\partial p^n}+u^m \frac{\partial \phi_m}{\partial p_n}\; , \qquad
 -\frac{\partial L}{\partial q^n} \biggr |_{\dot{q}} = \frac{\partial H}{\partial q^n}\biggl |_{p}+u^m \frac{\partial \phi_m}{\partial q^n}\; ,
 \eeqs
  which now defines a transformation between spaces of the same dimension $2N$.
 From $(q,\dot{q})$ space to the primary constraint surface of phase space, we have
 \beqs
 q^n=q^n\; , \qquad p_n=\frac{\partial L}{\partial \dot{q}^n}(q,\dot{q})\; , \qquad u^m=u^m(q,\dot{q})\; .
 \eeqs

 We have thus achieved our goal of restoring invertibility of the Legendre transformation by adding the extra independent variables $u^m$ thereby permitting a transformation between two spaces of same dimensionality.\\

 With this in hand, we can go from the Euler-Lagrange equations to Hamilton's equations and we get
 \beqs\label{Hameom}
\dot{q}^n = \frac{\partial H}{\partial p^n}+u^m \frac{\partial \phi_m}{\partial p_n} \; , \qquad 
\dot{p}^n = -\frac{\partial H}{\partial q^n}\biggl |_{p}-u^m \frac{\partial \phi_m}{\partial q^n} \; , \qquad 
\phi_m(p,q)=0\; ,
 \eeqs
 where the equations of motion can be written using the Poisson bracket as
 \beqs
 \dot{F}=[F,H]+u^m[F,\phi_m]\; .
 \eeqs
The first relation in (\ref{Hameom}) permits to recover the $\dot{q}^n$ when given the momenta (upon imposing $\phi_m=0$) and the $u^m$. Because the $\partial \phi_m / \partial p_n$ are assumed to be independent, two different sets of $u^m$ must give two different sets of $\dot{q}^n$. This also implies that the $u^m$ can in principle be expressed as $u^m(q,\dot{q})$. 

Note that these equations of motion could also be obtained from the variational principle
\beqs
\delta \int_{t_1}^{t_2} (\dot{q}^n p_n -H-u^m \phi_m)=0\; ,
\eeqs
under arbitrary variations $\delta q$, $\delta p$, and $\delta u^m$ with $\delta q_n(t_1)=\delta q_n(t_2)=0$.  Here, the $u^{m}$ appear as Lagrange multipliers enforcing the primary constraints. 

\subsubsection{The consistency algorithm: a full set of primary and secondary constraints}

If we look at the equations of motion (\ref{Hameom}) for a constrained system, we see that a necessary requirement is that every primary constraint also satisfies
\beqs
\dot{\phi}_m=[\phi_m,H]+u^{n} [\phi_m,\phi_{n}]=0\; .
\eeqs

Depending on the appearance of the parameters $u^m$  in the above expression, this requirement provides us with a secondary constraint $N(p,q)=0$, a relation involving only the $q$'s and the $p$'s and independent of the primary constraints, or with a relation involving the $u$'s and thus restricting these parameters. Also, as an iterative consistency algorithm, if there is a secondary constraint  $N(q,p)$ we also need to check that
\beqs
\dot{N}=[N,H]+u^{m} [N,\phi_m]=0\; ,
\eeqs
does not bring new secondary constraints.
In the end, we are left with a total sytem of $J$ constraints (primary and secondary) that we collectively denote by
\beqs
\Phi_j, \: j=1...M,M+1...M+K(=J)\; ,
\eeqs
where $K$ of them are secondary constraints. Note that we also assume in the following that the regularity conditions discussed above for the primary constraints apply to the full set of primary and secondary constraints. Remember that the constraints are assumed to be all independent such as to form an irreducible set of constraints.

\subsubsection{Towards a more fundamental classification of constraints: first and second class}

Having determined  the full set of constraints, the set of $J$ nonhomogeneous equations linear in the $M$ unknown ($M\leq J$) parameters $u^m$
\beqs\label{Lagparam}
[\Phi_j, H]+u^{m}[\Phi_j,\phi_m]\approx 0\; ,
\eeqs
should possess solutions. Otherwise, this would mean that the system desribed by the Lagrangian is inconsistent. In the last equation, the sign $\approx$  refers to an equality that is only true on the primary constraint surface. We say that the expression is "weakly vanishing". One important consequence of Theorem 1 is that two functions that are the same on the constraint surface should be related by
\beqs\label{rell}
F\approx G \rightarrow F-G= c^k(q,p) \Phi_k \; ,
\eeqs
where the sign $=$ denotes equality on the full phase space. A function on phase space that is $=0$ is said to be "strongly vanishing". Also, for three quantities $A$, $B$, and $C$ with $C\approx 0$, we have, using the Jacobi identity,
\beqs\label{property}
[A,BC]\equiv B[A,C]+[A,B]C\approx B[A,C] \rightarrow B[A,C]\approx [A,BC]\; .
\eeqs

The general solution to the non-homogeneous first order differential equation (\ref{Lagparam}) is given by
\beqs\label{gensoll}
u^m\approx U^m +v^a V_a^m\; ,
\eeqs
where $U^m$ is a particular solution and $V^m\equiv v^a V_a^m$ is the general solution of the associated homogeneous system
\beqs
V^m[\Phi_j,\phi_m]\approx 0 \; ,
\eeqs
 written in the basis\footnote{Note that  we actually  need to require $[\Phi_j,\phi_m]$ to be of constant rank so that $V_a^m$ is fixed on the constraint surface. }
 of linearly independent solutions $V_a^m$. Upon solving this system, we have achieved a split between what is fixed by the consistency conditions, the $U^m$, and what is left totally arbitrary, i.e. the coefficients $v^a$.

One particularly interesting thing to notice is that both the primary constraints $\phi_a\equiv V_{a}^{\:\:m} \phi_m$ and $H'\equiv H+U^m \phi_m$, defined using \eqref{gensoll}, Poisson commute with the general set of constraints $\Phi_j$
 \beqs
 [\Phi_j, H'] \approx 0\; , \qquad  [ \Phi_j,\phi_a]\approx 0 \; .
 \eeqs
Indeed, this can be seen by plugging the general solution \eqref{gensoll} into the consistency conditions \eqref{Lagparam} and using \eqref{property}. Doing so, we obtain
\beqs\label{ccc}
[\Phi_j, H']+v^a [\Phi_j,\phi_a]\approx 0\; .
\eeqs
 However, $\phi_a$ is also a  basis of primary constraints which are solutions of the homogeneous equation
 \beqs
 v^a[ \Phi_j,\phi_a]\approx 0 \; ,
 \eeqs
 implying that $[ \Phi_j,\phi_a]\approx 0$. To complete our proof, we just need to implement this last relation into \eqref{ccc}.

The fact that $\phi_a$ and $H'$ Poisson commute with the general set of constraints is a motivation to introduce a more interesting classification of constraints which makes direct use of the Poisson bracket.

 A function $F(q,p)$   is said to be first class if its Poisson bracket with every constraint vanishes weakly
\beqs
[F,\Phi_j]\approx 0\; .
\eeqs
If it is not first class we will call it second class. A first class function on phase space is said to Poisson commute with all the constraints. Our first examples of first class functions are thus $\phi_a$ and $H'$. In the following, we split the full set of constraints $\Phi_j$ into first class constraints $\gamma_a$ and second class constraints $\chi_{\alpha}$.

Before proceeding, let us define the total Hamiltonian as being the sum of the first class Hamiltonian $H'\equiv H+ U^m \phi_m$ and the first class primary constraints $\phi_a$ multiplied by arbitrary factors $v^a$
\beqs
H_{T}=H'+v^a \phi_a.
\eeqs
One can check that the equations of motion reduce to
\beqs
\dot{F}&&=[F,H]+u^m [F,\phi_m]\approx [F, H+u^m \phi_m] \approx [F,H_T]\; ,
\eeqs
upon using \eqref{property}.

An important property of the definition of first class constraints is that it is preserved under the Poisson bracket operation. The Poisson bracket of two first class constraints is first class. This is shown by making use of Theorem 1 and the Jacobi identity. This result is indeed crucial because we will now show that first class constraints are generators of symmetries and thus form an algebra. Also, by Noether's theorem, charges associated to it should form a closed algebra under the Poisson bracket operation.

\subsubsection{First class functions as generators of symmetries}

It is easy to see that no second class functions but only first class functions can be generating symmetries. Indeed, a symmetry of the equations of motion is a variation of the dynamical variables that leaves the equations invariant. As such they should map an allowed state to another (equivalent or non-equivalent) allowed state.  Because an allowed state sits on the constraint surface, we should have by consistency of the theory that $\delta \Phi_i\approx 0$. If this last statement is not true, then we would allow for symmetry transformations that bring us out of the constraint surface, an inconsistent statement. For a canonical transformation, we have
\beqs
\delta \Phi_i &=& \frac{\partial \Phi_i}{\partial q_n} \delta q^n + \frac{\partial \Phi_i}{\partial p_n} \delta p^n
=  \frac{\partial \Phi_i}{\partial q_n} [q,G]+ \frac{\partial \Phi_i}{\partial p_n} [p,G] \nonumber \\
&=&  \frac{\partial \Phi_i}{\partial q_n} \frac{\partial G}{\partial p^n}  + \frac{\partial \Phi_i}{\partial p_n}  \frac{\partial G}{\partial q^n}  \equiv [\Phi_i,G] \approx 0\; ,
\eeqs
which is precisely the requirement that the generating function be a first class function.  When looking at symmetries of the action, we will see that  first class functions that are constant of the motion generate global symmetries. Let us for now review how  first class constraints are understood as generators of gauge transformations. Later, we will recover these results through Noether's theorem. 

As already stated before, given an initial set of canonical variables describing a physical state at time $t_0$, we expect the equations of motion to fully determine the physical state at other times. However, we know that, in the presence of constraints, different sets of canonical variables can describe the same physical state as it is reflected in the definition of the total Hamiltonian by the set of arbitrary functions $v^a$.

What this means is that \textit{any ambiguity in the value of the canonical variables at a time $t_1$ should be a physically irrelevant ambiguity}, also called a gauge transformation. In mathematical language, by picking $t_1=t_0+\Delta t$, and using the time evolution of dynamical variables with two different choices of $v^a$, denoted $v^a $ and $\tilde{v}^a$, in the total Hamiltonian expression, we have
\beqs
\delta F &=& \Delta F(t_1,t_2,\tilde{v}^a)-\Delta F(t_1,t_2,v^a)\nonumber \\
&=& ([F,H']+\tilde{v}^a [F,\phi_a])\Delta t- ([F,H']+v^a [F,\phi_a])\Delta t \nonumber \\
&=&\epsilon^a [F,\phi_a]\; ,
\eeqs
where $\epsilon^a=(v^a-\tilde{v}^a) \Delta t$ is an \textit{arbitrary} function of time.
 We say that first class primary constraints generate gauge transformations. Here, we see that  this transformation will not modify the physical state at the later time $t_1$. One obvious question is "what about first class secondary constraints" ? 

There exists no general proof that first class secondary constraints do generate gauge transformations. At most, one can show  that in principle they could. This led Dirac to conjecture that all first class constraints generate gauge transformations (this is known as Dirac's conjecture). We will not discuss the fate of first class secondary constraints in details here. We will rather assume that all first class constraints generate gauge symmetries, although one should be aware that counter-examples do exist, see \cite{Henneaux:1992ig}.  

If we assume that all first class constraints (primary and secondary) are generators of gauge transformations, then  the most general physically permissible motion should also allow for general gauge transformations. To this end, we define the extended Hamiltonian as
\beqs
H_E\equiv H' + u^a \gamma_a\; ,
\eeqs
where $\gamma_a$ denote first class constraints.
From the extended action principle
\beqs
S_E=\int (p_i \dot{q}^i-H'-u^i \Phi_i) dt\; ,
\eeqs
where the sum is understood over all the constraints, 
we get the equations of motion for the extended formalism, see \cite{Henneaux:1992ig},
\beqs
\dot{F}\approx [F,H_E]\; , \qquad  \Phi_j\approx 0\; .
\eeqs
Let us mention that the extended formalism is really a new feature of the Hamiltonian formalism that takes into account all the gauge freedom of the theory while the Lagrangian (or equivalently total Hamiltonian) just restricts to the gauge freedom introduced by the primary constraints. Indeed, when considering any physically relevant dynamical variable $O$, also called observable, which is by definition gauge-invariant, we should have $\delta O\approx 0$, which means that its Poisson bracket is weakly zero with all the first class constraints. Its evolution is thus the same when expressed with respect to $H_E$, $H_T$ or $H'$. But this is not true for gauge-variant dynamical variables where evolution should be described using  $H_E$ which takes into account all the gauge freedom of the theory.

\subsubsection{Solving the constraints}

Before moving to Noether's theorem, let us briefly comment on how constraints can be solved, as often implemented in the study of constrained Hamiltonian systems. In short, first class constraints can be gauge-fixed by introducing new ad-hoc constraints\footnote{Note that this procedure can not always be implemented  because there does not always exist gauge fixing conditions that are globally well defined. This phenomena is known as the "Gribov obstruction".} and second-class constraints are dealt with by reformulating the theory in terms of the Dirac bracket.

Solving the first class constraints, generators of gauge symmetries, permits to establish a one-to-one correspondence between physical states and values of the canonical variables, by avoiding a multiple counting of states. It also allows one to describe the true degrees of freedom of the theory under consideration. To implement this, one introduces new constraints, i.e. gauge fixing conditions. It can be checked that to obtain a satisfying set of gauge fixing conditions, that we denote by  $C_b(q,p)\approx 0$, we need a number of independent gauge conditions that is precisely  equal to the number of independent first class constraints. 

If such a set of gauge fixing conditions has been determined, one consistency requirement is that there must not exist gauge transformations other than the identity that preserve the set of gauge conditions and by this we mean
\beqs\label{blablabla}
\delta u^a [C_b,\gamma_a]\approx 0 \rightarrow \delta u^a=0\: .
\eeqs
This last statement is true when the above Poisson bracket defines an invertible matrix such that
\beqs
\text{det}( [C_b,\gamma_a] ) \neq 0\; ,
\eeqs
 which alternatively means that the introduced gauge fixing conditions are second class but that also our previously first class functions $\gamma_a$ have now become second class. In this case, we have completely fixed the gauge freedom and first class constraints have become second class. 
 
 Without entering into details, it was noticed by Dirac that to deal with second class constraints, one can just replace the Poisson bracket by a new one. This new bracket is known as the Dirac bracket
 \beqs
[F,G]^{\star}=[F,G]-[F,\chi_{\alpha}] C^{\alpha \beta} [\chi_{\beta},G]\; ,
\eeqs
 where $C_{\alpha \beta}$ is the Poisson bracket of second class constraints
 \beqs
C_{\alpha \beta}=[\chi_{\alpha},\chi_{\beta}] \; ,
\eeqs
and $C^{\alpha \beta}$ is the inverse matrix such that $C^{\alpha \beta} C_{\beta \gamma}=\delta^{\alpha}_{\gamma}$. By definition, the determinant of the antisymmetric matrix $C_{\alpha \beta}$ should not be zero. This implies that it has to be a  $n\times n$ matrix with $n$ even, i.e. second class constraints should always come in pairs.

To check that the introduction of the Dirac bracket permits to get rid of second class constraints, one readily checks that the Dirac bracket of any dynamical variable $F(q,p)$ with a second class constraint is strongly zero
\beqs
[F,\chi_{\gamma}]^{\star}&=&[F,\chi_{\gamma}]-[F,\chi_{\alpha}] C^{\alpha \beta} [\chi_{\beta},\chi_{\gamma}] = [F,\chi_{\gamma}]-[F,\chi_{\alpha}] C^{\alpha \beta} C_{\beta \gamma}
= 0 \; .
\eeqs
This means that we can always set the second class constraints to zero either before or after evaluating a Dirac bracket. The  equations of motion become
\beqs
\dot{F}\approx [F,H_E]\approx [F,H_E]^{\star}\; .
\eeqs

The most trivial example illustrating this procedure is a system with two second class constraints, i.e. constraints such that $p\approx 0$, $q\approx 0$ but $[p,q]=1$. One checks that the introduction of the Dirac bracket permits to completely forget about these coordinates.\\

Let us make a few additional comments:\\

Firstly, we have said that getting rid of the constraints is sometimes useful. One can choose to solve all constraints, and especially fix the gauge freedom of the theory, or only solve the second class constraints. However, it is also interesting to remember that while getting rid of first class constraints, we made them become second class. Second class constraints could thus be reformulated as gauge degrees of freedom before gauge fixation. This procedure can present some advantages as it permits to bypass the introduction of the Dirac bracket whose quantum realization may be highly non-trivial. Actually, even in the absence of second class constraints, the introduction of extra gauge degrees of freedom can be interesting, for example when one wants to make some hidden symmetry manifest.

We also said at the beginning that the solving the constraints permits to single out the true degrees of freedom of our theory. Indeed, we see that a constrained Hamiltonian of $2N$ independent canonical variables with $n$ first class constraints and $m$ second class constraints has $D$ degrees of freedom
\beqs
D=N-n-(m/2)\; ,
\eeqs
where $D$ is always an integer because $m$ is always even. From the above counting, one often says that first class constraints strike twice. This is understood from the fact that we had to introduce $n$ extra ad-hoc gauge conditions who implied that the set of first class constraints became second class, a ``new" set of $n$ constraints.

Eventually, let us mention that one can also study constrained Hamiltonians using symplectic manifolds, i.e. manifolds equipped with a closed non-degenerate differential two-form $\omega_{\alpha\beta}$, the symplectic form. A generic bracket between two functions $F$ and $G$ is defined as 
\beqs
[F,G]\equiv \omega^{\alpha\beta}\:  \frac{\partial F}{\partial y^{\alpha}} \:  \frac{\partial G}{\partial y^{\beta}} \; .
\eeqs
 In this context, one can give a geometrical meaning to the Dirac bracket. It is the bracket defined using as symplectic structure the pullback of the phase space symplectic structure onto the constraint surface.

\subsection{Noether's theorem}

Let us now formulate Noether's theorem for a generic set of symmetries of the extended action. Here, we show that the set of gauge and global symmetries of the extended action are generated by first class generating functions that are respectively the first-class constraints and the general functions on phase space that are constant of the motion. 

To show this, we consider the variation of the extended action
\beqs
S_E[q^n(t),p_n(t),u^a(t),u^{\alpha}(t)]=\int (p_n \dot{q}^n-H-u^a \gamma_a -u^{\alpha} \chi_{\alpha}) dt\; ,
\eeqs
 under an infinitesimal  transformation
\beqs
\delta q^n=Q^n\; , \qquad \delta p_n=P_n\; , \qquad \delta u^a=U^a\; , \qquad \delta u^{\alpha}=U^{\alpha}\; ,
\eeqs
where the $Q_n$, $P_n$,$U^a$, $U^{\alpha}$ are functions of $q,p,t$ and $u$ and their derivatives\footnote{For the same reason as in the unconstrained case, we can get rid of the dependence in $\dot{q}$ and $\dot{p}$ by means of redefinitions which are trivial symmetries of the equations of motion. Note that, technically, this can only be implemented under the assumption of locality, which we assume here.}.

The invariance of the action states that the variation of the Lagrangian for a given symmetry is zero up to a total derivative
\beqs\label{deltaLL}
\delta L&=& \dot{q}^n P_n + \frac{d}{dt} (p_n Q^n)-Q^n \dot{p}_n -\delta H-U^a \gamma_a -u^a \delta \gamma_a -U^{\alpha} \chi_{\alpha} -u^{\alpha} \delta \chi_{\alpha}  \nonumber \\
&=&\frac{df}{dt}\; .
\eeqs
Now, let us first introduce
\beqs\label{ddddt}
\frac{D}{Dt}=\frac{\partial }{\partial t}+\dot{u}^a \frac{\partial }{\partial u^a}+\ddot{u}^a \frac{\partial }{\partial \ddot{u}^a}+...+\dot{u}^{\alpha} \frac{\partial }{\partial u^{\alpha}}+\ddot{u}^{\alpha} \frac{\partial }{\partial \ddot{u}^{\alpha}}+...\; .
\eeqs
This definition permits to replace time derivatives in the above variation as
\beqs\label{ddt}
\frac{d}{dt}=\frac{D}{Dt}+\dot{q}^n \frac{\partial }{\partial q^n} + \dot{p}^n \frac{\partial }{\partial p^n}\; .
\eeqs
It is easy to see that the rhs of \eqref{deltaLL} will split, like in the unconstrained case, into three terms. The first two terms, i.e. the ones that multiply $\dot{q}$ and $\dot{p}$, will tell us that the transformation must be canonical
\beqs\label{canoo}
\delta q_n= Q^n=\frac{\partial{G}}{\partial p_n}=[q_n,G]\; , \qquad \delta p_n=P_n=-\frac{\partial G}{\partial q_n}=[p_n,G]\; ,
\eeqs
where $G=p_j Q^j-f$. Because $\gamma_a=\gamma_a(q,p)$, $\chi_{\alpha}=\chi_{\alpha}(q,p)$ and $H=H(p,q)$ are functions on phase space, this also implies that
\beqs\label{103}
\delta \gamma_a=[\gamma_a,G]\; , \qquad \delta \chi_{\alpha}=[\chi_{\alpha},G] \; , \qquad \delta H=[H,G]\; .
\eeqs
Replacing this in the third term, coming from the split of \eqref{deltaLL}, immediately gives us an additional condition on the generating function $G$
\beqs\label{GENERAL}
\frac{DG}{Dt}+[G,H] + u^a [G,\gamma_a]+ u^{\alpha} [G,\chi_{\alpha}]=U^a \gamma_a +U^{\alpha} \chi_{\alpha}\; .
\eeqs
As one can show, it is equivalent to deal with this equation after having set to zero the second class constraints which we will assume from now on, see \cite{Henneaux:1992ig}. With a lot of courage, one can show by solving the above equation, upon looking at its dependence in derivatives of $u$, that the general solution to the equation \eqref{GENERAL} is the sum of a particular solution $G_{part}=\bar{G}(q,p,t)$ of the non-homogeneous equation with the general solution of the homogeneous part of this equation \cite{Henneaux:1992ig}
\beqs\label{generalsolution}
G=g^a(q,p,t,u,\dot{u},\ddot{u},...) \gamma_a + \bar{G}(q,p,t)\; ,
\eeqs
where the particular solution $G_{part}=\bar{G}(q,p,t)$ must be first class and a constant of motion on the constraint surface
\beqs
[\bar{G},\gamma_a]\approx 0 \; , \qquad \frac{\partial \bar{G}}{\partial t}+[\bar{G},H]\approx 0\; .
\eeqs

The first term in \eqref{generalsolution} represents the charges associated to gauge transformations while the second term represents the charges associated to global symmetries. 

In the end, we have shown that global symmetries of the extended action are generated by functions $\bar{G}(q,p,t)$ which are first class and constants of the motion. Noether associates to it the conserved charge $\bar{G}$.
We have also checked, as previously announced, that gauge symmetries are generated by (linear combination of) first class constraints. 

One important consequence of this last result is that gauge symmetries have trivial associated conserved charges as the generators of these symmetries are on-shell vanishing. After a brief review of the Lagrangian and Hamiltonian formulations of general relativity, we will see that this is not always true for gauge field theories, such as general relativity.

\setcounter{equation}{0}
\section{Einstein's theory of gravity}
\label{einsteinn}

General Relativity is a theory for a spin 2 field that was designed to describe the gravitational interaction at the classical level. 
The celebrated equations of Einstein can be written as
\beqs
R_{\mu\nu}-\frac{1}{2} R g_{\mu\nu} = 8 \pi G T_{\mu\nu}\; .
\eeqs
For the vacuum equations, in the absence of sources $T_{\mu\nu}=0$, the Lagrangian formulation was established by Hilbert. The Lagrangian is known as the Einstein-Hilbert Lagrangian 
\beqs
S=\int d^4 x \: \sqrt{g} \:  R.
\eeqs
Asking stationarity of the Einstein-Hilbert action, one recovers Einstein's equations in the vacuum. Actually, this last statement is only true when boundary contributions are discarded. Alternatively, one says that the variational principle is valid, i.e. it gives the Einstein's equations, only when boundary contributions can be neglected. We show, in section \ref{sec:RT}, that this is not always the case. As we will see in the following, the importance of these boundary terms are primordial to the construction of non-trivial conserved charges for general relativity. 

The formulation of the Hamiltonian theory for classical systems, constrained or not, implicitly required a specification of time. The theory of general relativity is a theory of space and time where there is a priori no preferred time direction. To be able to cast it into an Hamiltonian form by understanding how one can single out a time direction, we first review the closely related problem of establishing that general relativity has a well-posed initial value formulation. Indeed, a positive answer was provided by considering a splitting of space and time as we now review. 

\subsubsection{A well-posed initial value formulation}

A system possesses a well-posed initial value formulation if an allowed state of that system can be unambiguously described at a future time $t_1$, upon using the equations of motion, given a set of initial conditions at time $t_0$ and if small fluctuations of these initial conditions at $t_0$ do not alter drastically the state at time $t_1$. One of the successes of general relativity is that it has a well-posed initial value formulation when considering, as we explain below, globally hyperbolic spacetimes. In here, we do not pretend to be rigorous as we are only interested in reviewing some concepts we will use in the following.

Given a manifold $\mathcal{M}$, a Cauchy surface $\Sigma$ is a space-like surface (meaning that all points on this surface are space-like separated) such that
\beqs
\mathcal{M}=\mathcal{D}^{-}(\Sigma)\;  \cup \;   \Sigma \;   \cup \;  \mathcal{D}^{+}(\Sigma)\; ,
\eeqs
where $\mathcal{D}^{\pm}(\Sigma)$ represent respectively the future and past regions of that surface $\Sigma$. The future region of the surface $\Sigma$ is the region of space-time that can be reached from any point lying on the surface $\Sigma$ when going along time-like or null directions. A spacetime which possesses a Cauchy surface is a globally hyperbolic spacetime. To understand the importance of such a definition, let us comment on the case of AdS which is not a globally hyperbolic spacetime. In AdS, the information at a given point of spacetime may not be characterized by the information coming from a specified three-surface as information arriving from infinity may also contribute. However, Anti de Sitter spacetime has what is called a Cauchy horizon, a region of spacetime where one can define a Cauchy surface. 

The Arnowitt-Deser-Misner \cite{Arnowitt:1962hi} (ADM) splitting of space-time precisely achieves this initial value formulation for globally hyperbolic spacetimes (see also chapter 10 of \cite{Wald:1984rg}) of general relativity through  the consideration of Cauchy surfaces. Indeed, the ADM decomposition of spacetime is a 3+1 space-like slicing of space and time. It permits to consider the information that lies on a Cauchy surface to be the dynamical information and study its evolution through the introduction of a preferred arrow of time. The four dimensional metric is split into a three dimensional metric $\g_{ij}$ which is the induced metric on the Cauchy surface. The other components of the metric are seen to describe deformations of $\Sigma$. They are known as the lapse $N$ and the shift $N_i$. In terms of the four-dimensional metric $g_{\mu\nu}$, we have
\beqs\label{ADMcoord}
\g_{ij}\equiv g_{ij}\; , \qquad N_i \equiv g_{0i} \; , \qquad N\equiv(-g^{00})^{-1/2}\; .
\eeqs
Eventually, to describe the evolution of the Cauchy surface, we need a notion of its embedding into spacetime. This is the r\^ole of the extrinsic curvature $K_{ij}$, which measures the difference between a normal vector to $\Sigma$ at a point $p$ and a normal vector (at a point $q$) that has been parallel transported to $p$.

It has been proved that the initial conditions can be described by the triplet ($\Sigma$, $\g_{ij}$, $K_{ij}$) when subject to additional initial value constraints. These additional constraints are the first class constraints obtained from  varying the shift and the lapse in the Hamiltonian action of general relativity, which we now review.

\subsubsection{Hamiltonian formulation}

Because general relativity is a theory invariant under diffeomorphisms, its Hamiltonian should be constrained. 
Using ADM coordinates \eqref{ADMcoord}, one constructs the Hamiltonian of general relativity from its Lagrangian formulation. One checks that it depends on the three-dimensional metric $\g_{ij}$, its associated conjugate momenta $\pi^{ij}$ (which can be expressed in terms of the extrinsic curvature), a lapse $N$ and a shift $N_i$. Note that the lapse and shift functions do not have associated conjugates as the action does not contain time derivatives of these functions. They are thus non-dynamical quantities. The Hamiltonian is
\beqs\label{H0}
H_{0}(\g_{ij}, \pi^{ij}, N, N^i)=\int d^3 x \Big( N(x) \mathcal{H}(x) +N^i(x) \mathcal{H}_i (x) \Big)\; ,
\eeqs
where
\beqs
\mathcal{H}&=& G_{ijkl} \pi^{ij} \pi^{kl}- \sqrt{\g}\:  \mathcal{R}  \; ,\nonumber \\
\mathcal{H}_i&=& -2 \:  \pi_{i\;\;\;|j}^{\:\: j}= -2 \: \g_{ik} \:  \pi^{kj}_{\;\;\; ,j} -(2 \: \g_{ki,j}-  \g_{kj,i}) \pi^{kj}\; ,
\eeqs
and where the column denotes the covariant derivative associated to $\g_{ij}$, $\mathcal{R}$ is the three-dimensional Ricci scalar associated to $\g_{ij}$ and $G_{ijkl}$ is the DeWitt supermetric
\beqs
G_{ijkl}&=&\frac{1}{2} \: \g^{-1/2} \: (\g_{ik}\g_{jl}+\g_{il} \g_{jk}- \g_{ij}\g_{kl})\; ,\nonumber \\
G^{ijkl}&=&\frac{1}{2} \: \g^{1/2} \: (\g^{ik}\g^{jl}+\g^{il} \g^{jk}-2 \g^{ij} \g^{kl}) \; ,\nonumber \\
G^{ijkl} G_{klmn} &=& \delta^{ij}_{mn}=\frac{1}{2} (\delta^i_m \delta^j_n+\delta^i_n \delta^j_m) \; .
\eeqs
Hamilton's equations take the generic form
\beqs\label{ABABAB}
\dot{\g}_{ij}(x)&=&\delta (\text{Hamiltonian})/\delta \pi^{ij}(x)\equiv A_{ij} , \nonumber \\
 \dot{\pi}^{ij}(x)&=&-\delta (\text{Hamiltonian})/\delta \g_{ij}(x)\equiv -B^{ij}\;.
\eeqs
In our case, one finds
\beqs
\dot{\g}_{ij}&=& 2N \g^{-1/2} (\pi_{ij}-\frac{1}{2} \g_{ij} \pi)+ N_{i|j}+N_{j|i}\; , \nonumber\\
\dot{\pi}^{ij}&=& -N \sqrt{\g} (\cR^{ij}-\frac{1}{2} \cR \g_{ij}) +\frac{N}{2} \g^{-1/2} \: \g^{ij} (\pi^{mn} \pi_{mn}-\frac{1}{2} \pi^2 )     +(\pi^{ij} N^m)_{|m}  \nonumber \\
&& -2 N \g^{-1/2}  (\pi^{im} \pi_m^j-\frac{1}{2} \pi \pi^{ij})  +\sqrt{\g} (N^{ij} -\g^{ij} N^{|m}_{\;|m})  -N^i_{\;|m} \pi^{mj} -N^j_{\;|m} \pi^{mi}\; , \nn\\
\eeqs
while variations with respect to the lapse and the shift give the Hamiltonian and momentum constraints
\beqs
\mathcal{H}(x) = 0 \; , \qquad      \mathcal{H}_i (x) = 0\; .
\eeqs
These last equations are constraints as they do not describe the time evolution of some quantity. The lapse and shift are seen as Lagrange multipliers that enforce the constraints. In our previous notation, we say that $\mathcal{H}(x)\approx 0$ and $\mathcal{H}_i (x)\approx 0$. One can check that the bracket of any two of these constraints gives another such constraint, implying that the bracket is also weakly vanishing.  By definition, this implies that all the constraints are first class. 

In their work, Arnowitt, Deser and Misner were motivated to cast the theory in an Hamiltonian form as a first step towards the quantization\footnote{Here, we do not want to enter into any discussion about the problems of a canonical quantization of general relativity and refer the reader to the book of Wald \cite{Wald:1984rg}, see especially chapter 14. } of general relativity. Far from technicalities, to implement this they introduced gauge conditions to solve for the constraints, i.e. to fix the gauge freedom. In particular, their analysis recovered the fact that the theory is a theory of a spin 2. This is so because the metric only has two dynamical degrees of freedom when all the redundancy of the theory has been eliminated.  From another perspective, their approach also permitted to give the first expressions for the energy and momentum \cite{Arnowitt:1961zz} for a specific class of spacetimes we will discuss in the following. 

General considerations about the definition of conserved charges for field  theories and especially for general relativity is the topic of the next section.

\setcounter{equation}{0}
\section{Conserved charges for general relativity}
\label{conservedy}

We have seen for classical mechanics that one can associate to each symmetry of the action, a conserved charge. In the case of a gauge symmetry, the generator of the symmetry is vanishing on account of the equations of motion.  

When looking at a specific field theory with a given set of gauge symmetries, one could also try to apply Noether's theorem. The generalization of our previous result to field theories provides us with the conservation of a current which is vanishing  on-shell up to the divergence of a superpotential. We will not reproduce the derivation here but it can be found, for example, in the introduction section of \cite{Barnich:2001jy} (see also exercise 3.4. of \cite{Henneaux:1992ig}). Roughly speaking, we have
\beqs
&&\textit{Classical Mechanics} \qquad \qquad \qquad \qquad \textit{Field theories} \qquad \qquad \nn\\
 \nn \\
&&\:\:\:\: \:\: \frac{dQ}{dt}=0    \; ,        \qquad \qquad    \qquad  \rightarrow   \qquad \qquad   \qquad \partial_{\mu} j^{\mu}=0 \; . \qquad
\eeqs
$ $\\
The Poincar\'e  Lemma actually tells us that the current $j^{\mu}$ is on-shell vanishing up to the divergence of this ``arbitrary" superpotential $k^{[\nu\mu]}$. We can write
\beqs
j^{\mu} \approx \partial_\nu k^{\nu \mu} \; ,
\eeqs
such that $\partial_{\mu} j^{\mu} =0$ by antisymmetry of the superpotential.

For field theories, the charges are thus ill-defined as they can be expressed in terms of an arbitrary superpotential upon using Stoke's theorem
\beqs\label{puzzle}
Q= \int_\Sigma \star j=\oint_{\partial \Sigma} \star k\; ,
\eeqs
where $\partial \Sigma$ is the boundary of $\Sigma$. This phenomena is known as the Noether puzzle. 

From \eqref{puzzle}, one however realizes that non-trivial conserved charges could be defined as the flux of the superpotential through the boundary $\partial \Sigma$. It thus only depends on the properties of $k$ near that boundary. This is a hint that, for gauge field  theories, one should define charges through ``surface" integrals by picking the right superpotential. One necessary requirement is that this superpotential be asymptotically, i.e. close to the boundary, a conserved quantity.

The construction of charges associated to gauge symmetries, through the determination of the right superpotential, is however a very difficult problem. Actually, we believe it is fair to say that there is currently no general understanding of how this can be implemented for a completely generic gauge field theory, i.e. where no assumptions have been made at first. 
The most generic well-established statement about the definition of conserved charges associated to gauge symmetries in field theories can be stated as follows \footnote{This formulation is mainly inspired from the results of Barnich and Brandt stated in \cite{Barnich:2001jy}, see also references in that paper. Note that there does exist a quite recent work by Barnich and Compere \cite{Barnich:2007bf} which discusses the removal of the condition of asymptotic linearity.  }\\

\textit{Under the assumption of asymptotic linearity, every non-trivial asymptotically conserved superpotential $k$ is related to an asymptotic symmetry of the background fields.} \\

Altough we do not want to enter into specific details right away, let us just comment on two important points. 

Firstly, the asymptotic symmetries, also known as large gauge or improper gauge transformations, can be understood as specific gauge transformations which act asymptotically like global transformations, i.e. they act on the physical state of the system.  In classical mechanics, we have seen that conserved charges associated to gauge symmetries are trivial. However, for field theories, one sees that it is possible to associate possibly \textit{non-trivial} conserved charges to some asymptotic, ``global",  transformations.  Indeed, as we have reviewed previously, charges associated to global transformations may turn out to be non-trivial. 

Secondly, in the above statement, asymptotic linearity means that the charges can be constructed from the linearized theory and are expressed in terms of linear combinations of the fields. The above statement is thus obviously generic for gauge theories that are linear such as electromagnetism. However, it clearly imposes restrictions if one deals with non-linear theories. Let us insist here on the fact that the assumption of asymptotic linearity is not a necessary requirement to the construction of charges. Indeed, conserved charges that are non-linear in the fields have already been considered in the literature. As we already pointed out, what this assumption really reflects is our lack of a completely general treatment of non-trivial conserved charges associated to gauge symmetries of non-linear field theories. In this thesis, we will only discuss the case of general relativity. As it is a non-linear theory, we will pay much attention to the possible restrictions this assumption may impose.

\subsubsection{The case of general relativity: Abbott-Deser charges}

For general relativity, one understands that Noether's theorem states that charges associated to diffeomorphisms are zero on-shell up to a divergence of a superpotential, and that conserved charges should thus be associated to asymptotic Killing vectors of a background metric. 
The characterization of the charges of a given solution of Einstein's equations will thus require much more work. With all we have said, our task in the rest of this chapter can be split into the following three steps
\begin{itemize}
\item[(1)] Define what we mean by asymptotic Killing vectors of a background metric. 
\item[(2)] Understand how a solution is said to approach a background metric asymptotically. 
\item[(3)] Determine the form of the conserved superpotential, i.e. the expressions of the conserved charges, and show they are generated by the asymptotic Killing vectors of a background metric. 
\end{itemize}

Because the aspects related to asymptotics are probably less trivial to introduce, we propose to start the discussion by a general construction of conserved charges, when asymptotic linearity holds, in a way that somehow evades the precise formulation of asymptotic notions, as referred in (1) and (2).  We will see that this construction partially answers step (3).

The construction of  Abbott and Deser \cite{Abbott:1981ff} is indeed a quite generic construction of charges associated to \textit{exact} Killing vectors, i.e. isometries, of a given background metric. As such, it evades considerations about the asymptotics of the Killing vectors to which we will return in the following. However, it is of great interest as we will see that it reproduces the correct expressions of conserved charges associated to asymptotic Killing vectors when one assumes asymptotic linearity. 

In their paper, Abbott and Deser started by writing the four-dimensional metric in terms of fluctuations $h_{\mu\nu}$ around a fixed background metric
\beqs
g_{\mu\nu} = \bar{g}_{\mu\nu} + h_{\mu\nu}\;,
\eeqs 
where the background metric $\bar{g}_{\mu\nu}$ is understood to be a solution of Einstein's equations. 
The Killing vectors $\bar{\xi}_{\mu}$ of a background metric $\bar{g}_{\mu\nu}$ are the vectors which satisfy
\beqs
\mathcal{L}_{\bar{\xi}} \bar{g}_{\mu\nu}\equiv \bar{D}_{\mu} \bar{\xi}_{\nu} +\bar{D}_{\nu} \bar{\xi}_{\mu}=0 \; ,
\eeqs
where $\bar{D}_{\mu}$ is the covariant derivative associated to $\bar{g}_{\mu\nu}$ and $\mathcal{L}$ is the Lie derivative.
To construct their conserved charges, they started from Einstein's equations with a cosmological constant
\beqs
R_{\mu\nu}-\frac{1}{2} R g_{\mu\nu}+\Lambda g_{\mu\nu}= 0\; ,
\eeqs
and linearized them such that the left hand side becomes
\begin{eqnarray}
R_{\mu\nu\: L}-\frac{1}{2} \bar{R} h_{\mu\nu}- \frac{1}{2} R_L \: \bar{g}_{\mu\nu} +\Lambda h_{\mu\nu}+O(h^2).
\end{eqnarray}
By simplifying this last expression with $\bar{R}=+4 \Lambda$ and  relabeling all non-linear terms  $O(h^2)$ by $T_{\mu\nu}$, understood as the energy-momentum tensor density of the gravitational field, Einstein's equations are
\beqs\label{einsteinnn}
R_L^{\mu\nu}-\frac{1}{2} R_L \bar{g}^{\mu\nu} -\Lambda h^{\mu\nu}= (-\bar{g})^{-1/2} T^{\mu\nu}\; .
\eeqs
From this, the conserved charges associated to the Killing vectors of the background are defined as
\beqs\label{chargesss}
Q[\bar{\xi}]\equiv \frac{1}{8\pi G} \int d^3 x \; T^{0\nu} \bar{\xi}_{\nu}\; .
\eeqs
These are the right quantities to be considered as \textit{conserved} charges because the Bianchi identity imposes $\bar{D}_{\mu}T^{\mu\nu}=0$ and thus
\beqs
\bar{D}_{\mu} (T^{\mu\nu} \bar{\xi}_{\nu})= (\bar{D}_{\mu} T^{\mu\nu}) \bar{\xi}_{\nu} + \frac{1}{2} T^{\mu\nu} (\bar{D}_{\mu} \bar{\xi}_{\nu}+\bar{D}_{\mu} \bar{\xi}_{\nu} )=0, 
\eeqs
but also because $T^{\mu\nu} \bar{\xi}_{\nu}$ is a vector density such that
\beqs
\bar{D}_{\mu} (T^{\mu\nu} \bar{\xi}_{\nu})=0\qquad \rightarrow \qquad \partial_{\mu} (T^{\mu\nu} \bar{\xi}_{\nu})=0\; .
\eeqs
The conservation is thus really understood as a conservation with respect to the partial derivative, and not the covariant one, of a contravariant density. 

Using \eqref{einsteinnn}, it is quite straightforward to show that \eqref{chargesss} can actually be written as a surface integral \cite{Abbott:1981ff}
\beqs\label{chargessss}
Q[\bar{\xi}] = \frac{1}{8\pi G}  \oint d^2 S_i \sqrt{-\bar{g}} \Big(\bar{D}_{\mu} K^{0i \nu \mu}-K^{0 j \nu i} \bar{D}_j \Big) \bar{\xi}_{\nu}\; ,
\eeqs
where
\begin{eqnarray}
K^{\mu \sigma \nu \kappa}&=&\frac{1}{2} \biggr [  \eta^{\mu\kappa} H^{\nu \sigma} + \eta^{\nu \sigma} H^{\mu \kappa} -\eta^{\mu\nu} H^{\sigma \kappa}- \eta^{\sigma \kappa} H^{\mu\nu} \biggl ] \; , \\
H^{\mu\nu}&=&h^{\mu\nu}-\frac{1}{2} \eta^{\mu\nu} h\; .
\end{eqnarray}

The charges \eqref{chargessss} are known as the Abbott-Deser (AD) conserved charges. What the work of Abbott-Deser has achieved can be stated as follows\\

\textit{Under the assumption of asymptotic linearity, one can derive a set of (potentially) non-trivial conserved charges related to the exact Killing vectors of a given background metric. }\\

The only difference with our previous statement resides in the fact that we have derived charges associated to the \textit{exact} Killing vectors of the background instead of the asymptotic Killing vectors. However, this also means that if one is given a set of asymptotic Killing vectors and assumes asymptotic linearity, the expressions of the conserved charges associated to the asymptotic Killing vectors should be equivalent to the expressions \eqref{chargessss} when evaluated on each asymptotic Killing vector.  We will see that this is indeed the case when the background metric under consideration is Minkowski. 

\subsubsection{Asymptotic Killing vectors versus exact Killing vectors}

As we have already said, we want to describe the set of charges that characterize an ``asymptotic state". By asymptotic state, we mean a specific solution of Einstein's equations, as seen from infinity, that approaches a given background metric. 
What we want to emphasize at this point, as it may not be enough clear from the above discussions, is that the charges that describe an ``asymptotic state" are really obtained by taking into account all the charges constructed from the \textit{asymptotic} Killing vectors of the background metric and not just the ones associated to \textit{exact} background Killing vectors. The difference stands in the fact that \\

\textit{The asymptotic Killing vectors of a given background metric may not be expressed as exact background Killing vectors of any background metric.} \\

To appreciate the difference between charges associated to exact or asymptotic symmetries, it is interesting at this point to introduce the notion of the asymptotic symmetry group. The asymptotic symmetry group is defined as the group of asymptotic symmetries associated to non-trivial conserved charges modulo the (trivial) asymptotic symmetries that are associated to trivial charges. 
Given this definition, one can be faced to three different possibilities whether the asymptotic symmetry group is generated by the exact Killing vectors of \textit{the} background metric, of \textit{a} background metric, or if it is generated by Killing vectors that are not isometries of \textit{any} background metric.

To illustrate our discussion, let us briefly comment on the very famous case of $AdS_3$. In that case, the group generated by the exact Killing vectors of the background is  $O(2,2)$. However, for a specific class of spacetimes that approach the one of AdS at infinity, it was found by J.D. Brown and M. Henneaux in \cite{Brown:1986nw} that the asymptotic symmetry group can be extended to the full conformal group. This asymptotic symmetry group can not be generated by the isometries of any given background metric. Note that this study also led to the discovery of a central extension of the algebra of asymptotic symmetries. The presence of a central charge has been an important clue of the celebrated AdS/CFT correspondence. We hope we have convinced the reader that the study of the asymptotic symmetries \textit{is} of physical relevance and is not just a technicality.

\setcounter{equation}{0}
\section{A path to the infinity of Minkowski spacetime}
\label{pathy}

In the rest of this thesis, we will be concerned with asymptotically flat spacetimes that are spacetimes approaching, at large distances, the boundary metric of Minkowski. We know from special relativity that the exact Killing vectors of Minkowski generate the Poincar\'e group. So, if the asymptotic symmetry group is just the Poincar\'e group, then one can say that the class of asymptotically flat spacetimes are characterized by the charges associated to the exact Killing vectors of the Minkowski background. If one moreover assumes asymptotic linearity, these charges should be equivalent to the AD expressions.

We will see in the following that the determination of the asymptotic symmetry group is a rather complicated problem as it strongly depends on the determination of the asymptotic Killing vectors of the background metric and the choice of a good set of so-called boundary conditions. Note that these specifications precisely correspond to the first two steps of the program depicted in the previous section. 

To obtain the asymptotic Killing vectors of the background, one first needs to know how to reach infinity. This will be the subject of this section. It will allow us to answer the step (1) of our program for the particular case of Minkowski spacetime. 

To know what are the \textit{allowed} asymptotic symmetries and the asymptotic symmetry generators which form part of the asymptotic symmetry group (assuming that such charges can be constructed), one needs to specify a set of boundary conditions that defines our class of asymptotically flat spacetimes. The boundary conditions thus refer to the set of conditions that defines ``spacetimes that approach asymptotically the asymptotic form of the background metric", in our case the boundary of Minkowski. Boundary conditions usually amount to the specification of the asymptotic form of a class of metrics. In some cases, additional restrictions may also be imposed on the boundary fields. The \textit{allowed} asymptotic symmetries are the transformations that map an allowed state to another allowed state, i.e. a state that satisfies the boundary conditions. A specific construction of an asymptotic symmetry group, given a set of asymptotically Killing vectors and specific boundary conditions, will be reviewed in the next section.

For the moment, let us focus on the description of the asymptotic region of the Minkowski spacetime, i.e. flat spacetime. 

\subsubsection{The boundary of Minkowski spacetime}

The structure of Minkowski spacetime at infinity is best understood through its so-called Carter-Penrose diagram. To characterize infinity in a more mathematical framework, one would like to possess concepts such as ``the neighborhood at infinity". As pointed out by R. Penrose in \cite{Rpen}:  "from the point of view of the metric structure of space-time, there is no such thing as a point \textit{at} infinity, since such a point would be an infinite distance from its neighbors". In \cite{Rpen}, Penrose  evades these issues by proposing to work with the conformal structure of space-time, which implies that only ratios of neighboring infinitesimal distances are to have significance. The idea can be summed up as follows. We bring what we have called ``infinity" to a finite distance by considering a new ``unphysical" metric $g_{\mu\nu}$ which is related to the physical metric $\hat{g}_{\mu\nu}$ by
\beqs
g_{\mu\nu}=\Omega^2 \hat{g}_{\mu\nu},
\eeqs
where $\Omega$ is the conformal factor. One says that both metrics are conformally related to each other. The ``infinity part", or boundary $\mathcal{J}$ of our spacetime $\mathcal{M}$, has been brought to a finite distance. It lies at $\Omega=0$. Note that $\Omega_{;\mu}\neq 0$. 

For Minkowski spacetime, one can check that on $\mathcal{J}$ we have
\beqs
\Omega_{;\rho} \Omega^{;\rho}=0,
\eeqs
telling us that the boundary is a null hypersurface. As illustrated on Fig.1.1., one distinguishes five disjoint parts  on this null hypersurface : the three points  $I^-$, $I^+$, $I^0$ which represent past, spatial and future infinity and the two null hypersurfaces $\mathcal{J}^-$ and  $\mathcal{J}^+$ which represent the past and future null infinities. The conformal structure of a space-time is often represented by a small diagram called the Carter-Penrose diagram (see Fig.1.1.).
\begin{figure}[!hbt]
  \centering \label{fig1}
\includegraphics[width=0.3\textwidth]{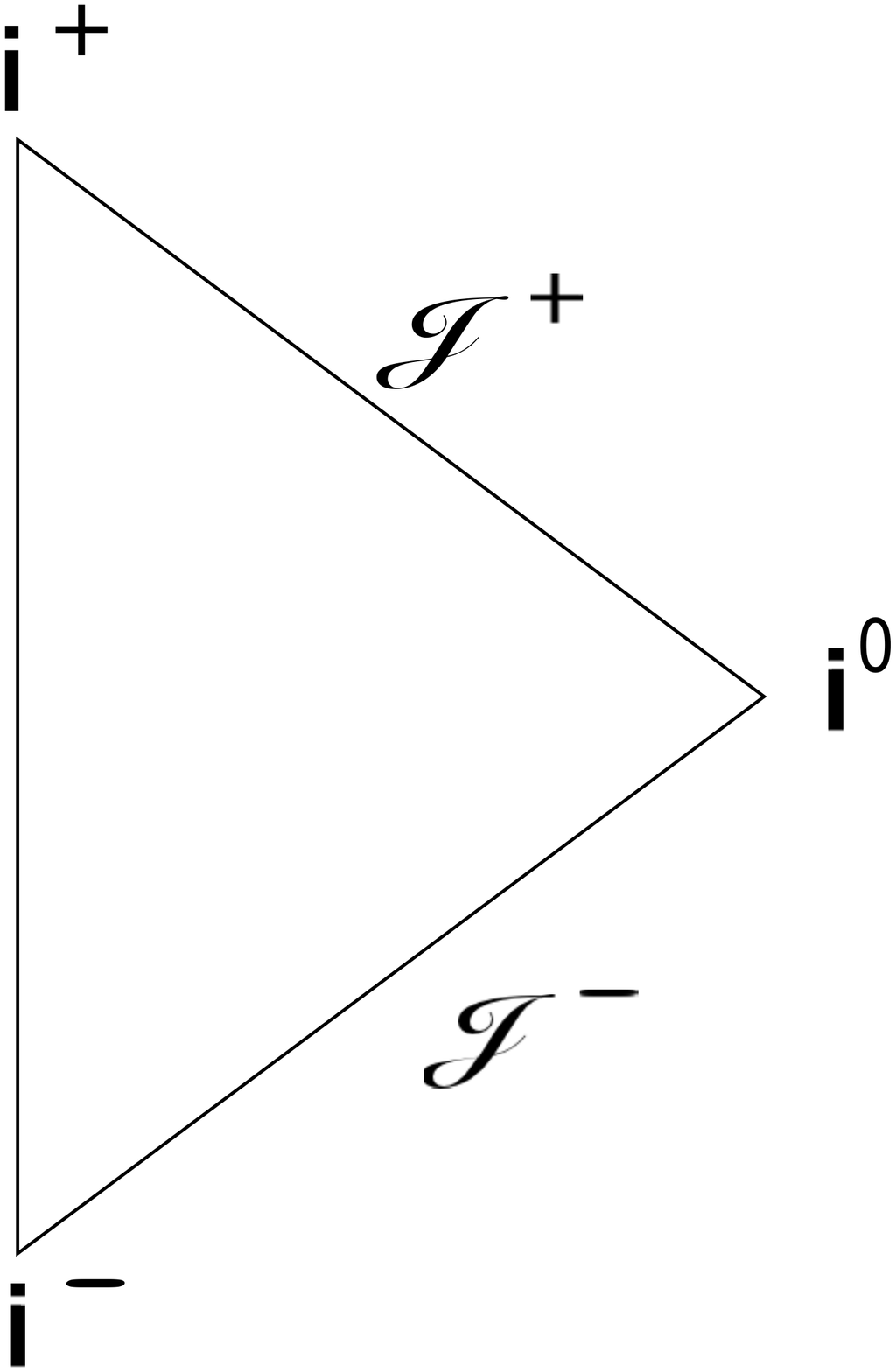}
  \caption{Carter-Penrose diagram of Minkowski spacetime.}
\end{figure}

In this approach, a spacetime is thus specified by the bulk metric, the metric describing the geometry of the inside, and the boundary, an hypersurface attached to the manifold.  The method just described is known as the conformal (or Penrose) completion of spacetime.

From this analysis, we thus see that Minkowski is a quite non-trivial case as there exists two different ways one can reach infinity. Indeed, to characterize a given solution, that we may understand as a source, we make the distinction between
\begin{itemize}
\item Null infinity $\mathcal{J}^{\pm}$: region situated at very large null separations from the source.\\
\item Spatial infinity: region situated at very large spacelike distances from the source.
\end{itemize}

Null infinity was first considered for the study of gravitational waves. Indeed, as compared to spatial infinity, gravitational radiation can escape through this boundary. The study of asymptotic symmetries revealed a much richer structure than expected as first discovered by Bondi, Metzner and Sachs (BMS). The group of asymptotic symmetries at null infinity is known as the BMS group (see \cite{Bondi:1962px} and \cite{Sachs:1962wk}). It is the semi-direct product of the group of globally defined conformal transformations of the unit 2-sphere, which is isomorphic to the orthochronous homogeneous Lorentz group, times the abelian normal subgroup of so-called supertranslations. This result was revisited in \cite{Barnich:2009se} (see also \cite{Barnich:2010eb,Barnich:2011ct,Barnich:2011mi}) where it was shown that local conformal transformations can actually be considered.

In this thesis we will restrict ourselves to considerations about spatial infinity. Let us just mention that for null infinity, one reaches it by considering null three-dimensional hypersurfaces. However, on the Carter-Penrose diagram, we just saw that spatial infinity is a point. If we want to approach spatial infinity as a limiting procedure of data given on 3-surfaces, we are faced with the dilemma of choosing whether the 3-surface used to describe spatial infinity should be spacelike or timelike.  Let us now comment on these two approaches.

\subsubsection{Hamiltonian description of spatial infinity: Cylindrical (ADM) slicing}

The first description of spatial infinity was established by Arnowitt, Deser and Misner \cite{Arnowitt:1961zz}. As we have discussed previously, they considered a splitting of space and time so that one can formulate Einstein's theory in terms of a well-posed initial formulation. In their work, spatial infinity is described by taking $r\rightarrow \infty$ on a specific Cauchy slice of spacetime. 

Going through the Hamiltonian formalism, after fixation of the first class constraints, they obtained expressions for the charges associated to energy and momentum. We do not want to enter into details of their procedure as we will review the Regge-Teitelboim construction which recover the ADM expressions, and agree with the Abbott-Deser expressions, in the next section. 

Criticisms (see for example \cite{Ashtekar:1978zz}) towards this approach to spatial infinity rely on the fact that the formalism is not covariant and does not permit comparison with conserved quantities defined at null infinity, such as the Bondi energy \cite{Bondi:1962px}.

\subsubsection{Covariant description of spatial infinity: Hyperbolic slicing}

A way to deal with these drawbacks was first formulated by Ashtekar and Hansen  \cite{Ashtekar:1978zz} who developed a formalism, known as the $i^0$ formalism. This formalism deals with spatial infinity as the vertex of the light cone representing future and past null infinities $\mathcal{J}^{\pm}$. As such, they were able to compare quantities defined at null and spatial infinity. The comparison with the 3+1 description was described by Ashtekar and Magnon in  \cite{Ashtekar:1984aa}. However, the Ashtekar-Hansen formalism considers spatial infinity as a point and as such awkward differentiability conditions have to be imposed at $i^0$.

A way to overcome these awkward differentiability conditions,  in a coordinate-dependent way, was provided by the formalism\footnote{We refer the reader to the next chapter for a description of this formalism.} of Beig and Schmidt \cite{BS}.  This led Ashtekar and Romano \cite{Ashtekar:1991vb} to formulate spatial infinity, in a coordinate-indepedent way, as a limit of timelike 3-surfaces. 
Their formalism is a reformulation of Penrose's conformal approach (of null infinity) to deal with spatial infinity. In this sense, they gave a more geometrical formulation of spatial infinity. 

In \cite{Ashtekar:1991vb}, the way spatial infinity is described is through a specific compactification that permits to deal with spatial infinity in terms of a limiting procedure of timelike hypersurfaces. The first thing to notice is that if one starts with Minkowski spacetime in cartesian coordinates
\beqs
d\hat{s}^2=-dt^2 +dx^2 +dy^2 +dz^2,
\eeqs
one can introduce, in the region of Minkowski spacetime exterior to the light cone at the origin, the standard hyperbolic coordinates
\beqs\label{Ashcoord}
t= \rho \sinh \tau , \qquad x=\rho \cosh \tau \sin \theta \cos \phi , \nonumber \\
y= \rho \cosh \tau \sin \theta \sin \phi , \qquad z= \rho \cosh \tau \cos \theta .
\eeqs
In this chart, the metric takes the form
\beqs
d\hat{s}^2=\hat{\eta}_{\mu\nu} dx^{\mu} dx^{\nu}= d\rho^2 + \rho^2 h^{(0)}_{ab} d\phi^a d\phi^b , \eeqs
where $h^{(0)}_{ab}$ is the unit time like hyperboloid metric
\beqs
h^{(0)}_{ab} d\phi^a d\phi^b = - d\tau^2 +\cosh^2 \tau (d\theta^2 +\sin^2 \theta d\phi^2).
\eeqs
Following Penrose's approach described above, since spatial infinity is at $\rho \rightarrow \infty$, one defines $\Omega=1/\rho$. Doing this, spatial infinity is at $\Omega=0$.  In these new coordinates, the physical metric is such that
\beqs\label{ita}
d\hat{s}^2= \hat{\eta}_{\mu\nu} dx^{\mu} dx^{\nu}=\Omega^{-4} \nabla_{\mu} \Omega \nabla_{\nu} \Omega \;  dx^{\mu} dx^{\nu} +\Omega^{-2} h^{(0)}_{ab}\; d\phi^a d\phi^b ,
\eeqs
and it is singular at $\Omega=0$. 

It is obvious from \eqref{ita} that the usual conformal completion we described above will not work. Indeed, it would tell us to introduce an unphysical metric as a conformal rescaling of the physical one such as $\hat{\eta}_{\mu\nu}=\Omega^4 \eta_{\mu\nu}$. As such, the surface $\Omega=0$ will have zero volume with respect to the unphysical metric. Spatial infinity would then be described by a single point, which is not what we were aiming at. 

Looking again at \eqref{ita}, and because we want something like $g_{\mu\nu}=n_{\mu} n_{\nu} + h_{\mu\nu}$, the solution to this problem is to rescale the 3-metric and the normals to the $\Omega=$cst 3-surfaces by different powers of $\Omega$. Indeed, the induced metric on this timelike hypersurface is
\beqs
\hat{h}_{ab}=\Omega^{-2} q_{ab}=\hat{\eta}_{ab}-l^{-1} \nabla_a \Omega \nabla_b \Omega\; ,
\eeqs
where $l\equiv \hat{\eta}^{ab} \nabla_a \Omega \nabla_b \Omega$ and $q_{ab}$ is a general 3-metric which reduces to $h^{(0)}_{ab}$ for Minkowski. As we just said, this scaling can not be applied to the full metric as $\Omega^2 \hat{\eta}_{ab}$ does not admit a smooth extension to the 3-surface $\Omega=0$. However, the rescaled 3-metric
\beqs
h_{ab}\equiv \Omega^2 \hat{h}_{ab}\; ,
\eeqs
is well defined on the boundary. Now, the contravariant normal to these 3-surfaces 
\beqs
\hat{\eta}^{ab}\nabla_b \Omega=\Omega^4 \Big(\frac{\partial}{\partial \Omega}\Big)^a\; ,
\eeqs
needed to extract information off the $\Omega=$cst surfaces can be rescaled such that
\beqs
n^a\equiv \Omega^{-4}\hat{\eta}^{ab}\nabla_b \Omega\; ,
\eeqs
because $(\partial/\partial \Omega)^a$ is well defined on the boundary. By rescaling the 3-metrics and the normals to $\Omega=$cst surfaces by different powers of $\Omega$, we have achieved a description of spatial infinity in terms of timelike 3-surfaces.

Asymptotically flat spacetimes at spatial infinity are understood as spacetimes that resemble Minkowski spacetime sufficiently so as to admit a completion in which the fields $h_{ab}$ and $n^a$ have smooth limits to the boundary. 

We will not discuss how asymptotic symmetries and charges can be constructed in this geometrical way and refer the reader to the original paper. However, we will study in the next two chapters the Beig-Schmidt formalism which is understood as a coordinate-dependent formulation of the Ashtekar-Romano formalism. This will be motivated by the fact that this formalism is easier to deal with when concerned with solutions of equation of motions.

\setcounter{equation}{0}
\section{The Regge-Teitelboim approach}
\label{sec:RT}

 For general relativity, we have reviewed the construction of the Abbott-Deser conserved charges associated to exact Killing vectors of a background metric. However, to describe the charges of a specific class of spacetimes, one needs to consider the charges associated to their asymptotic symmetries. In this section, we review the work of Regge and Teitelboim who recovered the Poincar\'e group as the asymptotic symmetry group of a class of asymptotically flat spacetimes at spatial infinity.

 To find the set of asymptotic symmetries, we said that one needs to: (1) describe how one reaches infinity and (2) give a set of  boundary conditions.  In \cite{Regge:1974zd}, T. Regge and C. Teitelboim considered asymptotically flat spacetimes at spatial infinity, using an ADM slicing of spacetime.  The class of spacetimes they considered are asymptotically of the form
\beqs\label{AERT}
\g_{ij} &\underset{r \rightarrow \infty}{\rightarrow}& \delta_{ij} + \frac{h^{(1)}_{ij}(\textbf{n})}{r}+  \frac{h^{(2)}_{ij}(\textbf{n})}{r^2} +O(r^{-(2+\epsilon)})\; ,  \nn\\
\pi^{ij} &\underset{r \rightarrow \infty}{\rightarrow}&  \frac{\pi^{(2) \: ij}(\textbf{n})}{r^2}+  \frac{\pi^{(3)\: ij}(\textbf{n})}{r^3} +O(r^{-(3+\epsilon)})\; ,
\eeqs
where $\g_{ij}$ is the three-dimensional metric on the Cauchy surface and $\pi^{ij}$ is the conjugate momenta to the three-metric $\g_{ij}$. Note that $\pi^{ij}$ is a tensorial density of weight +1. They also restricted their study to spacetimes which also obey the following parity conditions
\beqs\label{parityy}
h^{(1)}_{ij}(-\textbf{n})=h^{(1)}_{ij}(\textbf{n}) \; , \qquad \pi^{(2) \: ij}(-\textbf{n})=-\pi^{(2) \: ij}(\textbf{n})\; , \qquad \textbf{n}=r^{-1}(x,y,z)\; ,
\eeqs
i.e. $h^{(1)}_{ij}$ must be of even parity and $\pi^{(2)\: ij}$ of odd parity. The specifications \eqref{AERT} and \eqref{parityy} are referred as the boundary conditions. 

One can now derive the group of asymptotic symmetries or allowed diffeomorphisms at infinity, i.e. diffeomorphisms that map an allowed configuration at infinity to another allowed one. It was shown in \cite{Regge:1974zd} that the most general deformations of the hypersurface, on which the state is defined (see also section \ref{einsteinn}), that leave the form of \eqref{AERT} and \eqref{parityy} invariant are
\beqs\label{transfoo}
N^{\mu} \underset{r \rightarrow \infty}{\sim}\alpha^{\mu}+\beta^{\mu}_{\:\:i} x^i +\xi^{\mu}(n)+ O(1/r), \qquad \beta_{\mu i}=-\beta_{i \mu}\; , \qquad \xi^{\mu}(-n)=-\xi^{\mu}(n)\; .
\eeqs
The parameters $\alpha^{\mu}$ are the four translations, the $\beta_{\mu i}$ are six parameters associated to boosts and rotations, while the  $\xi^{\mu}$ are (parity odd) supertranslations. As we have explained in the previous section, this is again a  manifestation that asymptotic symmetries are not always equivalent to background Killing vectors. Indeed, the background Minkowski metric only has ten Killing vectors associated to translations, rotations and boosts, i.e. the ten Poincar\'e transformations, while our class of asymptotically flat spacetimes at spatial infinity admits a richer set of transformations as an infinite class of (parity-odd) supertranslations are also allowed transformations.

 In section \ref{lagvshamm}, we have seen that Noether associates a conserved charge to a symmetry of the action. If the Abbott-Deser charges associated to asymptotic symmetries are really the generators of these asymptotic symmetries, one should be able to associate them to symmetries of an action. Actually, even before the Abbott-Deser construction, this is precisely how Poincar\'e charges for asymptotically flat spacetimes at spatial infinity were constructed by  Regge and Teitelboim. Starting from the Hamiltonian action,  they studied the definition of a good variational principle for asymptotically flat spacetimes and obtained the conserved charges as generators of the asymptotic symmetries of the Hamiltonian action. Doing so, they recovered the previous expressions given by ADM.

Regge and Teitelboim showed in \cite{Regge:1974zd} that starting from the Hamiltonian \eqref{H0}, and allowing asymptotic transformations of the form \eqref{transfoo} for the class of spacetimes described by \eqref{AERT}, Hamilton's principle is not a good variational principle as it is not stationary under variations including all these allowed trajectories. Actually, a good variational principle can be achieved if and only if specific surface integrals are supplemented to the original Einstein-Hilbert action. These surface integrals are understood as the generators of the asymptotic symmetries, i.e. the Noether charges, when evaluated on solutions of the constraint equations.
 
To have a well-defined variational principle, it is necessary that the variation of the Hamiltonian, under any variation allowed in our phase space, takes the form
\beqs
\delta H_0= \int d^3 x \: A^{ij} \delta \g_{ij} +B_{ij} \delta \pi^{ij}\; ,
\eeqs
where $A^{ij}$ and $B_{ij}$ were defined in \eqref{ABABAB}. However, one can compute this variation from (\ref{H0}) and realize that this is true up to surface integral terms. By using
\beqs
\delta \sqrt{\g}=\frac{1}{2} \sqrt{\g} \g^{ij} \delta \g_{ij} \; ,  \qquad \delta \g^{-1/2}= -\frac{1}{2} \g^{-1/2} \: \g^{ij} \delta \g_{ij} \; ,
\eeqs
and the following quantities
\beqs
\int d^3 x \: \delta (N^i \cH_i)&=&  -\oint d^2 s_l \: \Big ( 2N_i \delta \pi^{il} +(2 N^k \pi^{jl}- N^l \pi^{jk})  \delta \g_{jk} \Big) \nonumber \\
&& + \int d^3 x \: (N_{i|j}+N_{j|i}) \delta \pi^{ij} + (N^i_{\;|m} \pi^{mj} +N^j_{\;|m} \pi^{mi}) \delta \g_{ij}\; ,\nonumber \\
\int d^3 x \: \delta (N \: \cH)&=&  \int d^3x\: \delta( N G_{ijkl} \pi^{ij} \pi^{kl}) + N \sqrt{\g} \: \g^{ij} \Big(  (\cR^{ij}- \frac{1}{2} \cR)  \delta \g_{ij} -\delta \cR_{ij}) \; , \nonumber \\
\eeqs
we find
\beqs
&& \int d^3 x \: \delta( N G_{ijkl} \pi^{ij} \pi^{kl}) = \int d^3 x \: \Big( 2 N \g^{-1/2} (\pi_{ij}-\frac{1}{2} \pi \: \g_{ij})   \Big)  \delta \pi^{ij} \nonumber \\
&&\qquad \qquad \qquad + \Big(  -\frac{N}{2} \g^{-1/2} \: \g^{ij} (\pi^{mn} \pi_{mn}-\frac{1}{2} \pi^2 ) +2 N \g^{-1/2}  (\pi^{im} \pi_m^j-\frac{1}{2} \pi \pi^{ij})\Big) \delta \g_{ij},  \nonumber \\
 &&\int d^3 x \: N \sqrt{\g} \: \g^{ij} \delta(\cR_{ij})=  \oint d^2 s_{l} \:G^{ijkl}  (N \: \delta \g_{ij|k} - N_{|k} \delta \g_{ij}) \nonumber \\
 &&\qquad \qquad \qquad \qquad \qquad \qquad +\int d^3 x \: \sqrt{\g} (N^{ij} -\g^{ij} N^{|m}_{\;|m}) \delta \g_{ij}  \; .
\eeqs
In the end, one rapidly obtains
\beqs\label{delH0}
\delta H_0&=& \int d^3 x \: A^{ij} \delta \g_{ij} +B_{ij} \delta \pi^{ij} \nonumber \\
&& -\oint d^2 s_{l} \:G^{ijkl}  (N \: \delta \g_{ij|k} - N_{|k} \delta \g_{ij}) \nonumber \\
&& - \oint d^2 s_l \: \Big(  2  N_k \delta \pi^{kl} +(2N^k \pi^{jl}-N^l \pi^{jk}) \delta \g_{jk} \Big)  \; .
\eeqs
From this, we see that Hamilton's equations can not be obtained from asking stationarity of the Hamiltonian action
\beqs
S_0=\int d^3 x \:  \g_{ij} \dot{\pi}^{ij}-H_0\; ,
\eeqs
 and reproduce Einstein's equations, unless the surface integrals are vanishing. This is obviously true for closed space-times and in this case Hamilton's principle is well-defined when one starts from the Hamiltonian \eqref{H0}. However, it is not the case for the class of asymptotically flat spacetimes under consideration and the Hamiltonian has thus to be supplemented with the non-vanishing parts of these surface integrals to give a good variational principle and reproduce Hamilton's equations.

One can now evaluate those surface integrals for each particular asymptotic transformation given in \eqref{transfoo}.
Plugging \eqref{AERT} and \eqref{transfoo} into those integrals, it was realized that some surface integrals might potentially linearly diverge. This is why Regge and Teitelboim imposed the additional parity conditions \eqref{parityy}. For the divergences to cancel, one should have $h^{(1)}_{ij}$ and $\pi^{(2)\:ij}$ of opposite parities. The fact that $h^{(1)}_{ij}$ is chosen to be parity even is only motivated by the fact that the Schwarzschild solution lies in the phase space of allowed metrics. Imposing these additional parity conditions, the linearly divergent parts of the surface integrals identically vanish as the integrands are of odd parity. Non-linear terms that would have appeared otherwise in the expressions for the Lorentz charges vanish for the same reasons. In the end, they find that the correct extended Hamiltonian should be
\beqs
H=H_{0}- \alpha^{\mu} P_{\mu} +\frac{1}{2} \beta^{\mu\nu} M_{\mu\nu}\; ,
\eeqs
where $P_{\mu}$ are the four momenta, generators of translations, and $M_{\mu\nu}$ are the six Lorentz charges associated to boosts and rotations. Note that charges associated to parity-odd supertranslations are always zero because of the parity conditions they imposed on the first order fields in the asymptotic expansion. Here, parity-odd supertranslations, associated to trivial charges, are true gauge transformations.

For the energy, associated with the invariance of the action under a time translation, we have $N_{\perp}=\alpha_{\perp}$, $N_{i}=0$, and only the first integral in \eqref{delH0} contributes. The associated conserved charge is
\beqs
P^{\perp}\equiv E=\oint d^2S_j  \Big( h^{(1)}_{ij,i}-h^{(1)}_{ii,j}\Big)\;.
\eeqs
For spatial translations, we have $N_i \sim \alpha_i$ and we find
\beqs
P^i \equiv -2 \: \oint d^2S_j \: \pi^{(2)\: ij}\; .
\eeqs
These expressions are the same as the expressions obtained by ADM. 
For rotations and boosts, we respectively find
\beqs
M^{ij}&\equiv& -2 \: \oint d^2S_j \: \pi^{(3)\: ij}\; , \\
M_{\perp r} &\equiv& \oint d^2 s_l \:G^{(0) ijkl} (x_r \g^{(2)}_{ij|k}-\delta_{rk}  \g^{(2)}_{ij}) \nn\\
&=&\oint d^2 S_l [x_r (\g^{(2)}_{sl,s} -\g^{(2)}_{ss,l} ) -\g^{(2)}_{rl} +\g^{(2)}_{ss} \delta_{rl}] \: .
\eeqs
The algebra of the above ten charges was shown to reproduce the Poincar\'e algebra. 

On can rapidly convince himself that the expressions of Abbott and Deser, when evaluated on each of the Killing vectors of Minkowski, agree with the above (Regge-Teitelboim) Poincar\'e charges. 

Regge and Teitelboim realized that, following Dirac's idea, the canonical phase space should be described by $(\g_{ij},\pi^{ij})$ supplemented with 10 additional independent canonical pairs, the allowed asymptotic lapse and shift and their canonical conjugates, which describe the Poincar\'e  transformations of the spacelike surface at infinity. The supertranslations $\xi^{\mu}$ should not be considered as extra variables as they are pure gauge transformations, i.e. their associated charges are trivial, and can thus be gauge-fixed.

\setcounter{equation}{0}
\section{Summary: Asymptotic linearity, parity conditions and equations of motion.}
\label{sec:summaryyy}

In this chapter, we applied Noether's theorem for global and local symmetries of the extended action of a constrained Hamiltonian system. We have seen that, for classical mechanics, gauge symmetries are associated to trivial charges. For gauge field theories, such as general relativity, Noether's theorem associates a current which is vanishing on-shell up to the divergence of a superpotential. 

These results have motivated the construction of  ``global" conserved charges for gauge field theories in terms of surface integrals.  For general relativity, we have reviewed the work of Abbott and Deser who have given a generic definition of charges that describe metrics which can be expressed as fluctuations around a given background metric $\bar{g}_{\mu\nu}$. From the linearized equations of motion, these charges are written as surface integrals. 

Although the Abbott-Deser construction seems to answer the problem of defining conserved charges for general relativity, we have pointed out that charges should actually be associated to asymptotic symmetries and that this often differs from the consideration of charges associated to exact background Killing vectors. To find what are the allowed asymptotic symmetries, one needs a specification of the asymptotic sector to be considered. 

For asymptotically flat spacetimes, spacetimes that approach Minkowski spacetime at infinity, we have seen that two different regimes may exist at infinity, i.e. null infinity and spatial infinity. Focusing on spatial infinity, we have then provided two different ways one approaches it  using either a family of Cauchy surfaces, a cylindrical slicing of spacetime, or a specific conformal completion of spacetime that introduces an hyperbolic slicing of spacetime. In the first of these frameworks, we have reviewed the work of Regge and Teitelboim who constructed, for a class of metrics specified by a given set of boundary conditions, the ten Poincar\'e surface charges as generators of asymptotic symmetries of the Hamiltonian action.  For this splitting of spacetime and these boundary conditions, the construction of charges is equivalent to a construction that would use the Abbott-Deser charges associated to the ten Poincar\'e Killing vectors. 

As a way to motivate the considerations presented in the next two chapters, let us finish here with some comments on the asymptotic linearity assumption of Abbott and Deser, the parity boundary conditions used by Regge and Teitelboim, and the relevance of a study of Einstein's equations in the process of determining a physically interesting set of boundary conditions. 

In the work of Abbott and Deser, conserved charges were constructed from the linearization of a metric around a given background and the subsequent linearization of Einstein's equations. The charges one obtains are thus linear in the fluctuations $h_{\mu\nu}$. This is referred as asymptotic linearity.  However, as we have already pointed out, general relativity is a non-linear theory, charges may thus contain non-linearities. It can thus not be the end of the story, unless general relativity is proved to be asymptotically linear. 

The construction of Regge and Teitelboim does not refer at all to asymptotic linearity. Their construction of charges relies on Noether's theorem applied to asymptotic symmetries of an action and, as such, they may well turn out to be non-linear. However, they do find charges which are linear in the fluctuations and which are equivalent to the Abbott-Deser charges. The fact that the charges are linear is achieved thanks to the parity conditions imposed on the first order fields. As we have quickly explained, this was implemented to cancel divergences present in the Lorentz charges. The presence of such parity conditions remains however quite obscure and, maybe for this reason, their discussion was  relegated to the appendices of their paper. 

As we will show in the next chapter, these parity conditions are sufficient to ensure that the spacetimes allowed by the boundary conditions are solutions of the Einstein's equations. They may however not be necessary conditions. One way to state about the necessity of these conditions would be to show that only such asymptotically flat spacetimes are solutions of Einstein's equations. The analysis of the next chapter will show that the Regge-Teitelboim class of spacetimes, where parity conditions would not have been imposed, are solutions of Einstein's equations only if a specific subset of conditions on the fields are imposed. From this perspective, the requirement of parity conditions is seen as a very stringent condition. 

The justification of Regge and Teitelboim for introducing parity conditions was that these conditions are sufficient to cancel the linear divergences present in the expressions of the charges. However, we will also see that these contributions vanish under the conditions imposed by Einstein's equations in the absence of parity conditions. However, parity conditions do have their importance as logarithmic divergences are still present as firstly noticed by Beig and o'Murchadha in \cite{Beig:1987aa}. Although these divergences have not been discussed by Regge and Teitelboim, we will see that it is connected to the status of logarithmic translations which can be, for a specific set of boundary conditions, considered as asymptotic symmetries in the absence of parity conditions. 

From the existing litterature, the fact that the asymptotic symmetry group at spatial infinity is always found to be the Poincar\'e group and that the theory should be seen as asymptotically linear is thus intimately connected with the choice of boundary conditions. To state about asymptotic linearity, we believe that one should impose the less possible stringent boundary conditions, but still enough stringent so that they describe solutions of Einstein's equations. In the last chapter of Part I, we propose a way to relax parity conditions.  As such, we find that the asymptotic symmetry group is larger than the Poincar\'e group and contain charges that can not be obtained from the linearized theory. This construction differs thus radically from the results presented in the litterature. Our analysis relies on the Beig-Schmidt formalism we describe in the next chapter.

%%%%%%%%%%%%%%%%%%%%%%%%%%%%%%%%%%%%%%%%%%%%%%%%%%%%%%%%%%%%%%%%%%%

\chapter{Generalized Beig-Schmidt formalism}\label{BS}

The Beig-Schmidt formalism is a coordinate dependent description of the covariant Ashtekar-Hansen formalism, which was presented in chapter \ref{chap:lagvsham}, to describe asymptotically flat spacetimes at spatial infinity. Our first aim in this chapter will consist in reviewing their formalism. This will enable us to discuss the unicity of conserved charges that one can construct when taking into account the equations of motion. The other aim is to extend this formalism such as to pave the road for the discussions in the following chapter.

We start in section \ref{BeigSchm} by reviewing the definition of asymptotically flat spacetimes given by R. Beig and B. Schmidt in \cite{BS}, discuss the asymptotic symmetries of their ansatz and motivate the consideration of a generalized ansatz that includes a logarithmic contribution at second order in a radial expansion.
 We continue, in section \ref{theeq}, by plugging our generalized ansatz in Einstein's equations to obtain the zeroth, first and second order equations in the radial expansion. In section \ref{sec:nice}, we rewrite the first and second order equations in terms of symmetric and divergenceless tensors that we have previously classified. In the meantime, we also discuss the solutions to these equations.  After reviewing the conditions imposed by Einstein's equations at second order in section \ref{sec:LS}, which we recognize as linearization stability constraints, and describing the properties of tensors and Killing vectors on de Sitter space, in section \ref{sec:conservedd} we  construct charges associated to translations, rotations and boosts and show that they are unique within the Beig-Schmidt formalism. We also comment about these constructions in our general set up. 
 
 Our main result in this chapter resides in the fact that, within the Beig-Schmidt formalism, only ten independent non-trivial Poincar\'e charges can be defined from the analysis of the equations of motion. Also, we see that six of them, the Lorentz charges, can be written in two equivalent ways using either the electric or the magnetic part of the Weyl tensor. As we will see in the next chapter, this gives a generic proof of the equivalence, as shown in \cite{Mann:2008ay}, between the counterterm Lorentz charges of Mann and Marolf \cite{Mann:2005yr}, who use the electric part of the Weyl tensor, and the Ashtekar-Hansen Lorentz charges \cite{Ashtekar:1978zz} described with the magnetic part of the Weyl tensor.

As this chapter is rather technical, we end up in section \ref{sec:summ} with an extended summary of the results contained in this chapter. The reader familiar with this formalism, not interested with the details, or lost in the middle of technical details, may directly proceed to this summary section.

In Appendix \ref{App: BS}, we describe how the Schwarzschild solution can be brought to the Beig-Schmidt form. Appendix  \ref{app:properties} sums up a number of useful properties of tensors on the unit hyperboloid that are used throughout this chapter and the following one. Eventually, Appendix \ref{app:proofs} provides the proofs of the five Lemmae stated in the main text.

\setcounter{equation}{0}
\section{The Beig-Schmidt ansatz}
\label{BeigSchm}

R. Beig and B. Schmidt considered, in \cite{BS}, a class of spacetimes which are asymptotically flat at spatial infinity. Their considerations were motivated by previous results obtained at null infinity and their will to learn more about the structure at infinity, and in particular about solutions, satisfying specific boundary conditions, of the equations of motion. Their definition provides a framework where, in a neighborhood of spatial infinity, the new class of spacetimes admits an expansion in negative powers of a radial coordinate. Einstein's equations can be expressed as a hierarchy of equations for the coefficients in this expansion sourced by nonlinear terms of subleading orders. The first work of R. Beig and B. Schmidt in \cite{BS} proves that this hierarchy can be completely solved provided the initial data satisfies certain constraints.  The follow-up work of R. Beig \cite{B} proves that the system can actually be solved under the milder assumption that the first order field in the expansion satisfies six conditions, that he refers to as integrability conditions. 

Before discussing Einstein's equations, let us discuss the form of the ansatz for the metric. The ansatz for the metric will be part of the definition of our boundary conditions.

\subsection{Asymptotically flat spacetimes at spatial infinity}

In \cite{BS}, asymptotically flat spacetimes at spatial infinity are defined as space-times admitting a radially smooth Minkowskian spacelike infinity. The coordinate $\rho$ defined in the following is the same as the one previously defined in \eqref{Ashcoord}. 

\begin{flushleft}
\textit{Definition} : ($\mathcal{M},g$) is radially smooth of order $m$ at spatial infinity, if the following holds:\\
$ $\\
(1) For a part of $\mathcal{M}$, a chart ($x^{\mu}$) exists which is defined for
\beqs
\rho_0 < \rho < \infty , \qquad \rho^2=\eta_{\mu\nu} x^{\mu} x^{\nu} \; .
\eeqs
(2) The components of the metric in this chart satisfy
\beqs\label{genericm}
g_{\mu\nu} =\eta_{\mu\nu} +\sum_{n=1}^{m} \frac{1}{\rho^n} \: l^n_{\mu\nu} \Big( \frac{x^{\s}}{\
\rho}\Big) +f^{m+1}_{\mu\nu} ,
\eeqs
where \\
$ $\\
(3) $l^n_{\mu\nu}$ is $C^{\infty}$  in $x^{\s}/\rho$ and $|f^m_{\mu\nu}|\leq \frac{const}{\rho^m}$, $|f^{m+1}_{\mu\nu}|\leq \frac{const}{\rho^{m+1}}$,...
\end{flushleft}
Apart from technicalities, this definition is readily the same as was previously presented by Ashtekar in \cite{AshRev}.

The first important thing to remark is that there exists a large freedom of performing changes of coordinates such that \eqref{genericm} holds. Indeed, this is true for example for $s\geq m-1$ with
\beqs
x^{\mu}=\bar{x}^{\mu}+\sum_{n=1}^s \frac{a^{\mu}(\bar{x}^\nu/\rho)}{\bar{\rho}^n}, \qquad \bar{\rho}^2=\eta_{\mu\nu} \bar{x}^{\mu} \bar{x}^{\nu}.
\eeqs
There are also other transformations known as supertranslations or logarithmic translations which preserve the form of the metric \eqref{genericm}. They read respectively
\beqs
x^{\mu}&=&\bar{x}^{\mu}+S^{\mu}(\bar{x}^{\nu}) \; ,\label{supp}\\
x^{\mu}&=&\bar{x}^{\mu}+C^{\mu} \ln \bar{\rho}\; , \qquad C^{\mu}=const, \label{log}
\eeqs
where supertranslations are direction-dependent shifts of the origin.

In summary, we see that the set of diffeomorphisms preserving the form of the metric \eqref{genericm}, i.e. the asymptotic symmetries, can be written as
\beqs
\bar{x}^{\mu}=L^{\mu}_{\:\:\nu} x^{\nu} +T^{\mu} + S^{\mu} (x^{\nu}) +C^{\mu} \ln \rho +o(\rho^0)\; ,
\eeqs
where $L^{\mu}_{\:\:\nu} $ are the Lorentz transformations, $T^{\mu}$ are the translations, $S^{\mu}$ are the supertranslations , and $C^{\mu}$ are the four logarithmic translations.

If one consider that ($\phi^a$) is a local chart on the manifold of directions $x^\mu/\rho$, it implies that there exists functions $w^{\mu}(\phi^a)$ such that
\beqs
\frac{x^{\mu}}{\rho}=w^{\mu}(\phi^a), \qquad dx^{\mu}=w^\mu d\rho + \rho \: w^\mu_{,a} \: d\phi^a.
\eeqs
Given this and also
\beqs
&&\eta_{\mu\nu} dx^{\mu} dx^{\nu} = d\rho^2 + \rho^2 h^{(0)}_{ab} d\phi^a d\phi^b , \\
&&\tilde{\s}^n\equiv l^n_{\mu\nu} w^{\mu} w^{\nu}, \qquad h^{n}_{ab}\equiv l^n_{\mu\nu} w^{\mu}_{,a} w^{\nu}_{,b}, \qquad A^{n}_{a}\equiv l^n_{\mu\nu} w^{\mu} w^{\nu}_{,a},
\eeqs
we see that the metric \eqref{genericm} can be written as
\begin{eqnarray}\label{BG1}
ds^2&=& d\rho^2 \biggr [ (1+\sum_{n=1}^{m} \frac{ \sigma^{(n)}}{ \rho^n} )^2+O ( 1 / \rho^{m+1} ) \biggr ] + 2\rho d\rho d\phi^a  \biggl [ \sum_{n=1}^{m} \frac{A^{(n)}}{\rho^n} + O(\rho^{m+1}) \biggr ] \nonumber \\
&& +\rho^2 d\phi^a d\phi^b \biggl [ h^{(0)}_{ab} + \sum_{n=1}^{m} \frac{h^{(n)}_{ab}}{\rho^n }+ O(1/\rho^{m+1}) \biggr ] ,
\end{eqnarray}
where we have set
\beqs\label{311}
\Big( 1 + \sum_{n=1}^{m} \frac{\tilde{\s}^n}{\rho^n} \Big) = \Big( 1 + \sum_{n=1}^{m} \frac{\s^n}{\rho^n}\Big)^2 + O(1/\rho^{m+1}).
\eeqs

\subsection{The Beig-Schmidt ansatz}

As we already said, Beig and Schmidt were motivated by the study of Einstein's equations in a radial expansion. In \cite{BS}, they pointed out that solutions of $\Box \Phi=0$, for a certain field $\Phi$ admitting a radial expansion of the form
\beqs
\Phi=\frac{\Phi^1(\tau, \theta, \phi)}{\rho}+\frac{\Phi^2(\tau, \theta, \phi)}{\rho^2
}+...\: ,
\eeqs
can be obtained, due to the choice of coordinates, from solving a decoupled system of equations for various terms in the expansion
\beqs
h^{(0)\:\:ab} \DD_a \DD_b \Phi^n + n(n-2) \Phi^n=0\; .
\eeqs
This is what motivated their will to write a generic ansatz for the metric as
\begin{eqnarray}\label{wanted}
ds^2=(1+\frac{\tilde{\sigma}^1}{\rho} + \frac{\tilde{\sigma}^2}{\rho^2}+...) d\rho^2 +\rho^2 \Big( h^{(0)}_{ab} +\frac{h^{(1)}_{ab}}{\rho}+... \Big) d\phi^a d\phi^b ,
\end{eqnarray}
where $h^{(0)}_{ab}$ is the metric on the unit hyperboloid $\mathcal{H}$. Let us now see how they obtained such an ansatz for the metric starting from \eqref{BG1} and fixing specific supertranslations and higher order transformations.

\subsubsection{Beig-Schmidt algorithm}

Starting from \eqref{BG1}, there is always a
change of coordinates such that
\begin{eqnarray}
\sigma^{(2)}&=&\sigma^{(3)}=\sigma^{(4)}=...=\sigma^{(m)}=0  ,\nonumber \\
A^{(1)}_a&=&A^{(2)}_a=A^{(3)}_a=...=A^{(m)}_a=0 ,
\end{eqnarray}
which allows to bring the metric into the desired form
\begin{eqnarray}
ds^2=(1+\frac{\sigma}{\rho})^2 d\rho^2 + h_{ab} d\phi^a d\phi^b ,
\end{eqnarray}
where we have set $\sigma^{(1)}=\sigma$ and
\begin{eqnarray}
h_{ab}= \rho^2 h^{(0)}_{ab}+ \rho h^{(1)}_{ab}+ h^{(2)}_{ab}+... + + \frac{1}{\rho^{n}} h^{(n)}_{ab}+ O(1/ \rho^{n+1}) .
\end{eqnarray}
This was proven by the following iterative procedure. The terms $A^{(1)}_a$ and $\s^{(2)}$ can be cancelled by making the following transformations: first act with a supertranslation of the form
\beqs
\phi^a=\bar{\phi}^a +\frac{1}{\rho} G^{(1)\:a}(\bar{\phi}^b) , \qquad G^{(1)\:b}=h^{(0)\:ab} A^{(1)}_a,
\eeqs
 such that the mixed term $d\bar{\phi}^a d\rho (A^1_a -G^{(1)\:b} h^{(0)}_{ab})$ cancels, and then with the higher order transformation
 \beqs
 \rho=\bar{\rho} +\frac{F^{(2)}}{\bar{\rho}}, \qquad F^{(2)}=\s^2 \: ,
 \eeqs
 such that the $1/\bar{\rho}^2$ in $d\bar{\rho}^2$ also cancels. By iteration, a transformation of the form
 \beqs
 \phi^a=\bar{\phi}^a +\frac{1}{\bar{\rho}^{n-1}} G^{(n-1)\:a}, \qquad   \rho=\bar{\rho} +\frac{F^{(n)}}{\bar{\rho}^{n-1}} \: ,
 \eeqs
 removes the terms $\s^{(n)}$ and $A^{(n-1)}_a$. In a similar manner, one removes $A^{(n)}_a$. Following this procedure, one arrives at the requested form \eqref{wanted} that we write as
 \beqs
 ds^2= \Big(1+\frac{\s}{\rho}\Big)^2 d\rho^2 +  h_{ab} d\phi^a d\phi^b, \qquad h_{ab}= \rho^2 h^{(0)}_{ab}+\rho h^{(1)}_{ab}+...\: ,
 \eeqs
where we have used \eqref{311} and then set $\s=\s^1$. This ansatz for the metric is known as the Beig-Schmidt ansatz. 

In this Part I,  we will mainly be concerned with metrics up to second order in $h_{ab}$. To cast a metric into Beig-Schmidt coordinates, we only need to perform the change of coordinates that brings Minkowski metric into the unit hyperboloid and then follow the previous detailed algorithm such that
\begin{eqnarray}
A^{(1)}_a=A^{(2)}_a=0 , \qquad \sigma^{(2)}=0 .
\end{eqnarray}
Starting from the general metric up to second order
\begin{eqnarray}\label{BG2bis}
ds^2&=& d\rho^2\biggl [1+\frac{2\sigma^{(1)}}{\rho} + \frac{(\sigma^{(1)})^2+2\sigma^{(2)}}{\rho^2}+O(1/\rho^3) \biggl ] + 2 \rho d\rho d\phi^a \biggl [ \frac{A^{(1)}_{a}}{\rho} + \frac{A^{(2)}_{a}}{\rho^2} +O(1/\rho^3) \biggr ] \nonumber \\
&& +\rho^2 d\phi^a d\phi^b \biggl [ h^{(0)}_{ab}+ \frac{1}{\rho} h^{(1)}_{ab}+ \frac{1}{\rho^2} h^{(2)}_{ab}+  O(1/\rho^3)\biggr ] ,
\end{eqnarray}
the Beig-Schmidt algorithm brings the metric into the form
\begin{eqnarray}\label{BSS}
ds^2&=& \biggr [ (1+\sigma)^2 +O(1/\rho^3) \biggl ] d\rho^2 + 2 \rho d\rho d\phi^a \biggr [ O(1/\rho^3) \biggl ] \nn\\
&&  +\rho^2 \biggl [ h^{(0)}_{ab}+ \frac{1}{\rho} h^{(1)}_{ab}+ \frac{1}{\rho^2} h^{(2)}_{ab} +O(1/\rho^3)\biggl ] . \nonumber \\
\end{eqnarray}

In Appendix \ref{App: BS}, we show how this procedure is implemented for the Schwarzschild black hole. It has been recently described in \cite{Virmani:2011gh} how to cast the Kerr-NUT black hole into Beig-Schmidt coordinates (see also \cite{Mann:2008ay} for results for the Kerr or the boosted Schwarzschild black holes).

\subsubsection{Gauge freedom and gauge fixing of the Beig-Schmidt ansatz}

Up to second order, apart from Lorentz transformations, we first see that the metric is now invariant under a \textit{subgroup} of the supertranslations \eqref{supp} that we write as
\beqs\label{supertrans}
\rho&=&\bar{\rho} +\omega (\bar{\phi}^a) +\frac{F^{(2)}(\bar{\phi}^a)}{\bar{\rho}}+... \: , \\
\phi^a&=& \bar{\phi}^a +\frac{1}{\bar{\rho}} h^{(0)\:ab} \omega_{,b}+\frac{G^{(2)\:a}}{\bar{\rho}^2}+... \: ,
\eeqs
and where $\omega$ is a function that can be chosen arbitrarily. Let us insist on the fact that it is a subgroup as we have partially fixed the gauge freedom when writing the metric into the Beig-Schmidt form. For reasons that will become clear later, Beig and Schmidt \cite{BS} only considered a restricted set of metrics where we can impose the additional condition
\begin{eqnarray}\label{kab}
k_{ab}\equiv h^{(1)}_{ab}+ 2\sigma h^{(0)}_{ab}=0.
\end{eqnarray}
For this specific class of solutions, which we will specify later, this can always be reached by  a supertranslation of the form \eqref{supertrans}. Indeed, applying the transformation to (\ref{BSS}), we see that it generates no new $A^{(1)}_{a}$ or $\sigma$ term. Also, one can see that under these supertranslations, we have
\begin{eqnarray}
h^{(1)}_{ab} \rightarrow h^{(1)}_{ab}+ 2 \DD_a \DD_b \omega + 2 \omega h^{(0)}_{ab}.
\end{eqnarray}
In the end, we see that $\omega$ must be fixed by imposing that
\begin{eqnarray}
h^{(1)}_{ab}+ 2 D_a D_b \omega + 2 \omega h^{(0)}_{ab}=-2 \sigma h^{(0)}_{ab}.
\end{eqnarray}
When it is possible, we see that, under a specific supertranslation, $h^{(1)}_{ab}$ is completely fixed by $\sigma$ and $h^{(0)}_{ab}$ up to the ambiguity of performing a supertranslation of the form $\DD_a \DD_b \omega + h^{(0)}_{ab} \omega=0$. These last supertranslations can however be recognized as translations. We see that imposing the additional condition \eqref{kab} completely removes the freedom of performing supertranslations, but not translations. Let us mention also that the first order terms in \eqref{supertrans} also bring in contributions to $\sigma^{(2)}$ and $A^{(2)}_{a}$ and thus interfere with the second step of the Beig-Schmidt algorithm. This is the reason we added the higher-order terms ($F^{(2)}$ and $G^{(2)\:a}$) in the supertranslation described here above. The functions $F^{(2)}$ and $G^{(2)}_a$ are precisely there to cancel the terms $\sigma^{(2)}$ and $A^{(2)}$ and are thus functions of $\omega$, $\s$ and their derivatives. As we restrict ourselves to an expansion at second order, we do not need to take care of higher order terms.

We said previously that there also exists logarithmic translations that leave the form of the Beig-Schmidt metric invariant. This is actually not completely true. Indeed, the logarithmic translation can be written in the form
\beqs
\rho&=&\bar{\rho}+ H(\bar{\phi})(\ln \bar{\rho}-1) +o(\bar{\rho^0}), \\
\phi^a&=& \bar{\phi}^a + H^a (\bar{\phi}) (\ln \bar{\rho})/\bar{\rho} +o(\bar{\rho}^{-1}),
\eeqs
where $H^a\equiv \DD^a H$ and $H$ satisfies $\DD_a \DD_b H+ H h^{(0)}_{ab}=0$ (which also implies $(\DD_c\DD^c+3)H=0$). We can check that it leaves the form of the metric invariant up to first order with
\beqs
\s \rightarrow \s +H, \qquad h^{(1)}_{ab} \rightarrow h^{(1)}_{ab}-2 H h^{(0)}_{ab}.
\eeqs
However, at second order, it introduces a logarithmic term in the expansion. To leave the form of the metric invariant, one sees after a quite lengthy computation that we also need to require (see \cite{B})
\beqs\label{conddd}
\DD_c(E^{(1)}_{ab} H^c)=H^c \DD_c E^{(1)}_{ab}-3 H E^{(1)}_{ab}=0,
\eeqs
where $E^{(1)}_{ab}=-\s_{ab} -\s h^{(0)}_{ab}$ and $\s_{ab}\equiv \DD_b \DD_a \s$. This condition was recognized in \cite{B} as the condition for the spacetime to admit an asymptotically translation Killing vector related to H. Indeed, one can check that asymptotically $\mathcal{L}_H g_{ab}=\DD_c(E^{(1)}_{ab} H^c) $.

\subsubsection{Generalized Beig-Schmidt ansatz}

Logarithmic translations are ambiguities in the choice of asymptotically cartesian coordinates and were first discovered by P. Bergmann in \cite{Bergmann:1961zz}. Following the work of R. Beig and B. Schmidt, Ashtekar studied more carefully those transformations in \cite{logambig}. With Ashtekar's definition of asymptotically flat spacetimes in the above described coordinates, which only differs from the Beig-Schmidt definition by the relaxed condition
\beqs\label{reff}
\lim \rho f^{2}_{ab}=0 \: ,
\eeqs
instead of the condition $|f^2_{\mu\nu}|\leq \frac{const}{\rho^2}$, logarithmic translations are now completely allowed transformations. Based on his results pointing out that logarithmic translations do not affect the definition of momenta or Lorentz charges, and are thus pure gauge, he showed that one can fix them appropriately by setting a parity condition on $\s$. Indeed, by imposing the further strict condition
\bea
\sigma(\tau , \theta, \phi ) = \sigma (-\tau , \pi - \theta ,\phi + \pi) \:,
\eea
one removes the freedom of performing logarithmic translations. Indeed, as we will see later in equation \eqref{transareodd}, the functions $H$ which are solutions of $\DD_a \DD_b H + H h^{(0)}_{ab}=0$ are parity odd functions on the hyperboloid. A logarithmic translation that sends $\s \rightarrow \s+H$ is thus not allowed anymore.

Actually, as we will try to explain in detail in the following chapter, we have found it interesting to consider a generalized form of the Beig-Schmidt ansatz that includes a logarithmic term at second order such that logarithmic translations are allowed transformations. This will be motivated by the construction of an enlarged phase space where logarithmic and specific supertranslations are allowed and associated to non-trivial charges. Our generalized class of asymptotically flat spacetimes are spacetimes whose metrics can be brought into the form up to second order\footnote{Let us mention here that the Beig-Schmidt algorithm was designed to deal with a series in powers of $\rho$. This is why we have to keep $\s$. Indeed, by acting with a logarithmic supertranslation (function $H$ defined above but that does not satisfy $H_{ab}+H h^{(0)}_{ab}=0$), one can get rid of $\s$ but at the cost of introducing a logarithmic term between $h^{(0)}_{ab}$ and $h^{(1)}_{ab}$. One could reasonably ask why we do not get rid of $\s$ in here by acting with such a transformation as we do deal with logarithmic terms. We have refrained from doing so as we wanted to compare our results with the existing ones in the litterature. It would however surely be interesting to see how logarithmic supertranslations can be dealt with. We would like to thank Kostas Skenderis from bringing our attention to this point.}
\bea
\label{metricsecondorder}
ds^2 &=& \left( 1 + \frac{2\s}{\rho}+ \frac{\s^2}{\rho^2}  + o(\rho^{-2})\right) d \rho^2 +  o(\rho^{-1}) d\rho dx^a \nn\\
&& + \rho^2\left( h^{(0)}_{ab} + \frac{h^{(1)}_{ab}}{\rho} + \ln\rho \frac{i_{ab}}{\rho^2} + \frac{h^{(2)}_{ab}}{\rho^2} + o(\rho^{-2}) \right) dx^a dx^b .
\eea

Although we have not explicitly checked it, we believe that this metric can be readily obtained from the definition of asymptotically flat spacetimes of Beig and Schmidt, where we allow for the relaxed condition \eqref{reff}, and after proceeding through the Beig-Schmidt algorithm as described above. 

As we want to be as general as possible, we will also not consider that supertranslations are fixed but rather allow for non-trivial values of $k_{ab}$. For reasons that will become clear in the next chapter, we consider that $k_{ab}$ is a symmetric, traceless and divergenceless (SDT) tensor. It thus fulfills
\beqs\label{kSDT}
k_{[ab]}=0\; , \qquad h^{(0)\: ab} k_{ab}=0 \; , \qquad \DD^b k_{ab}=0 \;.
\eeqs
Because $k_{ab}$ transforms under supertranslations as
\beqs
k_{ab} \rightarrow k_{ab} +2 (\omega_{ab} +\omega h^{(0)}_{ab}),
\eeqs
we see that only supertranslations that obey $(\DD_a \DD^a +3)\omega=0$ are allowed. When referring to our enlarged boundary conditions, we will always assume that the metric is written in the form \eqref{metricsecondorder} where $k_{ab}$ is an SDT tensor. The Beig-Schmidt boundary conditions assume moreover that $k_{ab}=i_{ab}=0$.

Let us now move to the study of the equations of motion and the conserved charges that can be defined from symmetric and divergence-free (SD) tensors contracted with some asymptotic symmetry. We will come back, in the next chapter, to the comparison between these charges and the ones obtained from the variational principle.

\setcounter{equation}{0}
\section{The equations of motion}
\label{theeq}

In this section we will see how Einstein's equations can be expanded in powers of $\rho$ for our enlarged boundary conditions \eqref{metricsecondorder}-\eqref{kSDT}.  We first start by splitting them into a set of three equations using a 3+1 split.

\subsection{The 3+1 split}
\label{subsec: gausscod}

The 3+1 split is achieved as soon as the relations between the three and four-dimensional Riemann tensors are established. These relations are known as the Gauss-Codazzi equations. Here, our objective is to expand the Einstein-equations into the (generalized) Beig-Schimdt form, so that the 3+1 split provides, contrarily to the usual ADM formulation, a split between a spatial coordinate and the coordinates on the hyperboloid. Note that this merely changes signs in the relations but not the overall form of the Gauss-Codazzi equations.

Before reviewing these important relations, we need to establish a few definitions and identities involving the extrinsic curvature.

\subsubsection{The extrinsic curvature}

The extrinsic curvature is defined as
\begin{eqnarray}
K_{ab} \equiv h_{a}^{\: \mu} h_{b}^{\: \nu} \nabla_{\mu} n_{\nu},
\end{eqnarray}
where $\nabla_{\mu}$ is the covariant derivative associated to $g_{\mu\nu}$ and the metric $h_{a}^{\:\mu}$  is understood here as a projector that projects directions normal to the hypersurface.

To rewrite $K_{ab}$ differently, let us now define the quantity
 \begin{eqnarray}\label{lieder}
 K_{\mu\nu} \equiv \frac{1}{2} (\mathcal{L}_n h)_{\mu\nu}= \frac{1}{2} (n_{\mu;\nu}+ n_{\nu;\mu} - n_{\mu} a_{\nu} - n_{\nu} a_{\mu}) ,
 \end{eqnarray}
and remind the reader that the Lie derivative of a general tensor $T^{a_1\cdots a_k}_{\qquad b_1 \cdots b_l}$ with respect to a vector field $v^a$ in a given coordinate basis is defined as
\begin{eqnarray}
(\mathcal{L}_v T)^{a_1 \cdots a_k}_{\qquad b_1 \cdots b_l} &\equiv&  v^c \nabla_{c} T^{a_1 \cdots a_k}_{\qquad b_1 \cdots b_l} - \sum_{i=1}^{k} T^{a_1 \cdots c \cdots a_k}_{\qquad \:\:\:\:\:\: b_1 \cdots b_l}\: \nabla_c v^{a_i} \nonumber \\
&& +\sum_{j=1}^{l}  T^{a_1 \cdots a_k}_{\qquad b_1 \cdots c \cdots b_l} \: \nabla_{b_j} v^c.
\end{eqnarray}
Given this, the Lie derivative of the three and four-dimensional metrics along the unit normal $n^{\mu}$ are
\begin{eqnarray}
(\mathcal{L}_n g)_{\mu\nu}&=&n^{\sigma} \nabla_{\sigma} g_{\mu\nu} +g_{\sigma \nu} \nabla_{\mu} n^{\sigma} + g_{\mu \sigma} \nabla_{\nu} n^{\sigma}= \nabla_{\mu} n_{\nu} + \nabla_{\nu} n_{\mu} , \nonumber \\
(\mathcal{L}_n h)_{\mu\nu}&=&(\mathcal{L}_n g)_{\mu\nu} - n_{\mu} (\mathcal{L}_n n)_{\nu}-n_{\nu}  (\mathcal{L}_n n)_{\mu} \nonumber \\
&=& \nabla_{\mu} n_{\nu} + \nabla_{\nu} n_{\mu}  - n_{\mu} a_{\nu} - n_{\nu} a_{\mu},
\end{eqnarray}
where in the last equation we used $h_{\mu\nu}=g_{\mu\nu}-n_{\mu}n_{\nu}$. We also introduced $a^{\mu}$ which is the curvature vector (4-acceleration) of the spacelike normal curves whose tangent field is $n^{\mu}$
\begin{eqnarray}
 a_{\nu}\equiv (\mathcal{L}_n n)_{\nu} = n^{\lambda} \nabla_{\lambda} n_{\nu} + n_{\sigma} \nabla_{\nu} n^{\s}=n^{\lambda} \nabla_{\lambda} n_{\nu} , \qquad a^{\mu}= g^{\mu\nu} a_{\nu} ,
\end{eqnarray}
where it is understood that $0=\nabla_{\nu} (n_{\s} n^{\s})= 2 n_{\sigma} \nabla_{\nu} n^{\sigma}$.
The definition \eqref{lieder} allows us to re-express our extrinsic curvature $K_{ab}$ as
\begin{eqnarray}
K_{ab}= h_{a}^{\: \mu} h_{b}^{\: \nu} K_{\mu\nu} =\frac{1}{2} h_{a}^{\: \mu} h_{b}^{\: \nu} (\mathcal{L}_n h)_{\mu\nu}= \frac{1}{2} (\mathcal{L}_n h)_{ab}=h_{a}^{\: \mu} h_{b}^{\: \nu} \nabla_{\mu} n_{\nu} ,
\end{eqnarray}
where we used $h_{a}^{\:\: \mu} n_{\mu}=0$. To summarize, we have
\begin{eqnarray}\label{defKK}
K_{ab} =h_{a}^{\: \mu} h_{b}^{\: \nu} \nabla_{\mu} n_{\nu}= \frac{1}{2} (\mathcal{L}_n h)_{ab}.
\end{eqnarray}
The trace of the extrinsic curvature is defined as
\begin{eqnarray}
K= h^{ab} K_{ab} .
\end{eqnarray}

For the following, let us now derive some identities involving the extrinsic curvature. By means of \eqref{defKK}, one easily sees that
\begin{eqnarray}\label{Kid}
 h_{a}^{\:\: \mu} h_{c}^{\:\: \sigma} \nabla_{\mu}  h_{\sigma}^{\:\:\lambda}&=& h_{a}^{\:\: \mu} h_{c}^{\:\: \sigma} \nabla_{\mu} (g_{\sigma}^{\:\:\lambda}-n_{\sigma} n^{\lambda})=- h_{a}^{\:\: \mu} h_{c}^{\:\: \sigma} (\nabla_{\mu} n_{\sigma})  n^{\lambda}\equiv -K_{ac}n^{\lambda} , \nonumber \\
 h_{b}^{\:\: \kappa} n^{\lambda}    \nabla_{\kappa}  \omega_{\lambda}&=&-h_{b}^{\:\: \kappa} \omega_{\lambda}   \nabla_{\kappa} n^{\lambda}\equiv-K_{b}^{\:\: \lambda} \omega_{\lambda} ,
 \end{eqnarray}
where $\omega_{\mu}$ is a dual vector field on the timelike hypersurface, so that $\omega_{\mu} n^{\mu}=0$.

The Lie derivative of $K_{ab}$ can be obtained by considering the Lie derivative of $K_{\mu\nu}$. Using (\ref{lieder}), we find
\begin{eqnarray}
\mathcal{L}_n K_{\mu\nu}=\frac{1}{2}\biggl [ \mathcal{L}_n n_{\mu;\nu} + \mathcal{L}_n n_{\nu;\mu}-2 a_{\mu} a_{\nu}-(\mathcal{L}_n a)_{\mu} n_{\nu} -(\mathcal{L}_n a)_{\nu} n_{\mu} \biggr ].
\end{eqnarray}
However, we also have
\begin{eqnarray}
\mathcal{L}_n n_{\mu;\nu} &\equiv& n^{\sigma} \nabla_{\sigma} \nabla_{\nu} n_{\mu} + \nabla_{\nu} n_{\sigma} \nabla_{\mu} n^{\sigma} +\nabla_{\sigma} n_{\mu} \nabla_{\nu} n^{\sigma} \nonumber \\
&=& n^{\sigma} ( \nabla_{\sigma} \nabla_{\nu}-  \nabla_{\nu} \nabla_{\sigma}) n_{\mu} +\nabla_{\nu} a_{\mu} + \nabla_{\nu} n_{\sigma} \nabla_{\mu} n^{\sigma} \nonumber \\
&=&  n^{\sigma} R_{\sigma \nu \mu}^{\:\:\:\:\:\:\:\:\lambda} n_{\lambda}+\nabla_{\nu} a_{\mu} + \nabla_{\nu} n_{\sigma} \nabla_{\mu} n^{\sigma} , \nonumber
\end{eqnarray}
so that
\begin{eqnarray}
\mathcal{L}_n K_{\mu\nu}= - R_{\mu \lambda  \nu  \sigma} n^{\lambda} n^{\sigma} + \nabla_{(\mu} a_{\nu)} +\nabla_{\mu} n_{\sigma} \nabla_{\nu} n^{\sigma}-a_{\mu} a_{\nu}-\frac{1}{2}(\mathcal{L}_n a)_{\mu} n_{\nu} -\frac{1}{2}(\mathcal{L}_n a)_{\nu} n_{\mu} . \nonumber \\
\end{eqnarray}
By projecting with the three-dimensional metric, we obtain
\begin{eqnarray}\label{LieofK}
\mathcal{L}_{n} K_{ab}=h_{a}^{\:\:\mu} h_{b}^{\nu} \mathcal{L}_{n} K_{\mu\nu} =  - h_{a}^{\:\:\mu} h_{b}^{\nu} R_{\mu \lambda  \nu  \sigma} n^{\lambda} n^{\sigma}+D_{(a}a_{b)}+K_{ac}K_{b}^{\:\:c}-a_{a}a_{b}.
\end{eqnarray}
Eventually, we also have
\beqs\label{lnK}
\mathcal{L}_n K \equiv \mathcal{L}_n (h^{ab} K_{ab})&=& h^{ab} \mathcal{L}_n K_{ab} +K_{ab} (n^c \nabla_c h^{ab} -h^{ac} \nabla_c n^b -h^{cb} \nabla_c n^a) \nonumber \\
&=& h^{ab} \mathcal{L}_n K_{ab} - 2 K^{ab}  h_a^{\:\:c} h_{b}^{\:\:d} \nabla_c n_d  \nonumber \\
&=& h^{ab} \mathcal{L}_n K_{ab} - 2 K_{ab} 
K^{ab}\; .
\eeqs

\subsubsection{The Gauss-Codazzi equations}

One way to derive the Gauss-Codazzi equations is to re-express the four dimensional Riemann tensor in terms of three dimensional quantities. In here, we will closely follow Wald's approach \cite{Wald:1984rg} who first defines three-dimensional objects and then connects them to their four dimensional counterparts.

To start with,  the three-dimensional Riemann tensor on the hypersuface is denoted by $\mathcal{R}_{abc}^{\:\:\:\:\:\: d}$ and is defined by
\begin{eqnarray}
\mathcal{R}_{abc}^{\:\:\:\:\:\: d}\:  \omega_{d} \equiv  [D_{a},D_{b}] \omega_{c} =(D_a D_b-D_b D_a) \omega_c,
\end{eqnarray}
where $\omega_{\sigma}$ is a dual vector field defined on the hypersurface.  In the following, $\nabla_{\mu}$ and $D_{\mu}$ are the covariant derivatives associated respectively with $g_{\mu\nu}$ and $h_{\mu\nu}$. \\

To link $\mathcal{R}_{abc}^{\:\:\:\:\:\: d}$ to the Riemann tensor $R_{\mu\nu\sigma}^{\:\:\:\:\:\: \:\:\lambda}$ in four dimensions, we first see that
\begin{eqnarray}
D_{a} D_{b} \omega_c &=& D_{a} (h_{b}^{\:\: \kappa} h_{c}^{\:\:\lambda} \nabla_{\kappa} \omega_{\lambda})= h_{a}^{\:\: \mu} h_{b}^{\:\: \nu} h_{c}^{\:\: \sigma} \nabla_{\mu} (h_{\nu}^{\:\: \kappa} h_{\sigma}^{\:\:\lambda} \nabla_{\kappa} \omega_{\lambda}) \nonumber \\
&=&  \biggl [ h_{a}^{\:\: \mu} h_{b}^{\:\: \kappa} h_{c}^{\:\: \lambda} \nabla_{\mu} \nabla_{\kappa}  +K_{ac} K_{b}^{\:\: \lambda} -K_{ab} h_{c}^{\:\:\lambda} n^{\kappa} \nabla_{\kappa} \biggr ] \omega_{\lambda} ,
\end{eqnarray}
where in the last equality we made use of the identities (\ref{Kid}). This means that
\begin{eqnarray}
\mathcal{R}_{abc}^{\:\:\:\:\:\: d}\:  \omega_{d} = \biggl [ h_{a}^{\:\: \mu} h_{b}^{\:\: \kappa} h_{c}^{\:\: \lambda} (\nabla_{\mu} \nabla_{\kappa}-\nabla_{\kappa} \nabla_{\mu} )  +K_{ac} K_{b}^{\:\: \lambda}-K_{bc} K_{a}^{\:\: \lambda}  \biggr ] \omega_{\lambda} ,
\end{eqnarray}
which immediately gives us the first Gauss-Codazzi equation
\begin{eqnarray}
\mathcal{R}_{abc}^{\:\:\:\:\:\: d}=  h_{a}^{\:\: \mu} h_{b}^{\:\: \nu} h_{c}^{\:\: \sigma}  h^{d}_{\:\:\lambda} R_{\mu\nu\sigma}^{\:\:\:\:\:\:\:\:\lambda}+K_{ac} K_{b}^{\:\:d}-K_{bc} K_{a}^{\:\:d} .
\end{eqnarray}
With this result in hand, it is interesting to look at the relation between the three-dimensional Ricci tensor or scalar and the four dimensional Riemann tensor. We have
\begin{eqnarray}\label{helpforcompuR}
\mathcal{R}_{ab}&\equiv&\mathcal{R}_{acb}^{\:\:\:\:\:\:c}=h_{a}^{\:\: \mu}h_{b}^{\:\:\sigma} h^{\nu\lambda}R_{\mu\nu\sigma\lambda} + K_{ab} K  - K_{bc}K_{a}^{\:\:c} \: ,  \nonumber \\
\mathcal{R} &\equiv& h^{ab}\mathcal{R}_{ab}= h^{\mu\sigma} h^{\nu\lambda} R_{\mu\nu\sigma\lambda}- K_{ab}K^{ab}+K^2 \;.
\end{eqnarray}
One can simplify the expression for the three dimensional Ricci scalar by realizing that
\begin{eqnarray}
R_{\mu\nu\sigma\lambda} h^{\mu\sigma} h^{\nu\lambda}&=&R_{\mu\nu\sigma\lambda} (g^{\mu\sigma}-n^{\mu}n^{\sigma})(g^{\nu\lambda}-n^{\nu} n^{\lambda})
= R-2 R_{\mu\nu} n^{\mu} n^{\nu} \nonumber \\
&=& -2 G_{\mu\nu} n^{\mu} n^{\nu}\: ,
\end{eqnarray}
which also implies
\begin{eqnarray}\label{eq1}
\mathcal{R}=-2 G_{\mu\nu} n^{\mu} n^{\nu}- K_{ab}K^{ab}+K^2 .
\end{eqnarray}
If we look at the expression for the Ricci tensor, we can rewrite it as
\begin{eqnarray}
\mathcal{R}_{ab}&=& h_{a}^{\:\: \mu} h_{b}^{\:\: \sigma} (g^{\nu \lambda}-n^{\nu} n^{\lambda})  R_{\mu\nu\sigma\lambda} + K_{ab} K  - K_{bc}K_{a}^{\:\:c}  \nonumber \\
&=& h_{a}^{\:\: \mu} h_{b}^{\:\: \sigma} R_{\mu\sigma} - h_{a}^{\:\: \mu} h_{b}^{\:\: \sigma} R_{\mu\nu\sigma\lambda}  n^{\nu} n^{\lambda}+ K_{ab} K  - K_{bc}K_{a}^{\:\:c} .
\end{eqnarray}
The second term on the right hand side can be re-expressed using (\ref{LieofK}), and we eventually find
\begin{eqnarray}\label{eq3}
\mathcal{R}_{ab}= h_{a}^{\:\: \mu} h_{b}^{\:\: \nu} R_{\mu\nu} +\mathcal{L}_n K_{ab} -D_{(a} a_{b)} +a_a a_b  + K_{ab} K  - 2 K_{bc}K_{a}^{\:\:c}\; .
\end{eqnarray}

As for the second Gauss-Codazzi equation, one can check that
\begin{eqnarray}\label{eq2}
D_{b} K^{b}_{\:\: a}- D_{a} K^{b}_{\:\:b}=h^{\:\:\mu}_{a} n^{\nu} R_{\mu\nu} = h^{\:\:\mu}_{a} n^{\nu} G_{\mu\nu}.
\end{eqnarray}

\subsubsection{Einstein's equations in the 3+1 split}

As we said, the 3+1 split  sums up to a splitting of Einstein equations. In our case, it is obtained by projecting them either along, or perpendicular to, the hyperboloid of constant $\rho$ using either the projector $h_{\mu\nu}=g_{\mu\nu}-n_{\mu} n_{\nu}$, which is also the metric on the timelike hypersurface, or the outward pointing unit (spacelike) normal $n_{\mu}$(such that $n_{\mu} n^{\mu}=1$). The equations we obtain are a set of equations depending on the lapse $N=1+\sigma^{(1)}$ and the three-dimensional metric $h_{ab}$. They take the form
\begin{eqnarray}\label{seteq}
H&:=& -2 n^{\mu} n^{\nu} G_{\mu\nu}=0 , \nonumber \\
F_a&:=& h_{a}^{\:\: \mu} n^{\nu} G_{\mu\nu}=h_{a}^{\:\: \mu} n^{\nu} R_{\mu\nu}=0 , \nonumber \\
F_{ab}&:= &h_{a}^{\:\: \mu} h_{b}^{\:\: \nu} R_{\mu\nu} =0 .
\end{eqnarray}

Using respectively equations  (\ref{eq1}),(\ref{eq2}) and (\ref{eq3}), we directly obtain
\begin{eqnarray}
H &=& \mathcal{R}-K^2+K_{ab}K^{ab}=0 \: ,\nonumber \\
F_a &=& D_{b} K^{b}_{\:\: a}- D_{a} K^{b}_{\:\:b} =0 \: ,\nonumber \\
F_{ab} &=& \mathcal{R}_{ab} -\mathcal{L}_n K_{ab} +D_{(a} a_{b)} -a_a a_b  - K_{ab} K  + 2 K_{bc}K_{a}^{\:\:c} =0\: .
\end{eqnarray}
The first equation can be simplified by taking the trace of the third equation
\beqs
h^{ab} F_{ab}=0 \rightarrow \mathcal{R}= h^{ab} \mathcal{L}_n K_{ab} - D_{a} a^{a} +a_a a^a  + K^2  - 2 K_{ab}K^{ab},
\eeqs
which implies
\beqs\label{Hamilton}
H &=& -h^{ab} \mathcal{L}_n K_{ab} + D_{a} a^{a} -a_a a^a +  K_{ab}K^{ab} \nonumber \\
&=& - \mathcal{L}_n K + D_{a} a^{a} -a_a a^a - K_{ab}K^{ab} \: ,
\eeqs
where in the last equation we made use of \eqref{lnK}.

In analogy with the Arnowitt-Deser-Misner formalism, we will refer to the equation \eqref{Hamilton} as the Hamiltonian equation, and to the second and third equations of \eqref{seteq} as the momentum equation and the equation of motion.

\subsubsection{A simplified form of the equations}

For our ansatz, these equations can be even more simplified. Indeed, in our case, the shift is zero and the lapse $N$ is defined by
\beqs
N \equiv 1+\frac{\s}{\rho},
\eeqs
so that
\begin{eqnarray}
n^{\mu}=(1/N) \delta^{\mu}_{\rho}, \qquad g_{\mu\nu} n^{\mu} n^{\nu}=1 .
\end{eqnarray}

Now, using $D_{a}\equiv h_{a}^{\:\:\mu} \nabla_{\mu}$ the covariant derivative associated to $h_{ab}$ and the projector $h_{a}^{\mu}=g_{a}^{\mu}-n_{a} n^{\mu}=g_{a}^{\mu}$ because $n_a=0$, we rapidly obtain
\begin{eqnarray}
K_{ab}&=&\frac{1}{2} (\mathcal{L}_{n} h)_{ab}=\frac{1}{2} h_{a}^{\: \mu} h_{b}^{\: \nu}  (\mathcal{L}_{n} h)_{\mu\nu} = \frac{1}{2} h_{a}^{\: \mu} h_{b}^{\: \nu}  ( n^{\sigma} \nabla_{\sigma} h_{\mu\nu} + h_{\mu \sigma} \nabla_{\nu} n^{\sigma} +h_{\sigma \nu} \nabla_{\mu} n^{\sigma}) \nonumber \\
            &=&\frac{1}{2}  ( n^{\sigma} \nabla_{\sigma} h_{ab} + h_{a \sigma} D_{b} n^{\sigma} +h_{\sigma b} D_{a} n^{\sigma}) \nonumber \\
            &=& \frac{1}{2}   n^{\rho} \partial_{\rho} h_{ab} ,
\end{eqnarray}
where in the last line we used $h_{a\sigma} D_{b} n^{\sigma}=h_{a \rho} \partial_{b} n^{\rho} -h_{a c}  n^{d} \Gamma^{c}_{\:\: d b}=0 $ and
$n^{\sigma} \nabla_{\sigma} h_{ab} =n^{\rho} \partial_{\rho} h_{ab}$. We also have
\beqs
\mathcal{L}_n K_{ab} = N^{-1} \partial_{\rho} K_{ab} .
\eeqs
The 4-acceleration also simplifies as
\begin{eqnarray}
a^{\mu}=n^{\nu} \nabla_{\nu} n^{\mu} =n^{\nu} \biggr [  \partial_{\nu} n^{\mu} +  n^{\sigma} g^{\mu\kappa} (g_{\kappa \sigma,\nu} -\frac{1}{2} g_{\sigma \nu,\kappa}) \biggl ] ,
\end{eqnarray}
so we easily see that
\begin{eqnarray}
a^{a}=-\frac{1}{N} D^{a} N , \qquad a_{a}=-\frac{1}{N} D_{a} N .
\end{eqnarray}
We eventually have
\begin{eqnarray}
D_a a_{b}-a_a a_b=-\frac{1}{N} D_{a}D_{b}N .
\end{eqnarray}

Given all this, the equations take the much simpler form \cite{BS}
\beqs
H &\equiv&  -\mathcal{L}_{n} K - K_{ab}K^{ab}-N^{-1} h^{ab} D_a D_b N=0  \: , \nonumber \\
F_{a} &\equiv& D_{b} K^{b}_{\:\:a}-D_a K=0  \: , \nonumber \\
F_{ab} &\equiv &   \mathcal{R}_{ab}- N^{-1} \partial_{\rho} K_{ab}-N^{-1} D_a D_b N-K K_{ab}+ 2 K_{a}^{\:\:c} K_{cb}=0  \: .
\eeqs

\subsection{Radial expansion}
\label{subsec: ansatz}

We now expand these equations using our metric ansatz \eqref{metricsecondorder}. The results we obtain reduce at zeroth and first order in the expansion when $i_{ab}=0$ to the results presented in \cite{BS} and, at second order when $k_{ab}=i_{ab}=0$ to the results presented in \cite{B}. In several places, we simplify the computations by setting the trace and the divergence of $k_{ab}$ to zero. As already discussed these are additional boundary conditions that we will justify in the next chapter. 

The inverse metric is expanded as
\begin{eqnarray*}
h^{ab}(\rho,x^c)&=& \rho^{-2} h^{(0) ab}-\rho^{-3} h^{(1) ab}-\ln\rho\, \rho^{-4} i^{ab}-\rho^{-4} (h^{(2) ab}-h^{(1)a}_{\quad \: c} h^{(1)cb})+O(\rho^{-5}) \; . \nonumber
\end{eqnarray*}
The extrinsic curvature admits the simple expansion
\bea
K_{ab} = \rho \, h^{(0)}_{ab} + \left( \frac{1}{2} h^{(1)}_{ab} -\s h^{(0)}_{ab} \right) + \frac{1}{\rho} \left( \frac{1}{2}i_{ab} - \frac{1}{2} h^{(1)}_{ab} \s + \s^2 h^{(0)}_{ab}   \right)+O(\rho^{-2}) \, ,
\eea
and we also have
\bea
K^a_{\; b} &=& \frac{1}{\rho} \delta^a_b - \frac{1}{2\rho^2} k^a_{\; b}-\frac{\ln\rho}{\rho^3} i^a_{\; b} \nonumber \\
&&+\frac{1}{\rho^3} \left( -h_{(2)\; b}^a+\frac{1}{2}i^a_{\; b} +2\s^2 \delta^a_b + \frac{1}{2} k^a_{\; c}k^c_{\; b} - \frac{3}{2} \sigma k^a_{\; b}  \right) +O(\rho^{-4}) \; .
\eea
The covariant derivative requires an expansion of the Christoffel symbols
\begin{eqnarray}
\Gamma^a_{bc} = \Gamma^{(0)\:a}_{\quad\:\: bc}+ \rho^{-1} \Gamma^{(1)\:a}_{\quad\:\: bc}+\ln\rho\,  \rho^{-2}\Gamma^{(ln,2)\:a}_{\quad\:\: bc}+ \rho^{-2}\Gamma^{(2)\:a}_{\quad\:\: bc}+ O(\rho^{-3})  \: ,
\end{eqnarray}
where
\bea
\Gamma^{(1)\:a}_{\quad\:\: bc} &=& \frac{1}{2}\left( \DD_c h^{(1)a}_{\;\; b} +  \DD_b h^{(1)a}_{\;\; c} -  \DD^a  h^{(1)}_{bc}\right)  \: , \nn\\
\Gamma^{(ln,2)\:a}_{\quad\:\: bc} &=& \frac{1}{2}\left( \DD_c i^{a}_{\;\; b} +  \DD_b i^{a}_{\;\; c} -  \DD^a  i_{bc}\right)\, ,\\
\Gamma^{(2)\:a}_{\quad\:\: bc} &=& \frac{1}{2}\left( \DD_c h^{(2)a}_{\;\; b} +  \DD_b h^{(2)a}_{\;\; c} -  \DD^a  h^{(2)}_{bc}\right) - \frac{1}{2}h^{(1) ad}
\left( \DD_c h^{(1)}_{db} +  \DD_b h^{(1)}_{dc} -  \DD_d  h^{(1)}_{bc}\right)   \: .\nn
\eea

The expansion of the three-dimensional Ricci curvature tensor is
\begin{eqnarray}\label{riccitensor}
\mathcal{R}_{ab}= \mathcal{R}^{(0)}_{ab}+ \rho^{-1}\mathcal{R}^{(1)}_{ab}+\ln\rho  \rho^{-2}\mathcal{R}^{(ln,2)}_{ab}+ \rho^{-2}\mathcal{R}^{(2)}_{ab}  +O(\rho^{-3})  \: .
\end{eqnarray}
The zeroth order Ricci tensor is the one constructed with the metric $h^{(0)}_{ab}$. The first order Ricci tensor and the tensor $\mathcal{R}^{(ln,2)}_{\:\:\:\:ab}$ are
\begin{eqnarray*}
\mathcal{R}^{(1)}_{\:\:\:\:ab}&=& \DD_c \biggr [ \Gamma^{(1)\:c}_{\quad \:\:ab} \biggl ]-\DD_b \biggr [ \Gamma^{(1)\:c}_{\quad \:\:ac} \biggl ]
= \frac{1}{2} \biggr [ \DD^{c} \DD_{b} h^{(1)}_{ac}+ \DD^{c} \DD_{a} h^{(1)}_{bc} -\DD_{a} \DD_{b} h^{(1)} - \DD^{c} \DD_{c} h^{(1)}_{ab} \biggl ]   \: ,\\
\mathcal{R}^{(ln,2)}_{\:\:\:\:ab} &=&\frac{1}{2} \biggr [ \DD^{c} \DD_{b} i_{ac}+ \DD^{c} \DD_{a} i_{bc} -\DD_{a} \DD_{b} i - \DD^{c} \DD_{c} i_{ab} \biggr ] ,
\end{eqnarray*}
and the second order Ricci tensor reads as
\begin{eqnarray*}
\mathcal{R}^{(2)}_{\:\: ab} &=& \frac{1}{2} \biggr [ \DD^{c} \DD_{b} h^{(2)}_{ac}+ \DD^{c} \DD_{a} h^{(2)}_{bc} -\DD_{a} \DD_{b} h^{(2)} - \DD^{c} \DD_{c} h^{(2)}_{ab} \biggl ] +\frac{1}{2} \DD_b \biggr [ h^{(1)cd} \DD_a h^{(1)}_{cd} \biggl ]\nonumber \\
&& -\frac{1}{2} \DD_d \biggr [ h^{(1)cd} (\DD_a h^{(1)}_{bc}+\DD_b h^{(1)}_{ac}-\DD_c h^{(1)}_{ab} ) \biggl ]  +\frac{1}{4} \DD^{c} h^{(1)} \biggr [ \DD_a h^{(1)}_{bc}+\DD_b h^{(1)}_{ac}-\DD_c h^{(1)}_{ab}\biggl ]\nonumber \\
&&-\frac{1}{4} \DD_{a} h^{(1)}_{cd} \DD_{b} h^{(1)cd}  +\frac{1}{2} \DD_{c} h^{(1)}_{ad} \DD^{c} h^{(1)d}_{\:\:\:\:\:\: \:\:b} -\frac{1}{2} \DD_{c} h^{(1)}_{ad} \DD^{d} h^{(1)c}_{\:\:\:\:\:\: \:\:b}\, .
\end{eqnarray*}
Finally, the equations can be expanded as
\begin{eqnarray}
H &=& \rho^{-3} H^{(1)} +\ln\rho\, \rho^{-4} H^{(ln,2)} +\rho^{-4} H^{(2)} +O(\rho^{-5})  \: , \label{equation1bis} \nonumber \\
F_a &=& \rho^{-2} F^{(1)}_a+\ln \rho\, \rho^{-3} F^{(ln,2)}_a+\rho^{-3} F^{(2)}_a+O(\rho^{-4})   \: , \label{equation2}  \label{equation3}\\
F_{ab}&=& F^{(0)}_{ab} + \rho^{-1} F^{(1)}_{ab}+ \ln\rho \, \rho^{-2} F^{(ln,2)}_{ab}+\rho^{-2} F^{(2)}_{ab} +O(\rho^{-3})  \: .\nonumber
\end{eqnarray}

\subsubsection{At zeroth order}

At zeroth order, the Hamiltonian and momentum equations are trivial. We are left with the equation of motion
\beqs\label{zeroeq}
F^{(0)}_{ab}=\mathcal{R}^{(0)}_{ab}-2 h^{(0)}_{ab}=0 ,
\eeqs
 which implies that the boundary metric is the three-dimensional de Sitter spacetime
 \beqs
 ds^2_{\mathcal{H}}=h^{(0)}_{ab} dx^a dx^b= -d\tau^2 +\cosh^2 \tau (d\theta^2+\sin^2 \theta d\phi^2)\; .
 \eeqs
 The metric $h^{(0)}_{ab}$ is the unit hyperboloid metric on $\mathcal{H}$.

Indeed, if $h^{(0)}_{ab}$ is an unspecified Lorentz metric on the manifold $S^2\times R$, the equation  \eqref{zeroeq} and the vanishing of the Weyl tensor in three dimensions imply that
\beqs
\mathcal{R}^{(0)}_{abcd}=h^{(0)}_{ac}h^{(0)}_{bd}-h^{(0)}_{bc}h^{(0)}_{ad}.
\eeqs

\subsubsection{At first order}

At first order, the Hamiltonian equation $H^{(1)}=0$ is simply
\beqs\label{divs1}
(\Box+3)\s=0 .
\eeqs
The momentum equation $F^{(1)}_a = 0$ is
\beqs\label{divk1}
\DD^b k_{ab}=\DD_a k,
\eeqs
and the radial equation of motion $F^{(1)}_{ab}$ is
\bea\label{dive1}
(\Box - 3 )k_{ab} = \DD_a \DD_b k - k h^{(0)}_{ab}\: ,
\eea
which can also be written as
\beqs\label{momentumdef}
\Box k_{ab}-\DD^c \DD_a k_{bc}=0 \: .
\eeqs
If we further impose that $k_{ab}$ is a traceless and divergence-free tensor, the momentum equation is  trivial and the first order equations of motions are summarized as
\bea\label{eqfirstorder}\fbox{
$(\Box+3)\s=0 \; , \qquad (\Box - 3 )k_{ab} = 0\; ,$}
\eea

\subsubsection{At second order}

At second order we easily get for the logarithmic terms $ H^{(ln,2)} = 0$, $F^{(ln,2)}_a = 0$ and $ F^{(ln,2)}_{ab} = 0$
\begin{eqnarray}\label{eqh2a}
\fbox{
   $i = 0,\qquad \DD^b i_{ab}=0, \qquad (\Box-2) i_{ab}=0\; ,$
}
\end{eqnarray}
For the finite terms at second order we find
\begin{eqnarray}
H^{(2)}&=& -h^{(2)}+\frac{3}{2} i +\frac{1}{4}h^{(1)ab} h^{(1)}_{ab}+  \frac{1}{2} \sigma h^{(1)} \nonumber \\
&& +9 \sigma^{2}+\sigma \DD^2 \sigma+ h^{(1)ab} \DD_a \DD_b \sigma +  \DD^{b} \sigma \DD^{a} h^{(1)}_{ab}  -\frac{1}{2} \DD_a \sigma \DD^{a} h^{(1)} \: .
\end{eqnarray}
Using only $\sigma$ and $k_{ab} = h_{ab}^{(1)} + 2\s h_{ab}^{(0)}$, and also $k=i=0$, we obtain
\begin{eqnarray}\label{eqh2b}
\fbox{
   $ h^{(2)}=12 \sigma^2 + \sigma_c \sigma^c+\frac{1}{4} k_{cd} k^{cd}  + k_{cd} \sigma^{cd}\; ,$
}
\end{eqnarray}
where we also made use of the first order equations of motion. We also have
\be
F^{(2)}_a \equiv \DD_b K^{(2)b}_a - \DD_a K^{(2)} + \Gamma_{bc}^{(1)b}K^{(1)c}_{\; a} - \Gamma_{ab}^{(1)c}K^{(1)b}_{c}=0 \: ,
\ee
which amounts, after simplifications, to
\begin{eqnarray}\label{eqh2c}
\fbox{  $\DD^b h^{(2)}_{ab}= \frac{1}{2} \DD^b k_{ac} k_{b}^{\:\: c}  + \DD_a \left( \sigma_c \sigma^c +8 \sigma^2 -\frac{1}{8} k_{cd} k^{cd}  +k_{cd} \sigma^{cd} \right)\; , $ }
\end{eqnarray}
The radial equation of motion can be obtained after a straightforward computation and we find the quite intricate form
\beqs\label{eqh2d}
&\fbox{ $(\Box - 2) h^{(2)}_{ab} = 2 i_{ab} + \text{NL}_{ab}(\s,\s) + \text{NL}_{ab}(\s ,k)+\text{NL}_{ab}(k,k)$}
\eeqs
where the non-linear terms are given by
\beqs
\label{equation3bis}
\text{NL}_{ab}(\s ,\s )  &=& \DD_a \DD_b \left( 5 \sigma^2+\s_c \s^c  \right) + h_{ab}^{(0)} \left( -18 \s^2+4\s^c \s_c  \right) + 4 \s \s_{ab} \: , \nn \\
\text{NL}_{ab}(\s ,k)  &=& \DD_a \DD_b \left( k_{cd} \s^{cd}\right) -2k_{cd}\s^{cd}  h_{ab}^{(0)} + 4\s k_{ab}+4 \s^c (D_{(a}k_{b)c} - D_c k_{ab}) +4\s_{c(a} k^{c}_{\;\, b)} \: , \nn \\
\text{NL}_{ab}(k,k)  &=& k_{ac}k^c_{\; \, b} +k^{cd}( -\DD_{d}\DD_{(a}k_{b)c} +\DD_c \DD_d k_{ab}) \nonumber \\
&&  -\frac{1}{2}\DD_b k^{cd}\DD_a k_{cd}+\DD^d k_{c(a}\DD_{b)} k_{d}^{\; c}+\DD_c k_{ad}\DD^c k^d_{\; b}-\DD_c k_{ad}\DD^d k^{c}_{\; b}\, .\label{NLterms}
\eeqs
Using the relation $\s^c \s_{abc} = \s^c \s_{cab} + \s_a \s_b - h_{ab}^{(0)}\s_c \s^c$ (see also Appendix \ref{app:properties}), one can rewrite the $\text{NL}_{ab}(\s ,\s )$ non-linear terms as
\be
\text{NL}_{ab}(\s ,\s ) = 6 \s_c \s^c h_{ab}^{(0)} + 8\s_a \s_b +14 \s \s_{ab} -18 \s^2 h_{ab}^{(0)}+2\s_{ac} \s^{c}_{\; b}+2\s_{abc}\s^c \, .
\ee
The equations of motion reproduce the expressions of \cite{B} when $k_{ab}= i_{ab} = 0$.

\setcounter{equation}{0}
\section{Compact equations and their solutions}
\label{sec:nice}

In this section, we would like to put the previously derived equations into more compact forms that would allow us to study them efficiently. To achieve this, we first review the definition of the Weyl tensor and its decomposition into electric and magnetic parts. We then move to the classification of symmetric and divergence-free tensors (SD tensors) that can be built out of quadratic quantities in the first order fields $\s$ and $k_{ab}$ and their derivatives. Here, the reader should remember that we assume that $k_{ab}$ is a symmetric, traceless and divergence-free (SDT) tensor.

\subsection{The electric and magnetic parts of the Weyl tensor}
\label{subsec:weyl}

The Weyl tensor is the trace-free part of the Riemann tensor and it is defined by
\begin{eqnarray}
C_{\mu\nu\rho\sigma}=R_{\mu\nu\rho\sigma} - (g_{\mu [\rho} R_{\sigma] \nu}-g_{\nu [\rho} R_{\sigma] \mu}) + \frac{1}{3} R g_{\mu [\rho} g_{\sigma] \nu}.
\end{eqnarray}

It can be decomposed into its electric and magnetic parts, respectively denoted $E_{ab}$ and $B_{ab}$ which are defined as follows
\begin{eqnarray}\label{emweyl}
E_{ab}\equiv h_{a}^{\:\: \mu} h_{b}^{\:\:\nu} C_{\mu\lambda\nu\sigma} \: n^{\lambda} \: n^{\sigma} , \qquad
B_{ab}\equiv  \frac{1}{2}\epsilon^{cd}{}_{ae}C_{cdbf}n^e n^f  .
\end{eqnarray}

\subsubsection{Electric part of the Weyl tensor}

Starting from its definition \eqref{emweyl}, the electric part of the Weyl tensor can also be written as
\begin{eqnarray}
E_{ab}&=& h_{a}^{\:\: \mu} h_{b}^{\:\:\nu} C_{\mu\lambda\nu\sigma} \: n^{\lambda} \: n^{\sigma} \nonumber \\
             &=& h_{a}^{\:\: \mu} h_{b}^{\:\:\nu} R_{\mu\lambda\nu\sigma} \: n^{\lambda} \: n^{\sigma} - \frac{1}{2} h_{ab} R_{\sigma \lambda} n^{\lambda} n^{\sigma} -\frac{1}{2}  h_{a}^{\:\: \mu} h_{b}^{\:\:\nu} R_{\mu\nu} +\frac{1}{6} R h_{ab} ,
\end{eqnarray}
where in the second line we used the fact that
$g_{\mu\nu}=h_{\mu\nu} +n_{\mu} n_{\nu}$ and $h_{a}^{\:\: \mu} n_{\mu}=0$. This implies
$h_{a}^{\:\: \mu} n^{\mu} g_{\mu\nu}=0$, $h_{a}^{\:\: \mu} h_{b}^{\:\: \nu} g_{\mu\nu}=h_{ab}$ and also $n^{\mu} n^{\nu} g_{\mu\nu}=1$.
In the following, we will only deal with the on-shell Weyl tensor. Upon setting $R_{\mu\nu}=0$ and using (\ref{LieofK}), we have
\beqs
E_{ab}&=& h_{a}^{\:\: \mu} h_{b}^{\:\:\nu} R_{\mu\lambda\nu\sigma} \: n^{\lambda} \: n^{\sigma} = -\mathcal{L}_{n} K_{ab}+D_{(a}a_{b)}+K_{ac}K_{b}^{\:\:c}-a_{a}a_{b} \nonumber \\
&=& -\frac{1}{N}(\partial_\rho K_{ab}+D_a D_b N)+K^c_{\; a} K_{cb}.
\eeqs
Its asymptotic expansion is given by
\bea
E_{ab} =  \frac{1}{\rho} E_{ab}^{(1)}+ \frac{\ln\rho}{\rho^2} E_{ab}^{(ln,2)}  + \frac{1}{\rho^2} E_{ab}^{(2)} + O(\rho^{-3}) .
\eea
By computing the following quantities
\beqs
\mathcal{L}_n K_{ab}&=& \frac{1}{N} \partial_{\rho} K_{ab}= h^{(0)}_{ab}-\rho^{-1} \sigma h^{(0)}_{ab}+\rho^{-2} \frac{1}{2} ( \sigma k_{ab}-2 \s^2 h^{(0)}_{ab}-i_{ab})+O(\rho^{-3}) , \nonumber \\
 K_{ac}K^{c}_{\:\:b}&=&  h^{(0)}_{ab}+\rho^{-1} \biggr [ -2\sigma h^{(0)}_{ab}\biggl ]- \rho^{-2} \ln \rho \: i_{ab} \nonumber \\
&& +\rho^{-2} \biggr [ -h^{(2)}_{ab}  + i_{ab} +4 \s^2 h^{(0)}_{ab}+\frac{1}{4} k_{ac} k_b^{\:\:c}- \sigma k_{ab}\biggl ] +O(\rho^{-3}) , \nonumber \\
D_a a_{b}-a_a a_b &=& -\frac{1}{N} D_{a} D_{b}N \nonumber \\
                                   &=&  -\rho^{-1}  \sigma_{ab} + \rho^{-2} \biggr [ \s \s_{ab} -2 \s_a \s_b +\s_c \s^c h^{(0)}_{ab} +\s^c \DD_{(a} k_{b)c}-\frac{1}{2} \s^c \DD_c k_{ab}\biggl ]  \nonumber \\
                   &&+O(\rho^{-3}) ,
\eeqs
we find
\beqs
E^{(1)}_{ab} &=&   -\s_{ab} - \s h^{(0)}_{ab} , \label{E1}\\
E^{(ln,2)}_{ab} &=& - i_{ab} , \\
E^{(2)}_{ab} & =& - h^{(2)}_{ab} +\frac{1}{2}i_{ab}+ 5 \sigma^2 h_{ab}^{(0)}+\sigma_{ab} \sigma - 2\sigma_a \sigma_b +\s_c \s^c h_{ab}^{(0)} \nonumber \\
&&+ \frac{1}{4}k_{ac} k^{c}_{\; b} - \frac{3}{2} \sigma k_{ab} + \DD_{(a} k_{b)c} \s^c - \frac{1}{2} \DD_c k_{ab} \s^c \, .
\eeqs

\subsubsection{Magnetic Part of the Weyl Tensor}
The magnetic part of the Weyl tensor is defined as
\be
B_{ab} = \frac{1}{2}\epsilon^{mn}{}_{ae}C_{mnbf}n^e n^f  = - \underline{\epsilon}^{mn}{}_{a}D_{m} K_{nb}.
\ee
where $\underline{\eps}_{mna} \equiv \eps_{mnab}n^b$ admits the expansion, upon setting $k=0$,
\bea
\underline{\eps}_{mna} &=& \rho^3 \eps_{mna} \left(1-\frac{3\sigma}{\rho}+o(\rho^{-1})  \right),\\
\underline{\eps}^{mn}{}_a &=& \rho^{-1} \eps^{cd}{}_a \left( \delta^m_c \delta^n_d (1+\frac{\sigma}{\rho}) -\frac{1}{\rho} ( k^m_c \delta^n_d + k^n_d \delta^m_c   ) +o(\rho^{-1}) \right).
\eea
Also, we have
\bea
D_{m} K_{nb} &=&  \DD_m K^{(1)}_{nb} - \Gamma^{(1)}_{mn}{}^e h^{(0)}_{eb} - \Gamma^{(1)}_{mb}{}^e h^{(0)}_{ne}
+ \frac{\ln\rho}{\rho} \left(  - \Gamma^{(ln,2)}_{mn}{}^e h^{(0)}_{eb}- \Gamma^{(ln,2)}_{mb}{}^e h^{(0)}_{ne}  \right) \nn \\
& & +  \frac{1}{\rho} \left( \DD_m K^{(2)}_{nb} - \Gamma^{(1)}_{mn}{}^e K^{(1)}_{eb} - \Gamma^{(1)}_{mb}{}^e K^{(1)}_{ne} - \Gamma^{(2)}_{mn}{}^e h^{(0)}_{eb}- \Gamma^{(2)}_{mb}{}^e h^{(0)}_{ne} \right) \\
&& + o \left(\rho^{-1}\right),\\
&=&  -\frac{1}{2}  \DD_m k_{nb} \nn + \frac{\ln\rho}{\rho}\left( - \DD_m i_{bn} \right) + \frac{1}{\rho} \Bigg{(} -\DD_m h^{(2)}_{nb} +\frac{1}{2}\DD_m i_{nb} + 2 \s \s_m h^{(0)}_{nb}  \\
&& + \frac{1}{2} \s \DD_m h^{(1)}_{nb} - \frac{1}{2} \s_m h^{(1)}_{nb} + \frac{1}{4} \DD_m \left(h^{(1)}_{b}{}^{f} h^{(1)}_{nf}\right)    + \frac{1}{2} h^{(1)f}{}_{(b}\DD_{n)}h^{(1)}_{mf}  \nn \\ && -\frac{1}{2} h^{(1)f}{}_{(b}\DD_{|f|}h^{(1)}_{n)m} \Bigg{)} +o\left(\rho^{-1}\right),
\eea
where we used
\bea
K^{(1)}_{ab} = \frac{1}{2}k_{ab}-2\sigma h_{ab}^{(0)},\qquad K^{(2)}_{ab} = \frac{1}{2}i_{ab} -\frac{1}{2} \s k_{ab} +2\s^2 h_{ab}^{(0)}\, .
\eea
Finally, we obtain
\bea
B_{ab} = \frac{1}{\rho} B_{ab}^{(1)} + \frac{\ln\rho}{\rho^2} B_{ab}^{(ln,2)}+ \frac{1}{\rho^2} B_{ab}^{(2)}+O(\rho^{-3}) \: ,
\eea
where
\bea
B^{(1)}_{ab} &=& \frac{1}{2}\eps_a^{\; cd}\DD_c k_{db}\: , \label{B1}\\
B_{ab}^{(ln, 2)} &=& \eps_a^{\; cd} \DD_c   i_{db} \: , \\
B_{ab}^{(2)} &= & \eps_a^{\; cd} \biggr [ \DD_c \left(  h^{(2)}_{db} -\frac{1}{2} i_{db} -2 \s^2 h^{(0)}_{db} +\sigma k_{db} -\frac{1}{4} k_{de} k_{b}^{\;e}\right) \nonumber \\
&& \qquad  -\frac{1}{2}  k_{de} \sigma^e h^{(0)}_{bc} +k_{d}^{\;e} (\DD_{[e}k_{c]b} +\frac{1}{2} \DD_{[e}k_{b]c}) -\frac{1}{4} k_{b}^{\:e} \DD_{[d} k_{c]e} \biggl ] \nonumber \\
&=& \eps_a^{\; cd} \biggr [ \DD_c \left(  h^{(2)}_{db} -\frac{1}{2} i_{db} -2 \s^2 h^{(0)}_{db} +\sigma k_{db} -\frac{1}{4} k_{de} k_{b}^{\;e}\right)  -\frac{1}{2}  k_{de} \sigma^e h^{(0)}_{bc} \biggl ] \nonumber \\
&& + \frac{3}{2} k_{a}^{\;c} B^{(1)}_{bc} +k_{b}^{\:c} B^{(1)}_{ac}-\frac{1}{2} k^{cd} B^{(1)}_{cd} h^{(0)}_{ab}\: ,
\eea
where in the last equation we used the definition of $B^{(1)}_{ab}$ to write
\bea
\DD_{[e}k_{c]b}=-\eps^{f}_{\;ec} B^{(1)}_{fb} \; .
\eea

Just remark that assuming $k_{ab} =i_{ab} =0$, the expansion simplifies to
\bea\label{B2b}
B_{ab}&=& \frac{1}{\rho^2} \epsilon^{mn}{}_{a} \left(  D_{m} h^{(2)}_{nb} - 4 \s \s_{m} h^{(0)}_{nb} \right) + o \left( \frac{1}{\rho^2}\right), \label{B2simp}
\eea
which agrees with the equation (C.7) of \cite{Mann:2006bd}.

\subsection{Classification of symmetric and divergence-free tensors}
\label{appSD}
\label{sec:SD}

In this section, we prove that all  symmetric and divergence-free tensors (SD tensors) built out of quadratic terms in the first order fields $\s$ and $k_{ab}$ can be formed from symmetric tensors $M_{ab}$ obeying $\DD^b M_{ab} = \DD_a M$ and which we call tensor potentials. A complete set of SD tensors consists of SD tensors given by $\kappa_{ab}\equiv M_{ab}-M h^{(0)}_{ab}$ and of symmetric, traceless and divergence-free tensors (SDT tensors) obtained by acting with successive curls on $M_{ab}$ or equivalently by acting with successive symmetrized curls on $\kappa_{ab}$. Indeed, an SDT tensor can be constructed from $T_{ab} = \eps_a^{\;\, cd}\DD_c M_{d b }= \eps_{cd(a}\DD^c \kappa^{\:\:\: d}_{b)}$. As the curl of a tensor potential might be trivially zero, the SD and SDT tensors, whose curls are non-zero, can be classified using the equivalence of classes of tensor potentials where two tensor potentials are equivalent if their difference has a trivial curl. We will refer to one representative of such equivalence class of non-trivial tensor potentials as a RNT tensor potential.

Because the first order fields $\sigma$ and $k_{ab}$ obey decoupled linear equations as it is obvious from \eqref{eqfirstorder}, we can consider separately the quadratic combinations $(\sigma,\sigma)$, $(k,k)$ and $(\sigma,k)$. What we show in the following is that any SD or SDT tensor whose curl is non-zero, let us denote it $X_{ab}$, can be written up to the addition of SD tensors with trivial curls as
\beqs
X_{ab}=\sum_{i} a^{(0)}_{\:\:\:(i)} \kappa^{(i)}_{ab} + a^{(1)}_{\:\:\:(i)} \curl(\kappa^{(i)})_{(ab)} + a^{(2)}_{\:\:\:(i)} \curl^2 (\kappa^{(i)})_{ab} + ... +  a^{(j)}_{\:\:\:(i)} \curl^j (\kappa^{(i)})_{ab} , \nonumber \\
\eeqs
where $a^{(j)}_{\:\:\:(i)}$ are arbitrary real coefficients and $\kappa^{(i)}_{ab}$ is an SD tensor given by one of the following SD tensors
 \beqs
\kappa^{[\s,\s,I]}_{ab} &=& (2 \s^2 -2\s_c \s^c ) h_{ab}^{(0)} + 4\s \s_{ab} \label{k1} ,\\
\kappa_{ab}^{[\s,k, I]} &=& \s k_{ab} +\s_{c(a} k_{b)}^{\:\:\:c}  -\frac{1}{2} \s_{cd} k^{cd} h^{(0)}_{ab} -\s^c \DD_c k_{ab} + \s^c \DD_{(a} k_{b)c} \: , \label{k3}\\
\kappa_{ab}^{[\s,k,II]} &=& Z^{(3)}_{ab} + 2 B^{(1)}_{c(a} \s_{b)}^{\:\:\:c} -\s^{cd} B^{(1)}_{cd} \: h^{(0)}_{ab} +2 \s B^{(1)}_{ab} \label{k4}\, , \\
\kappa^{[k,k]}_{ab}&=& M^{[1,k,k]}_{ab}- M^{[1,k,k]} h^{(0)}_{ab}= -4 B^{(1)}_{c(a} k_{b)}^{\:\:c} +2 B^{(1)}_{cd} k^{cd} h^{(0)}_{ab}, \label{kkkk} \\
\kappa^{[k,k,I]}_{ab}  &=& \frac{3}{4} k_{cd} k^{cd} h_{ab}^{(0)} -  k_{ac}k^c_{\; b} + \frac{1}{4} \DD_c k_{de}\DD^c k^{de} h_{ab}^{(0)}-\frac{1}{2} \DD_a k_{cd}\DD_b k^{cd} \: , \label{k6} \\
\kappa^{[k,k,II]}_{ab}&=&  -\frac{1}{2} B^{(1)\: cd} B^{(1)}_{cd} h^{(0)}_{ab} +  B^{(1)}_{c(a} B^{(1)\: c}_{b)} \label{k7}\: ,
\eeqs
which form a complete basis of SD tensors whose symmetrized curls are non-trivial. In \eqref{k4}, $Z^{(3)}_{ab}$ is an SDT tensor
\beqs
Z^{(3)}_{ab}&=& 8 \sigma B^{(1)}_{ab} +\frac{1}{2} \epsilon_{cd(a} \sigma_{b)}^{\;\:ce} k^d_{\;e} + 5 \sigma^c_{(a} B^{(1)}_{b)c}  \nonumber \\
&&-\frac{1}{2} \epsilon_{cd(a} \sigma^{ce} \DD_{b)} k^d_{\;e} - 2 h^{(0)}_{ab} \sigma^{cd} B^{(1)}_{cd} -\sigma^{c} \DD_c B^{(1)}_{ab} \: .
\eeqs
From \eqref{kkkk}, one can use the curl of the tensor potential $M^{[1,k,k]}_{ab}$ to construct the SDT tensor that we denote $Y^{(2)}_{ab}$ 
\beqs
Y^{(2)}_{ab} &\equiv& \epsilon_{cda} \DD^c M^{[1,k,k]\:d}_{b} =\epsilon_{cd(a} \DD^c \kappa^{[k,k]\:d}_{b)}\nonumber \\
&=&-4 \:  B^{(1)}_{c(a}   B^{(1) \: c}_{b)}  -2\: \epsilon_{cd(a}  \DD^c k_{b)}^{\;\:e}  B^{(1) \: d}_{e}  -2\: \epsilon_{cd(a}   \DD_{b)} B^{(1) \: c}_{e}   k^{de} \: . \label{k5} 
\eeqs

There exists obviously also an infinite list of SD tensors that have a trivial symmetrized curl. We will not consider them further in the rest of this thesis. Indeed, we will be mainly concerned with charges and we will see that regular SD tensors with trivial symmetrized curl are associated to trivial charges. For the sake of completeness, and because such tensors will appear in the following constructions, let us just present two such tensors
\beqs
\kappa^{[\s,\s,II]}_{ab}&=& 2 \s_a \s_b + 2 \s \s_{ab} + h^{(0)}_{ab} \Big( 4 \s^2 -2 \s_c \s^c \Big) \label{k2} , \\
\kappa^{[k,k,III]}_{ab} &=& (-\frac{1}{8} \DD_c k_{de} \DD^c k^{de}  -\frac{1}{2} k_{cd} k^{cd} ) h^{(0)}_{ab}  +\frac{1}{8} \DD_{(a} k^{cd} \DD_{b)} k_{cd}\nonumber \\
&& +\frac{1}{8} k^{cd} \DD_{(a} \DD_{b)} k_{cd} \: .
\eeqs
We start our analysis by explaining the general procedure we have followed in order to prove that we have listed all RNT tensor potentials needed to construct any SD or SDT tensor whose curl is non-zero. We then move on to the specific classification for each class: ($\s,\s$), ($\s,k$) and ($k,k$). Remember that we will always consider $k_{ab}$ to be SDT in the following.

\subsubsection{Algorithm for classification}

For each case ($\s,\s$), ($\s,k$) or ($k,k$), we use the following procedure

\begin{enumerate}

\item We start by listing a basis  symmetric tensors of rank two with $m$ derivatives built out of quadratic terms which are independent on-shell. It exists a number of derivatives $m^\star$ such that for all $m \geq m^\star$ the number of terms in that basis is maximal. At lower values $m < m^\star$, not all possible tensor structures can appear due to a lack of derivatives. We find $m^\star = 3$ for $(\s,\s)$, $m^\star = 3$ for $(k,k)$ and $m^\star = 4$ for $(k,\s)$ tensors. Due to the presence or the absence of the epsilon tensor, depending on whether $m$ is even or odd, the general form of a symmmetric tensor of rank two built out of linear combinations of the basis takes a different form. We denote this tensor as $Q^{(2n)}_{ab}$ or $Q^{(2n+1)}_{ab}$ and we provide its general form.

\item We continue by deriving a bound on the possible SDT tensors that one can build at a fixed number $m \geq m^\star$ of derivatives. We simply compute the number $H$ of linearly independent tensors $Q^{(m)}_{ab}$ which obey both $\DD^b Q^{(m)}_{ab}=0$ and $Q^{(m)a}_a=0$ where equalities here are valid up to terms with lower derivatives. At this stage, the number $H$ is only a bound on the number of SDT tensors at order $m$ because none of them has been fully yet constructed. We obtain that $H=1$ in the  $(\sigma, \sigma)$ case, $H=3$ in the $(k,k)$ case and $H=2$  in the $(\sigma, k)$ case, for both $m$ even or odd.

\item We then derive the explicit form of all RNT potentials, SD tensors and SDT tensors at each low value $m \leq m^\star$ of derivatives by enumeration. We write a basis of symmetric tensors of rank two built out of quadratic terms which are independent on-shell with at most $m$ derivatives for each $m \leq m^\star$ and impose the RNT or SDT conditions. The SD tensors $\kappa_{ab}$, whose curls are non-zero, are obtained from the RNT potentials by the correspondence $\kappa_{ab} = M_{ab} - h_{ab}^{(0)} M^c_c$.

\item We finally observe that there are exactly $H$ SDT tensors that have at most $m^\star$ derivatives and at least one term with $m^\star$ derivatives. This provides a proof that each candidate SDT tensor exists at order $m^\star$. We then note that the SDT tensors obtained by acting with the curl operator on these tensors form a basis for SDT tensors at order $m^\star + 1$ and by successive iterations at each order $m \geq m^\star$. Since there are $H$ SDT tensors at each order $m \geq m^\star$, there cannot be any other SD tensor which is not traceless but whose curls are non-zero or equivalently any RNT tensor at order $m$. Otherwise, there would be one additional SDT tensor at order $m+1$ by applying the curl operator but this would raise the number of SDT at level $m+1$ to $H+1$, which is not the case.

\item We conclude that all RNT potentials and SD tensors whose curls are non-zero are classified by the explicit tensors that we build out of terms with up to $m^\star$ derivatives. At higher order $m > m^\star$ in derivatives, all SD tensors, whose curls are non-zero, are traceless and can be obtained by applying curls on the RNT potentials.

\end{enumerate}

\subsubsection{$(\sigma,\sigma)$ SD tensors}

The analysis in the $(\s,\s)$ case, first performed in \cite{Compere:2011db}, is rather straightforward. One can rapidly realize that, for an odd number ($2n+1$) of derivatives, there is only one independent structure such that
\beqs\label{q2n1}
Q^{(2n+1)}_{ab}= \eps^{e f }{}_{(a} \s_{b) e c_1 c_2 \dots c_{n-1}} \s_f^{\; c_1 c_2 \dots c_{n-1}},
\eeqs
while for an even number (2n) of derivatives, we have
\beqs\label{q2n}
Q^{(2n)}_{ab}= a h_{ab}^{(0)}\s_{c_1c_2 \dots c_n}\s^{c_1 c_2 \dots c_n}+b \s_{ab c_1 \dots c_{n-1}}\s^{c_1 \dots c_{n-1}}+c \s_{a c_1 \dots c_{n-1}}\s_b^{\; c_1\dots c_{n-1}} .
\eeqs
This is so because we take into account that $\s_c^c=-3\s$  on shell and also that a structure such as
\beqs
\s_{ c_1 \dots c_{n-1} ab} \: \s^{c_1 \dots c_{n-1}} ,
\eeqs
is not an independent structure. Indeed, one can always bring it back to a form identical to the second term in \eqref{q2n} by commuting derivatives. This operation will only add terms with lower derivatives, terms that we neglect for this argument.

From \eqref{q2n1} and \eqref{q2n}, we immediately see that $m^{\star}=3$. Imposing divergence-free and traceless conditions on \eqref{q2n} provides us with the following requirements on the parameters
\bea
3 a +c = 0, \qquad  2a+b+c = 0.
\eea
Also, one can check that \eqref{q2n1} is always SDT. All in all, this tells us that $H=1$.

Let us now derive the explicit form of all RNT potentials, SD tensors and SDT tensors for $m \leq m^\star$. A generic symmetric tensor containing zero, one or two derivatives has the form
\bea
(a \sigma^2 +b \s_c \s^c )h_{ab}^{(0)} + c\s_a \s_b + d\s \s_{ab},
\eea
for some coefficients $a,b,c,d$. One can easily show that there are no SDT tensors in this class. However, there are two independent tensor potentials
\beqs\label{M2ss}
 M^{[2,\s,\s, I]}_{ab}&=&  (5 \s^2 +\s_c \s^c ) h_{ab}^{(0)} + 4\s \s_{ab} ,\nonumber \\
 M^{[2,\s,\s,II]}_{ab} &=&  (\DD_a \DD_b + h^{(0)}_{ab}) \: \s^2\, .
 \eeqs
The first one is a RNT potential while the curl of the second term is trivial. From this first potential, one can build an SDT tensor with three derivatives 
\bea\label{X3}
X^{(3)}_{ab} \equiv \eps_{a}^{\;\,\; cd}\DD_{c} M^{[2,\s,\s,I]}_{db} = 4 \eps_{cd(a}\s^c \s^d_{\;\, b)} = -4 \eps_{cd(a}\s^c E^{(1)}{}^d{}_{b)}\, ,
\eea
where $E^{(1)}_{ab}$ is the first order electric part of the Weyl tensor given in \eqref{E1}.

By a recursive application of the curl operator, one can build one SDT tensor at each order in derivatives. At the next order, we have
\beqs
X^{(4)}_{ab}&=& \curl X^{(3)}_{ab}= (\square-3)M^{[2,\s,\s,I]}_{ab}-\DD_a \DD_b M^{[2,\s,\s,I]} + M^{[2,\s,\s,I]} h^{(0)}_{ab} \nonumber \\
&=& 2\sigma_c\sigma^c h^{(0)}_{ab} +2\sigma_{cd} \sigma^{cd} h^{(0)}_{ab} +2\sigma_{abc} \sigma^c -18 \sigma^2 h^{(0)}_{ab}-18 \sigma \sigma_{ab} -6 \sigma^c_{\;(a} \sigma_{b)}^{\:\;c}. \nonumber \\
\eeqs

This ends the classification for the $(\s,\s)$ case. Indeed, we have found one RNT potential whose successive curls generate the unique SDT tensor at each order $m > m^\star$. The two SD tensors associated with the tensor potentials \eqref{M2ss} are given by
\beqs
\kappa^{[\s,\s,I]}_{ab} &=&  M^{[2,\s,\s, I]}_{ab}-  M^{[2,\s,\s, I]} h^{(0)}_{ab}  \nonumber \\
&=& (2 \s^2 -2\s_c \s^c ) h_{ab}^{(0)} + 4\s \s_{ab} \label{k2ss1} ,\\
\kappa^{[\s,\s,II]}_{ab}&=&  M^{[2,\s,\s,II]}_{ab}-  M^{[2,\s,\s,II]} h^{(0)}_{ab}  \nonumber \\
&=& 2 \s_a \s_b + 2 \s \s_{ab} + h^{(0)}_{ab} \Big( 4 \s^2 -2 \s_c \s^c \Big) \label{k2ss2} .
\eeqs
The second SD tensor has a zero symmetrized curl. We also see from \eqref{q2n1} that at each odd order $n \geq 3$ any SD tensor is also SDT.  At each even order $n \geq 4$, any SD tensor is a linear combination of the unique SDT tensor and a tensor with vanishing curl. If it was not the case, then a new independent SD tensor with at most four derivatives would generate an additional independent tensor with at most five derivatives that we do not observe. 
The general SD tensor whose curl is non-zero, up to the addition of SD tensors with trivial curls, has the form
\bea\label{XnSTDapp}
a_{(1)} \kappa^{[\s,\s,I]}_{ab}  + a_{(2)} X^{[3]}_{ab} + a_{(3)} \curl(X^{[3]})_{ab} + \cdots +  a_{(n)}  \curl^n(X^{[3]})_{ab} + \dots \; ,
\eea
for some arbitrary coefficients $a_{(i)}$ and $X^{[3]}_{ab}= (\curl \kappa^{[\s,\s,I]})_{(ab)}$.

\subsubsection{$(k,k)$ SD tensors}

The classification of $(k,k)$ structures is considerably more tedious than the classification of $(\s,\s)$ structures. In order to simplify the identification of a basis of independent tensors on-shell, we will make an efficient use of the relations
\beqs\label{ddqq}
\DD_{[a}B^{(1)}_{b]c}=0 \: , \qquad \DD_{[a}k_{b]c}=-\epsilon_{abd} \: B^{(1)\: d}_{c} \: .
\eeqs
where $B^{(1)}_{ab}$ is the first order magnetic part of the Weyl tensor given in \eqref{B1}.
We start by listing the independent structures quadratic in $B^{(1)}$ and its derivatives. Then, we add an independent subset of structures of the form $(B^{(1)}, k)$ such that no linear combinations are of the form $(B^{(1)}, B^{(1)})$ and eventually we add a subset of independent $(k,k)$ structures such that no linear combinations are of the form $(B^{(1)}, B^{(1)})$ or $(B^{(1)}, k)$. To look if such linear combinations exist, we just need to take into account the equations (\ref{ddqq}).

For an odd number $(2n+1)$ of derivatives we find the general form
\beqs\label{Qkkodd}
&&  Q^{(2n+1)}_{ab}=  a \; \epsilon_{cd(a} \DD^{i_1}...\DD^{i_{n-1}} \DD_{b)} B^{(1) ce} \DD_{i_1}...\DD_{i_{n-1}} B^{(1)d}_{e}\nn\\
&& +  b \; \epsilon_{cd(a}  \DD^{i_1}...\DD^{i_{n-1}}  \DD^f \DD_{b)} k^{ce} \DD_{i_1}...\DD_{i_{n-1}}  \DD_{f} k^d_{\;e}   \nonumber \\
&& + c \:  \DD^{i_1}...\DD^{i_{n-1}}  \DD_c B^{(1)}_{d(a}  \DD_{i_1}...\DD_{i_{n-1}}  \DD^{c} k_{b)}^{\:\;d} \nn\\
&&+ d \: \DD^{i_1}...\DD^{i_{n-1}}  \DD_c B^{(1)}_{de}  \DD_{i_1}...\DD_{i_{n-1}}  \DD^{c} k^{de} h^{(0)}_{ab}  \nonumber \\
 &&+ e \: \DD^{i_1}...\DD^{i_{n-1}}   \DD_c \DD_d B^{(1)}_{ab}  \DD_{i_1}...\DD_{i_{n-1}}  k^{cd}  \nn\\
 && + f\:  \DD^{i_1}...\DD^{i_{n-1}}  B^{(1)}_{cd}  \DD_{i_1}...\DD_{i_{n-1}}  \DD_{c} \DD_{(a} k_{b)}^{\:\;d} \: ,
\eeqs
while for an even  number  ($2n$)  of derivatives we find
\beqs\label{Qkkeven}
   Q^{(2n)}_{ab} &=& a \: \DD_{i_1}...\DD_{i_{n-2}} \DD_{c} B^{(1)}_{de} \DD^{i_1}...\DD^{i_{n-2}} \DD^c B^{(1) \: de} \: h^{(0)}_{ab} \nn \\
&& +b\:   \DD_{i_1}...\DD_{i_{n-2}} \DD_c B^{(1)}_{d(a} \DD^{i_1}...\DD^{i_{n-2}} \DD^c B^{(1) \: d}_{b)} \nn\\
&&+ c\:  \DD_{i_1}...\DD_{i_{n-2}} \DD_c \DD_d B^{(1)}_{ab} \DD^{i_1}...\DD^{i_{n-2}} B^{(1) \: cd} \nonumber \\
&&+ d\: \epsilon_{cd(a} \DD^{i_1}...\DD^{i_{n-2}} \DD^f \DD^c k_{b)}^{\;\:e}  \DD_{i_1} ...\DD_{i_{n-2}} \DD_f B^{(1) d}_{e}  \nonumber\\
&&  + e\: \epsilon_{cd(a} \DD^{i_1}...\DD^{i_{n-2}} \DD^f  \DD_{b)} B^{(1) \: c}_{e}  \DD_{i_1} ...\DD_{i_{n-2}}  \DD_f k^{de}\nn\\
&& +f\: \DD_{i_1}...\DD_{i_{n-2}} \DD_e \DD_{(a} k^{cd} \DD^{i_1}...\DD^{i_{n-2}} \DD^e \DD_{b)} k_{cd} \nonumber \\
 && +g\:  \DD_{i_1}...\DD_{i_{n-2}} \DD^e \DD_{(a} \DD_{b)} k^{cd} \DD^{i_1}...\DD^{i_{n-2}} \DD_e  k_{cd}\nn\\
 && +h\:   \DD_{i_1}...\DD_{i_{n-2}}  \DD_{c} \DD_{d} k_{ef} \DD^{i_1}...\DD^{i_{n-2}} \DD^c  \DD^d  k^{ef} h^{(0)}_{ab} \: .
\eeqs
We deduce that $m^\star = 3$. Indeed, when $m=2$ the third term in $Q^{(2)}_{ab}$ does not exist while when $m=3,4,$ or higher, all terms in $Q^{(m)}_{ab}$ exist. Looking at $m \geq m^\star$ and imposing the SDT condition, we find after a straightforward analysis that there can be at most three independent SDT tensors. We thus have $H=3$.

We now need to construct RNT potentials and SDT tensors at each order $m \leq m^\star$. At $m=0$, there are no tensor potentials and no SDT tensors. At $m=1$, we have
\beqs
Q^{(1)}_{ab}= a \:\epsilon_{cd(a}  \DD_{b)} k^{ce}  k^{d}_{\;\;e} + b \: B^{(1)}_{c(a} k_{b)}^{\:\;c} + c \:  B^{(1)}_{cd} k^{cd} h^{(0)}_{ab} \: ,
\eeqs
and one easily checks that there are no SDT tensors but there is one RNT potential 
\beqs
M^{[1,k,k]}_{ab}= -4 B^{(1)}_{c(a} k_{b)}^{\:\:c} +B^{(1)}_{cd} k^{cd} h^{(0)}_{ab}.
\eeqs
At $m=2$, we have
\beqs
Q^{(2)}_{ab}&=& a \:   B^{(1)}_{cd}  B^{(1) \: cd} \: h^{(0)}_{ab} +b\:  B^{(1)}_{c(a}   B^{(1) \: c}_{b)}  + c\: \epsilon_{cd(a}  \DD^c k_{b)}^{\;\:e}  B^{(1) \: d}_{e}  + d\: \epsilon_{cd(a}   \DD_{b)} B^{(1) \: c}_{e}   k^{de}  \nonumber \\
 && +e\:  \DD_{(a} k^{cd} \DD_{b)} k_{cd}  +f\:  k^{cd}  \DD_{(a} \DD_{b)}    k_{cd}   +g\:   \DD_{c} k_{de}  \DD^c   k^{de} h^{(0)}_{ab}  \nonumber \\
 && + h\: k^c_{\: (a} k_{b)c} + i \: k_{cd} k^{cd} h^{(0)}_{ab} \:,
\eeqs
where we also introduced the  structures with $m=0$ derivatives. Here, we get
\beqs
Q^{(2)}&=& [3a+b-2c] B^{(1)\:cd} B^{(1)}_{cd} + \DD_b k_{cd} \DD^b k^{cd} [e+3g] +[3f+h+3i] k_{cd} k^{cd} \: ,\nonumber \\
\DD_a Q^{(2)}&=&[6a+2b-4c] B^{(1)\:cd} \DD_a B^{(1)}_{cd}+ [2e+6g] \DD_a \DD_b k_{cd} \DD^b k^{cd} \nonumber \\
&&  + [6f+2h+6i] k_{cd} \DD_a k^{cd} \: , \nonumber \\
\DD^b Q^{(2)}_{ab}&=& [ 2a+b-2d] B^{(1)\:cd} \DD_a B^{(1)}_{cd} + [\frac{c}{2} -\frac{d}{2}] \epsilon_{cda}\DD^c k_{b}^{\:\:e} \DD^b B^{(1)\:d}_e \nonumber \\
&&+ [2c-2d-4e+4f+2h] \:  \epsilon_{cda} k^{ce} B^{(1)\:d}_e  +[e+f+2g]  \DD_a \DD^b k_{cd} \DD_b k_{cd} \nonumber\\
&&+k^{cd} \DD_a k_{cd} [e+7f+h+2i] \: ,
\eeqs
where we made use of the relations
\beqs
&& k^{cd} \square \DD_a k_{cd} = 5 k^{cd} \DD_a k_{cd} + 4 k^{cd} \DD_c k_{ad} , \qquad k^{cd} \DD^b \DD_c \DD_d k_{ab}=7 k^{cd} \DD_c k_{ad}  \: , \nonumber \\
&& k^{cd} \DD^b \DD_a \DD_b k_{cd}= k^{cd} \square \DD_a k_{cd}- 2k^{cd} \DD_c k_{ad} ,\qquad  k^{cd} \DD_c k_{ad}=2 \epsilon_{cda} k^c_{\:\:e} B^{(1)\:d}_{e}+k^{cd} \DD_a k_{cd}\, .\nn
\eeqs

We obtain that tensors $Q^{(2)}_{ab}$ satisfying $\DD^b Q^{(2)}_{ab}=\DD_a Q^{(2)}$ are of the form
\beqs
 m_1 Y^{(2)}_{ab} + m_2 M^{[2,k,k,I]}_{ab}+ m_3 M^{[2,k,k,II]}_{ab} +m_4 M^{[2,k,k,III]}_{ab} \: ,
\eeqs
where
\beqs
Y^{(2)}_{ab} &=& -4 \:  B^{(1)}_{c(a}   B^{(1) \: c}_{b)}  -2\: \epsilon_{cd(a}  \DD^c k_{b)}^{\;\:e}  B^{(1) \: d}_{e}  -2\: \epsilon_{cd(a}   \DD_{b)} B^{(1) \: c}_{e}   k^{de}= \epsilon_{cd(a} \DD^c M^{(1)\:d}_{b)} \: , \label{Y2aa}\nonumber \\
M^{[2,k,k,I]}_{ab}&=& \frac{1}{8}k_{cd} k^{cd} h_{ab}^{(0)} - k_{ac}k^c_{\; b} + \frac{1}{8}\DD_c k_{de}\DD^c k^{de} h_{ab}^{(0)}-\frac{1}{2}\DD_a k_{cd}\DD_b k^{cd} \: , \nonumber \\
M^{[2,k,k,II]}_{ab}&=&  -\frac{1}{4}B^{(1)}_{cd}  B^{(1) \: cd} \: h^{(0)}_{ab} +\:  B^{(1)}_{c(a}   B^{(1) \: c}_{b)}  \: ,\nonumber \\
M^{[2,k,k,III]}_{ab}&=& \frac{1}{8} k^{cd} \DD_{(a} \DD_{b)} k_{cd} +\frac{1}{8} \DD_{(a} k^{cd} \DD_{b)} k_{cd} +  \frac{1}{16} k_{cd} k^{cd}h^{(0)}_{ab} \: .
\eeqs
We thus see that $Y^{(2)}_{ab}$  is the unique SDT tensor and it is obtained from the RNT potential $M^{[1,k,k]}_{ab}$, that $M^{[2,k,k,I]}_{ab}$ and $M^{[2,k,k,II]}_{ab}$ are two new RNT potentials and that $M^{[2,k,k,III]}_{ab}$ is a tensor potential whose curl iss vanishing as it is of the form $(\DD_a \DD_b +h^{(0)}_{ab})k_{cd} k^{cd}$.
From the three tensor potentials found, we can define 3 SD tensors
\beqs
\kappa^{[k,k,I]}_{ab} &=& (M^{[2,k,k,I]}_{ab}- M^{[2,k,k,I]} h^{(0)}_{ab})  \nonumber \\
        &=& \frac{3}{4} k_{cd} k^{cd} h_{ab}^{(0)} -  k_{ac}k^c_{\; b} + \frac{1}{4} \DD_c k_{de}\DD^c k^{de} h_{ab}^{(0)}-\frac{1}{2} \DD_a k_{cd}\DD_b k^{cd} \: , \nonumber \\
\kappa^{[k,k,II]}_{ab}&=& (M^{[2,k,k,II]}_{ab}- M^{[2,k,k,II]} h^{(0)}_{ab}) = -\frac{1}{2} B^{(1)\: cd} B^{(1)}_{cd} h^{(0)}_{ab} +  B^{(1)}_{c(a} B^{(1)\: c}_{b)} \: , \nonumber \\
\kappa^{[k,k,III]}_{ab} &=&  (M^{[2,k,k,III]}_{ab} -M^{[2,k,k,III]} h^{(0)}_{ab}) \nonumber \\
                             &=& (-\frac{1}{8} \DD_c k_{de} \DD^c k^{de}  -\frac{1}{2} k_{cd} k^{cd} ) h^{(0)}_{ab}+\frac{1}{8} \DD_{(a} k^{cd} \DD_{b)} k_{cd} +\frac{1}{8} k^{cd} \DD_{(a} \DD_{b)} k_{cd} \: .\nn\\
\eeqs

Now, at $m=3$, we obtain
\beqs
Q^{(3)}_{ab}&=& a \: \epsilon_{cd(a}  \DD_{b)} B^{(1)\: ce}  B^{(1)\:d}_{e}  +  b \: \epsilon_{cd(a}   \DD^f \DD_{b)} k^{ce}   \DD_{f} k^d_{\;e}    + c \:   \DD_c B^{(1)}_{d(a}   \DD^{c} k_{b)}^{\:\;d} \nonumber \\
&& + d \:  \DD_c B^{(1)}_{de}  \DD^{c} k^{de} h^{(0)}_{ab} + e \:    \DD_c \DD_d B^{(1)}_{ab}    k^{cd}  + f\:   B^{(1)}_{cd}   \DD_{c} \DD_{(a} k_{b)}^{\:\;d} \: ,
\eeqs
up to terms with lower derivatives. One can explicitly construct three independent SDT tensors which can also be obtained as curls of the three previous RNT potentials. We have thus completed our algorithm. We have three towers of SDT tensors generated by the three RNT potentials $M^{[1,k,k]}_{ab}$  (which leads to $Y^{(2)}_{ab}$), $M^{[2,k,k,I]}_{ab}$ and $M^{[2,k,k,II]}_{ab}$.

\subsubsection{$(\sigma,k)$ SD tensors}

For the $(\sigma,k)$ case, we find that a generic tensor with $2n+1$ derivatives is of the form
\beqs
Q^{(2n+1)}_{ab}&=& a \; \DD_{i_1}...\DD_{i_{n}} B^{(1)}_{ab} \sigma^{i_1...i_{n}} + b \; \DD_{i_1}...\DD_{i_{n-1}} B^{(1)}_{e(a} \sigma_{b)}^{\;\:e i_1 ...i _{n-1}}  \nonumber \\
&&+ c \; \epsilon_{cd(a} \DD^{i_1} ...\DD^{i_{n-1}} \DD_{b)} k^c_{\;e} \sigma^{de}_{\;\;\;i_1 ... i_{n-1}} +  d\; \epsilon_{cd(a} \DD^{i_1} ...\DD^{i_{n-1}} k^{ce} \sigma_{b)\;\;\;e i_{1} ...i_{n-1}}^{\;\;d}\nonumber \\
&& + e \; \DD_{i_1}...\DD_{i_{n-1}} B^{(1)}_{cd} \sigma^{cd i_1...i_{n-1}} h^{(0)}_{ab}  + f \; \DD_{i_1}...\DD_{i_{n-2}} B^{(1)}_{cd} \sigma_{ab}^{\;\;\;cd i_1...i_{n-2}} \: ,
\eeqs
while, for an even number $2n$ of derivatives, it is of the form
\beqs
Q^{(2n)}_{ab}&=& a \; \sigma^{c d i_1...i_{n-2}} \DD_{i_1}...\DD_{i_{n-2}} \DD_c \DD_{(a}k_{b)d}    +\: b \; \sigma^{c d  i_1...i_{n-2}} \DD_{i_1}...\DD_{i_{n-2}} \DD_{(a} \DD_{b)} k_{cd} \nonumber \\
             &&   +\: c \; \sigma^{ cd  i_1...i_{n-2}} \DD_{i_1}...\DD_{i_{n-2}} \DD_c \DD_d  k_{ab} + \:  d \; \sigma_{c d i_1...i_{n-2} (a} \DD^{i_1}...\DD^{i_{n-2}} \DD_{b)}k^{cd} \nonumber \\
             && +\:  e \;  \sigma_{c d  i_1...i_{n-2} (a} \DD^{i_1}...\DD^{i_{n-2}} \DD^c k_{b)}^{\:\:\:d} + \: f \; \sigma_{a b c d  i_1...i_{n-2}} \DD^{i_1}...\DD^{i_{n-2}} k^{cd} \nonumber \\
             && + \: g \; h^{(0)}_{ab} \: \sigma_{c d  i_1...i_{n-1}} \DD^{i_1}...\DD^{i_{n-1}} k^{cd} \: .
\eeqs
We find that $m^\star =4$. Indeed, for $m=3$ derivatives, the last term in $Q^{(3)}_{ab}$ does not exist while for $m=4,5,\dots$ all terms in $Q^{(m)}_{ab}$ exist. For any $m \geq 4$, one can check that there are at most 2 SDT tensors. We therefore find $H=2$. At lower levels $m=0$ or $m=1$, we see that there are no SDT tensors and no tensor potentials. At $m=2$, we have
\beqs
Q^{(2)}_{ab}&=& a \:  \sigma k_{ab} +
b \: \epsilon_{cd(a} k^d_{\:\: b)} \sigma^c+ c \:   \sigma B^{(1)}_{ab}
+ d \:  \sigma^c \DD_c k_{ab}  \nonumber \\
&&+ e \:  \sigma^c \DD_{(a} k_{b)c} + f \:  \sigma_{c(a} k_{b)}^{\:\:\:c} + g \:  \sigma_{cd} k^{cd} h^{(0)}_{ab}\label{2der} \: .
\eeqs
There is no SDT tensor, but there is one RNT potential
\bea
M_{ab}^{[2,\s,k]} =   \sigma k_{ab} - \:  \sigma^c \DD_c k_{ab} +  \:  \sigma^c \DD_{(a} k_{b)c} +  \:  \sigma_{c(a} k_{b)}^{\:\:\:c} -\frac{1}{4} \:  \sigma_{cd} k^{cd} h^{(0)}_{ab}  \label{Mprimitive2} \, .
\eea
Its curl gives an SDT tensor
\beqs\label{Z3}
Z^{(3)}_{ab}&=& 8 \sigma B^{(1)}_{ab} +\frac{1}{2} \epsilon_{cd(a} \sigma_{b)}^{\;\:ce} k^d_{\;e} + 5 \sigma^c_{(a} B^{(1)}_{b)c}  \nonumber \\
&&-\frac{1}{2} \epsilon_{cd(a} \sigma^{ce} \DD_{b)} k^d_{\;e} - 2 h^{(0)}_{ab} \sigma^{cd} B^{(1)}_{cd} -\sigma^{c} \DD_c B^{(1)}_{ab} \: .
\eeqs
At $m=3$, the general SDT tensor or RNT potential can be written as a linear combination of a basis of terms with 1 and 3 derivatives
\beqs
Q^{(3)}_{ab}&=&  a \: B^{(1)e}_{(a}\sigma_{b)}^{\;\:e} + b \: h^{(0)}_{ab} \sigma^{cd} B^{(1)}_{cd} + c \: \eps_{cd(a }\s_{b)}^{\; c e}k^d_{\; e}  + d \:  \eps_{cd(a}\DD_{b)} k_{\; e}^{d}\s^{ce}+ e \: \sigma^e \DD_e B^{(1)}_{ab} \nonumber \\
&& + f \epsilon_{cd(a} k^d_{\;b)} \sigma^c + g \: \sigma B^{(1)}_{ab}\, .
\eeqs
We find one SDT tensor which is obviously $Z^{(3)}_{ab}$ and a new RNT potential $M^{[3,\s,k]}_{ab}$
\beqs
 M^{[3,\s,k]}_{ab}&=& 7 B^{(1)}_{c(a} \sigma^c_{b)}  - \frac{5}{2} h^{(0)}_{ab} \sigma^{cd} B^{(1)}_{cd} +\frac{1}{2} \epsilon_{cd(a} \sigma_{b)}^{\;\:ce} k^d_{\;e} \nn \\
 &&  -\frac{1}{2} \epsilon_{cd(a}  \DD_{b)} k^d_{\;e} \sigma^{ce} -\sigma^{c} \DD_c B^{(1)}_{ab}+ 10 \sigma B^{(1)}_{ab}\, .
\eeqs
 As expected, at $m=4$, one can check that we have 2 SDT tensors. The algorithm is therefore completed. The two RNT potentials that generate the two independent towers of SDT tensors are $M^{[2,\s,k]}_{ab}$ and $M^{[3,\s,k]}_{ab}$.
To each of these potentials corresponds a unique SD tensor
\bea
\kappa_{ab}^{[\s,k, I]} &=& M_{ab}^{[2,\s,k]}- h_{ab}^{(0)} M^{[2,\s,k]} \nonumber \\
&=& \s k_{ab} +\s_{c(a} k_{b)}^{\:\:\:c}  -\frac{1}{2} \s_{cd} k^{cd} h^{(0)}_{ab} -\s^c \DD_c k_{ab} + \s^c \DD_{(a} k_{b)c} \: , \label{k2sk}\\
\kappa_{ab}^{[\s,k,II]} &=& M_{ab}^{[3,\s,k]}- h_{ab}^{(0)} M^{[3,\s,k]} \nonumber \\
&=& Z^{(3)}_{ab} + 2 B^{(1)}_{c(a} \s_{b)}^{\:\:\:c} -\s^{cd} B^{(1)}_{cd} \: h^{(0)}_{ab} +2 \s B^{(1)}_{ab} \label{k3sk}
 \, .
\eea
All other SDT tensors, with $m \geq m^{*}$, are then constructed from linear combinations of successive curls of those SD tensors. This ends the classification of RNT and SDT $(\s,k)$ tensors.

\subsection{First order equations}
\label{sec:firstorder}

Now that we have established all this, we start by rewriting the first order equations of motion and study their solutions.

\subsubsection{Compact form}
From the definitions of the first order electric $E^{(1)}_{ab}$ and magnetic $B^{(1)}_{ab}$ parts of the Weyl tensor that were given in \eqref{E1} and \eqref{B1}, we see that these tensors satisfy the properties
\beqs
h^{(0)\:ab} B^{(1)}_{ab}=0, \qquad E^{(1)}_{[ab]}=0,\qquad \DD_{[c} E^{(1)}_{a]b}=0.
\eeqs
The first two properties are trivial, while the third one stating that $E^{(1)}_{ab}$ is curl-free can be easily proved using (see also Appendix \ref{app:properties})
\beqs
[\DD_a, \DD_b] \s_c= 2 h^{(0)}_{c[a} \s_{b]} .
\eeqs
Now, it is also easy to see that the first order equations of motion can be summarized by
\beqs
h^{(0)\:ab} E^{(1)}_{ab}=0 , \qquad B^{(1)}_{[ab]}=0  , \qquad \DD_{[c} B^{(1)}_{a]b}=0.
\eeqs
Indeed, the first one is a trivial rewriting of the first order Hamiltonian equation $(\Box+3)\s=0$ using the definition of $E^{(1)}_{ab}$. The second one states that
\beqs
\epsilon^{cab} \: B^{(1)}_{ab}=0,
\eeqs
and by definition of $B^{(1)}_{ab}$, we recover the first order momentum equation
\beqs
2 \epsilon^{cab} \: B^{(1)}_{ab}=\epsilon^{cab} \: \epsilon_{a}^{\:\:de} \DD_d k_{eb}= \DD^c k-\DD^b k^c_{\:\:b}=0 .
\eeqs
Eventually, the third equation implies that
\beqs
2 \epsilon_{a}^{\:\:cd} \: \DD_c B^{(1)}_{db}= \Box k_{ab}-\DD^c \DD_a k_{bc}=0,
\eeqs
which is just the first order equation of motion obtained in \eqref{momentumdef}. One can check that all these properties also imply that
\beqs
&&\DD^b E^{(1)}_{ab}=0 , \qquad \DD^b B^{(1)}_{ab}=0, \\
&&(\Box-3)E^{(1)}_{ab}=0 , \qquad (\Box-3)B^{(1)}_{ab}=0. \label{boxxx}
\eeqs

The first order electric and magnetic parts of the Weyl tensor are thus, on shell, SDT tensors which are moreover curl-free and satisfy \eqref{boxxx}

\subsubsection{Solutions to the equations}

Let us characterize the solutions to these first order equations. As we will see just after, it is sufficient, to specify the physical content of the tensors $E^{(1)}_{ab}$ and $B^{(1)}_{ab}$, to solve the hyperbolic equation
\beqs
\square \Phi +3\Phi =0.
\eeqs
The general solution to this equation is the sum of a function of $\tau$ times a spherical harmonic,  $f_{lm}(\tau)Y_{lm}(\theta,\phi)$, $l = 0,1,\dots$, $m=-l,\dots,l$. However, two independent solutions for $f_{lm}(\tau)$, which we denote as $f_{lm}(\tau)$ and $\hat{f}_{lm}(\tau)$, exist since the hyperbolic equation is second order. We express each independent solution of that equation as a hyperbolic harmonic. One can then write
\be
\Phi = \sum_{i=0}^3 \left( \alpha_{(i)}  \zeta_{(i)}+ \hat \alpha_{(i)} \hat \zeta_{(i)}  \right) + \sum_{l \geq 2} \sum_{m=-l}^l \left( a_{lm} f_{lm}(\tau) +\hat a_{lm} \hat f_{lm}(\tau) \right) Y_{lm}(\theta,\phi) ,\label{decompphi}
\ee
where the first set of lowest harmonics are given by the four solutions
\bea\label{transareodd}
\zeta_{(0)} = \sinh\tau,\qquad \zeta_{(k)} = \cosh\tau f_{(k)}, \quad k=1,2,3,\label{zetas}
\eea
of $\DD_a \DD_b \zeta_{(a)} + h_{ab}^{(0)}\zeta_{(a)}= 0,$ which are odd under parity $(\tau, \theta, \phi) \rightarrow (-\tau, \pi-\theta, \phi+\pi)$ while the other set of lowest harmonics are
\begin{eqnarray}
\hat \zeta_{(0)} &=& \frac{\cosh 2\tau}{\cosh \tau}, \qquad \hat \zeta_{(k)} = \left(2\sinh \tau+\frac{\tanh \tau}{\cosh \tau}\right)f_{(k)}\label{zetahat} ,\quad k=1,2,3,
\end{eqnarray}
which are even under parity. Here,
\bea
f_{(1)} &=& \cos \theta, \qquad f_{(2)} =  \sin \theta \cos \phi, \qquad f_{(3)} =  \sin \theta \sin \phi , \label{harmosphere}
\eea
are the three $l=1$ harmonics on the two-sphere.

This analysis clearly classifies the possible solutions for $E^{(1)}_{ab}$ since it is expressed in terms of a scalar that fulfills this hyperbolic equation. Actually, it is also sufficient to classify solutions of $B^{(1)}_{ab}$ as we now see with the following lemma that we prove in Appendix \ref{app:proofs}.

\begin{lemma}[Ashtekar-Hansen]\label{Ash} On the three-dimensional hyperboloid, any traceless curl-free divergence-free symmetric tensor $T_{ab}$ such that $\square T_{ab}= 3 T_{ab}$ can be written as
\begin{equation}
T_{ab} = \DD_a \DD_b \Phi + h^{(0)}_{ab}\Phi,
\end{equation}
with $\square \Phi + 3\Phi = 0$. The scalar $\Phi$ is determined up to the ambiguity of adding a combination of the four functions \eqref{zetas}. 
\end{lemma}
Following this lemma, one can express the first order magnetic part of the Weyl tensor as
\beqs\label{B111}
B^{(1)}_{ab}=-\s^D_{ab}-\s^D h^{(0)}_{ab}, \qquad (\Box+3)\s^D=0.
\eeqs
Let us note, as also proved in Appendix \ref{app:proofs}, that

\begin{lemma}[Beig-Schmidt] \label{BS} On the three dimensional hyperboloid, any scalar $\Phi$ satisfying $\square \Phi + 3\Phi = 0$ and such that it does not contain the four lowest hyperbolic harmonics \eqref{zetahat} defines a symmetric, traceless, curl-free and divergence-free tensor $T_{ab} = \DD_a \DD_b \Phi + h^{(0)}_{ab}\Phi$ that can be written as
\begin{equation}
T_{ab} = \epsilon_{a}^{\;\, cd}\DD_c P_{d b},
\end{equation}
where $P_{ab}$ is a symmetric, traceless and divergence-free tensor. This tensor is defined up to the ambiguity $P_{ab} \rightarrow P_{ab} + \DD_a \DD_b \omega + h^{(0)}_{ab}\omega$ where $\omega$ is an arbitrary scalar obeying $\square \omega +3\omega =0$.
\end{lemma}

This second lemma tells us that, equivalently, the first order electric part can be rewritten as
\beqs
E^{(1)}_{ab}=\frac{1}{2}  \epsilon_{a}^{\;\, cd}\DD_c k^D_{d b} .
\eeqs

\subsection{Second order equations}
\label{subsec:second}
Let us now move to the second order equations. In \cite{B}, Beig showed that in the case $k_{ab}=i_{ab}=0$, the system could be written in terms of the second order part of the magnetic Weyl tensor. Here, we show that for our enlarged boundary conditions, the second order equations of motion can actually be written in terms of two equivalent systems constructed from two SDT tensors that are mutually conjugate in a way that we precise below. In the case $k_{ab}=i_{ab}=0$, we see that one of the SDT tensors reduces to the result of Beig and that the other can be expressed in terms of the electric part of the Weyl tensor.

\subsubsection{Linearized equations and their solutions}

In an attempt to present the material in a more pedagogical way, we start by discussing the second order equations in the linear case  when the quadratic terms in $(\s,\s),(\s,k)$ and ($k,k$) are set to zero. We also set $i_{ab}=0$. The equations reduce to
\beqs\label{linea}
h^{(2)\:a}_a=0, \qquad \DD^b h^{(2)}_{ab}=0, \qquad (\Box-2)h^{(2)}_{ab}=0\; .
\eeqs
Let us define
\bea
V_{ab} &\equiv & - h_{ab}^{(2)}, \label{defV0}\qquad W_{ab} \equiv  \curl h_{ab}^{(2)} \label{defW0},
\eea
where the curl operator is $(\curl T)_{ab} \equiv  \eps_{a}^{\;\, cd}D_{c} T_{db}$. Let us emphasize here that the curl operator obeys remarkable properties. The curl of a symmetric tensor $T_{ab}$ satisfying $\DD^bT_{ab} = \DD_a T_{b}{}^{b}$ is symmetric: $(\curl T)_{[ab]}=0$. Moreover, the curl is the square root of the operator $\square - 3$ when acting on an SDT tensor $T_{ab}$
\be
\curl(\curl(T))_{ab} = (\square - 3 )T_{ab}.
\ee
The latter property also implies that the square of the curl operator when acting on an SDT tensor $T_{ab}$ obeying $(\square - 2)T_{ab} = 0$ is minus the identity. This shows that this operator is invertible when acting on $T_{ab}$ satisfying $(\square - 2)T_{ab} = 0$. Finally, one has $[\square - 2, \curl] T_{ab} = 0$ for an SDT tensor $T_{ab}$.

With these properties in mind, we see that the above defined quantities enjoy the duality properties
\bea
W_{ab} = - \curl V_{ab},\qquad V_{ab} = \curl W_{ab}.
 \eea
Moreover, the linearized second order equations  \eqref{linea} can then be written in two equivalent ways
\begin{align}
V^a_{\;a} &= 0,& \DD^b V_{ab} &= 0,&  (\square - 2)V_{ab} &= 0,\label{eq:001a}
\end{align}
or
\begin{align}
W^a_{\;a} &= 0,&  \DD^b W_{ab} &= 0,&  (\square - 2)W_{ab} &= 0\label{eq:001b}.
\end{align}
The two sets of equations are related by the curl operator. Given a solution to one of these systems, one can reconstruct the metric using definitions \eqref{defV0}. We have thus shown the equivalence of linearized Einstein's equations at second order (with $i_{ab}=0$) to any one of the above two systems of equations.

Let us remember and insist on the fact that tensor fields that derive from a scalar potential, like the first order electric part of the Weyl tensor $E_{ab}^{(1)}= - \DD_a \DD_b \sigma - h^{(0)}_{ab}\sigma$, have vanishing curl,
\bea
T_{ab} = \DD_a \DD_b \Phi + h^{(0)}_{ab} \Phi \quad \Rightarrow \quad (\curl T)_{ab} = 0.
\eea
These tensors therefore obey $(\curl^2 \,T)_{ab} = (\square - 3)T_{ab} = 0$ as opposed to $(\square - 2)T_{ab} = 0.$ As a consequence, none of the lemmae used in the first order analysis, see also \cite{BS, Ashtekar:1978zz}, capture tensor structures appearing at second order. The general form of the solutions to the equations
\bea
T^a_a = \DD^b T_{ab} = (\square - 2)T_{ab} = 0 ,\label{eq:Taa2}
\eea
is summarized in the following lemma, proved in Appendix \ref{app:proofs}, stating that
\begin{lemma}\label{lemma8}
On the hyperboloid, any regular symmetric divergence-free traceless tensor $T_{ab}$ obeying $(\square -2) \, T_{ab}=0$ can be uniquely decomposed as
\bea
T_{ab} = \sum_{i=1}^3 \left( v_{(i)} V_{(i)ab}+ w_{(i)} W_{(i)ab} \right)+ J_{ab},
\eea
where the three tensors $V_{(i)ab}$ and the three tensors $W_{(i)ab}$, $i=1,2,3$ are given by
\begin{align}
V_{(i)\tau\tau} &= 2 \sech ^5 \tau \zeta_{(i)},& V_{(i)\tau i} &= \sech ^3 \tau \tanh \tau \partial_i \zeta_{(i)},& V_{(i)ij}&=\eta_{ij} \sech ^3\tau\zeta_{(i)}, \\
W_{(i)\tau \tau} &= 0,&   W_{(i) \tau i} &=  \sech ^3\tau  \eps_{i}^{\;\, j}\partial_j \zeta_{(i)},&   W_{(i)ij} &=0.
\end{align}
These tensors are dual to each other in the sense that
\bea
\eps_a^{\;\, cd}\DD_c V_{(i)d b} = -W_{(i)ab},\qquad \eps_a^{\;\, cd}\DD_c W_{(i)d b}=  V_{(i)ab}.
\eea
These tensors also obey the orthogonality properties
\begin{align}
\int_{S}  V_{(k)ab}\zeta^a_{(l)} n^b d^2S    =  \int_{S} W_{(k)ab}\zeta^a_{(l)} n^b d^2S  = 0,\\
\int_{S}  V_{(k)ab}\xi^a_{\rom{rot}(l)} n^b d^2S    =   \int_{S}  W_{(k)ab}\xi^a_{\rom{boost}(l)}n^b d^2S = 0,\\
\int_{S}  V_{(k)ab}\xi^a_{\rom{boost}(l)}n^b d^2S  = \int_{S}  W_{(k)ab}\xi^a_{\rom{rot}(l)} n^b d^2S    = \frac{8\pi}{3} \delta_{(k)(l)},
\end{align}
where $\delta_{(i)(j)} = 1$ if $i=j$, and $\zeta^a_{(l)} = D^a \zeta_{(l)}$ are the four translation Killing vectors (conformal Killing vectors on $dS_3$) with $\zeta_{(l)}$ given in \eqref{zetas}. The tensor $J_{ab}$ is a symmetric traceless divergence-free tensor obeying
\bea
\int_{S} d^2S J_{ab}\xi_\rom{rot}^a n^b  = \int_{S} d^2S J_{ab} \xi_\rom{boost}^a n^b =\int_{S} d^2S J_{ab}\zeta^a_{(l)} n^b  = 0, \qquad (\square -2)J_{ab}= 0.\nn\\
\eea
\end{lemma}

\subsubsection{Full non-linear equations}
\label{equivsys}

Let us now consider the full non-linear second order equations. Based on the previous analysis, we would like to rewrite them in terms of two tensors  $V_{ab}$ and $W_{ab}$
\bea
V_{ab} &\equiv& - h_{ab}^{(2)}+\frac{1}{2} i_{ab}+ Q_{ab}^V \: ,\label{defnlVb} \\
W_{ab} &\equiv& \eps_a^{\;\, cd}D_c \left( h^{(2)}_{d b }-\frac{1}{2} i_{db} + Q_{db}^W \right) \: , \label{defnlWb}
\eea
where $Q_{ab}^{V,W}$ are appropriate quadratic terms in $(\s,\s)$, $(\s,k_{ab})$ or $(k_{ab},k_{ab})$ that we will construct herebelow. We require that $V$ and $W$ are SDT tensors that  obey the following duality properties
\bea\label{dualiprop}
W_{ab} + \eps_a^{\;\, cd}D_c V_{d b } = K^W_{ab},\qquad V_{ab} - \eps_a^{\;\, cd}D_c W_{d b } = -2 i_{ab} + K^V_{ab}\, ,\label{nl03b}
\eea
where $K^{V,W}_{ab}$ are non-linear terms quadratic in $\sigma$ and $k_{ab}$, which are also SDT. Applying the curl operator on both equations \eqref{nl03b} we obtain that $V_{ab}$ and $W_{ab}$ obey
\bea
(\square - 2)V_{ab} = -2i_{ab} + K^V_{ab}+ \eps_a^{\;\, cd}D_c K^W_{d b } , \label{rhs01} \\
(\square - 2)W_{ab} = - 2 j_{ab} + K^W_{ab} - \eps_a^{\;\, cd}D_c K^V_{d b } ,  \label{rhs02}
\eea
where $j_{ab} \equiv -curl(i)_{ab}$. Our construction of the non-linear tensors $Q^{V,W}$, $K^{V,W}$ goes as follows. In order for $W_{ab}$ to be traceless, we require $Q^{W}_{ab}$ to be symmetric. Using the Hamiltonian and  momentum equation of motion, one can rewrite the symmetry condition of $W_{ab}$ as the following equation on $Q^{W}_{ab}$
\bea
\DD^b Q^W_{ab} - \DD_a Q^{W\, b}_b =  -\frac{1}{2}k^{bc} \DD_b k_{ac} + \DD_a \left( 4 \sigma^2 + \frac{3}{8} k_{cd} k^{cd} \right)  . \label{eqQ}
\eea
The divergence-free conditions of $W_{ab}$ can then be rewritten as
\bea
\eps_a^{\; \, cd} \DD_c \left( \DD^b Q^W_{d b} +\frac{1}{2} \DD_e k_{df} k^{ef} \right)= 0 \: ,
\eea
which is a consequence of the previous equation. The equation \eqref{eqQ} can be solved up to the ambiguity of adding to $Q^W_{ab}$ tensors obeying $\DD^b M_{ab} = \DD_a M$. We will fix the ambiguity in defining $Q_{ab}^W$ since we would like to find one equivalent formulation of the equations of motion, not all possible formulations. By choosing a $Q_{ab}^W$ with the smallest possible number of derivatives, we obtain
\bea\label{QW}
Q^W_{ab} = \left( - 2\sigma^2 +\frac{1}{16} k_{cd} k ^{cd} \right) h_{ab}^{(0)} - \frac{1}{2}k_{ac}k^{c}_{\; b}\, .\label{defQW}
\eea
Using again the Hamiltonian and momentum equations of motion, one can rewrite the traceless and divergence-free conditions of $V_{ab}$ as the following equations on $Q^{V}_{ab}$
\bea
Q_{a}^{V\; a} &=& 12 \sigma^2 + \sigma_c \sigma^c+\frac{1}{4} k_{cd} k^{cd}  + k_{cd} \sigma^{cd} \: , \\
\DD^b Q_{ab}^V &=& \frac{1}{2} \DD^b k_{ac} k_{b}^{\:\: c}  + \DD_a \left( \sigma_c \sigma^c +8 \sigma^2 -\frac{1}{8} k_{cd} k^{cd}  +k_{cd}\sigma^{cd} \right) \,.
\eea
This system has a unique solution up to the ambiguity of adding an SDT tensor to  $Q^{V}_{ab}$. We will make a specific choice for the ambiguity in defining $Q_{ab}^V$ as well. In fact, the SDT tensor $K^V_{ab}$ can be computed using \eqref{nl03b} and the equations of motion \eqref{equation3bis} as
\bea
K^V_{ab} &=& -NL_{ab}(\s,\s)-NL_{ab}(\s,k)-NL_{ab}(k,k) + Q^V_{ab} - (\Box - 3)Q^W_{ab} \nonumber \\
&& +\DD_a \DD^c (h^{(2)}_{bc} +Q^W_{bc})-h_{ab}^{(0)}(h^{(2)}+Q^{W \, a}_{a})\; .
\eea
We will choose to fix the ambiguity  of adding SDT tensors to  $Q_{ab}^V$ by requiring
\bea
K^V_{ab} = 0 \: .\label{KVzero}
\eea
After a tedious computation, we obtain simply
\bea\label{QV}
Q_{ab}^V &=& (6\s^2 + \s^c \s_c+ \frac{1}{8} k_{cd} k^{cd}+\frac{1}{8} \DD_c k_{de} \DD^c k^{de} )  h_{ab}^{(0)} +2\s \s_{ab} -2\s_a \s_b \nonumber \\
&& + 4\:  \sigma k_{ab} - 4\:  \sigma^c \DD_c k_{ab} + 4 \:  \sigma^c \DD_{(a} k_{b)c} +  \:  4 \sigma_{c(a} k_{b)}^{\:\:\:c} - \:  \sigma_{cd} k^{cd} h^{(0)}_{ab} \nonumber\\
&& -\frac{1}{2} k_{ac} k_{b}^{\:\:c}  - \frac{3}{8} \DD_a k_{cd} \DD_b k^{cd} +\frac{1}{8} k^{cd} \DD_ {(a} \DD_{b)}  k_{cd} \, +Y^{(2)}_{ab} \: , \label{defQV}
\eea
where $Y^{(2)}_{ab}$ is an SDT tensor given in \eqref{Y2aa}.
Remark that to perform this computation, one can separate the analysis of non-linear terms for each set of quadratic terms $(k,k)$, $(\s,k)$ or $(k,k)$ independently since those terms never mix in the equations.

Using then the definition of $K^W_{ab}$ in \eqref{nl03b} we find
\bea
K^W_{ab} &=& curl(M)_{ab} =  \eps_a^{\;\, cd}  \DD_c M_{db} \: ,
\eea
where
\bea
M_{ab} \equiv Q^W_{ab}+Q^V_{ab} \; ,
\eea
is a tensor obeying $\DD^b M_{ab} = \DD_a M$. Using the classification of such tensors described in section \ref{sec:SD}, we have explicitly
\bea
M_{ab} = M^{[2,\s,\s,I]}_{ab}- M^{[2,\s,\s,II]}_{ab} + 4 M^{[2,\s,k]}_{ab} + Y^{(2)}_{ab}+ M^{[2,k,k,I]}_{ab}+ M^{[2,k,k,III]}_{ab} \: ,
\eea
where
\beqs
M^{[2,\s,\s, I]}_{ab}&=&  (5 \s^2 +\s_c \s^c ) h_{ab}^{(0)} + 4\s \s_{ab}, \nonumber \\
 M^{[2,\s,\s,II]}_{ab} &=&  (\DD_a \DD_b + h^{(0)}_{ab}) \: \s^2 ,  \nonumber \\
M_{ab}^{[2,\s,k]} &=&   \sigma k_{ab} - \:  \sigma^c \DD_c k_{ab} +  \:  \sigma^c \DD_{(a} k_{b)c} +  \:  \sigma_{c(a} k_{b)}^{\:\:\:c} -\frac{1}{4} \:  \sigma_{cd} k^{cd} h^{(0)}_{ab}   ,\nonumber \\
M^{[2,k,k,I]}_{ab}&=& \frac{1}{8}k_{cd} k^{cd} h_{ab}^{(0)} - k_{ac}k^c_{\; b} + \frac{1}{8}\DD_c k_{de}\DD^c k^{de} h_{ab}^{(0)}-\frac{1}{2}\DD_a k_{cd}\DD_b k^{cd} , \nonumber \\
M^{[2,k,k,III]}_{ab}&=& \frac{1}{16} \Big( \DD_a \DD_b + h^{(0)}_{ab}\Big)  k^{cd} k_{cd} \, .
\eeqs
In summary, the equations of motion can be written in the form
\bea
W^a_a &=& \DD^b W_{ab}=0 \, ,\nonumber \\
(\square - 2)W_{ab} &=& \curl(2 i + M )_{ab} \, ,\label{eqWfb}\\
i^a_a &=& \DD^b i_{ab} = 0 \, ,\nonumber\\
(\square -2)i_{ab} &=& 0\, .\label{eqib}
\eea
Using the curl operator and the definition $j_{ab} \equiv -(\text{curl} \: i)_{ab}$, one can derive an equivalent form of those equations in terms of $V_{ab}$ as
\bea
V^a_a &=& \DD^b V_{ab}=0 ,\nonumber \\
(\square - 2)V_{ab} &=& \curl(-2j + \curl(M))_{ab},\label{eqVf} \\
j^a_a &=& \DD^b j_{ab} = 0, \nonumber\\
(\square -2)j_{ab} &=& 0.\label{eqj}
\eea
Since the curl of the latter set of equations lead to \eqref{eqWfb}-\eqref{eqib}, the two sets are equivalent. Once the set of equations \eqref{eqWfb}-\eqref{eqib} is solved, one can reconstruct $h_{ab}^{(2)}$ from the definitions \eqref{defnlVb} or \eqref{defnlWb}. Einstein's equations at second order are therefore equivalent to either set of the above systems of equations.

As a last remark, we could also use in the above equations the symmetrized curl of $\kappa_{ab}\equiv M_{ab}-M h^{(0)}_{ab}$ instead of $M_{ab}$. Indeed, the symmetrized curl of $\kappa_{ab}$ is equivalent to the curl of the tensor potential $M_{ab}$.

\subsubsection{Non-linear equations with $k_{ab}=i_{ab}=0$}

If we restrict ourselves to $k_{ab}=i_{ab}=0$, we see from the definitions of $V_{ab}$ and $W_{ab}$ given in \eqref{defnlVb} and  \eqref{defnlWb}, but also \eqref{QV} and \eqref{QW}, that
\beqs
V_{ab} &=& -h^{(2)}_{ab} + 6\s^2 h_{ab}^{(0)} +2\s \s_{ab} -2\s_a \s_b +\s^c \s_c   h_{ab}^{(0)} \: ,\\
W_{ab} &=& \eps_a^{\;\, cd}D_c \left( h^{(2)}_{d b } -2 \s^2 h^{(0)}_{db} \right) \: .
\eeqs
From the expansions of the electric and magnetic parts of the Weyl tensor given in \eqref{E1}, \eqref{B1}, and \eqref{B2b}, one easily realizes that
\beqs
V_{ab}=E^{(2)}_{ab}-\s E^{(1)}_{ab}, \qquad W_{ab}=B^{(2)}_{ab}\: .
\eeqs

The equations of motion can be reformulated in terms of $B^{(2)}_{ab}$,  as firstly derived in \cite{B},
\bea
B^{(2)}_{a}{}^{a} &=& 0, \\
D^{a}B^{(2)}_{ab} &=& 0,\\
(\Box - 2)B^{(2)}_{ab} &=& \curl(M)_{ab}=- 4 \epsilon_{cd(a}\s^cE^{(1)}_{b)}{}^{d}, \label{B23eq}
\eea
but they can also be written in terms of $E^{(2)}_{ab} - \s E^{(1)}_{ab}$
\bea
(E^{(2)} - \s E^{(1)})_{a}{}^{a} &=& 0, \\
D^{a}(E^{(2)}_{ab} - \s E^{(1)}_{ab}) &=& 0,\\
(\Box - 2)(E^{(2)}_{ab} - \s E^{(1)}_{ab}) &=&  \curl [ -4  \epsilon_{cd(a}\s^cE^{(1)}_{b)}{}^{d}]. \label{E23eq}
\eea
Remark that the right hand side of \eqref{B23eq} is precisely the SDT tensor $X^{[3]}_{ab}$ defined in \eqref{X3}.
The duality properties of the tensors $V_{ab}$ and $W_{ab}$ written in \eqref{dualiprop} reduce to
\be\label{dualipropki}
E^{(2)}_{ab} - \s E^{(1)}_{ab} = \curl  B^{(2)}_{ab}, \qquad - \curl (E^{(2)}_{ab} - \s E^{(1)}_{ab}) = B^{(2)}_{ab} + 4  \epsilon_{cd(a}\s^cE^{(1)}_{b)}{}^{d}.
\ee

\setcounter{equation}{0}
\section{Linearization stability constraints}
\label{sec:LS}

Solutions of linearized equations are not always linearizations of solutions of non-linear equations. This phenomenon is well-known as a linearization instability \cite{Deser:1973zza,Moncrief:1975aa,Moncrief:1976un}.
The main result of \cite{BS} is that Einstein's equations can always be solved order by order provided a subpart of them, the Hamiltonian and momentum equations, are satisfied. The follow-up paper \cite{B} aims at removing this provision, i.e. solve the constraints. It is shown that\\
$ $\\
{\textit{
Given a spacetime of the Beig-Schmidt form with $k_{ab}=i_{ab}=0$, Einstein's vacuum equations can be solved to all orders if and only if the field $\s$ satisfies the field equation $(\Box+3)\s=0$, and is such that the six charges associated to Killing vectors 
\beqs
\mathcal Q [\xi_{(0)}] \equiv \oint_S d^2S\, \eps_{cd(a}\s^c E^{(1) \: d}_{\; \, b)} \xi^a_{(0)} n^b \; ,
\eeqs
where the integrand is a tensor that is conserved at infinity and built from quadratic terms of the first order field $\s$,  vanish.
}}\\

In the following, we refer to these six additional conditions imposed by Einstein's equations as the linearization stability constraints. They were named integrability conditions in \cite{B}.

This section is devoted to review how these six conditions arise and are generalized for our enlarged ansatz. These conditions will turn out to be crucial when discussing unicity of conserved charges in the next section. To derive these conditions, we start by reviewing properties of Killing vectors on $dS_3$ and by showing an important result, that we will use throughout our argumentation, which states that a charge constructed by contracting an SD tensor with a certain Killing vector is equivalent to a charge constructed by contracting the curl of this tensor with another particular Killing vector. This is the main result we will use in the next section to show that there exists two equivalent forms to describe conserved charges associated with boosts or rotations. It also tells us that charges associated to SD tensors that have a zero symmetrized curl are trivial.

\subsection{Properties of Killing Vectors on $dS_3$}
\label{vecKilling}

Three-dimensional de-Sitter space admits six Killing vectors. Three of them are rotations and the other three correspond to four-dimensional Lorentz boosts when interpreted in the asymptotically flat context. The rotations are
\bea
\xi^{a}_{\rom{rot}(1)} \partial_a &=& \partial_\phi, \\
\xi^{a}_{\rom{rot}(2)} \partial_a &=& - \sin \phi \partial_\theta - \cot \theta \cos \phi \partial_\phi,  \\
\xi^{a}_{\rom{rot}(3)} \partial_a &=& \cos \phi \partial_\theta - \cot \theta \sin \phi \partial_\phi\, .
\eea
These Killing vectors are precisely the three Killing vectors of the round two-sphere. On the round two-sphere, Killing vectors satisfy a special property: they can be written as
\be \xi^{m}_{\rom{rot}(k)} = \epsilon_{(2)}{}^{mn} D_{n}^{(2)} f_{(k)}, \label{rotexpl}
\ee
where $f_{(k)}$ are the three scalar $l = 1$ harmonics on the two-sphere
\bea
(\square^{(2)} + 2)f_{(k)} = 0,\qquad \int_S d^2 S f_{(k)} f_{(l)} = \frac{4\pi}{3}\delta_{(k)(l)},
\eea
given explicitly in \eqref{harmosphere}. We use the conventions $\epsilon_{(S^2)}{}_{\theta \phi} = \sin \theta$, $D_{a}^{(2)}$ is the unique torsion free covariant derivative on $S^2$,  $dS = \sin\theta d\theta \wedge d\phi$, and $\square^{(2)}$ is the scalar Laplacian on $S^2$.

The boost Killing vectors of the three-dimensional de-Sitter space can be written as
\bea
\xi^{a}_{\rom{boost}(i)} = f_{(i)} n^{a} + \cosh\tau \sinh\tau h_{(0)}^{ab}\partial_b f_{(i)} , \label{boostexpl}
\eea
or, explicitly, as
\bea
\xi^{a}_{\rom{boost}(1)} \partial_a &=& \cos \theta \partial_\tau - \tanh \tau \sin \theta  \partial_\theta, \\
\xi^{a}_{\rom{boost}(2)} \partial_a &=& \sin \theta \cos \phi \partial_\tau +  \tanh \tau \cos \theta \cos \phi  \partial_\theta - \tanh \tau  \csc \theta \sin \phi \partial_\phi,\\
\xi^{a}_{\rom{boost}(3)} \partial_a &=&  \sin \theta \sin \phi \partial_\tau +  \tanh \tau \cos \theta \sin \phi  \partial_\theta + \tanh \tau  \csc \theta \cos \phi \partial_\phi.
\eea
The unit vector normal to the two sphere in $dS_3$ is $n^{a}\partial_a = \partial_\tau$.

The boost Killing vectors are intimately related to the rotational Killing vectors by the following relation
\bea
\xi^{a}_{\rom{boost}(i)} &=& -\frac{1}{2}\eps^{abc} \DD_b \xi_{\rom{rot}(i)c},\\
\xi^{a}_{\rom{rot}(i)} &=& \frac{1}{2}\eps^{abc} \DD_b \xi_{\rom{boost}(i)c},
\eea
where $\DD_a$ is the covariant derivative on the hyperboloid and $\eps_{abc}$ the totally anti-symmetric tensor normalized as $\eps_{\tau \theta \phi} = +\cosh^2\tau \sin\theta$. The latter relation implies
\bea\label{Kill}
(\square +2)\xi^{a}_{\rom{rot}(i)} =0,\qquad (\square +2)\xi^{a}_{\rom{boost}(i)} =0,
\eea
where $\square = \DD^a\DD_a$.

Now, let us see how these relations imply that a charge constructed by contracting an SD tensor with a rotational (respectively boost) Killing vector is equivalent to the charge constructed by contracting the curl of this SD tensor with a boost (respectively rotational) Killing vector, or alternatively said how the relations between boost and rotational Killing vectors imply two equivalent forms for the conserved charges associated with these vectors.

Given a tensor  $T_{ab}$ without special properties, one can show that on the $\tau = 0$ slice of de Sitter space,
\bea
T_{ab}\xi_{\rom{rot} (i)}^a n^b &=& \eps_a^{\; cd}\DD_c T_{db}\xi^a_{\rom{boost}(i)}n^b+\DD^{(S^2)}_c ( T_{ab}\eps_{(S^2)}^{ac} f_{(i)} n^b)\, .\label{proprotboost}
\eea
For any symmetric and divergence-free tensor, one has $$\DD^b (T_{ab}\xi_{\rom{rot}(i)}^a) = 0, \qquad \qquad \DD^b ((\curl T)_{(ab)}\xi_{\rom{rot}(i)}^a) = 0.$$ Note that $(\curl T)_{ab}$ is symmetrized in the second equation, as it is not necessarily symmetric. Therefore, for any regular symmetric and divergence-free tensor, the conserved charges
\bea
Q[T_{ab},\xi_{\rom{rot}(i)}^a]\equiv \int_S d^2 S \:  n^b T_{ab}\xi_{\rom{rot}(i)}^a\: ,
\eea
can be expressed in two equivalent ways as follows
\bea
Q[T_{ab},\xi_{\rom{rot}(i)}^a]= \int_S d^2 S \:  n^b T_{ab}\xi_{\rom{rot}(i)}^a = \int_S d^2 S \:  n^b \curl(T)_{(ab)} \xi_{\rom{boost}(i)}^a \, . \label{eq:T1}
\eea

Replacing $T_{ab}$ by the curl of $T_{ab}$ in identity \eqref{proprotboost}, we get on the $\tau = 0$ slice
\be
T_{ab}\xi_{\rom{boost}(i)}^b n^a =- \curl( T)_{ab}\xi^b_{\rom{rot}(i)}n^a+\DD^{(S^2)}_c (2\DD_{[d}T_{c]b} n^d n^b f_{(i)})+(\curl(\curl T) +T)_{ab}n^a \xi_{\rom{boost} (i)}^b. \nonumber
\ee
For any  symmetric and divergence-free tensor, we have $(\curl(\curl T) +T )_{ab}= (\square - 2)T_{ab} + h_{ab}^{(0)}T$, and
\bea
\left( (\square - 2)T_{ab} + 2 h_{ab}^{(0)} T \right) \xi_{(0)}^c = 2 \DD^a \left( \xi_{(0)}^c \DD_{[a} T_{b]c}+ T_{c[a}\DD_{b]}\xi_{(0)}^c \right).
\eea
Therefore, one obtains
\bea
\int_S d^2 S \: n^b T_{ab}\xi_{\rom{boost}(i)}^a =- \int_S d^2 S \:  n^b \curl(T)_{ab} \xi_{\rom{rot}(i)}^a - \int_S d^2 S \,  T \xi^{\rom{boost}(i)}_a n^a  \, .
\eea
For a tensor $T_{ab}$ whose trace is non-vanishing, $\curl(T)_{ab}$ is not symmetric in general. Decomposing into symmetric and anti-symmetric parts and using integrations by parts, the following conserved charges
\bea
Q[T_{ab},\xi_{\rom{boost}(i)}^a]\equiv \int_S d^2 S \:  n^b T_{ab}\xi_{
\rom{boost}(i)}^a , 
\eea
can also be expressed in two equivalent forms
\bea
Q[T_{ab},\xi_{\rom{boost}(i)}^a]= \int_S d^2 S \:  n^b T_{ab}\xi_{
\rom{boost}(i)}^a =- \int_S d^2 S \:  n^b \curl(T)_{(ab)} \xi_{\rom{rot}(i)}^a  \, .\label{eq:T2}
\eea
In establishing this, we used $\curl(T)_{[ab]}= -\frac{1}{2} \epsilon_{abc} \DD^c T$.

To summarize, we have shown in equations (\ref{eq:T1}) and (\ref{eq:T2}) that charges associated with any symmetric and divergence-free tensor are equivalent to charges associated with the symmetrized curl of this tensor. In particular, this means that charges constructed using an SDT tensor are equivalent to charges constructed using the curl of this SDT tensor. Also, charges constructed with symmetric and divergence free tensors that have zero symmetrized curl are automatically zero.

\subsection{Linearization stability constraints when $k_{ab}=i_{ab}=0$}

In the case where $k_{ab}=i_{ab}=0$, we have seen in section \ref{subsec:second} that the second order equations  can be written in the following compact form
\bea
B^{(2)}_{a}{}^{a} &=& 0, \\
D^{a}B^{(2)}_{ab} &=& 0,\\
(\Box - 2)B^{(2)}_{ab} &=& \curl(M)_{ab}=- 4 \epsilon_{cd(a}\s^cE^{(1)}_{b)}{}^{d} \; .\label{B23eqb}
\eea

The presence of six necessary conditions, or obstructions, to the existence of non-linear solutions, constructed from given linear solutions, can be seen as follows. Contracting equation \eqref{B23eqb} with a Killing vector on $dS_3$, one can rewrite the l.h.s of the expression, upon using the equations of motion and \eqref{Kill}, as
\be
(\Box-2)B^{(2)}_{ab}  \xi_{(0)}^{a} =2\DD^{a}\left( \xi_{(0)} ^{c}\DD_{[a}B^{(2)} _{b]c}+B^{(2)} _{c[a}\DD_{b]}\xi_{(0)} ^{c}\right),
\label{Beig30}
\ee
which is a total divergence and vanishes when integrated on a Cauchy surface $S$ on the unit hyperboloid. For consistency, we must require that the integrals on the sphere of the r.h.s of \eqref{B23eqb} contracted with $\xi_{(0)}^{a}$ are also zero. These requirements are precisely Beig's integrability conditions
\bea
\mathcal Q [\xi_{(0)}] \equiv \oint_S d^2S\, \eps_{cd(a}\s^c E^{(1) \: d}_{\; \, b)} \xi^a_{(0)} n^b =0\; , \label{intO}
\eea
 where $\xi_{(0)}^a$ are the six Killing vectors on the hyperboloid, $S$ is a Cauchy surface in the unit hyperboloid, and $n^{a}$ is a unit timelike vector normal to $S$ in the unit hyperboloid.

Before proceeding, let us present a new way of looking at these integrability conditions. It is clear that the equation for the mass aspect $\sigma$, which was given by
\beqs
(\Box+3)\s=0,
\eeqs
can be derived from the free scalar Lagrangian $L^{(\sigma)}$
\bea
L^{(\sigma)} = \sqrt{-h^{(0)}} \Big( -\frac{1}{2} \partial_a \s \partial^a \s +\frac{3}{2}\s^2 \Big),\label{Ls}
\eea
with mass $m^2 = -3$ on three-dimensional de Sitter space. Now, it is interesting to note that Beig's integrability conditions are precisely the conditions that all six Noether charges derived from this Lagrangian vanish. Indeed, one has $ \eps_{cd(a}\s^c \s^d_{\;\, b)}  = -  (\curl \kappa)_{(ab)} $ where $\kappa_{ab}$ is precisely the stress-tensor of $L^{(\sigma)}$
\bea
T^{(\s)}_{ab} &\equiv& - \frac{2}{\sqrt{-h^{(0)}}} \frac{\delta L^{(\s)}}{\delta h^{(0)\: ab}} =\kappa_{ab} = - \frac{1}{2} \s^c
\s_c h^{(0)}_{ab} + \s_a \s_b + \frac{3}{2} \s^2 h^{(0)}_{ab} .
\eea
Note that this tensor is also an SD tensor and that following our classification it can be identified as a linear combination of the ($\s,\s$) SD tensors defined in \eqref{k1} and \eqref{k2}
\beqs
\kappa_{ab}=-\frac{1}{4} \kappa^{[\s,\s,I]}_{ab}+\frac{1}{2} \kappa^{[\s,\s,II]}_{ab}  .
\eeqs
Because the charges constructed using  the symmetrized curl of an SD tensor contracted with a Killing vector or with the SD tensor himself are equivalent as we have just shown in the previous section, the linearization stability constraints reduce to
\be
\oint_S d^2S \: T^{(\s)}_{ab}\: \xi_{(0)}^a n^b= \oint_S d^2S \, \kappa^{[\s,\s,I]}_{ab}\xi_{(0)}^a n^b =   0.\label{BeigIntk}
\ee
where we also used the fact that $\kappa^{[\s,\s,II]}_{ab}$ has a trivial symmetrized curl and is thus associated to trivial charges.

Let us mention that it was also understood in \cite{B} that charges constructed with $X_{ab}\equiv \eps_{cd(a}\s^c \s^d_{\;\, b)} $ contracted with a conformal Killing vector $\omega^a$, a translation which satisfies $\omega^{ab}+\omega h^{(0)\:ab}=0$, also vanish as
\beqs\label{Xabb}
X_{ab} \omega^a =\DD^d \biggr [ \epsilon_{cdb} \:  \sigma  \: E^{(1)\:c}_a \omega^a-\frac{1}{4} \: \epsilon_{bad} (\sigma_c \sigma^c +\sigma^2) \omega^a \biggl ],
\eeqs
can be written as a total divergence. 

At this point, we should warn the reader that the above construction only presents the linearization stability constraints as  \textit{necessary} conditions. It was shown in \cite{B} that these conditions are also \textit{sufficient} to solve Einstein's equations to all orders in the expansion. The general idea of this construction is to split the linear part in a $2+1$ decomposition and keep non linear terms general. Then, a study of the harmonic decomposition of those equations revealed that only six conditions are to be imposed on these general non-linear terms, conditions that appear at second order in the expansion. These conditions for the system of equations to have a solution are thus equivalent to the necessary conditions previously reviewed. Note that there are minor typos in equations (25) and (43) of \cite{B}. In (25), the indices are incoherent and have been corrected in our formula \eqref{Xabb}, while in (43) the coefficient multiplying $\beta_a$ is $(n-1)^2$ instead of $(n+1)^2$.

\subsection{Generalized linearization stability constraints}

The generalization of these constraints to the case where $k_{ab}$ and $i_{ab}$ are non-zero is straightforward. Indeed, from the general second order equation
\bea
(\square - 2)W_{ab} &=& \curl(2 i + \kappa )_{(ab)} \, ,
\eea
we see that the integrability conditions are
\beqs\label{integrability1}
\oint_S d^2S \: i_{ab} \xi^a_{(0)} n^b=-\frac{1}{2} \oint d^2 S \:  \kappa_{ab} \: \xi^{a}_{(0)} n^b ,
\eeqs
 where
 \beqs
\kappa_{ab} &=& \kappa^{[\s,\s,I]}_{ab} - \kappa^{[2,\s,\s,II]}_{ab}
+ 4 \kappa^{[\s,k, I]}_{ab} + Y^{(2)}_{ab}+ \kappa^{[k,k,I]}_{ab}  \label{defkappa}
+ \kappa^{[2,k,k,III]}_{ab} \: .
\eeqs

Let us try to simplify those constraints as much as we can by identifying the currents, constructed out of SD tensors contracted with Killing vectors, that can be written as total derivatives. 
We can already get rid of  $\kappa^{[2,\s,\s,II]}_{ab}$ and $\kappa^{[2,k,k,III]}_{ab} $ as these SD tensors have a vanishing symmetrized curl\footnote{One could also check that these tensors contracted with Killing vectors can be written as total derivatives.} and are thus associated to trivial charges. Now, we see that the two currents associated with the two independent $(\s,k)$ SD tensors are total divergences.  To prove this efficiently, one first needs to check that the current $\kappa^{[\s,k,I]}_{ab} \xi^b$ can be expressed as a total divergence
\beqs\label{k1sk2}
\kappa^{[\s,k,I]}_{ab} \xi^b =\DD^b \Big( -\xi_{[a} \: k_{b]}^{\:\:\:c} \s_c +\DD^c \xi_{[a} \: \s k_{b]c} +\xi^c \: \s \DD_{[a} k_{b]c} +\xi^c \: \s_{[a} k_{b]c}   \Big) \; .
\eeqs
This also implies that its symmetrized curl, the SDT tensor  $Z^{(3)}_{ab}$, defined in \eqref{Z3}, will not contribute either. Eventually, from the definition of $\kappa^{[\s,k,II]}_{ab}$ given in (\ref{k3sk}) and the following result
\beqs
\Big(  2 B^{(1)}_{c(a} \s_{b)}^{\:\:\:c} -\s^{cd} B^{(1)}_{cd} \: h^{(0)}_{ab} +2 \s B^{(1)}_{ab} \Big) \xi^b =  2    \DD^b \Big( \DD^c \xi_{[a} \: \s B^{(1)}_{b]c} -\xi_{[a} \: B^{(1) \: c}_{b]}  \s_c +\xi^c \: \s_{[a} B^{(1)}_{b]c}   \Big) ,\nn\\
\eeqs
we see that $\kappa^{[\s,k,II]}_{ab}$ contracted with a Killing vector can also be expressed as a total divergence.

In the same line of thoughts, for the terms quadratic in $(k,k)$, one can show by inspection that
\beqs\label{int1}
&&\DD^b \Big( 2 \xi_c \: k_{d[a} \DD^d k_{b]}^{\:\:\:c} -\DD_c \xi_d \: k^{c}_{\:\:[a} k_{b]}^{\:\:\:d} \Big) \nonumber \\
&&\qquad = \Big( \DD_c k_{d(a} \DD^d k_{b)}^{c}+4 k_{c(a} k_{b)}^{\:\:\:c}-k^{cd} \DD_c \DD_d k_{ab} \Big) \xi^b \nonumber \\
&&\qquad = \Big( 7 \:   B^{(1)}_{cd}  B^{(1) \: cd} \: h^{(0)}_{ab} - 6\:  B^{(1)}_{c(a}   B^{(1) \: c}_{b)}  + 4\: \epsilon_{cd(a}  \DD^c k_{b)}^{\;\:e}  B^{(1) \: d}_{e}  + 4\: \epsilon_{cd(a}   \DD_{b)} B^{(1) \: c}_{e}   k^{de}  \nonumber \\
 && \qquad \qquad   + \:  \DD_{(a} k^{cd} \DD_{b)} k_{cd}  - \:  k^{cd}  \DD_{(a} \DD_{b)}    k_{cd}   + 4 \: k^c_{\: (a} k_{b)c} +  \: k_{cd} k^{cd} h^{(0)}_{ab}  \Big) \xi^b \nonumber \\
 &&\qquad = \Big( - 2 Y^{(2)}_{ab}- 4 \kappa^{[k,k,I]}_{ab} - 14  \kappa^{[k,k,II]}_{ab} -8 \kappa^{[k,k,III]}_{ab}\Big)  \xi^b \: ,
\eeqs
and also
\beqs\label{int2}
&&\DD^b \Big( \DD_c \xi_{[a} \: k_{b]}^{\:\:\:d} k^c_{\:\:d} -\xi_{[a} \: \DD^c k_{b]}^{\:\:\:d} k_{cd} -\xi^c  \: k^d_{\:\:[a} \DD_{b]} k_{cd} +\xi^c \: k_{cd} \DD_{[a} k_{b]}^{\:\:\:d}\Big) \nonumber \\
&&\qquad = \Big( -2  \: k^c_{\: (a} k_{b)c} -\frac{3}{2}  \: k_{cd} k^{cd} h^{(0)}_{ab} -\frac{1}{2} \DD_c k_{de} \DD^d k^{ce} h^{(0)}_{ab} \nonumber \\
 && \qquad \qquad + \DD^c k_{(a}^{d} \DD_{b)} k_{cd} -\DD_c k_{d(a} \DD^c k_{b)}^{d}+k^{cd} \DD_c \DD_{(a} k_{b)d} \Big) \xi^b \nonumber \\
&&\qquad = \Big(-5 \:   B^{(1)}_{cd}  B^{(1) \: cd} \: h^{(0)}_{ab} +6 \:  B^{(1)}_{c(a}   B^{(1) \: c}_{b)}  -2 \: \epsilon_{cd(a}  \DD^c k_{b)}^{\;\:e}  B^{(1) \: d}_{e}  -2 \: \epsilon_{cd(a}   \DD_{b)} B^{(1) \: c}_{e}   k^{de}  \nonumber \\
 && \qquad \qquad  + \:  k^{cd}  \DD_{(a} \DD_{b)}    k_{cd}   -\frac{1}{2}\:   \DD_{c} k_{de}  \DD^c   k^{de} h^{(0)}_{ab}   -2\: k^c_{\: (a} k_{b)c} -\frac{5}{2} \: k_{cd} k^{cd} h^{(0)}_{ab}  \Big) \xi^b \nonumber \\
&&\qquad = \Big( Y^{(2)}_{ab}+ 2 \kappa^{[k,k,I]}_{ab}+ 10 \kappa^{[k,k,II]}_{ab} + 8 \kappa^{[k,k,III]}_{ab} \Big)  \xi^b \: .
\eeqs
These two equations show that  the current constructed out of $\kappa^{[k,k,II]}_{ab}$ can be written as a total divergence and that the equality
\beqs\label{papapa}
Y^{(2)}_{ab} \: \xi^{(0)\:b}=-2 \kappa^{[k,k,I]}_{ab} \: \xi^{(0)\:b} ,
\eeqs
is true up to a total divergence.

With all these results in hand, it is now  easy to see that the linearization stability constraints \eqref{integrability1} reduce to
\beqs
\oint_S d^2S \: i_{ab} \xi^a_{(0)} n^b=-\frac{1}{2} \oint d^2 S \: \Big( \kappa^{[\s,\s,I]}_{ab} + \frac{1}{2}  Y^{(2)}_{ab} \Big) \: \xi^{a}_{(0)} n^b .
\eeqs

In comparison with the analysis presented in the previous section, one realizes that the equations of motion for $k_{ab}$ can be derived from the Lagrangian $L^{(k)}$
\beqs
L^{(k)}=\sqrt{-h^{(0)}} \Big( \frac{1}{4} \: B^{(1)}_{ab} B^{(1)\:ab}\Big),
\eeqs
whose six associated Noether charges are
\beqs
T^{(k)}_{ab} \equiv -  \frac{2}{\sqrt{-h^{(0)}}} \frac{\delta L^{(k)}}{\delta h^{(0) \: ab}}= \frac{1}{2} \kappa^{[k,k,II]}_{ab} -\frac{1}{8} Y^{(2)}_{ab} . \label{Tk}
\eeqs
The linearization stability constraints can thus be written in the more elegant form
\bea
\int_S d^2S\,  i_{ab} \: \xi_{(0)}^a n^b =  2\:  \int_S d^2S  \, \Big(  T^{(\s)}_{ab} + T^{(k)}_{ab} \Big) \xi_{(0)}^a n^b , \label{intFF}
\eea
where
\bea
T^{(\s)}_{ab} &\equiv& - \frac{2}{\sqrt{-h^{(0)}}} \frac{\delta L^{(\s)}}{\delta h^{(0)\: ab}} = -\frac{1}{4} \kappa^{[\s,\s,I]}_{ab} +\frac{1}{2} \kappa^{[\s,\s,II]}_{ab} , \label{Ts} \\
T^{(k)}_{ab} &\equiv& -  \frac{2}{\sqrt{-h^{(0)}}} \frac{\delta L^{(k)}}{\delta h^{(0) \: ab}}
= \frac{1}{2} \kappa^{[k,k,II]}_{ab} -\frac{1}{8} Y^{(2)}_{ab} , \label{Tk}
\eea
are the stress-tensors associated to  $L^{(\s)}$ and $L^{(k)}$.

Although the realization that the integrability conditions can be expressed using the stress-tensors of specific actions for the first order fields may sound like a curiosity at this stage, we will show in section \ref{paritycond} that they play an important role in the discussion of a good variational principle.

One important thing that comes out of this analysis is that, when $i_{ab}$ and $k_{ab}$ are non-zero, we can build twelve possibly non-trivial and independent Noether charges out of quadratic expressions of the first order quantities. If we set $i_{ab}$ to zero, the integrability conditions impose that only six of them are independent.

\setcounter{equation}{0}
\section{Conserved charges from the equations of motion}
\label{sec:conservedd}

In this section we would like to consider all the possible conserved charges that can be constructed from SD tensors contracted with a Killing vector or SDT tensors with a conformal Killing vector. As we have seen in the previous sections, some charges are trivial by construction.  Indeed, charges constructed from an SD tensor with a trivial symmetrized curl contracted with a Killing vector are vanishing. Similarly, some currents built out of a specific SD tensor, respectively SDT tensor, contracted with a Killing vector, respectively a conformal Killing vector, can be expressed as total divergences.   Also, we have seen that some quantities should be restricted to zero when one imposes the equations of motion. All these results will obviously severely restrict the possible independent conserved charges one can construct.

We will see in the following that, in the particular case where $k_{ab}=i_{ab}=0$, only ten independent charges can be defined when the equations of motion are taken into account. These can be identified as the ten Poincar\'e charges. In this section, we will focus on the case $k_{ab}=i_{ab}=0$ and just give some comments on the construction of charges for our enlarged boundary conditions. This analysis in the general case will be presented in the next chapter.

\subsection{First order:  momenta and dual momenta}
\label{subsec:lemma45}
Since
\beqs
\DD^b(T_{ab} \zeta^a_{(i)})= \DD^b T_{ab} \zeta^a_{(i)} - T_a^a \zeta_{(i)},
\eeqs
for a symmetric tensor $T_{ab}$ and a conformal Killing vector $\zeta^a_{(i)}$, i.e. a translation, we need to consider SDT tensors $T_{ab}$ if we want to define charges associated to translations that are conserved.  
 
To start our analysis, the first thing to realize is that, given a symmetric tensor $U_{ab}$, we have
\bea
\curl(U)_{ab} \zeta_{(i)}^a n^b = \DD_c \left( \eps_b{}^{\; cd}U_{ad} \zeta_{(i)}^a \right) n^b. \label{curlzeta}
\eea
The r.h.s of this equation is a total divergence on the two-sphere. Let us now state the two following lemmas that we prove in Appendix \ref{app:proofs}.

\begin{lemma} \label{newlemma} On the three dimensional hyperboloid, any symmetric traceless and divergence-free tensor can be decomposed as
\bea
T_{ab} = \curl(\tilde T_{ab}) + \DD_a \DD_b \hat \zeta+h_{ab}^{(0)}\hat \zeta\; ,
\eea
where $\tilde T_{ab}$ is a symmetric, traceless and divergence-free tensor and $\hat\zeta$ is a combination of the four functions \eqref{zetahat}.
\end{lemma}

\begin{lemma} \label{higheroplemma} On the hyperboloid, any regular symmetric divergence-free traceless tensor $T_{ab}$ obeying $(\square +n^2 -2n-2 ) \, T_{ab}=0$ with $n$ any integer $n \geq 3$ also obeys
\bea
\int_{S}  T_{ab}\zeta^a_{(l)} n^b d^2S  = 0, \; \; l=0,1,2,3   , \qquad \int_{S}  T_{ab}\xi^a_{(0)} n^b d^2S = 0  \, ,
\eea
where  $\zeta^a_{(l)} = D^a \zeta_{(l)}$ are the four translation Killing vectors (conformal Killing vectors on $dS_3$) with $\zeta_{(l)}$ given in \eqref{zetas} and $\xi^a_{(0)}$ are the six Killing vectors on $dS_3$.
\end{lemma}

From Lemma \ref{newlemma} and \eqref{curlzeta}, it then follows that  charges constructed from SDT tensors $T_{ab}$ contracted with translations are simply associated with the coefficients of the four lowest harmonics $\hat \zeta_{(i)}$ given in \eqref{zetahat},
\bea\label{zetahatt}
Q[T_{ab},\zeta_{(i)}^a]\equiv\int d^2S T_{ab}\zeta_{(i)}^a n^b = \int d^2S (\DD_a \DD_b \hat \zeta + h_{ab}^{(0)}\hat \zeta ) \zeta_{(i)}^a n^b\, .
\eea
This readily means that there are only four such charges and the only possibility is\footnote{ Another way, to see this, is as follows. From Lemma \ref{lemma8} and the linearization stability constraints, we can check that any SDT tensor satisfying $(\Box+n^2-2n-2)T_{ab}=0$ with $n=2$, such as $E^{(2)}_{ab}-\s E^{(1)}_{ab}$ and $B^{(2)}_{ab}$, is associated to a trivial charge  when contracted with a conformal Killing vector. This is also due to the fact that any other SDT tensor built out of quadratic combinations in $\s$ can be expressed as the curl of $\kappa_{ab}^{[\s,\s,I]}$ and is thus not contributing because of \eqref{curlzeta}. Eventually, from Lemma \ref{higheroplemma}, we see that all higher order SDT tensors  including $h^{(3)}_{ab}$, $h^{(4)}_{ab}$ would also be associated to trivial charges. We are thus left considering tensors made out of linear quantities in $\s$ and the only possibility is $E^{(1)}_{ab}$}
\beqs\label{momentaa}
Q[\zeta^a_{(i)}]= -\frac{1}{8\pi G} \oint_S d^2 S \: E^{(1)}_{ab} \: n^a \: \zeta^b_{(i)}\: .
\eeqs
These expressions are precisely the ones derived by R. Geroch \cite{Geroch:1972up} (see also \cite{Geroch:1977jn}), Ashtekar-Hansen \cite{Ashtekar:1978zz}, Ashtekar-Romano \cite{Ashtekar:1991vb}. 

These were shown in \cite{Ashtekar:1984aa} (see also \cite{ABR}) to agree with the ADM momenta \cite{Arnowitt:1961zz}, which are also equivalent to the Regge-Teitelboim expressions. For the energy, using our Appendix \ref{app:BC} which establishes a link between our covariant boundary conditions and the boundary conditions of Regge-Teitelboim, we see that the ADM definition is just 
\beqs
E= \frac{1}{4\pi} \oint d^2 S \: \s\; ,
\eeqs
where $\s$ is our first order field which is often referred to as the mass aspect for this particular reason. One can now check that this expression is equivalent to 
\beqs
E=-\frac{1}{8\pi} \oint_S d^2 S \: E_{ab} \:  \xi^a_0 \:  n^b \; ,
\eeqs
where $\xi^a_0$ is a time-unit translation. Indeed, in \cite{Ashtekar:1984aa}, it it shown that if we pick $S$ to be an extremal slice where the extrinsic curvature vanishes, we have $\xi^a_0=n^a$ and
\beqs
E_{ab} \: \xi^a_0 \:  n^b= - n^a n^b\DD_a \DD_b \s -\s n_a n^a= -\partial_\tau^2 \s +\s = - \Box^{(2)} \s -2 \s \; ,
\eeqs
 where in the last equality we made use of the equation of motion of $\s$ to write $
 \Box^{(3)} \s +3 \s= (-\partial_\tau^2  +\Box^{(2)}+3) \s=0$. 
Then, the result follows immediately 
 \beqs
E=-\frac{1}{8\pi} \oint E_{ab} \:  \xi^a_0 \:  n^b  d^2S= \frac{1}{4\pi} \oint (\frac{1}{2} \Box^{(2)} \s + \s) d^2S=  \frac{1}{4\pi} \oint  d^2S \: \s \; .
\eeqs
The same could be done for the space translations along the same lines.

 Before moving to the classification of charges associated to Killing vectors, let us comment on the case where $k_{ab}$ is allowed to take non-trivial values.  Indeed, at first sight, one could enjoy constructing charges of the form (see \cite{ramaswamy,ashtekar})
 \beqs\label{dualmomenta}
 Q[\zeta^a_{(i)}]= -\frac{1}{8\pi G} \oint_S d^2 S \: B^{(1)}_{ab} \: n^a \: \zeta^b_{(i)}\: .
 \eeqs
From Lemma \ref{Ash}, we know that $B^{(1)}_{ab}$ can always be expressed as $
B^{(1)}_{ab}=-\s^D \s^D_{ab}-\s^D h^{(0)}_{ab}$.
From Lemma \ref{BS}, it can also be derived from a regular potential $k_{ab}$ if $\s^D$ does not contain the four lowest harmonics $\hat{\zeta}^{a}_{(i)}$. However, we just saw that these are the only harmonics that can contribute to \eqref{dualmomenta}. These charges are thus trivial for a regular $k_{ab}$. The following Lemma is a generalization of Lemma \ref{BS} that circumvents the restriction on the four lowest hyperbolic harmonics to define $B^{(1)}_{ab}$ in terms of a tensor potential. However, we see that it implies that $k_{ab}$ should develop wire singularities if it has to reproduce the correct value of $B^{(1)}_{ab}$ using $\s^D=\hat{\zeta}^a$. These singularities are of the same type as the Misner-string singularities we discuss in Part II of this thesis. 

\begin{lemma}\label{newlemma2} On the three dimensional hyperboloid, any scalar $\Phi$ satisfying $\square \Phi + 3\Phi = 0$ defines a symmetric, traceless, curl-free and divergence-free tensor $T_{ab} = D_a D_b \Phi + h^{(0)}_{ab}\Phi$ which can be written as
\begin{equation}
T_{ab} = \epsilon_{a}^{\;\, cd}D_c P_{d b},\label{eqkkk}
\end{equation}
where $P_{ab}$ is a symmetric, traceless tensor of the form
\bea
P_{ab} = \sum_{\mu=0}^3 N_{(\mu)} k^{(\mu)}_{ab} + P_{ab}^{reg},
\eea
where $P_{ab}^{reg}$ is regular and $k^{(\mu)}_{ab}$ are four singular tensors listed here below.
\end{lemma}
The regular tensors $P_{ab}^{reg}$ are the same as the ones described in Lemma 2 (see also \cite{BS}).
The four singular tensors $k_{ab}^{(\mu)}$ can be derived by integrating equation \eqref{eqkkk} for $\Phi = \hat \zeta_{(\mu)}$. 
They can be written in the traceless gauge $h_{(0)}^{ab}k_{(\mu)ab} = 0$ as
\beqs
k_{(0)ab} &=& \left( \begin{array}{ccc}  0&0 & 2 \frac{{\hat k}-\cos\theta}{\cosh\tau}\\ 0 &0 & \sinh\tau \frac{\cos{2\theta}-4{\hat k} \cos\theta +3}{2\sin\theta} \\ 2 \frac{{\hat k}-\cos\theta}{\cosh\tau} &  \sinh\tau \frac{\cos{2\theta}-4{\hat k} \cos\theta +3}{2\sin\theta}&0 \end{array}\right),\nonumber \\
k_{(1)ab} &= &\left( \begin{array}{ccc}  0&0 & -3 \frac{\tanh\tau}{\cosh\tau}\sin^2\theta\\ 0 &0 & \frac{a}{4\sin\theta} \cosh\tau\\ -3 \frac{\tanh\tau}{\cosh\tau}\sin^2\theta &  \frac{a}{4\sin\theta} \cosh\tau  &0 \end{array}\right),\label{kiab} \\
k_{(2)ab} &= &\left( \begin{array}{ccc} 0& 3 \frac{\tanh\tau}{\cosh\tau}\sin\phi   & 3 \frac{\tanh\tau}{\cosh\tau}\cos\theta \sin\theta \cos\phi \\ 3 \frac{\tanh\tau}{\cosh\tau}\sin\phi & -\frac{a}{2\sin^3\theta } \cosh\tau \sin\phi & \frac{b}{\sin^2\theta}\cosh\tau\cos\phi \\ 3 \frac{\tanh\tau}{\cosh\tau}\cos\theta \sin\theta \cos\phi &   \frac{b}{\sin^2\theta}\cosh\tau\cos\phi & \frac{a}{2\sin\theta}\cosh\tau \sin\phi \end{array}\right),\nonumber \\
k_{(3)ab} &= &\left( \begin{array}{ccc} 0& -3 \frac{\tanh\tau}{\cosh\tau}\cos\phi   & 3 \frac{\tanh\tau}{\cosh\tau}\cos\theta \sin\theta \sin\phi \\ -3 \frac{\tanh\tau}{\cosh\tau}\cos\phi & \frac{a}{2\sin^3\theta } \cosh\tau \cos\phi & \frac{b}{\sin^2\theta}\cosh\tau\sin\phi \\ 3 \frac{\tanh\tau}{\cosh\tau}\cos\theta \sin\theta \sin\phi &   \frac{b}{\sin^2\theta}\cosh\tau\sin\phi & -\frac{a}{2\sin\theta}\cosh\tau \cos\phi \end{array}\right)\; ,\nonumber
\eeqs
where $a=-8{\hat k}+9\cos\theta - \cos 3\theta , b= \cos^4\theta-4{\hat k} \cos\theta +3$.
These tensors are regular in the north patch upon choosing ${\hat k} = +1$ and in the south patch upon choosing ${\hat k} = -1$. They are transverse and obey the equation
\bea
(\square -3)k_{(\mu) ab} = 0\; ,
\eea
outside of the singularities. The singular transition functions between the south and north patches can be written as

\begin{eqnarray}
\delta k_{(0)ab} \equiv k_{(0)ab}|_{South} - k_{(0)ab}|_{North}  &=& \left( \begin{array}{ccc}  0&0 & -\frac{4}{\cosh\tau} \\ 0&0 &4\cot\theta \sinh\tau \\ -\frac{4}{\cosh\tau} & 4\cot\theta \sinh\tau & 0 \end{array}\right),\nonumber \\
\delta k_{(1)ab} \equiv k_{(1)ab}|_{South} - k_{(1)ab}|_{North}  &=& \left( \begin{array}{ccc} 0 &0 & 0\\ 0&0 & 4\frac{\cosh\tau}{\sin\theta} \\ 0& 4\frac{\cosh\tau}{\sin\theta}& 0\end{array}\right),\nonumber \\
\delta k_{(2)ab} \equiv k_{(2)ab}|_{South} - k_{(2)ab}|_{North}  &=& \left( \begin{array}{ccc} 0 &0 &0 \\0 & -\frac{8}{\sin^3\theta}\cosh\tau \sin\phi & \frac{8 \cos\theta}{\sin^2\theta}\cosh\tau \cos\phi \\ 0&\frac{8 \cos\theta}{\sin^2\theta}\cosh\tau \cos\phi  &  \frac{8}{\sin\theta} \cosh\tau \sin\phi \end{array}\right),\nn \\
\delta k_{(3)ab} \equiv k_{(3)ab}|_{South} - k_{(3)ab}|_{North}  &=& \left( \begin{array}{ccc} 0 &0 &0 \\0 & \frac{8}{\sin^3\theta}\cosh\tau \cos\phi & \frac{8 \cos\theta}{\sin^2\theta}\cosh\tau \sin\phi \\ 0&\frac{8 \cos\theta}{\sin^2\theta}\cosh\tau \sin\phi  &  -\frac{8}{\sin\theta} \cosh\tau \cos\phi \end{array}\right).\nonumber
\end{eqnarray}
These transition functions obey
\begin{eqnarray}
D_{[a} \delta k_{(\mu) b]c} = 0, \qquad (\square - 3) \delta k_{(\mu)ab} = 0,\qquad  h^{(0)\: ab}\delta k_{(\mu)ab}=0,\qquad D^b \delta k_{(\mu)ab}=0, \label{propdeltak}
\end{eqnarray}
on the hyperboloid outside the singular region $\theta =0$ and $\theta =\pi$ and obey the normalized orthogonality relations
\bea
\int_0^{2\pi} d\phi \;\delta k_{(\mu) \phi a}D^a \zeta_{(\nu)} = - 8\pi \; \delta_{(\mu)(\nu)},\qquad \mu,\nu=0,\dots 3\, ,\label{orthok}
\eea
where $\zeta_{(\mu)}$ are the four solutions of $\DD_a \DD_b \zeta_{(\mu)} + h_{ab}^{(0)}\zeta_{(\mu)}=0$ 
which are odd under parity-time reversal and normalized such that $\zeta_{(\mu)}\p_\rho + \rho^{-1}\p^a \zeta_{(\mu)} \p_a +o(\rho^{-1})= \p_\mu$ where $\p_\mu = \p_t,\, \p_i$.

\subsection{Second order: Lorentz charges}

In this section, we present a general construction of conserved Lorentz charges. Our approach is to construct these charges using SD tensors as 
\beqs
\DD^b (T_{ab} \xi^a)= 0 \; ,
\eeqs
for Killing vectors $\xi^a$ which satisfy $\DD^{(a} \xi^{b)}=0$, when $T_{ab}$ is an SD tensor which does not need to be moreover traceless. Since Killing vectors corresponding to asymptotic Lorentz transformations are larger at infinity than translations, corresponding conserved tensors are constructed both from the leading and the next-to-leading terms in the Beig-Schmidt expansion.  These tensors are linear in the next-to-leading terms and quadratic in the leading terms. What we readily show is that for $k_{ab}=i_{ab}=0$, only six independent non-trivial conserved charges can be associated to Killing vectors. These six charges must thus agree with the six Lorentz charges.  

The first thing to realize is that the SDT tensors $E^{(1)}_{ab}$ and $B^{(1)}_{ab}$, linear in the first order fields, have a trivial symmetrized curl and are thus associated to trivial charges. Following previous results, SD tensors quadratic in the first order field $\s$ will not contribute either.  This is obviously true for SD tensors whose symmetrized curls are zero. For SD tensors with non-trivial symmetrized curl, this is proved using the same argument as before and using the linearization stability constraints,  which state that charges associated to $\kappa^{[\s,\s,I]}_{ab}$ must be set to zero. 

We are thus left considering the tensors $E^{(2)}_{ab}-\s E^{(1)}_{ab}$ and $B^{(2)}_{ab}$ that could a priori define 12 independent charges
 \beqs\label{twelvec}
\mathcal J_{(i)} &\equiv & \frac{1}{8\pi G}\oint_S d^2S (E^{(2)}_{ab} - \s E^{(1)}_{ab})\xi^a_{\rom{rot}(i)}n^b\; , \nn\\ 
\mathcal K_{(i)} &\equiv & \frac{1}{8\pi G}\oint_S d^2S (E^{(2)}_{ab} - \s E^{(1)}_{ab})\xi^a_{\rom{boost}(i)}n^b \; ,\nn\\
\tilde{\mathcal J}_{(i)}&\equiv& - \frac{1}{8\pi G}\oint_S d^2S  B^{(2)}_{ab} \xi^a_{\rom{boost}(i)}n^b \; , \nn \\
\tilde{\mathcal K}_{(i)} &\equiv& \frac{1}{8\pi G} \oint_S d^2S B^{(2)}_{ab} \xi^a_{\rom{rot}(i)}n^b  \label{eq:Kb}\: .
 \eeqs
However, we have seen in \eqref{dualipropki} that these tensors are mutually conjugate in the sense that
\be\label{blabla}
E^{(2)}_{ab} - \s E^{(1)}_{ab} = \curl  B^{(2)}_{ab}, \qquad - \curl (E^{(2)}_{ab} - \s E^{(1)}_{ab}) = B^{(2)}_{ab} - X^{[3]}_{ab} \: ,
\ee
where $X^{[3]}_{ab} =- 4  \epsilon_{cd(a}\s^cE^{(1)}_{b)}{}^{d}$ is an SDT tensor. From the first of these relations, we see that charges associated with $E^{(2)}_{ab} - \s E^{(1)}_{ab}$  contracted  with either a rotational or a boost Killing vector are equivalent to charges associated with $B^{(2)}_{ab}$ contracted  with either a boost or a rotational Killing vector. This could also have been derived from the second relation in \eqref{blabla} as $X^{[3]}_{ab}$, or successive curls of this tensor, are associated to trivial charges by means of the linearization stability constraints. This eventually shows that only six charges are independent as
\beqs\label{sixc}
\mathcal J_{(i)}=\tilde{\mathcal J}_{(i)}\: , \qquad \mathcal K_{(i)}=\tilde{\mathcal K}_{(i)} \: .
\eeqs
Note that this result is in agreement with Lemma \ref{lemma8} which basically states that the general solution for SDT tensors $T_{ab}$ satisfying $(\Box-2)T_{ab}=0$ consists of two sets, related by the curl operator, of three tensors that capture the six charges $\int_S T_{ab} \xi_{(0)}^a n^b$ associated with Lorentz transformations, supplemented by higher harmonic tensors that do not contribute to the charges.

In the case $k_{ab}$ and $i_{ab}$ are non-trivial, one can define twelve additional boundary charges constructed from tensors quadratic in the first order fields
\beqs
Q^{[\s]}_{\text{bdr}}[\xi^a_{(0)}]\equiv \oint d^2 S \: T^{(\s)}_{ab} \: \xi^a_{(0)} n^b \; ,      \qquad Q^{[k]}_{\text{bdr}}[\xi^a_{(0)}]\equiv \oint d^2 S \: T^{(k)}_{ab} \: \xi^a_{(0)} n^b   \; .
\eeqs
These tensors represent the two classes of equivalence of SD tensors quadratic in $(\s,\s)$, $(\s,k)$ or $(k,k)$ that do not have a trivial symmetrized curl and that can not be written as total divergences when contracted with a Killing vector. All charges constructed with curls of these tensors are equivalent to the above charges by means of our previous results.  

One possible way to define the Lorentz charges is
\bea
\mathcal J_{(i)}&\equiv&\frac{1}{8\pi G}\oint_S d^2S  \sqrt{-h^{(0)}} (V_{ab} +2 i_{ab}) \xi^a_{\rom{rot}(i)}n^b = - \frac{1}{8\pi G}\oint_S d^2S \sqrt{-h^{(0)}}   W_{ab} \xi^a_{\rom{boost}(i)}n^b ,\,\label{eq:J} \nonumber \\
\mathcal K_{(i)}&\equiv&\frac{1}{8\pi G}\oint_S d^2S  \sqrt{-h^{(0)}}  (V_{ab}+2 i_{ab}) \xi^a_{\rom{boost}(i)}n^b = \frac{1}{8\pi G} \oint_S d^2S  \sqrt{-h^{(0)}} W_{ab} \xi^a_{\rom{rot}(i)}n^b  \label{eq:K}.\nn\\
\eea
However, as we have just emphasized, these charges are not unique as one can add any linear combination of the boundary charges to them  
\bea
\Delta \mathcal Q[\xi^a_{(0)}]=  \oint_S d^2S  \sqrt{-h^{(0)}} ( \alpha_1 \: T^{(\s)}_{ab} + \alpha_2 \: T^{(k)}_{ab} ) \xi^a_{(0)} n^b\, ,
\eea
where $\alpha_1$ and $\alpha_2$ are arbitrary constants. No matter how we define them, if we set $k_{ab}=i_{ab}=0$, they reduce to the six uniquely defined Lorentz charges that were given in \eqref{twelvec}-\eqref{sixc}.

To finish this section, let us just remark that singular contributions of $k_{ab}$, as discussed in Lemma \ref{newlemma2}, have not been considered here for Lorentz charges but would most certainly render the analysis much more complicated. As a first difficulty, the very definition of a charge that is conserved is not  straightforward anymore.

\setcounter{equation}{0}
\section{Summary of the results}
\label{sec:summ}

Choosing an hyperbolic slicing of spacetime to describe spatial infinity, which basically amounts to make the change of coordinates outside the light cone at large distances
\beqs
\rho=r \sqrt{1-\frac{t^2}{r^2}} \: , \qquad  \tau=\text{arctanh}(\frac{t}{r})\: ,
\eeqs
where $r$ is the usual radial coordinate from the spherical coordinates, we have defined asymptotically flat spacetimes as spacetimes whose metrics can be cast into the asymptotic form up to second order in the new radial coordinate $\rho$
\bea
\label{metric1}
ds^2 &=&   \left( 1 + \frac{2\s}{\rho}+ \frac{\s^2}{\rho^2}  + o(\rho^{-2})\right) d \rho^2 +  o(\rho^{-1}) d\rho dx^a \nn\\
&& + \rho^2\left( h^{(0)}_{ab} + \frac{h^{(1)}_{ab}}{\rho} + \ln\rho \frac{i_{ab}}{\rho^2} + \frac{h^{(2)}_{ab}}{\rho^2} + o(\rho^{-2}) \right) dx^a dx^b ,
\eea
where $h^{(0)}_{ab}$ is the metric on the unit hyperboloid, denoted $\mathcal{H}$,
\beqs
ds^2= h^{(0)}_{ab} d\phi^a d\phi^b= -d\tau^2 +\cosh^2 \tau (d\theta^2 +\sin^2 \theta d\phi^2)\: ,
\eeqs
and where $\s$, $h^{(1)}_{ab}$, $h^{(2)}_{ab}$ and $i_{ab}$ are fields defined on $\mathcal{H}$, i.e. on 3-dimensional de Sitter space. Note that $h_{ab}=\rho^2 h^{(0)}_{ab} + \rho \: h^{(1)}_{ab} +  \ln\rho \: i_{ab} +  h^{(2)}_{ab} +...$ denotes the full three-dimensional metric, the induced metric on the three-dimensional timelike hypersurface. We have referred to the ansatz $\eqref{metric1}$ as the generalized Beig-Schmidt ansatz. Indeed, it generalizes spacetimes considered by R. Beig and B. Schmidt in \cite{B} and \cite{BS} where only spacetimes where $B^{(1)}_{ab}=0$ were  considered, i.e. where one can set $k_{ab}\equiv h^{(1)}_{ab}+2\s h^{(0)}_{ab}$ and $i_{ab}$ to zero. 

Our metric \eqref{metric1} is invariant under Lorentz transformations but also under supertranslations of the form
\beqs\label{supertransb}
\rho=\bar{\rho} +\omega (\bar{\phi}^a) +\frac{F^{(2)}(\bar{\phi}^a)}{\bar{\rho}}+... \: , \qquad \phi^a= \bar{\phi}^a +\frac{1}{\bar{\rho}} h^{(0)\:ab} \omega_{,b}+\frac{G^{(2)\:a}}{\bar{\rho}^2}+... \: ,
\eeqs
where translations are singled out as the four supertranslations that satisfy $\omega_{ab}+\omega h^{(0)}_{ab}=0$ with $\omega_{ab}\equiv \DD_b \DD_a \omega$, and eventually under logarithmic translations
\beqs
\rho&=&\bar{\rho}+ H(\bar{\phi})(\ln \bar{\rho}-1) +o(\bar{\rho^0}), \qquad \phi^a= \bar{\phi}^a + H^a (\bar{\phi}) (\ln \bar{\rho})/\bar{\rho} +o(\bar{\rho}^{-1}),
\eeqs
where the functions $H$ are required to satisfy $H_{ab}+H h^{(0)}_{ab}=0$ so that no logarithmic terms are generated at first order. 

Our generalized ansatz was justified by our will to consider logarithmic translations and particular supertranslations as allowed transformations. Indeed, considering logarithmic transformations, that will generate a logarithmic term at second order, has driven us to include a non-trivial field $i_{ab}$. Supertranslations are allowed even when $i_{ab}=0$, they do not act on $\s$ but they do transform $k_{ab}$ as 
\beqs
k_{ab} \rightarrow k_{ab}+2 (\omega_{ab}+\omega h^{(0)}_{ab}).
\eeqs
The usual attitude in the literature is to consider only spacetimes for which $k_{ab}$ can be set to zero so that supertranslations are gauge-fixed. We have (and we will do so in the rest of this Part I)  relaxed this condition by considering spacetimes for which $k_{ab}$ can be non-trivial with the restriction that $k_{ab}$ must be an SDT tensor. This will be justified, in the next chapter, when discussing the variational principle. Note that this last condition implies that we restrict to these supertranslations $\omega$ that satisfy
\beqs
(\Box+3)\omega=0\: ,
\eeqs 
where $\Box \equiv \DD_a \DD^a$. Let us insist on the fact that, in the case $k_{ab}=i_{ab}=0$, logarithmic translations and supertranslations are not allowed transformations anymore as they have been gauge-fixed. Actually, the status of the logarithmic translations in this case is less trivial as they may be allowed when $\DD_c(E^{(1)}_{ab} H^c)=0$.

The section \ref{theeq} was devoted to the study of the equations of motion for our specific class of spacetimes. To study these equations, given the form of the metric \eqref{metric1}, we have reviewed in section \ref{subsec: gausscod} how the vacuum Einstein equations can be projected along, or  perpendicular to, the hyperboloid of constant $\rho$ using the projector $h_{ab} = g_{ab} -n_a n_b$, or the outward-pointing unit normal $n^a$, respectively.  This is known as the 3+1 split although it is in our case a split along the radial coordinate as compared to time in the Arnowitt-Deser-Misner formalism. This has provided us with Hamiltonian and momentum equations of motion (these equations contain time derivatives and therefore are not constraints) and equations of motion on the 3-dimensional hypersurface which respectively read \cite{BS}
\beqs\label{3equationssimp}
H &\equiv& R_{\mu\nu} n^{\mu} n^{\nu}=  -\mathcal{L}_{n} K - K_{ab}K^{ab}-N^{-1} h^{ab} D_a D_b N=0, \nonumber \\
F_{a} &\equiv& h_a^{\:\:\mu} n^\nu G_{\mu\nu}=h_a^{\:\:\mu} n^\nu R_{\mu\nu}= D_{b} K^{b}_{\:\:a}-D_a K=0, \nonumber \\
F_{ab} &\equiv & h_a^{\:\:\mu} h_b^{\:\:\nu} R_{\mu\nu}=  \mathcal{R}_{ab}- N^{-1} \partial_{\rho} K_{ab}-N^{-1} D_a D_b N-K K_{ab}+ 2 K_{a}^{\:\:c} K_{cb}=0, \nonumber \\
\eeqs
where $D$ is the covariant derivative compatible with the full metric $h_{ab}$ on the hyperboloid, $K_{ab}$ is the extrinsic curvature, $N = 1 + \sigma/ \rho$ is the lapse function,  $\mathcal{L}_{n}$ is the Lie derivative in the direction of $n^a$, and $\mathcal{L}_{n}K = \mathcal{L}_{n}(h^{ab}K_{ab})$.  In here, indices are raised and lowered with $h_{ab}$. 

In section \ref{subsec: ansatz}, we have plugged our general ansatz \eqref{metric1} for the metric into the set of equations \eqref{3equationssimp} and imposed $k_{ab}$ to be SDT. We have obtained the respective radial expansions of these equations at zeroth, first and second order in terms of the fields $h^{(0)}_{ab}, \s, k_{ab}, h^{(2)}_{ab}$ and $i_{ab}$. We have seen that the equations of motion at zeroth order imply that $h^{(0)}_{ab}$ is locally the metric on the unit hyperboloid as firstly assumed because we have locally that
\beqs
\mathcal{R}^{(0)}_{abcd}= h^{(0)}_{ac} h^{(0)}_{bd} -h^{(0)}_{bc}  h^{(0)}_{ad} .
\eeqs
The first order equations take the form  \cite{BS} 
\beqs\label{eqfirst}
(\Box-3)\s=0\: , \qquad (\Box+3)k_{ab}=0 ,
\eeqs
while the second order equations are of the generic form
\beqs\label{eqh2d}
i_a^{\:\:a} &=& 0,\qquad \DD^b i_{ab}=0, \qquad (\Box-2) i_{ab}=0\; , \nonumber \\
h^{(2)\:a}_a&=& \text{NL}^{(1)}_{ab} \: , \qquad 
\DD^b h^{(2)}_{ab}=  \text{NL}^{(2)}_{ab}\: , \qquad 
(\Box - 2) h^{(2)}_{ab} = 2 i_{ab} + \text{NL}^{(3)}_{ab}\: ,
\eeqs
where  $\text{NL}^{(i)}_{ab}$ for $i=1,2,3$ are non-linear terms, each given in  terms of $(\s,\s)$, ($\s,k$) and $(k,k)$ quadratic quantities. We have also seen how these second order equations for $h^{(2)}_{ab}$ reduce to the expressions obtained by Beig in \cite{B} when $k_{ab}=i_{ab}=0$.

In section \ref{sec:nice}, we have studied in more details the first and second order equations. To do that, we started by reviewing in \ref{subsec:weyl} the split of the Weyl tensor into electric $E_{ab}$ and magnetic $B_{ab}$ parts and gave their respective asymptotic expansions up to second order. This enabled us to rewrite in section \ref{sec:firstorder} the first order equations of motion using the electric and magnetic first order parts of the Weyl tensor
\bea
E^{(1)}_{ab} &\equiv& -\DD_a \DD_b \sigma - h^{(0)}_{ab} \sigma \, ,\qquad
B^{(1)}_{ab} \equiv  \frac{1}{2}\epsilon_a^{\;\; c d} \DD_c k_{d b} \, .\label{dBk}
\eea
Indeed, we saw that by construction, these two tensors enjoy the following properties
\bea
h^{(0)ab} B^{(1)}_{ab} = 0\; , \qquad  E^{(1)}_{[ab]}  = 0\; , \qquad  \DD_{[c} E^{(1)}_{a] b} = 0\; ,
\eea
and that the first order equations of motion \eqref{eqfirst} are equivalent to
\bea\label{EOM1bis}
h^{(0)ab} E^{(1)}_{ab} = 0\; ,\qquad B^{(1)}_{[ab]}  = 0\; , \qquad  \DD_{[c} B^{(1)}_{a] b} = 0.
\eea
Note that these quantities also satisfy
\beqs
&& \DD^b E^{(1)}_{ab}=0 \;, \qquad
(\Box-3)E^{(1)}_{ab}=0\; , \\
&& \DD^b B^{(1)}_{ab}=0 \;, \qquad 
(\Box-3)B^{(1)}_{ab}=0 \; ,
\eeqs
 telling us that the first order parts of the Weyl tensor are on-shell curl-free SDT tensors. Solutions to the equation $(\Box+3)\s=0$ were then studied. 

To express the second order equations of motion, we have first reviewed in section \ref{sec:SD} the classification of SD tensors $\kappa_{ab}$ constructed out of quadratic quantities in the first order fields. We then considered, in section \ref{subsec:second}, two SDT tensors, $V_{ab}$ and $W_{ab}$, which are defined in the generic form
\bea
V_{ab} &\equiv& - h_{ab}^{(2)}+\frac{1}{2} i_{ab}+ Q_{ab}^V \; , \label{defnlV} \\
W_{ab} &\equiv& \eps_a^{\;\, cd}D_c \left( h^{(2)}_{d b }-\frac{1}{2} i_{db} + Q_{db}^W \right) \; , \label{defnlW}
\eea
where $Q^V_{ab}$ and $Q^{W}_{ab}$ are specific tensor potentials satisfying
$\DD^b Q^{V,W}_{ab}=\DD_a Q^{V,W}$ and $Q^{V,W} \equiv  h^{(0)\: ab} Q^{V,W}_{ab}$. The tensors $V_{ab}$ and $W_{ab}$ are set to be mutually dual, or conjugate, in the sense that they obey the following duality properties
\bea
W_{ab} = - (\curl V)_{ab}  +  (\curl \kappa)_{(ab)} ,\qquad V_{ab} = (\curl W)_{ab} -2 i_{ab} \, .\label{nl03}
\eea
We eventually showed that the equations at second order can be recast into the equivalent form
\bea
W^a_a &=& \DD^b W_{ab}=0, \qquad
(\square - 2)W_{ab} = \curl(2 i + \kappa )_{(ab)} ,\label{eqWf}\\
i^a_a &=& \DD^b i_{ab} = 0 ,\qquad
(\square -2)i_{ab} = 0, \label{eqi}
\eea
or with the help of the curl operator, and the definition $j_{ab} \equiv -(\text{curl} \: i)_{ab}$, in the form
\bea
V^a_a &=& \DD^b V_{ab}=0 , \qquad (\square - 2)V_{ab} = \curl(-2j + \curl(\kappa ))_{ab},\label{eqVf} \\
j^a_a &=& \DD^b j_{ab} = 0, \qquad (\square -2)j_{ab} = 0.\label{eqj}
\eea
In the case where $k_{ab}=i_{ab}=0$, we have seen that the tensors $V_{ab}$ and $W_{ab}$ just reduce to $E^{(2)}_{ab}-\sigma E^{(1)}_{ab}$ and $B^{(2)}_{ab}$, where $E^{(2)}_{ab}$ and $B^{(2)}_{ab}$ are the second order electric and magnetic parts of the Weyl tensor.

In section \ref{sec:LS}, we have reviewed the important result of \cite{B} showing that Einstein's vacuum equations can be solved to all orders, in the case $k_{ab}=i_{ab}=0$, if and only if the field $\s$ satisfies the field equation $(\Box+3)\s=0$ and the following six conditions
\beqs
\mathcal Q [\xi^a_{(0)}] \equiv \oint_S d^2S\, \eps_{cd(a}\s^c E^{(1) \: d}_{\; \, b)} \xi^a_{(0)} n^b =0 \: ,
\eeqs
where $\xi^a_{(0)}$ are Killing vectors on $dS_3$. This result was established by first integrating the equation of motion \eqref{eqWf} contracted with a Killing vector and showing that it can be written as a total divergence. The integration of the r.h.s. of the equation of motion contracted with any Killing vector is precisely the above condition. These integrability conditions were understood as linearization stability constraints.  We then moved to the generalization of these conditions in the case where $k_{ab}$ and $i_{ab}$ are non-trivial. We found that, for the equations to have a solution at least up to second order, we need to require the following necessary conditions
\bea\label{linsumm}
\oint_S d^2S\,  i_{ab} \: \xi_{(0)}^a n^b =  2\:  \oint_S d^2S  \, \Big(  T^{(\s)}_{ab} + T^{(k)}_{ab} \Big) \xi_{(0)}^a n^b , \label{intFF}
\eea
where
\bea
T^{(\s)}_{ab} &\equiv& - \frac{2}{\sqrt{-h^{(0)}}} \frac{\delta L^{(\s)}}{\delta h^{(0)\: ab}} = -\frac{1}{4} \kappa^{[\s,\s,I]}_{ab} +\frac{1}{2} \kappa^{[\s,\s,II]}_{ab} , \label{Ts} \\
T^{(k)}_{ab} &\equiv& -  \frac{2}{\sqrt{-h^{(0)}}} \frac{\delta L^{(k)}}{\delta h^{(0) \: ab}}
= \frac{1}{2} \kappa^{[k,k,II]}_{ab} -\frac{1}{8} Y^{(2)}_{ab} , \label{Tk}
\eea
are the stress-tensors, expressed as linear combinations of SD tensors $\kappa_{ab}$, associated to the actions
\bea
L^{(\sigma)} &=& \sqrt{-h^{(0)}} \Big( -\frac{1}{2} \partial_a \s \partial^a \s +\frac{3}{2}\s^2 \Big),\label{Lsb} \qquad L^{(k)}= \sqrt{-h^{(0)}} \Big( \frac{1}{4} \: B^{(1)}_{ab} B^{(1)\:ab}\Big), \label{Lkb}
\eeqs
 which were found by requiring that their variations respectively reproduce the equations of motion for $\s$ and $k_{ab}$.
 
Eventually, in section \ref{sec:conservedd}, we have looked at all possible conserved charges that one can associate to conformal Killing vectors, i.e. translations, or Killing vectors, i.e. boosts or rotations. To do so, we have made extensive use of the results established in the previous sections such as the properties of Killing vectors on $dS_3$, the classification of SD tensors, the proof that charges constructed by contracting an SD tensor with a certain Killing vector is equivalent to a charge constructed with the curl of this tensor with another particular Killing vector, and the linearization stability constraints.  In the case where $k_{ab}=i_{ab}=0$, we have seen that only four charges, which agree with the ADM momenta, can be associated to translations
 \beqs
 Q[\zeta^a_{(i)}]= -\frac{1}{8\pi G} \oint_S d^2 S \: E^{(1)}_{ab} \: n^a \: \zeta^b_{(i)}\: \:,
 \eeqs
 and that six charges, the Lorentz charges, can be associated to the Killing vectors 
 \beqs
\mathcal J_{(i)} &\equiv & \frac{1}{8\pi G}\oint_S d^2S (E^{(2)}_{ab} - \s E^{(1)}_{ab})\xi^a_{\rom{rot}(i)}n^b = - \frac{1}{8\pi G}\oint_S d^2S  B^{(2)}_{ab} \xi^a_{\rom{boost}(i)}n^b ,\,\label{eq:Jb} \\
\mathcal K_{(i)} &\equiv & \frac{1}{8\pi G}\oint_S d^2S (E^{(2)}_{ab} - \s E^{(1)}_{ab})\xi^a_{\rom{boost}(i)}n^b = \frac{1}{8\pi G} \oint_S d^2S B^{(2)}_{ab} \xi^a_{\rom{rot}(i)}n^b  \label{eq:Kb},
 \eeqs
 using either the electric or magnetic parts of the Weyl tensor. As we will review in the following chapter, this last result is also a proof of the uniqueness of the Lorentz charges defined by Ashtekar-Hansen in \cite{Ashtekar:1978zz,Ashtekar:1991vb} using the magnetic part of the Weyl tensor and the counter-term charges of Mann and Marolf  \cite{Mann:2005yr} expressed in terms of the electric part of the Weyl tensor. This also trivially proves their equivalence as established in \cite{Mann:2008ay}. We also mentioned that, only in the case where $k_{ab}$ is allowed to develop singularities, one can define non-trivial dual momenta 
  \beqs
 Q[\zeta^a_{(i)}]= -\frac{1}{8\pi G} \oint_S d^2 S \: B^{(1)}_{ab} \: n^a \: \zeta^b_{(i)}\: .
 \eeqs
For regular $k_{ab}$, the above dual momenta are identically zero. 

We also pointed out that Lorentz charges are a priori not unique in the case where $k_{ab}$ and $i_{ab}$ are non-trivial. Indeed, one possible choice is
\bea
\mathcal J_{(i)}&\equiv&\frac{1}{8\pi G}\oint_S d^2S  \sqrt{-h^{(0)}} (V_{ab} +2 i_{ab}) \xi^a_{\rom{rot}(i)}n^b = - \frac{1}{8\pi G}\oint_S d^2S \sqrt{-h^{(0)}}   W_{ab} \xi^a_{\rom{boost}(i)}n^b ,\,\label{eq:J} \nonumber \\
\mathcal K_{(i)}&\equiv&\frac{1}{8\pi G}\oint_S d^2S  \sqrt{-h^{(0)}}  (V_{ab}+2 i_{ab}) \xi^a_{\rom{boost}(i)}n^b = \frac{1}{8\pi G} \oint_S d^2S  \sqrt{-h^{(0)}} W_{ab} \xi^a_{\rom{rot}(i)}n^b  \label{eq:K}.\nn\\
\eea
However, there is an infinite number of possible choices as one can add to them any linear combination of the twelve boundary charges 
\bea
\Delta \mathcal Q[\xi^a_{(0)}]=  \oint_S d^2S  \sqrt{-h^{(0)}} ( \alpha_1 \: T^{(\s)}_{ab} + \alpha_2 \: T^{(k)}_{ab} ) \xi^a_{(0)} n^b\, ,
\eea
where $\alpha_1$ and $\alpha_2$ are arbitrary constants. In all cases, they reduce to the Lorentz charges \eqref{eq:Jb}-\eqref{eq:Kb} for $k_{ab}=i_{ab}=0$ upon taking into account the linearization stability constraints \eqref{linsumm}. We will explore these issues in greater detail in the next chapter.

%%%%%%%%%%%%%%%%%%%%%%%%%%%%%%%%%%%%%%%%%%%%%%%%%%%%%%%%%%%%%%%%%%%

\chapter{Conserved charges from the variational principle}\label{actionprinc}

Asymptotically flat spacetimes are defined by boundary conditions at infinity. The class of diffeomorphisms which preserve the boundary conditions are the allowed diffeomorphisms. Allowed diffeomorphisms associated with non-trivial conserved charges - the large diffeomorphisms -  modulo diffeomorphisms associated with zero charges - the pure gauge transformations -  define the asymptotic symmetry group. For asymptotically flat spacetimes at spatial infinity with specific parity conditions imposed on the first order fields, the construction of the asymptotic symmetry group has been worked out both with Hamiltonian and covariant phase space methods. As we have reviewed, in the Hamiltonian framework, the Regge-Teitelboim (RT) construction \cite{Regge:1974zd}  shows that the asymptotic symmetry group is just the Poincar\'e group. In that framework, parity odd supertranslations also preserve the boundary conditions but they are associated with vanishing Hamiltonian generators. In covariant phase space, we have seen in the previous chapter that the asymptotic symmetry group is the Poincar\'e group when truncating the phase space, by setting to zero a part of the first order fields, i.e. the first order magnetic part of the Weyl tensor is set to zero. This last condition also fixes all supertranslations \cite{Ashtekar:1978zz,Ashtekar:1984aa}. We have however not discussed yet how parity conditions show up in this setup. 

In this chapter, we will deal with conserved charges associated to asymptotic symmetries that can be constructed from a Lagrangian which provides a good variational principle. Our analysis also relies on the covariant phase space (see \cite{ABR}) which is the space of dynamically allowed histories, i.e. solutions of Einstein's equations. We refer the reader to \cite{ABR} for more details about the construction of charges as Noether charges in this set-up and the precise meaning of a symplectic structure on the covariant phase space. 

We  start by reviewing in section \ref{sec:cov} the Crnkovic-Witten-Lee-Wald symplectic structure, constructed from the Einstein-Hilbert Lagrangian, the Barnich-Brandt-Comp\`ere symplectic structure, and the ambiguity in their definitions. We point out that for asymptotically flat spacetimes that can be cast into the Beig-Schmidt form, parity conditions must be imposed on the first order field $\s$ to obtain a finite symplectic structure, and thus a good definition of the covariant phase space, as first realized by Ashtekar, Bombelli and Reula (ABR) \cite{ABR}. 
Comparing the ABR phase space with the work of Regge-Teitelboim, where supertranslations are allowed at first, we show the existence of a larger covariant phase space which allows the inclusion of a traceless and divergence-free $k_{ab}$ with a fixed parity (see also comments in \cite{Beig:1987aa}). Here, when logarithmic translations are not allowed and parity even conditions have been imposed, we also recover the Poincar\'e group as asymptotic symmetry group by computing  the charges that can be constructed from the symplectic structure. In comparison with the ABR phase space,  particular parity even supertranslations are allowed, namely the covariant equivalent of the Hamiltonian odd supertranslations. These are associated to trivial charges in agreement with the RT analysis.

In section \ref{sec:MM}, we review the Mann-Marolf construction \cite{Mann:2005yr} of a good Lagrangian variational principle for asymptotically flat spacetimes at spatial infinity, their construction of a boundary stress-energy tensor and its associated counter-term charges. Based on the results presented in the previous chapter, we revisit the equivalence of the counter-term charges with the expressions of Ashtekar and Hansen (see also \cite{Mann:2008ay}). We finish this section by pointing out the relation between parity conditions and boundary terms at future and past  infinities in the variational principle.

In section \ref{paritycond}, we propose a new way of regulating the covariant phase space without imposing parity conditions. This is implemented by adding counter-terms to the action so that divergences cancel in the variational principle and makes it well-defined. The charges associated to the symplectic structure, when the bulk part is regulated by the boundary counter-term contribution, are thus also shown to be finite and non-trivial for all asymptotic transformations allowed by our enlarged Beig-Schmidt ansatz.  

We relegate to Appendix \ref{app:BC} a comparison between covariant and 3+1 boundary conditions.

\setcounter{equation}{0}
\section{The covariant symplectic structure and associated charges}
\label{sec:cov}

In this section, we review two different definitions of the symplectic structure and see how they are related by a boundary term  manifesting the ambiguity in its definition. We then study conservation and finiteness of the symplectic structure, when using our enlarged Beig-Schmidt ansatz, which are the crucial properties the symplectic structure should satisfy to define a good covariant phase space. Our computations review the result of Ashtekar-Bombelli-Reula \cite{ABR} which shows that the symplectic structure is conserved and logarithmically divergent. It can however be made finite, for the Beig-Schmidt ansatz, by imposing parity conditions on the field $\s$. Here, we also show that a generalized covariant phase space can be considered when both $\s$ and $k_{ab}$ with specific parity conditions are allowed. In the last part, we see how the symplectic structure can be used to define conserved charges, that are finite and reduce to previous expressions known in the litterature for the Beig-Schmidt ansatz.  Although we will not discuss it here, we refer the reader to \cite{Lee:1990nz,Wald:1999wa,Barnich:2001jy,Barnich:2007bf} for the understanding of the charges, constructed from the symplectic structure, as generators of  (Poincar\'e)  asymptotic symmetries.

\subsection{Two symplectic structures}

The symplectic structure is a phase space 2-form that is defined as the integral over a Cauchy slice of a spacetime 3-form that we refer to as the integrand for the symplectic structure.

As for conventions, we denote a $3$-form $\mathbf{\Theta}$ and a 2-form $\mathbf{k}$ as
\beqs
\mathbf{\Theta}&=& \frac{1}{3!} \: \Theta_{\mu\nu\rho} \: dx^\mu \wedge dx^\nu \wedge dx^\rho \equiv \tilde{\Theta}^{\mu} (d^{3}x)_{\mu}  \; , \label{thetar}\\
 \mathbf{k}&=& \frac{1}{2} k_{\mu\nu} dx^{\mu} \wedge dx^{\nu}=\tilde{k}^{\mu\nu} (d^2 x)_{\mu\nu} \; ,
\eeqs
where
\beqs
(d^{n-p} x)_{\mu_1 \cdots \mu_p}= \frac{1}{p! (n-p)!} \epsilon_{\mu_1 \cdots \mu_n} dx^{\mu_{p+1}} \cdots dx^{\mu_{n}} \; ,
\eeqs
with $n$ being the number of dimensions.

For Stokes' theorem to hold, we have set conventions (see Appendix B of \cite{Wald:1984rg}) that imply 
\beqs
\epsilon_{\tau \rho \theta \phi}=\epsilon_{\rho \theta \phi}=-\epsilon_{\tau \theta \phi}=\epsilon_{\theta \phi}.
\eeqs
 This also tells us that 
 \beqs
 &&(d^3x)_{\tau}= -d\rho (d^2 x)_{\tau}, \qquad (d^3 x)_{\rho}=-d\tau (d^2 x)_{\rho}\; , \nn\\
  && (d^2x)_{\tau}=-(d^2x)_{\rho}=2(d^2x)_{\rho\tau}=-d^2S\; ,
 \eeqs
 where $d^2S=\frac{1}{2} \epsilon_{\zeta \iota} dx^{\zeta} \wedge dx^{\iota}$ with $\zeta$ and $\iota$ being coordinates on the sphere. 

\subsubsection{The Crnkovic-Witten-Lee-Wald symplectic structure}

The Crnkovic-Witten-Lee-Wald integrand \cite{Crnkovic:1986ex,Lee:1990nz} for the covariant phase space symplectic structure is given by
\bea\label{integrand}
\omega [\delta_1 g ,\delta_2 g ] &=& \frac{1}{32\pi G}(d^{3}x)_{\mu}\sqrt{-g } \Big(
\delta_1 g^{\alpha\beta}\nabla^\mu \delta_2 g_{\alpha \beta}+\delta_1 g \nabla^\alpha \delta_2 g^\mu_{\alpha}+\delta_1 g^\mu_{\alpha}\nabla^\alpha \delta_2 g  \nn \\
&&  - \delta_1 g \nabla^\mu \delta_2 g - 2 \delta_1 g_{\alpha\beta}\nabla^\alpha \delta_2 g^{\mu\beta}
- (1 \leftrightarrow 2)
\Big),
\eea
where $\delta_1 g_{\mu\nu}$, $\delta_2 g_{\mu\nu}$ are perturbations around a general asymptotically flat spacetime $g_{\mu\nu}$ and we use the convention $\delta g^{\mu\nu} \equiv g^{\mu \kappa} g^{\nu \lambda} \delta g_{\kappa \lambda}$.

To recover this expression, we follow closely  the approach given by Burnett and Wald in \cite{waldbur}. Starting from the Einstein-Hilbert action
\beqs
S_{EH}=\frac{1}{16 \pi G} \int_\mathcal{M} d^4 x \: \sqrt{-g} \: R \:, \qquad L_{EH}= \sqrt{-g} \: R \: ,
\eeqs
one computes the on-shell variation of that action while keeping the boundary terms. We find
\beqs
\delta L=\delta(R_{\mu\nu}) g^{\mu\nu} \sqrt{-g} + R_{\mu\nu} \delta(g^{\mu\nu} \sqrt{-g}) \: .
\eeqs
By explicit computation, we check that 
\beqs
&& \delta R_{\mu\nu}=\nabla_{\s} \Lambda^{\s}_{\:\:\mu\nu} \: , \qquad 
\Lambda^{\s}_{\:\:\mu\nu}\equiv \Upsilon^{\s}_{\:\:\mu\nu}- \Upsilon^{\kappa}_{\:\:n(\mu} \delta_{\nu)}^{\:\:\s} \; , \qquad \Upsilon^{\s}_{\:\:\mu\nu}\equiv g^{\s \kappa} \Big( \nabla_{(\mu} \: \delta g_{\nu)\kappa} -\frac{1}{2} \nabla_{\kappa} \: \delta g_{\mu\nu}\Big)  ,\nn\\
&& \delta(\sqrt{-g}) = -\frac{1}{2} g_{\mu\nu} \delta g^{\mu\nu} \sqrt{-g}\: , \qquad \delta g^{\mu\nu}= -g^{\mu\kappa} \: g^{\nu \s} \: \delta g_{\kappa \s}\: ,
\eeqs
and we obtain
\beqs
\delta L= (\nabla_{\s} \Lambda^{\s}_{\:\:\mu\nu}) \: g^{\mu\nu} \sqrt{-g} +\sqrt{-g} \: G_{\mu\nu} \: \delta g^{\mu\nu}\: .
\eeqs
On shell, we have $G_{\mu\nu}=0$ so that the second term vanishes. The first term is a boundary term. Let us re-express it as
\beq
\frac{1}{16 \pi  G} \int_\mathcal{M}  d^4 x \: \delta L =  \frac{1}{16 \pi  G} \int_\mathcal{M}  d^4 x \: \partial_{\s}( \sqrt{-g} \: \Lambda^{\s}_{\:\:\mu\nu} \: g^{\mu\nu}) \equiv \int_\mathcal{M}  d  \mathbf{\Theta} \; ,
\eeq
where $\mathbf{\Theta}$ is a 3-form, as defined in \eqref{thetar}. Its exterior derivative is
\beqs
d\mathbf{\Theta}= \frac{1}{3 !}  \nabla_{\s} \: \Theta_{\mu\nu\rho} \:dx^\s \wedge dx^\mu \wedge dx^\nu \wedge dx^\rho \; . \label{dtheta}
\eeqs
The Hodge dual of this 3-form is a 1-form $\mathbf{\tilde{\Theta}}=\star \mathbf{\Theta}$
\beqs
\star \mathbf{\Theta}= \frac{1}{3! (4-3)!} \: \Theta^{\mu\nu\rho} \epsilon_{\mu\nu\rho\sigma} \: dx^\s \; ,\qquad 
\mathbf{\tilde{\Theta}}= \tilde{\Theta}_\mu \: dx^\mu\; .
\eeqs
This gives us
\beqs\label{thetad}
\tilde{\Theta}^\mu=\frac{1}{6} \:  \epsilon^{\nu\rho\s\mu} \: \Theta_{\nu\rho\s}\: , \qquad \Theta_{\nu\rho\s}=\tilde{\Theta}^\mu \: \epsilon_{\mu\nu\rho\s}\; ,
\eeqs
and implies that
\beqs
d\mathbf{\Theta}= d(  \: \tilde{\Theta}^\mu  \:  (d^3 x)_\mu )\; .
\eeqs
In our case, we immediately see that
\beqs
\mathbf{\Theta}=\tilde{\Theta}^\mu \: (d^{3}x)_\mu&=& \frac{\sqrt{-g}}{16\pi G} \: \Lambda^\mu_{\:\:\nu\s} \: g^{\nu\s} \: (d^{3}x)_\mu \nn \\
&=& \frac{\sqrt{-g}}{16\pi G}\: g^{\mu\kappa} \: g^{\nu\sigma} \: (\nabla_{\s} \delta g_{\nu\kappa}-\nabla_\kappa \delta g_{\sigma\nu}) \: (d^{3}x)_\mu \; .
\eeqs
The pre-symplectic structure is obtained from the pre-symplectic potential $\mathbf{\Theta}$ as
\beqs
\omega(\delta_1 g, \delta_2 g)\equiv\delta_1 \mathbf{\Theta}(\delta_2 g)-\delta_2 \mathbf{\Theta}(\delta_1 g)\; ,
\eeqs
and the bulk symplectic structure is
\beqs\label{symplecticc}
\Omega^{\text{bulk}}=\int_{\Sigma} \omega \; .
\eeqs
We find the pre-symplectic structure associated to the Einstein-Hilbert Lagrangian to be given by
\beqs\label{omega}
\omega(\delta_1 g, \delta_2 g)=\frac{\sqrt{-g}}{16\pi G } R^{\mu\nu\alpha\beta\gamma\delta} \Big( \delta_2 g_{\nu \alpha} \nabla_{\beta} \delta_1 g_{\gamma \delta} -(1\leftrightarrow 2) \Big) (d^3 x)_\mu \; ,
\eeqs
where
\beqs
R^{\mu\nu\alpha\beta\gamma\delta}\equiv g^{\mu \gamma} g^{ \delta \nu} g^{\alpha \beta}-\frac{1}{2} g^{\mu \beta} g^{\nu \gamma} g^{\delta \alpha} -\frac{1}{2} g^{\mu \nu} g^{\alpha \beta} g^{\gamma \delta}- \frac{1}{2} g^{\nu \alpha} g^{\mu \gamma} g^{\delta \beta}+\frac{1}{2} g^{\nu \alpha} g^{\mu \beta} g^{\gamma \delta}\; .
\eeqs
This is exactly the expression given by  Wald and Zoupas in \cite{Wald:1999wa}. It is completely equivalent to our expression \eqref{integrand} as soon as one rewrites it using our convention
$\delta g^{\mu\nu}\equiv g^{\mu \sigma} \: g^{\nu \kappa} \: \delta g_{\sigma \kappa}$.
In short, one usually says that the Crnkovic-Witten-Lee-Wald integrand \eqref{integrand} is obtained by varying a second time the boundary term $\Theta [\delta g]$ obtained after a variation of the Einstein-Hilbert Lagrangian as $\omega[\delta_1 g,\delta_2 g] = \delta_1 \Theta[\delta_2 g] - \delta_2 \Theta[\delta_1 g]$.

\subsubsection{Barnich-Brandt-Comp\`ere symplectic structure}
In  \cite{Barnich:2001jy,Barnich:2007bf}, the integrand for the symplectic structure was obtained from Einstein's equations. We refer the reader to these papers for more details about its construction. The Barnich-Brandt-Comp\`ere symplectic structure is given by
\bea
W [\delta_1 g ,\delta_2 g ] &=& \frac{1}{32 \pi G} (d^3x)_{\mu} \: \sqrt{-g}\: P^{\mu\nu\alpha\beta\gamma\delta} \Big( \delta_1 g_{\alpha \beta} \n_\nu \delta_2 g_{\gamma \delta} -\delta_2 g_{\alpha \beta} \n_\nu \delta_2 g_{\gamma \delta} \Big) \nonumber \\
&=& \frac{1}{32\pi G}(d^{3}x)_{\mu}\sqrt{-g } \Big(
\delta_1 g^{\alpha\beta}\nabla^\mu \delta_2 g_{\alpha \beta}+\delta_1 g \nabla^\alpha \delta_2 g^\mu_{\alpha}+\delta_1 g^\mu_{\alpha}\nabla^\alpha \delta_2 g \nn \\
&&  - \delta_1 g \nabla^\mu \delta_2 g - \delta_1 g_{\alpha\beta}\nabla^\alpha \delta_2 g^{\mu\beta} - \delta_1 g^{\mu\alpha}\nabla^\beta \delta_2 g_{\alpha \beta}
- (1 \leftrightarrow 2)
\Big),
\eea
where
\beqs
P^{\mu\nu\alpha \beta \gamma \delta}&\equiv&g^{\mu\nu} g^{\gamma(\alpha} g^{\beta)\delta} + g^{\mu(\gamma} g^{\delta)\nu} g^{\alpha \beta}+ g^{\mu(\alpha} g^{\beta)\nu} g^{\gamma \delta} \nonumber \\
 &&-g^{\mu\nu} g^{\alpha \beta} g^{\gamma \delta}-g^{\mu (\gamma} g^{\delta)(\alpha} g^{\beta)\nu}-g^{\mu(\alpha} g^{\beta)(\gamma} g^{\delta)\nu}.
 \eeqs

\subsubsection{Comparison of symplectic structures and the ambiguity}

One can check that the integrands of the two symplectic structures defined above differ by the boundary term
\bea\label{diffsymp}
W [\delta_1 g ,\delta_2 g ] - \omega [\delta_1 g ,\delta_2 g ]  = \frac{1}{32\pi G}(d^{3}x)_{\mu}\sqrt{-g } \nabla_\nu \Big( \delta_1 g^\nu_{\; \beta} \delta_2 g^{\mu\beta} - (\mu \rightarrow \nu)\Big).
\eea
This result reflects the well-known fact that there is an ambiguity in the definition of the symplectic structure. Indeed, if one adds a boundary term  $d\mathbf{K}$ to the Lagrangian, then the equations of motion are unaffected. However, the presymplectic potential will be modified by a term $\delta \mathbf{K}\equiv d\mathbf{Y}[\delta g]$. Then, the integrand for the symplectic structure becomes
\beqs
\omega \rightarrow \omega + d \Big( \delta_1 Y[\delta_2 g]- (1\leftrightarrow 2)\Big)\; .
\eeqs

In the following, we will be interested by their expressions for a specific ansatz of the Beig-Schmidt form. In this case, we see that this difference vanishes on constant $\rho$ and $\tau$ surfaces when the metric and perturbations are expanded in the Beig-Schmidt expansion with
\bea
g_{\rho a}=0,\qquad \delta g_{\rho a} = 0\, .
\eea
We can therefore use interchangeably $\omega$ and $W$ in what follows. Note that this does not change the fact that there is still an ambiguity in the definition of the symplectic structure. 

\subsubsection{Conservation and finiteness of the symplectic structure}

Before looking at the construction of charges, let us discuss conservation and finiteness of the symplectic structure. 

Conservation is established as soon as the symplectic structure given in \eqref{symplecticc} is shown to be independent of the specific choice of the three surface $\Sigma$. Because the symplectic structure is closed by definition, to analyze its conservation it is sufficient to evaluate the integrand  on a constant $\rho$ hypersurface and see if it is of order $o(\rho^0)$. Indeed, if this is the case, the integration on a region delimited by $\Sigma_1$, $\Sigma_2$ and $\partial \Sigma$, representing two Cauchy surfaces joining the boundary of these two surfaces at fixed $\rho$ will be vanishing. The conservation then automatically follows as
\beqs
\int_{\mathcal{M}} d\omega=0=-\int_{\Sigma_1} \omega+\int_{\Sigma_2} \omega +\int_{\partial \Sigma} \omega \; , \qquad \int_{\Sigma_1} \omega=\int_{\Sigma_2} \omega \;.
\eeqs
The finiteness of the symplectic structure can be studied by integrating on a constant $\tau$ hypersurface. It is finite if 
\beqs
\int_{\Sigma,  \tau=\textbf{fixed}} \omega \: d\rho < \infty \: .
\eeqs
Note that the above argumentation for conservation of the integrand implicitly assumes that it is also finite.

To compute these quantities, let us first give some relevant information. Let us first remember our generalized Beig-Schmidt ansatz
\beqs
\label{metric}
ds^2 = \left( 1 + \frac{\s}{\rho} \right)^2 d \rho^2 + \rho^2\left( h^{(0)}_{ab} + \frac{h^{(1)}_{ab}}{\rho} +\ln\rho \frac{i_{ab}}{\rho^2}+ \frac{h^{(2)}_{ab}}{\rho^2} + \ldots \right) dx^a dx^b\; ,
\eeqs
meaning that
\beqs
g^{\rho\rho} &=& 1 - \frac{2\sigma}{\rho} +\frac{3\sigma^2}{\rho^2}+O(\rho^{-3})\, , \qquad  g^{\rho a} = 0\; ,\nn\\
g^{ab} &=& \rho^{-2} h^{(0)\: ab} - \rho^{-3} h^{(1) \: ab} -\rho^{-4} \; \ln \rho \; i^{ab} -\rho^{-4} (h^{(2)\:ab} -h^{(1)\;ac} h^{(1)\;b}_{c})+ o(\rho^{-4}) \; , \nn\\
\p_\rho g_{\rho\rho} &=& -\frac{2\s}{\rho^2} -\frac{2\s^2}{\rho^3}+O(\rho^{-4})\; , \qquad \p_\rho g_{ab} = 2\rho h_{ab}^{(0)} + h_{ab}^{(1)}+\rho^{-1} i_{ab}+o(\rho^{-1})\; ,\nn\\
\sqrt{-g} &=&  (1+\frac{\s}{\rho}) \rho^3 \sqrt{-h^{(0)}} (1+\frac{1}{2} h +\frac{1}{8} h^{\mu}_{\mu} h^{\nu}_{\nu}  -\frac{1}{4} h_{\mu\nu} h^{\mu\nu})\nonumber \nn\\
&=&  \sqrt{-h^{(0)}} \rho^3 \biggr [ 1+ \frac{(h^{(1)}+2\s)}{2\rho}+  \frac{1}{2\rho^2} (h^{(2)}+\frac{1}{4} h^{(1)\:2} -\frac{1}{2} h^{(1)}_{ab} h^{(1)\:ab} +\sigma h^{(1)}) \biggr ] \nonumber \\
&=&  \sqrt{-h^{(0)}} \biggr [ \rho^3 - 2 \s \rho^{2} +\frac{\rho}{2} ( h^{(2)}-3\s^2 -\frac{1}{2} k_{ab} k^{ab})+o(\rho^2) \biggr ]\; \label{start},
\eeqs
where we work in the gauge $k^a_a=0$. The variation of the metric is
\beqs
\label{deltametric}
\delta g_{\mu\nu}dx^\mu dx^\nu &=& \left( \frac{2\delta \s}{\rho} + \frac{2\sigma \delta \sigma}{\rho^2} \right) d \rho^2 \nn\\
&&+ \left( \rho ( \delta k_{ab}-2\delta \sigma h_{ab}^{(0)})  + \ln\rho \delta i_{ab} + \delta h^{(2)}_{ab}  + o(\rho^0)  \right) dx^a dx^b\; ,
\eeqs
where
\beqs\label{deltametricdet}
\delta g &=& g^{\mu\nu}\delta g_{\mu\nu} = \frac{-4\delta \sigma}{\rho}+\frac{1}{\rho^2}\left( \delta h_{(2)} -14 \s \delta \s - k^{ab} \delta k_{ab} \right)+o(\rho^{-2})\; , \\
\delta g^{ab} &=& g^{ac}g^{bd}\delta g_{cd} = \rho^{-3}(\delta k^{ab}-2\delta \s h_{(0)}^{ab})+\nn\\
&&\rho^{-4}(\ln \rho \delta i^{ab} + \delta h^{ab}_{(2)}-8 \s \delta \s h_{(0)}^{ab} + 4 k^{ab}\delta \s +4 \s \delta k^{ab} - k^{ac}\delta k_c^{b}- k^{bc}\delta k_c^{a})\; , \nn\\
\delta g^{\rho \rho} &=& (g^{\rho\rho})^2 \delta g_{\rho\rho} = \frac{2 \delta \sigma}{\rho}-\frac{6 \s \delta \s}{\rho^2} \; .\\
\eeqs
Remember our notation $\delta g^{\mu\nu}\equiv g^{\mu\kappa} g^{\nu \lambda} \delta g_{\kappa \lambda}$.

The covariant derivative requires an expansion of the Christoffel symbols. One checks that
\bea
\Gamma^\rho_{\rho\rho} &=& \frac{1}{2}g^{\rho\rho}g_{\rho\rho,\rho}=-\frac{\s}{\rho^2} + \frac{\s^2}{\rho^3}+O(\rho^{-4}) \; ,\qquad \Gamma^\rho_{\rho a} = \frac{1}{2} g^{\rho \rho} g_{\rho\rho,a}= \frac{\s_a}{\rho}-\frac{\s \s_a}{\rho^2}+O(\rho^{-3}) \; ,\nn\\
\Gamma^a_{\rho \rho} &=& -\frac{1}{2} g^{ab} g_{\rho\rho,b}= -\frac{\s^a}{\rho^{3}} + \frac{k^{ab}\s_b -3 \s \s^a}{\rho^4}+o(\rho^{-4}) \; ,\nn\\
\Gamma^\rho_{a b} &=& -\rho h_{ab}^{(0)} - \frac{1}{2} k_{ab}+3\s h_{ab}^{(0)} -\frac{1}{2\rho} (i_{ab}-2\s k_{ab}+10\s^2h^{(0)}_{ab}) + O(\rho^{-2})\; ,\nn\\
\Gamma^a_{\rho b} &=& \frac{1}{2} g^{ac} g_{cb,\rho}=\rho^{-1} \delta^a_b + \rho^{-2} \left( -\frac{1}{2} k^a_b + \s \delta^a_b \right) +\rho^{-3}\log\rho (- i ^a_{b} ) \nonumber \\
&& \qquad \qquad  \qquad+\rho^{-3} (-h^{(2)\:a}_{b} +\frac{1}{2} i^a_b+\frac{1}{2} k^{ac} k_{bc} - 2\s k^a_b+2\s^2 h^{(0)\:a}_{b})+o(\rho^{-3}) \; ,\nn\\
\eea
and also 
\begin{eqnarray}
\Gamma^a_{bc} = \Gamma^{(0)\:a}_{\quad\:\: bc}+ \rho^{-1} \Gamma^{(1)\:a}_{\quad\:\: bc}+o(\rho^{-1})\; ,
\end{eqnarray}
where
\bea
\Gamma^{(1)\:a}_{\quad\:\: bc} &=& \frac{1}{2}\left( \DD_c k^{a}_{\;\; b} +  \DD_b k^{a}_{\;\; c} -  \DD^a  k_{bc}\right) -\s_c \delta^a_b - \s_b \delta^a_c +\s^a h_{bc}^{(0)}\; . \label{finish}
\eea

Now, on the one hand, using the expansion \eqref{metric} only up to first order, the symplectic structure integrand evaluated on a hypersurface $\rho =$ constant gives
\bea
W [\delta_1 g ;\delta_2 g ]|_{fixed\; \rho} &=& \frac{1}{32\pi G}(d^{3}x)_{\rho}\sqrt{-g } \Big(
\delta_1 g^{\alpha\beta}\nabla^\rho \delta_2 g_{\alpha \beta}+\delta_1 g \nabla^\alpha \delta_2 g^\rho_{\alpha}+\delta_1 g^\rho_{\alpha}\nabla^\alpha \delta_2 g \nn \\
&&  - \delta_1 g \nabla^\rho \delta_2 g - \delta_1 g_{\alpha\beta}\nabla^\alpha \delta_2 g^{\rho\beta} - \delta_1 g^{\rho\alpha}\nabla^\beta \delta_2 g_{\alpha \beta}
- (1 \leftrightarrow 2)
\Big) \nn\\
&=& \frac{1}{32\pi G}(d^{3}x)_{\rho}\sqrt{-g } \Big(
\delta_1 g^{ab}\nabla^\rho \delta_2 g_{ab}+ (\delta_1 g - \delta_1 g^\rho_\rho)(\nabla_\rho \delta_2 g^\rho_\rho - \nabla_\rho \delta_2 g) \nn \\
&&- (1 \leftrightarrow 2) \Big)\:,
\eea
where we used $g_{\rho a}=0$ and  $\delta g_{\rho a} = 0$ in the second equation. Computing the following terms
\bea
\delta_1 g^{ab}\nabla^\rho \delta_2 g_{ab} &=& \rho^{-3} \; (12 \delta_1 \s \delta_2 \s +\delta_1 k^{ab} \delta_2 k_{ab})\;,\\
(\delta_1 g - \delta_1 g^\rho_\rho)&=& -6 \delta_1 \s \rho^{-1} \;,\\
(\delta_1 g - \delta_1 g^\rho_\rho)(\nabla^\rho \delta_2 g^\rho_\rho - \nabla^\rho \delta_2 g)  &=& -36 \rho^{-3} \delta_1 \s \delta_2 \s\;,
\eea
we see that
\bea
W [\delta_1 g ,\delta_2 g ]|_{fixed\; \rho}  = o(\rho^{0}).
\eea
This shows that our boundary conditions ensure that the symplectic structure is conserved. 

On the other hand, the integrand for the symplectic structure evaluated on a Cauchy slice $\Sigma$ asymptotic to a constant $\tau$ hypersurface reads
\bea
W [\delta_1 g ;\delta_2 g ]|_{fixed\; \tau} &=& \frac{1}{32\pi G}(d^{3}x)_{\tau}\sqrt{-g } \Big(
\delta_1 g^{\rho\rho}\nabla^\tau \delta_2 g_{\rho\rho} +\delta_1  g^{ab} (\nabla^\tau \delta_2 g_{ab} - \nabla_b \delta_2 g^\tau_a) \nn \\
&& +\delta_1 g (\nabla^a \delta_2 g^\tau_a - \nabla^\tau \delta_2 g)+\delta_1 g^{\tau a} (\nabla_a \delta_2 g - \nabla^b \delta_2 g_{ab}) - (1 \leftrightarrow 2)
\Big)\; .\nn\\
\eea
We find
\bea
\delta_1 g^{\rho\rho}\nabla^\tau \delta_2 g_{\rho\rho}  &=&  +4 \rho^{-4} \delta_1 \s \DD^\tau \delta_2 \s \; ,\nn\\
\delta_1  g^{ab} (\nabla^\tau \delta_2 g_{ab} - \nabla_b \delta_2 g^\tau_a)
&=& \rho^{-4} \: \Big( 8  \delta_1 \s \DD^\tau \delta_2 \s -2 \epsilon^{\tau c e }\:  \delta_1 k_c^{\:\:d} \: \delta_2 \:  B^{(1)}_{ed} + 2 \delta_1 k^{\tau b} \delta_2 \sigma_b \Big) \; , \nn\\
\delta_1 g (\nabla^a \delta_2 g^\tau_a - \nabla^\tau \delta_2 g) &=&  -8 \rho^{-4} \delta_1 \s \DD^\tau \delta_2 \s \; ,\nn\\
\delta_1 g^{\tau a} (\nabla_a \delta_2 g - \nabla^b \delta_2 g_{ab}) &=& \rho^{-4} \Big( 4  \delta_1 \s \DD^\tau \delta_2 \s - 2 \delta_1 k^{\tau b} \delta_2 \s_b \Big)\; .
\eea
Therefore, we obtain
\beqs
W [\delta_1 g ;\delta_2 g ]|_{\Sigma} 
&=& \frac{\rho^{-1}}{4\pi G} (d^3 x)_a \sqrt{-h^{(0)} } \Big( \delta_1 \sigma \DD^a \delta_2 \sigma  -\frac{1}{4} \epsilon^{a c e } \delta_1 k_c^{\:d} \: \delta_2   B^{(1)}_{ed} - (1 \leftrightarrow 2) \Big) + o(\rho^{-1}) \; . \label{Weval} \nn \\
\eeqs
The bulk symplectic structure
\bea\label{bulksymp}
\Omega_{bulk}[\delta_1 g,\delta_2 g] \equiv \int_\Sigma W [\delta_1 g ,\delta_2 g ]\; ,
\eea
is therefore logarithmically divergent for generic $\sigma$ and $k_{ab}$. 

This result was originally obtained by Ashtekar, Bombelli and Reula in \cite{ABR}, under the restrictive assumption that $k_{ab}=0$, where it was suggested that one should impose that $\s$ satisfies the following (even) parity condition
\bea
\sigma(\tau , \theta, \phi ) &=&  \sigma (-\tau , \pi - \theta ,\phi + \pi) \:,
\eea
so that the symplectic structure be finite. Actually, one can make the symplectic structure finite even in the presence of $k_{ab}$  by imposing
\bea
\sigma(\tau , \theta, \phi ) &=& s_\sigma\, \sigma (-\tau , \pi - \theta ,\phi + \pi)\label{par5} \:,\\
k_{ab}(\tau , \theta, \phi ) &=& s_k \, k_{ab} (-\tau , \pi - \theta ,\phi + \pi)\, .\label{par52}
\eea
Here, $s_\sigma$ and $s_k$ are two signs which define the phase space with parity conditions. Indeed, these conditions are sufficient as they impose the integrand to be of odd parity. Following the dictionary between cylindrical and hyperbolic coordinates presented in Appendix \ref{app:BC}, the RT parity conditions, reviewed in chapter 1, amount to $s_\sigma = s_k =+1$. Remember that in RT work, the parities were chosen so that Schwarzschild is allowed as a solution. Note that a necessary and sufficient condition to allow Schwarzschild is actually only $s_\s = +1$. For the sake of simplicity, we will not discuss the case of mixed parities here but restrict the analysis to even parities. The conclusions concerning mixed parities can be straightforwardly obtained from the results presented in the following.  

The covariant phase space where even parity conditions on $\s$ and $k_{ab}$ are imposed is more general than the one considered by ABR in \cite{ABR} since here we do not impose $k_{ab}=0$ but only require that $\DD^a k_{ab} = k_a^a=0$. In the latter case, the parity even supertranslations fulfilling $(\Box+3)\omega=0$, such that $k_{ab}$ remains an SDT tensor with even parity, are allowed while logarithmic translations are not, because we do not allow here for $i_{ab}$ and $\sigma$ of odd parity. As we will see in the following, the conserved charges associated with these allowed supertranslations are vanishing while the Poincar\'e generators are non-vanishing. Therefore, this enlarged phase space is a consistent phase space where the asymptotic symmetry group is still the Poincar\'e group. This consistent phase space generalizes the one defined in \cite{ABR} by allowing the field $k_{ab}$ and therefore the first order part of the magnetic Weyl tensor $B^{(1)}_{ab}= \frac{1}{2}\epsilon_a^{\;\; c d} \DD_c k_{d b} $ to be non-vanishing while still keeping the Poincar\'e group as asymptotic symmetry group. Note that the four lowest harmonics in $B_{ab}^{(1)}$ are zero because we impose that $k_{ab}$ has to be regular. 

Let us eventually mention that this enlarged covariant phase space we have just discussed is a subset of the phase space described by Regge and Teitelboim in \cite{Regge:1974zd}.  There is a one-to-one mapping of the Poincar\'e transformations between the 3+1 formalism and the covariant formalism. To see the mapping between supertranslations, let us decompose the odd supertranslations $\xi^i(\textbf{n})$ of RT as temporal supertranslations $\xi^{\perp}$, radial supertranslations $\xi^r$ and angular supertranslations $\xi^{\iota}$ where $\iota$ is a coordinate on the two-sphere. Following our Appendix \ref{app:BC}, there is a one-to-one mapping between canonical temporal and radial supertranslations and even parity covariant supertranslations satisfying $(\Box+3)\omega=0$. However, our phase space is a subset of the RT phase space as angular supertranslations generate mixed components $g_{\rho a}$ in hyperbolic coordinates. These were already discarded, i.e. gauge-fixed, when putting the metric into its Beig-Schmidt form.

\subsection{Charges constructed with the symplectic structure}
\label{subsec:chargessymp}
To discuss conserved charges, it turns out to be much more easy to work with the integrand for the symplectic structure $W$, the one derived from Einstein's equations. Indeed, when contracted with the Lie derivative of the metric, it becomes a boundary term
\bea\label{result1}
W[\delta g , \mathcal L_\xi g ] &=& d k_\xi [ \delta g ; g ],\label{propsym}
\eea
where the last equality has to be understood up to Einstein's equations of motion for the metric and the linearized perturbations (see also \cite{Barnich:2007bf}). The boundary term $k_\xi [ \delta g ; g ]$ is exactly given by the Abbott-Deser expression \cite{Abbott:1981ff} (see also chapter 1) for the surface charge constructed from a linear perturbation $\delta g_{\mu\nu}$ around a solution $g_{\mu\nu}$
\bea
k_\xi [\delta g ; g ] &=&  \frac{2}{3} \: \frac{1}{16\pi G} \: \sqrt{-g}\: (d^2 x)_{\mu\alpha} P^{\mu\nu\alpha\beta\gamma\delta} \Big( 2 \xi_{\nu} \n_{\beta} \delta g_{\gamma \delta} -\delta g_{\gamma \delta} \n_{\beta} \xi_{\nu} \Big) \nonumber \\
&=& \frac{\sqrt{-g}}{16\pi G}(d^{2}x)_{\mu\nu}\sqrt{-g } \Big( \xi^\nu (\D^\mu \delta g - \D_\sigma \delta g^{\s \mu}) +\xi_\sigma \D^\nu \delta g^{\s \mu} \nonumber \\
&&+\frac{1}{2} \delta g \D^\nu \xi^\mu +\frac{1}{2}\delta g^{\mu \sigma}\D_\sigma \xi^\nu +\frac{1}{2} \delta g^{\nu \sigma}\D^\mu \xi_\sigma - (\mu \leftrightarrow \nu)
\Big) .
\label{AD}
\eea
The equality \eqref{result1} can be checked rapidly as follows. The two-form $k_{\xi}$ is 
\beqs
k= k_{\mu\nu} dx^{\mu} \wedge dx^{\nu}= \tilde{k}^{\mu\nu} (d^2x)_{\mu\nu}\; , \qquad 
k_{\mu\nu}= \frac{1}{2} \epsilon_{\mu\nu\sigma\kappa} \tilde{k}^{\sigma\kappa}\; , \qquad  \tilde{k}^{\alpha \beta}=-\frac{1}{2} \epsilon^{\alpha \beta \mu \nu}\: k_{\mu\nu} \; ,
\eeqs
which gives, from \eqref{AD},
\beqs
\tilde{k}^{\mu\nu}=  \frac{1}{3} \: \frac{1}{16\pi G}\: (P^{\mu\alpha\nu\beta\gamma\delta}-P^{\nu\alpha\mu\beta\gamma\delta})\Big( 2 \xi_{\alpha} \n_{\beta} \delta g_{\gamma \delta} -\delta g_{\gamma \delta} \n_{\beta} \xi_{\alpha} \Big)\; ,
\eeqs
and we also have by definition
\beqs
dk&=&\nabla_{\nu} \:  \tilde{k}^{\mu\nu} \: (d^3x)_{\mu} \nonumber \\
    &=&  \frac{1}{3} \: \frac{1}{16\pi G}\: (d^3 x)_{\mu} \: (P^{\mu\alpha\nu\beta\gamma\delta}-P^{\nu\alpha\mu\beta\gamma\delta}) \nn\\
    && \Big( 2 \n_{\nu} \xi_{\alpha} \n_{\beta} \delta g_{\gamma \delta}+ 2  \xi_{\alpha} \n_{\nu} \n_{\beta} \delta g_{\gamma \delta} - \n_{\nu} \delta g_{\gamma \delta} \n_{\beta} \xi_{\alpha}-\delta g_{\gamma \delta} \n_{\nu}  \n_{\beta} \xi_{\alpha} \Big)\; .
\eeqs
One then needs to compute this last expression by discarding terms which are zero  on-shell. For example, one obtains
\beqs
&&   \frac{1}{3} \: (P^{\mu\alpha\nu\beta\gamma\delta}-P^{\nu\alpha\mu\beta\gamma\delta}) \Big( 2 \n_{\nu} \xi_{\alpha} \n_{\beta} \delta g_{\gamma \delta} - \n_{\nu} \delta g_{\gamma \delta} \n_{\beta} \xi_{\alpha} \Big) = P^{\mu\nu\alpha\beta\gamma\delta} \: \n_\alpha \xi_{\beta} \: \n_\nu \delta g_{\gamma \delta}\; . \nn\\
\eeqs
In the end, one finds
\beqs
W[ \delta g, \mathcal{L}_{\xi} g] &\equiv&  - \frac{1}{32 \pi G} (d^3x)_{\mu} \: P^{\mu\nu\alpha\beta\gamma\delta} \Big( (\n_{\alpha} \xi_{\beta} +\n_{\beta} \xi_{\alpha}) \n_\nu \delta g_{\gamma \delta} -\delta g_{\alpha \beta} \n_\nu (\n_{\gamma} \xi_{\delta} +\n_{\delta} \xi_{\gamma})\Big) \nonumber \\
&=& dk_{\xi} \;,
\eeqs
where the first equality was obtained by definition of the Lie derivative $\mathcal{L}_{\xi} g_{\mu\nu}\equiv \nabla_{\mu} \xi_{\nu}+\nabla_{\nu} \xi_{\mu}$.

The covariant phase space infinitesimal charges associated with a diffeomorphism tangent to the phase space are defined from the symplectic structure as
\bea
\slash \hspace{-5pt} \delta Q_\xi [g] = \Omega [\delta g , \mathcal L_\xi g  ] \, .
\eea
Here, we use the symbol $\slash\hspace{-5pt} \delta Q_\xi[g]$ to remind the reader that the infinitesimal charge between the solution $g$ and $g+\delta g$ can be considered as a one-form in field space which is not necessarily exact. When $\slash\hspace{-5pt} \delta Q_\xi[g]$ is an exact form in phase space, as will turn out to be the case for each diffeomorphism we consider, the charges can be defined. We denote the integrated charge as $Q_\xi[g ; \bar g]$ (we have $\delta Q_\xi[g ; \bar g] = \slash\hspace{-5pt} \delta Q_\xi[g]$) and fix the integration constant so that $Q_\xi[\bar g ; \bar g] = 0$ for Minkowski spacetime $\bar g$. See also \cite{Barnich:2007bf} for more details. 

Using the definition of the bulk symplectic structure \eqref{bulksymp} and its property \eqref{propsym}, the charge one-form $\slash\hspace{-5pt} \delta Q_\xi[g]$ can be written as a surface integral
\bea
\slash\hspace{-5pt} \delta Q_\xi[g] = \int_S k_\xi [\delta g ; g ] , \label{totkkb}
\eea
evaluated on the sphere $S$ at constant time $t$ and at $\rho = \Lambda$.  The expression for the charges can be rewritten in the alternative form
\bea
\hspace{-6pt}k_\xi [\delta g ; g ] &=& - \delta k^K_\xi +  \frac{1}{8\pi G}(d^{2}x)_{\mu\nu}\sqrt{-g } \Big( \xi^\nu (\D^\mu \delta g - \D_\sigma \delta g^{\s \mu})+\D^\mu \delta \xi^\nu \Big) - E[\mathcal L_\xi g, \delta g] \label{kk1b}, \nn\\
\eea
where
\bea
k^K_\xi[g] =  \frac{1}{16\pi G}(d^{2}x)_{\mu\nu}\sqrt{-g } \Big( \D^\mu \xi^\nu - \D^\nu \xi^\mu \Big)\, ,
\eea
is the Komar term and
\bea
E[\mathcal L_\xi g, \delta g] =   \frac{1}{16\pi G}(d^{2}x)_{\mu\nu}\sqrt{-g } \Big( \frac{1}{2} g^{\mu \alpha} \delta g_{\alpha \beta }g^{\beta \gamma}\mathcal L_{\xi}g_{\gamma \sigma}g^{\sigma\nu} - (\mu \leftrightarrow \nu)  \Big)\, , \label{kk3}
\eea
is a term linear in the Killing equations that might not vanish in general for asymptotic symmetries.  Here $\delta$ acts on the metric and on the asymptotic Killing vector, as  $\xi(g)$ might depend on the metric. Note that in \eqref{kk1b}, the first term is the exact variation of the Komar term and the third term is zero when evaluated on constant $\rho$ and $\tau$ surfaces since
\bea
g_{\rho a}=0,\qquad \delta g_{\rho a} = 0,\qquad \mathcal L_\xi g_{\rho a} = 0.
\eea
Therefore the bulk surface charge one-form \eqref{totkkb} is given by the generalization of the Iyer-Wald expression \cite{Iyer:1994ys} when the asymptotic Killing vector is allowed to depend on the metric
\bea
k_\xi [\delta g ; g ] = - \delta k^K_\xi[g]  +  \frac{1}{8\pi G} (d^{2}x)_{\mu\nu}\sqrt{-g } \Big( \xi^\nu (\D^\mu \delta g - \D_\sigma \delta g^{\s \mu})+\D^\mu \delta \xi^\nu \Big) .\label{kk2b}
\eea

We say that surface charges are integrable if and only if
\bea
\int_S \Big( \delta_2 k_\xi [\delta_1 g ; g ] - (1 \leftrightarrow 2) \Big) = 0 \; .
\eea
The integrable charges are then defined by integrating in phase space as 
\bea\label{charge}
\mathcal Q[\xi ; g ] = \int_S \int_\gamma k_\xi [\delta g ; g ] \; ,
\eea
where $\gamma$ is a path in the space of symmetric configurations, see \cite{Barnich:2007bf} for more details. 
The charges are asymptotically linear, and thus reduce to the Abbott-Deser expressions (see also \cite{Barnich:2001jy}), if
\bea
k_\xi [\delta g ; g ]  = k_\xi [\delta g ; \bar g ] \; =k_{\xi}[g-\bar{g} ; \bar{g}] .
\eea

\subsubsection{Evaluating the charges for our Beig-Schmidt ansatz}
Let us now review the explicit evaluation of the charge  \eqref{charge} for each asymptotic Killing vector (translations, supertranslations satisfying $(\Box+3)\omega=0$, rotations, boosts and logarithmic translations). Although we have not considered yet in this chapter a phase space where logarithmic translations or generic supertranslations are allowed, we produce the generic (divergent) results here as these will be useful in the next section. The main goal of this section is to establish that in the case where logarithmic translations are discarded but even parity supertranslations satisfying $(\Box+3)\omega=0$ can act on the phase space, i.e. $\s$ and $k_{ab}$ obey even parity conditions and $i_{ab}=0$, we recover the usual expressions for the Poincar\'e charges derived in previous works on asymptotically flat spacetimes in Lagrangian framework \cite{Geroch:1977jn,Ashtekar:1978zz,AshRev,Abbott:1981ff,Ashtekar:1991vb}. The only difference with these works, when parity conditions are imposed, is that we have allowed for even parity supertranslations by allowing the SDT tensor $k_{ab}$ to be non-zero.  In agreement with the work of Regge and Teitelboim, we show that the charges associated to these supertranslations are always zero.

In order to do  the computation only once, we consider the vector field
\bea\label{vectorr}
\xi^\rho &=& \log\rho H(x) + ( \omega(x) - H(x) ) + \frac{1}{\rho} F^{(2)}(\rho, x)+ o(\rho^{-1}),\nn\\
\xi^a &=& \xi_{(0)}^a+\frac{\log\rho}{\rho}H^a(x) +\frac{1}{\rho}\omega^a(x)+\frac{1}{\rho^2}G^{(2)a}(\rho , x)+o(\rho^{-2})\; ,
\eea
where $\omega$ stands for supertranslations (or translations), $H$ for logarithmic translations, and $\xi^a_{(0)}$ for the Killing vectors on $dS_3$. Remember that the functions $F^{(2)}$ and $G^{(2)\:a}$ are fixed in terms of $\omega$, $H$ and the first order fields such that the form of the generalized Beig-Schmidt ansatz is preserved. 

Let us now plug  the vector field \eqref{vectorr} and  our generalized ansatz \eqref{metric} into  \eqref{kk2b} and make intensive use of the results stated between \eqref{start} and \eqref{finish}.

For the first term in \eqref{kk2b}, the Komar term, we find
\bea
D^\rho \xi^a &=& \frac{1}{\rho}\xi_{(0)}^a + \frac{1}{\rho^2} (-\frac{1}{2}k^a_{\; b}\xi_{(0)}^b -\s \xi_{(0)}^a +H^a ) + \frac{\ln \rho}{\rho^3} \Big( -i^a_b \xi^b_{(0)} - H \s^a +\s H^a - \frac{1}{2} k^a_b H^b \Big) \nn \\
&& + \frac{1}{\rho^3} \Big( -G^{(2)\: a} +\rho \p_\rho G^{(2)\: a} - \s^a \omega +\s^a H - 2\s H^a  -\frac{1}{2} k^a_b \omega^b +  \s \omega^a  \nn\\
&& + (-h_b^{(2)\:a} + \frac{1}{2} i^a_b +\frac{1}{2}k^{a c}k_{cb}-\s k^a_{b}+3\s^2 h^{(0)\:a }_b)\xi_{(0)}^b \Big)\; , \\
D^a \xi^\rho &=& - \frac{1}{\rho}\xi_{(0)}^a + \frac{1}{\rho^2} (\frac{1}{2}k^a_{\; b}\xi_{(0)}^b +\s \xi_{(0)}^a - H^a ) + \frac{\ln \rho}{\rho^3} ( i^a_c \xi^c_{(0)} +H \s^a -\frac{1}{2} k^{ab} H_b + 3 \s H^a ) \nn \\
&& + \frac{1}{\rho^3} \Big( h_{(2)}^{ac}\xi_c^{(0)}+F_{(2)}^a - G^{(2)a} +k^{ab} H_b -H\s^a - 2 \s H^a  +\s^a \omega - \frac{1}{2}k^a_b \omega^b +3 \s \omega^a \nn\\
&&  - \frac{1}{2}i^a_b \xi^b_{(0)}+\s k^a_b \xi^b_{(0)}-3\s^2 \xi_{(0)}^a - \frac{1}{2}k^{ac}k_{cb}\xi_{(0)}^b \Big) \; .
\eea
The functions $F^{(2)}$ and $G^{(2)\:a}$ are fixed by computing  
\bea
\sqrt{-g} \left( D^\rho \xi^a+D^a \xi^\rho \right) &=& \sqrt{-h_{(0)}} \Big( \log\rho (-k^{a b}H_b + 4 \s H^a )  + F^{(2)\: a} - 2 G^{(2)\:a} + \rho \p_\rho G^{(2)\:a} \nn\\ 
&&  -4\s H^a - k^a_b \omega^b + 4\s \omega^a +k^{ab}H_b \Big)+o(\rho^0) \; .
\eea
Indeed, if we want the vector to be Killing up to order $o(\rho^0)$, we should impose 
\bea
 F^{(2)\:a} - 2 G^{(2)\: a} + \rho \p_\rho G^{(2)\:a} +(\log\rho-1 ) (-k^{a b}H_b + 4 \s H^a )   - k^a_b \omega^b + 4\s \omega^a  = 0 \; .
\eea
Note that this is just another equivalent way of saying that we stay in the Beig-Schmidt frame under such transformations. 

In the end, for the Komar term, we find
\beqs
-16 \pi G \int_{\gamma} \delta K^K_{\xi} \equiv  -\mathcal{A}[g]+\mathcal{A}[\bar{g}]\; ,
\eeqs
where
\bea\label{AAA}
\mathcal{A}[g] &=& \sqrt{-g} \left( D^\rho \xi^a - D^a \xi^\rho \right) = \sqrt{-h_{(0)}} \Big( 2 \xi^a_{(0)} \rho^2 + \left( - 6 \s \xi^a_{(0)} -k^a_b \xi^b_{(0)} +2H^a  \right) \rho \nn \\
&&+ \ln \rho ( - 2 i^a_b \xi^b_{(0)} -2 H \s^a - 2 \s H^a ) +  \Big( \rho \p_\rho G^{(2)\: a} - F^{(2)\:a}  -2 \s^a \omega -2\s \omega^a  +2 \s^a H \nn \\ 
&& - 4 \s H^a -k^{a b}H_b + (-2h^{(2)\:a}_b +i^a_b + k^{a c}k_{cb}  )\xi^b_{(0)}  + (h^{(2)} + 7\s^2 - \frac{1}{2}k_{ab}k^{ab} )\xi^a_{(0)} \Big) \nn\\
&& + o(\rho^0) \; .
\eea
The last term in \eqref{kk2b} is 
\bea
\mathcal B \equiv \int_\gamma \sqrt{-g} \left( D^\rho \delta \xi^a - D^a \delta \xi^\rho \right) &=& \sqrt{-h_{(0)}} \Big( \rho \p_\rho G^{(2)\:a} - F^{(2)\:a}  \Big) + o(\rho^0) \; .
\eea
To compute the last two terms of \eqref{kk2b}, we see that
\bea
D^\rho \delta g - D_\sigma \delta g^{\s \rho} &=& -\frac{6\delta \s}{\rho^2}+\frac{1}{\rho^3}(-\delta h^{(2)}+18 \s \delta \s + \frac{1}{2}k^{ab}\delta k_{ab})\; , \\
D^a \delta g - D_\sigma \delta g^{\s a} &=& - \frac{2\delta \s^a}{\rho^3}+o(\rho^{-3}) \; ,
\eea
which gives
\bea
\sqrt{-g}\xi^a ( D^\rho \delta g - D_\sigma \delta g^{\s \rho}) &=& - 6 \sqrt{-h_{(0)}}\delta \s \xi^a_{(0)} \rho - 6 \sqrt{-h_{(0)}}\delta \s H^a  \log\rho \nn \\ 
&& +\sqrt{-h_{(0)}} \Big( ( -\delta h^{(2)}+30 \s \delta \s + \frac{1}{2}k^{ab}\delta k_{ab})\xi^a_{(0)}  -6\delta \s \omega^a \Big)\nn\\
&& +o(\rho^0)\; , \nn \\
\sqrt{-g}\xi^\rho ( D^a \delta g - D_\sigma \delta g^{\s a}) &=& \sqrt{-h_{(0)}} \Big(-2 H \delta \s^a \log\rho -2 ( \omega-H) \delta \s^a \Big) +o(\rho^0)\; .
\eea
The integral in phase space turns out to be trivial
\bea
\mathcal C &\equiv&\int_\gamma \sqrt{-g}\xi^\rho ( D^a \delta g - D_\sigma \delta g^{\s a} - \xi^a ( D^\rho \delta g - D_\sigma \delta g^{\s \rho}) ) \nn \\
&&= \sqrt{-h_{(0)}} \Big[ (6 \s \xi^a_{(0)} )\rho + \log\rho (-2 H \s^a + 6 \s H^a ) + \nn \\
&& ( h_{(2)} -15\s^2 - \frac{1}{4}k_{ab}k^{ab} )\xi_{(0)}^a  +6 \s \omega^a - 2\omega \s^a +2 H \s^a  \Big] \; .
\eea
The final answer is 
\bea
(16 \pi G) \int_{\gamma} k_\xi = (- \mathcal A[g] +\mathcal A [\bar g]+  \mathcal B -\mathcal C ) \: (2 (d^{2}x)_{\rho a} )\; .
\eea
The quadratically divergent term, present in \eqref{AAA}, cancels between $ -\mathcal A[g]$ and $\mathcal A [\bar g]$. The linear divergent term proportional to $\sigma$ in the Komar integral $\mathcal A$ cancels the one in $\mathcal C$.  By convention, we express the charges in terms of $\hat \xi^a_{(0)} = - \xi^a_{(0)}$, so that the linear divergent term reduces to 
\bea
\int_S \int_{\bar g}^g k_{\hat \xi_{(0)}}  = \frac{1}{8\pi G}\int_S d^2 S \sqrt{-h_{(0)}} \left(  \frac{1}{2}k^\tau_a \hat \xi_{(0)}^a \rho \right) \; .
\eea
It is however identically zero after using the equations of motion for $k_{ab}$ and properties of integrals. In the end, we see that the final result has a log divergent piece and a finite piece 
\bea\label{generalcharges}
Q[\xi ; g] &=& \int_S  \int_{\gamma} k_\xi [\delta g ; g] \nn\\
&=& \frac{1}{16\pi G} \int_S d^2 S \, n_a \Big( \log\rho (- 2 i^a_b \hat{\xi}_{(0)}^b + 4 H \s^a - 4 \s H^a ) \nn \\ 
&& + 4 \s^a \omega - 4 \s \omega^a  - 4 \s^a H + 4\s H^a + k^{a b}H_b \nn \\
&& - 2 \hat{\xi}_{(0)}^b (h^{(2)\: a }_b - \frac{1}{2}i^a_b - \frac{1}{2} k^{a c }k_{cb} + h^{(0)\: a}_b (-h^{(2)} +4 \s^2 + \frac{3}{8}k_{cd}k^{cd}) ) \Big) \; ,
\eea
 as expected from the logarithmic divergence of the symplectic structure. 
 
 Let us now study in more detail the charges associated to translations, supertranslations and Lorentz transformations. We will not consider logarithmic translations as these are not allowed in the phase space where even parity conditions are imposed. For the other symmetries, we try to stay as general as possible in the following, anticipating the considerations to be presented in section \ref{paritycond}.

\subsubsection{Translations and supertranslations}

Even without assuming parity conditions, the charges associated to translations and supertranslations are finite and asymptotically linear $k_\xi [\delta g ; g ] = k_\xi [\delta g ; \bar g ]$. They reduce to
\beqs\label{superr}
Q[\xi ; g] &=& \frac{1}{4\pi G} \int_S d^2 S\: \sqrt{-h^{(0)}} \, n_a \Big(   \s^a \omega -  \s \omega^a   \Big)\; .
\eeqs
One can check that the charges are also  conserved as we restrict ourselves to supertranslations satisfying $(\Box+3)\omega=0$, so that $\DD_a (  \s^a \omega -\s \omega^a   ) =  \square \s \omega-\s \square \omega  = 0$.

For translations, the expression for the charges can be further simplified by noticing that, when $\omega_{ab}+\omega h^{(0)}_{ab}=0$, we have
\beqs
2 \omega \s_b - 2 \omega_b \s   = -E^{(1)\:a}_b \omega_a - 2 \DD^a (\omega_{[a}\s_{b]})\; .
\eeqs
One recovers the well-known expression for the four-momenta, already obtained in \eqref{momentaa},
\bea
Q_{(\mu)}[g ; \bar g ] = - \frac{1}{8 \pi G} \int_S d^2S \:  E_{ab}^{(1)} n^a \DD^b \zeta_{(\mu)}\, ,\label{ch:tr}
\eea
where the four scalars $\zeta_{(\mu)}$, $\mu=0,1,2,3$ are the four solutions of $\DD_a \DD_b \zeta_{(\mu)} + h_{ab}^{(0)}\zeta_{(\mu)}= 0$.  The vector $\p/\p t$ in the 3+1 asymptotic frame $(t,r,\theta,\phi)$ corresponds to the translation $\zeta_{(0)} = -\sinh \tau$. One can check \cite{Mann:2008ay} that the charge $Q_{(0)}$ for the Schwarzschild black hole of mass $m$ is $+m$ after identifying $\sigma = G m \cosh 2\tau \sech \tau$, as it should.

For supertranslations, when parity even conditions are imposed on both $\sigma$ and $k_{ab}$, $\omega$ also has to be parity even. One realizes from expression  \eqref{superr} that supertranslation charges identically vanish in this case. Note that this does not apply to translations $\omega = \zeta_{(\mu)}$, which are of odd parity, as these are still allowed and are associated with non-trivial charges.

 \subsubsection{Lorentz charges}

As we can see from \eqref{generalcharges}, the expression for the bulk charge admits a logarithmic divergence and a finite piece. 
The logarithmically divergent piece of the Lorentz charges takes the form
\bea
k_\xi [\delta g ; g ] = - \frac{\log\Lambda}{8\pi G} \int_S d^2S \sqrt{-h^{(0)}}\:  i_{ab} \: \xi^a_{(0)} n^b +O(\Lambda^0)\, .
\eea
It can be expressed using the linearization stability constraints \eqref{intFF} as
\bea
k_\xi [\delta g ; g ] =   - \frac{\log\Lambda}{4\pi G} \int_S d^2S \sqrt{-h^{(0)}} \Big(  T^{(\s)}_{ab}+ T^{(k)}_{ab} \Big)  \xi^a_{(0)} n^b +O(\Lambda^0),\label{infLo}
\eea
where $T^{(\s)}_{ab}$ and $T^{(k)}_{ab}$ are the stress-tensors of the actions \eqref{Lkb}. 
The finite part of the Lorentz charges  can be written as
\bea\label{finitepart}
\mathcal Q_{-\xi_{(0)}}[g ;\bar  g ] &=& \frac{1}{8\pi G}\int_S d^2 S \sqrt{-h^{(0)}} \Big(
 - h^{(2)}_{ab} +\frac{1}{2} i_{ab} +\frac{1}{2} k_a^{c}k_{cb} \nn \\
&&+h_{ab}^{(0)} (8\s^2 +\s^c \s_c -\frac{1}{8}k_{ab}k^{ab}+k_{cd}\s^{cd} ) \Big) \xi_{(0)}^a n^b .\label{finalQrotb}
\eea
where we used the second order equation $h^{(2)}=12 \s^2 +\s_c \s^c +\frac{1}{4} k_{cd} k^{cd} +k_{cd} \s^{cd}$.
The charges are also conserved as a consequence of the momentum equation \eqref{eqh2c}. 

Let us remark here that, following considerations presented in the previous chapter, the finite part of the charges given in \eqref{finitepart} can be written in two equivalent ways as
\bea
\mathcal J_{(i)}&\equiv&\frac{1}{8\pi G}\int_S d^2S  \sqrt{-h^{(0)}} (V_{ab} +2 i_{ab}) \xi^a_{\rom{rot}(i)}n^b = - \frac{1}{8\pi G}\int_S d^2S \sqrt{-h^{(0)}}   W_{ab} \xi^a_{\rom{boost}(i)}n^b ,\,\label{eq:J} \nonumber \\
\mathcal K_{(i)}&\equiv&\frac{1}{8\pi G}\int_S d^2S  \sqrt{-h^{(0)}}  (V_{ab}+2 i_{ab}) \xi^a_{\rom{boost}(i)}n^b = \frac{1}{8\pi G} \int_S d^2S  \sqrt{-h^{(0)}} W_{ab} \xi^a_{\rom{rot}(i)}n^b \; , \label{eq:K} \nn\\
\eea
where $V_{ab}$, $W_{ab}$ are defined in \eqref{defnlV}-\eqref{defnlW}. 

When parity conditions hold, the integral of all quadratic pieces vanish\footnote{This is a non-trivial statement. The reason for this has been previously explained in \cite{Mann:2008ay}.} and so does the divergent part (the integral of $i_{ab}$ vanishes as a consequence of the linearization stability constraints). It then implies that asymptotic linearity \eqref{al8} holds and the charges thus agree with the Abbott-Deser formula
\bea
Q_{-\xi_{(0)}}[g ; \bar g] = \int_S  k_{-\xi_{(0)}} [ g-\bar g ; \bar g ] \, .
\eea
One can check \cite{Mann:2008ay} that for the Kerr black hole, one gets the standard result $\mathcal J_{(3)} = + m a$ for $\xi_{(0),rot}=\frac{\p}{\p \phi}$.

When parity conditions are not imposed on $\sigma$ and $k_{ab}$, the charges have a divergent contribution and contain a finite part with  quadratic terms in the fields that cannot be obtained from the linearized theory alone. In other words, the property of asymptotic linearity
\bea
k_\xi [\delta g ; g ] = k_\xi [\delta g ; \bar g ], \label{al8}
\eea
would not hold for Lorentz transformations in this case. 

\subsubsection{An extended covariant phase space}

We have thus seen that from the symplectic structure, when imposing $i_{ab}=0$ and $\s$, $k_{ab}$ to be parity even, one can define Poincar\'e charges that are asymptotically linear and thus reduce to standard results known in the litterature. Also, we have generalized the phase space of ABR \cite{ABR} by allowing for an SDT parity even $k_{ab}$. We have thus allowed for even parity supertranslations that obey $(\Box+3)\omega=0$. In agreement with the work of Regge and Teitelboim, we have seen that the associated charges are trivial.

\setcounter{equation}{0}
\section{Mann-Marolf counter-term charges and parity conditions}
\label{sec:MM}

As we have seen in the previous section, the Einstein-Hilbert action is stationary, upon varying it, up to a boundary term. It was realized by G. Gibbons and S. Hawking in \cite{Gibbons:1976ue} that when one considers variations with $\delta h_{ab}=0$ on the boundary, one should add a boundary term to the Einstein-Hilbert action to have a good variational principle. This term is known as the Gibbons-Hawking term
\beqs\label{GH}
S=S_{EH} + S_{GH} \; ,  \qquad S_{GH}\equiv 2 \int_{\partial \mathcal{M}} K\: ,
\eeqs
where $K$ is the trace of the extrinsic curvature. In a more general way, one would like to consider an action principle which is stationary under the full class of variations corresponding to the space of paths under which the integral is performed. The boundary terms should then vanish under any allowed variations, and not just say $\delta h_{ab}=0$. The action \eqref{GH} is actually well defined for spatially compact geometries but may diverge for non-compact ones (see also \cite{Hawking:1995fd}). To define the action in this latter case, one should choose a reference background $g_0$ and write
\beqs
S'=S[g]-S[g_0]=S_{EH}+ 2 \int_{\partial \mathcal{M}} (K-K_0) \;.
\eeqs

The Mann-Marolf counterterm is a prescription for a good variational principle for asymptotically flat spacetimes which are described by metrics of the Beig-Schmidt form.
 Inspired by this ``reference background approach" just described, see also \cite{Gibbons:1976ue,Hawking:1995fd,Brown:1992br,Kraus:1999di,Mann:1999pc,Mann:2002fg,deHaro:2000wj,Astefanesei:2005ad,Astefanesei:2006zd}, it was proposed in \cite{Mann:2005yr} to define the action for asymptotically flat spacetimes as
\begin{equation}
\label{covaction1} S_{MM} =  \frac{1}{16\pi G} \int_{\cal M} d^4 x \sqrt{-g}\,R + \frac{1}{8\pi G} \int_{\rho = \Lambda} d^3 x \sqrt{-h} \,(K - \hat K) ,
\end{equation}
where $\hat K \equiv h^{ab} \hat K_{ab}$ and $\hat K_{ab}$ is defined to satisfy
\begin{equation}
\label{Khat}
{\cal R}_{ab} = \hat K_{ab} \hat K - \hat K_a{}^{c} \hat K_{cb},
\end{equation}
where ${\cal R}_{ab} $ is the Ricci tensor of the boundary metric $h_{ab}$ at a cut-off $\rho=\Lambda$. This equation, because it is quadratic in $\hat K_{ab}$, admits more than one solution for $\hat{K}_{ab}$. The prescription of \cite{Mann:2005yr} consists of choosing the solution that asymptotes to the extrinsic curvature of the boundary of Minkowski space as the cut-off $\rho=\Lambda$ is taken to infinity. It was then shown that the action is finite on-shell and also that asymptotically flat solutions are stationary points under all variations preserving asymptotic flatness. Indeed, one starts by computing the variation of the action for asymptotically flat spacetimes that are defined as spacetimes admitting a Beig-Schmidt expansion. One can check that this variation is equal on-shell to (see also \cite{Virmani:2011gh} and \cite{Mann:2008ay} for computational details)
\begin{eqnarray}\label{deltaS}
\delta S_{MM} = - \frac{1}{16 \pi G} \int_{\mathcal{H}} d^3 x \sqrt{-h^{(0)}}\:  E^{(1)\: ab} \delta k_{ab} ,
\end{eqnarray}
where $E_{ab}^{(1)} = -\s_{ab} - \s h_{ab}^{(0)}$ is the first order electric part of the Weyl tensor and boundary terms at the future and past boundaries, let us call them $C_{\pm}$, have been neglected. We will study in more detail the importance of these contributions at the end of this section. Here, Einstein's equations imply that $h_{ab}^{(0)}$ is locally the metric on the hyperboloid and therefore we set $\delta h_{ab}^{(0)} = 0$ by fixing the boundary metric to be the unit hyperboloid.

 Now, the variation of the action as obtained in \eqref{deltaS} is obviously zero if we consider, as is done in \cite{Mann:2005yr}, the phase space where $\delta k_{ab}=0$. However, after integrations by parts, one can also write \eqref{deltaS} as
 \beqs
 \delta S_{MM}= \frac{1}{16 \pi G} \int_{\mathcal{H}} d^3 x \sqrt{-h^{(0)}}\: \Big(-\s^a \delta(\DD^b k_{ab}) + \s\delta(h^{(0)\:ab} k_{ab}) \Big).
 \eeqs
We thus see that a good variational principle can be defined under the more general assumption that the symmetric tensor $k_{ab}$ is a traceless and divergenceless tensor
\beqs
k_a^{\:\:a}=0, \qquad \DD^b k_{ab}=0 .
\eeqs
This action provides thus also a good variational principle for asymptotically flat spacetimes defined by our generalized ansatz when $k_{ab}$ is restricted to be an SDT tensor\footnote{Actually, this is not completely true as parity conditions should also be imposed to ensure that the boundary contributions at future and past boundaries can be legitimately neglected. See the end of this section for more details about this. }. This is our justification for having considered solutions with $k_{ab}$ non-trivial but such that $k_{ab}$ is an SDT tensor. 

\subsubsection{Boundary stress-energy tensor}

Here, let us stick for a moment to the stronger assumption $k_{ab}=0$.
Based on this good variational principle, where the action has been supplemented by a counterterm, it was proven natural in \cite{Mann:2005yr} to define a renormalized, or boundary, stress-energy tensor. The boundary stress tensor is defined as the functional derivative of the on-shell action with
respect to $h_{ab}$
\begin{equation}\label{StressTensor}
T_{ab} \equiv -\frac{2}{\sqrt{-h}}\,\frac{\delta S}{\delta h^{ab}} = -\frac{1}{8 \pi G}\,\left(\pi_{ab} - \hat{\pi}_{ab} + \Delta_{ab}
\right),
\end{equation}
where $\pi^{ab}$ is the conjugate momenta defined as $\pi^{ab}=K h^{(0)\:ab}-K^{ab}$, but also $\hat{\pi}^{ab}=\hat{K} h^{(0)\:ab} - \hat{K}^{ab}$ and $\Delta_{ab}$ is an additional term that was first overlooked in \cite{Mann:2005yr} but later dealt with in \cite{Mann:2008ay}.  The formal expression for $\Delta_{ab}$ is not needed here but can be found in  appendix A of \cite{Mann:2008ay}, see also \cite{Virmani:2011gh}. The asymptotic expansion of the stress tensor is given by \cite{Mann:2008ay}
 \be
T_{ab} = \frac{1}{8 \pi G} \left( T^{(1)}_{ab} + \frac{ T^{(2)}_{ab} + \Delta^{(2)}_{ab}}{\rho} + o(\rho^{-1}) \right),\label{stress1}
 \ee
where \cite{Mann:2005yr, Mann:2006bd, Mann:2008ay}
\bea
T^{(1)}_{ab} &= & E^{(1)}_{ab} = -\s_{ab} -\s h^{(0)}_{ab},  \\
T^{(2)}_{ab} &=& -h^{(2)}_{ab} + \left( \frac{5}{2} \,\s^2 + \s_c \s^c +
\frac{1}{2} \,\s_{cd} \s^{cd} \right) h^{(0)}_{ab}  - 2\,\s_a
\s_b - \s \s_{ab} - \s_{a}{}^c\s_{cb}  \\
&=& E_{ab}^{(2)} - \gamma_{ab}, \\
\gamma_{ab}&=& 2 \s \s_{ab} + \frac{5}{2} \s^2 h^0_{ab} + \s_a^c \s_{cb} - \frac{1}{2} \s_{cd} \s^{cd} h^0_{ab},  \\
\label{DeltaPi2}
\Delta^{(2)}_{ab} &=&
-\frac{1}{4} \left[ 9 \s_c \s^c h^{(0)}_{ab} -29 \s_a \s_b + 63\s\s_{ab} + 24 \s_{ap}\s_{b}{}^{p} -5 \s_{cd}\s^{cd}h^{(0)}_{ab} + 45 \s^2 h^{(0)}_{ab}\right. \nn \\
& &
\left. \quad -  3\s_{mnp}\s^{mnp}h^{(0)}_{ab} + 9 \s_{pq(a}\s^{pq}{}_{b)} -3 \s^{pq}\s_{pq(ab)} - 2\s^e\s_{e(ab)} \right]. \label{Delta}
\eea
Here, $E_{ab}^{(1)}$ and $E_{ab}^{(2)}$ are the first and second order terms in the expansion of the electric part of the Weyl tensor.

\subsubsection{Counter-term charges and Ashtekar Hansen charges revisited}

The boundary stress tensor (\ref{StressTensor}) obtained from the action (\ref{covaction1}) can be used to define the conserved charges \cite{Mann:2005yr}
 \begin{equation}\label{GenericCharge}
 Q[\xi] = \frac{1}{8 \pi G} \int_{S_\rho} d^2 S \sqrt{-h} \,  T_{ab} \,  u^a  \xi^{b},
\end{equation}
for any asymptotic Killing field $\xi^{a}$, where $h_{ab}$ is the induced metric on the hyperboloid $\mathcal{H}_\rho$ defined as a constant $\rho$ slice, $S_\rho$ is a Cauchy surface in $\mathcal H_\rho$ and $u^{a}$ is a timelike unit vector in $\mathcal{H}_\rho$ normal to $S_\rho$. Expanding the expression in powers of $\rho$ for rotations and boosts, one notices a potentially linearly divergent term in $\rho$. However, since $E_{ab}^{(1)}$ admits $\s$ as its scalar potential, the divergent term is in fact zero \cite{Ashtekar:1978zz}.
Then, the finite part\footnote{Here, we do not consider translations as we readily find exactly the same definition as the one given by Ashtekar-Hansen \cite{Ashtekar:1978zz} that we derived in the previous chapter. } of \eqref{GenericCharge} reduces to \cite{Mann:2006bd, Mann:2008ay}
\bea
Q[\xi_{(0)}] = \frac{1}{8 \pi G} \int_{S} d^2 S \sqrt{-h_{(0)}}  (E_{ab}^{(2)}-\sigma E_{ab}^{(1)}- \gamma_{ab} -\Delta_{ab}^{(2)})\xi_{(0)}^a n^b \: ,\label{tough}
\eea
where $n^a$ is the leading order coefficient of $u^a$ in the asymptotic expansion and $S$ is the unit two-sphere. It is now easy to show that $\gamma_{ab}$ does not contribute to conserved charges as
\bea
\gamma_{ab} \xi_{(0)}^a &=&  D^a \left( \xi_{(0)}^c D_{[a} \kappa_{b]c} + \kappa_{c[a} D_{b]} \xi_{(0)}^c \right) +  D^a ( \s^2 D_{[a} \xi^{(0)}_{b]})  - 4 D^a ( \s \s_{[a} \xi^{(0)}_{b]}),
\eea
where $\kappa_{ab}\equiv T^{(\s)}_{ab}$, can be written as a total divergence on $S$. The tensor $\Delta^{(2)}_{ab}$ also does not contribute to conserved charges. Indeed, from our classification of SD tensors in section \ref{sec:SD}, one realizes that it can be written as
\bea
\Delta^{(2)}_{ab} &=& -\frac{7}{4} (\curl^2 \kappa)_{ab}-\frac{3}{4} (\curl^4 \kappa)_{ab}.
\eea
Imposing the linearization stability constraints stating that $\kappa_{ab}$ must be associated to trivial charges, the result immediately follows. Thus, expression \eqref{tough} simply reduces to our previous expression \eqref{eq:Jb}-\eqref{eq:Kb}
\bea
Q[\xi_{(0)}] = \frac{1}{8 \pi G} \int_{S} d^2 S \sqrt{-h_{(0)}}  (E_{ab}^{(2)}-\sigma E_{ab}^{(1)})\xi_{(0)}^a n^b \: .
\eea
We have thus shown that the linearization stability constraints, established in the previous chapter, are all that is needed to show the equivalence between the counter-term charges given in terms of the electric part of the Weyl tensor and the Ashtekar-Hansen charges given in terms of the magnetic part of the Weyl tensor. Note that this was expected as we proved in the previous chapter that only six non-trivial independent charges associated to Killing vectors can be defined. It would be interesting to see if the construction of a boundary stress-energy tensor can be generalized when $k_{ab}$ and/or $i_{ab}$ are non-trivial \footnote{Let us point out the recent work \cite{Virmani:2011pf} where generic supertranslations are considered in this set-up.}.

\subsubsection{Parity conditions and boundary contributions in the variational principle}

In \cite{Mann:2006bd, Mann:2008ay}, the equivalence between the counter-term charges and the Ashtekar-Hansen charges was established using the even parity condition on $\sigma$
\bea
\sigma(\tau , \theta, \phi ) = \sigma (-\tau , \pi - \theta ,\phi + \pi) \:.
\eea

We have reviewed in the two previous sections that, both in Hamiltonian framework or in covariant phase space methods, asymptotically flat spacetimes at spatial infinity have only been defined when parity conditions on the first order part of the boundary fields are imposed. We have seen that these conditions have been introduced so that Lorentz charges are finite and so that the canonical structure, or in Lagrangian formalism the covariant symplectic structure, is also finite. Now, as we have reviewed here with the Mann-Marolf construction \cite{Mann:2005yr}, see also \cite{Gibbons:1976ue,Hawking:1995fd}, it seems that asymptotically flat spacetimes at spatial infinity admit a variational principle whether or not parity conditions on the first order part of the boundary fields in the asymptotic cylindrical or hyperbolic radial expansion hold, at least when one neglects boundary terms at the past and future boundaries \cite{Gibbons:1976ue,Hawking:1995fd,Mann:2005yr}. It is however quite intriguing that even though the action is finite, the symplectic structure and the conserved charges are infinite when parity conditions do not hold. One would like to think that the action determines the entire dynamics and therefore in particular the symplectic structure and conserved charges\footnote{ We thank D. Marolf and A. Virmani for drawing our attention to this issue and A. Ashtekar for emphasizing the role of past and future boundary terms in the variational principle.}. This apparent puzzle was one of the motivations of the work presented in the next section. To understand the mismatch, one has to realize that \\

\textit{The boundary contributions in the variational principle of  Mann and Marolf} \cite{Mann:2005yr} \textit{ vanish if one imposes parity conditions but do not in general. These boundary contributions are responsible for the logarithmic divergence of the covariant symplectic structure when parity conditions are not imposed. }\\

 Although we have not stated so previously, the construction of the boundary stress-tensor thus implicitly assumes that parity conditions have been imposed. 
To see how these boundary contributions emerge and reproduce the logarithmic divergence of the symplectic structure, we  limit the domain where the variational principle is defined between an initial and a final spacelike hypersurface that we denote as $\Sigma_\pm$. Such temporal slices are relatively boosted with respect to each other close to spatial infinity. The spheres lying at the intersection of the boundary hyperboloid $\mathcal H$ with the hypersurfaces $\Sigma_\pm$ are denoted as $S_\pm$ and are defined at hyperboloid times $\tau = \tau_\pm$, see Figure~3.1.

\begin{figure}[!hbt]
  \centering \label{fig1}
\includegraphics[width=0.3\textwidth]{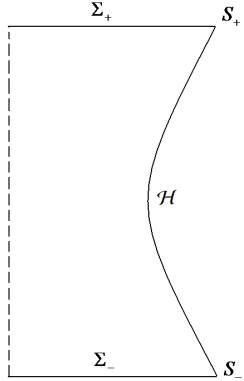}
  \caption{The variational principle is defined in the spacetime delimited by initial and final time slices $\Sigma_\pm$ and the hyperbolic cut-off $\mathcal H$.}
\end{figure}
Following previous discussions, we find
\begin{eqnarray}
\delta S_{MM} = \mathcal C_\pm \:  ,
\end{eqnarray}
where
\bea
\mathcal C_\pm = \pm \frac{1}{16 \pi G}\int_{\Sigma_\pm} (d^3 x)_\mu \Theta^\mu [\delta g] \; ,
\eea
and where
$\Theta^\mu[\delta g] (d^3 x)_\mu$ is the presymplectic form. Using the asymptotic expansion of the fields and $\int_{\Sigma} (d^3 x)_a \Theta^a = - \int^{\rho = \Lambda} d\rho \int_S (d^2 x)_a \Theta^a$, we obtain a linearly divergent term, a logarithmically divergent term and a finite term $\mathcal{F}_{\pm}$ 
\bea\label{loglog}
\mathcal C_\pm = \delta (\Lambda \mathcal{R}_{\pm}) \mp \frac{\log \Lambda}{16 \pi G} \int_{S_\pm} d^2 S \sqrt{-h_{(0)}}n_a \left( 4 \s \DD^a \delta \s + \frac{1}{2}k^{bc} \DD^a \delta k_{bc} \right)+\mathcal F_\pm \; ,  \label{varS2}
\eea
where we denote by $\mathcal F_\pm$ the finite terms that are obtained from integrating the presymplectic form in the bulk of $\Sigma_\pm$ after removing the logarithmic divergences. We expect that the finite terms $\mathcal F_\pm$, as well as the linearly divergent term, might be set to zero by imposing appropriate boundary conditions on $\Sigma_\pm$ and by adding appropriate boundary terms to the action on $\Sigma_\pm$. The derivation of the boundary conditions and boundary terms at $\Sigma_\pm$ would require a careful analysis that we will not perform here. 

Having derived the boundary contributions to the variational principle, we see that the logarithmically divergent term  in \eqref{loglog} cannot be set to zero for general variations $\delta \s$, $\delta k_{ab}$ unless additional boundary conditions are imposed on the first order fields. If one imposes that $\s$ and $k_{ab}$ satisfy specific parity conditions, then these boundary conditions vanish. For generic $\sigma$ and $k_{ab}$, one could compute the contribution of these boundary terms to the symplectic structure and would actually recover the divergent part we obtained before
\beqs
W[\delta_1 g; \delta_2 g]|_{\Sigma}
&=& \frac{\log \Lambda}{4\pi G} (d^3 x)_a \sqrt{-h^{(0)} } \Big( \delta_1 \sigma \DD^a \delta_2 \sigma  -\frac{1}{4} \epsilon^{a c e } \delta_1 k_c^{\:d} \: \delta_2   B^{(1)}_{ed} - (1 \leftrightarrow 2) \Big) + o(\rho^{-1}) \; .  \nn \\
\eeqs

To summarize, we have seen that the variational principle is actually ill-defined on future and past boundaries $\Sigma_\pm$ when parity conditions are not imposed. As a direct consequence of this, one sees that the contributions to the symplectic structure of the boundary terms \eqref{loglog} make the symplectic structure infinite.  The Mann-Marolf action we have described in this  section is thus only a valid variational principle, i.e. boundary contributions can be neglected so that $\delta S_{MM}=0$, if some other boundary conditions such as the parity conditions hold.  We review in the next section another way to regulate these infinities without imposing parity conditions, allowing us to deal with a phase space where logarithmic and generic supertranslations satisfying $(\Box+3)\omega=0$ are allowed transformations.

\setcounter{equation}{0}
\section{Relaxing parity conditions}
\label{paritycond}

Both in the usual covariant phase space and in the Hamiltonian framework, different logarithmic translation frames or generic parity supertranslation frames are not related by transformations associated with finite Hamiltonian or Lagrangian generators when one uses the standard canonical bracket or symplectic structure. There is therefore no covariant phase space or canonical space that encompasses spacetimes or initial data surfaces with  such different frames.
Now, either these frames are unphysical or they are physical. On the one hand, if they are unphysical, one would expect that the corresponding transformations are pure gauge (see \cite{logambig} for arguments that logarithmic translations are unphysical). The fact that logarithmic translations and generic parity supertranslations are not degenerate directions of the symplectic structure - they are not allowed directions of the symplectic structure in the first place - is however in tension with the intuition that pure gauge transformations should be degenerate directions of the symplectic structure. On the other hand, if the choice of frame has some implication for the dynamics of the theory, it is important to consider these transformations and study if the infinities can be suitably regularized. If an enlarged phase space exists where both logarithmic and generic supertranslations are allowed in the first place, it would allow us to settle these questions of frame-fixing. However, there does not exist in the literature a construction of a consistent phase space where parity conditions have not been imposed.

We have just seen that this relaxation will strongly rely on the regulation of the standard covariant phase space symplectic structure. In Hamiltonian formalism, this would equivalently rely on the canonical bracket defined from the canonical fields $\g_{ij}$ and $\pi^{ij}$. Now, it is important to remember from chapter 1 that the definition of the canonical structure depends on what fields are considered to be canonical. We reviewed that, in the work of \cite{Regge:1974zd}, the asymptotic values for the shift and lapse functions at infinity should be considered as additional canonical variables in addition to $\g_{ij}$ and $\pi^{ij}$. The consideration of additional canonical variables might then lead to a modification of the canonical structure. Also, we have seen above that the algebraic derivation of the covariant phase space symplectic structure from the bulk Lagrangian suffers from ambiguities \cite{Wald:1993nt,Iyer:1994ys} and that the Mann-Marolf variational principle is well-defined only under the assumption that future and past boundary terms can be dealt with without affecting the analysis at spatial infinity.

In this section, we show that four-dimensional asymptotically flat spacetimes at spatial infinity can be defined from first principles without imposing parity conditions or restrictions on the Weyl tensor. In section \ref{sec:action}, the Einstein-Hilbert action is shown to be a correct variational principle when it is supplemented by an anomalous counter-term which breaks asymptotic translation, supertranslation and logarithmic translation invariance. We also fix the ambiguity in the definition of the symplectic structure by looking at the contribution of the counterterms added to the action such that they minimally cancel the divergences that are present in the absence of parity conditions. We see in section \ref{LagCharges} that the contributions of these counter-terms to the charges make the Poincar\'e transformations as well as the supertranslations and the logarithmic translations finite, conserved and non-trivial. Lorentz charges are generally non-linear functionals of the asymptotic fields but reduce to well-known linear expressions when parity conditions hold. In section \ref{subsec:algebra}, we discuss how those transformations represent the asymptotic symmetry group. We see that Lorentz charges as well as logarithmic translations transform anomalously under a change of regulator.  Eventually, in section \ref{subsec:Hamiltonian}, we briefly discuss how this covariant construction can be  translated into the Hamiltonian formalism.

\subsection{Action principle and finite symplectic structure}
\label{sec:action}
As  another mechanism for canceling the logarithmic divergence, without assuming that $\s$ or $k_{ab}$ obey parity conditions, we simply propose to consider the action
\beqs
S= \frac{1}{16 \pi G} \int_\mathcal{M} d^4x \sqrt{-g} R +\mathcal{S}_{\Sigma_{\pm}} +\mathcal{S}_\mathcal{H},
\eeqs
where the spatial boundary counterterm $\mathcal{S}_\mathcal{H}$ is
\bea
\mathcal{S}_\mathcal{H} = \mathcal{S}^{\text{part}}_{\mathcal{H}} +\frac{\log\Lambda}{4 \pi G} \Big( S^{(\s)} + S^{(k)}\Big)   \label{action2},
\eea
and where $\mathcal{S}^{\text{part}}_{\mathcal{H}}$ is a counter-term action built algebraically from the boundary fields, i.e. such that its contribution in the variational principle contain no logarithmic divergence. For simplicity, we will consider $\mathcal{S}^{\text{part}}_{\mathcal{H}}$ to be the Mann-Marolf counterterm discussed earlier. In the above expression, we choose the actions $S^{(\s)}$ and $S^{(k)}$ to be
\bea
S^{(\s)} &=&   \int_{\mathcal{H}} d^3 x  \sqrt{-h^{(0)}} \: \Big( \frac{1}{2}  \: \s  (\Box +3) \s \Big) \pm \frac{1}{2} \int_{S_{\pm}} (d^2 x)_a \s \DD^a \s  \nn\\
 &=&   \int_{\mathcal{H}} d^3 x  \sqrt{-h^{(0)}} \: \Big( -\frac{1}{2}  \: \DD_a \s \DD^a \s +\frac{3}{2}\s^2 \Big) \pm \int_{S_{\pm}} (d^2 x)_a \s \DD^a \s   \label{Ss},\\
S^{(k)} &=& \int_{\mathcal{H}} d^3 x \sqrt{-h^{(0)}} \: \Big( \frac{1}{16} \: k^{ab} (\square - 3)k_{ab} \Big) \pm\frac{1}{16} \int_{S_\pm} (d^2 x)_a k^{cd}\DD^a k_{cd} \nn\\
&=& \int_{\mathcal{H}} d^3 x \sqrt{-h^{(0)}} \: \Big( \frac{1}{4} \: B^{(1)\: ab} B^{(1)}_{ab} \Big) \pm \frac{1}{8} \int_{S_\pm} (d^2 x)_a k^{cd}\DD^a k_{cd} \label{Sk} \; ,
\eea
for the scalar $\s$ and the SDT tensor $k_{ab}$ defined on the hyperboloid. In the above expression, we have defined
\beqs
\pm  \int_{S_{\pm}} (d^2 x)_a \s \DD^a \s &\equiv&+ \int_{S_{+}} (d^2 x)_a \s \DD^a \s -  \int_{S_{-}} (d^2 x)_a \s \DD^a \s \; ,\nn\\
\pm \int_{S_\pm} (d^2 x)_a k^{cd}\DD^a k_{cd}  &\equiv&+ \int_{S_+} (d^2 x)_a k^{cd}\DD^a k_{cd}- \int_{S_-} (d^2 x)_a k^{cd}\DD^a k_{cd}\; .
\eeqs
Note that, apart from boundary terms at $S_{\pm}$, these are precisely the actions for the fields $\s$ and $k_{ab}$ we already obtained in \eqref{Lkb} when discussing linearization stability constraints of Einstein's equations.  As we have said at that time, the variations of the actions $S^{(\s)}$ and $S^{(k)}$ reduce on-shell to a boundary term at $S_\pm$ because the equations of motion for $S^{(\s)}$ and $S^{(k)}$ are precisely the equations obeyed by $\s$ and $k_{ab}$ obtained from Einstein's equations. 

When parity conditions hold, the variations of the actions $S^{(\s)}$ and $S^{(k)}$ are identically zero off-shell.
In the absence of parity conditions,  when varying the action we see that the sum of the bulk and counter-term boundary terms at $S_\pm$ precisely cancel, for the choice of coefficients in \eqref{action2} and the choice of boundary terms in the actions \eqref{Ss}-\eqref{Sk}. 
We therefore obtained a variational principle without need for parity conditions. The terms that we added to the Mann-Marolf action at the spatial boundary are boundary terms which vanish on-shell and whose variations are also zero on-shell. In some sense, our approach is a refinement of \cite{Mann:2005yr} which consists in fixing the off-shell boundary value of the action at spatial infinity, which was left unfixed in \cite{Mann:2005yr}, in order to cancel the divergences at the boundary of $\Sigma_\pm$.

The action \eqref{action2} explicitly breaks translation, logarithmic translation and supertranslation invariance but does not break Lorentz invariance. The presence of logarithmic counter-terms is reminiscent of the Weyl anomaly \cite{Capper:1974ic,Deser:1976yx} in the holographic renormalization of anti-de Sitter spacetimes in odd spacetime dimensions \cite{Henningson:1998gx}. We will refer to the action
\bea
\mathcal A = \frac{1}{8 \pi G} \int_{\mathcal{H}} d^3 x  \sqrt{-h^{(0)}} \: \Big( \s  (\Box +3) \s +\frac{1}{8} k^{ab} (\square - 3)k_{ab} \Big)\; ,
\eea
as the (super/log-)translation anomaly. The anomaly is invariant under all symmetries that are broken. Indeed, translations do not act on the fields $\sigma$ and $k_{ab}$. Logarithmic translations act as $\delta_H \s= H$ and supertranslations as $\delta_{\omega} k_{ab}= 2 \omega_{ab} +2 \omega h^{(0)}_{ab}$ but the anomaly is invariant up to boundary terms at timelike boundaries (that we neglect for this argument). Therefore, the Noether charges of the anomaly associated with the (super/log)-translation symmetries represent the algebra of (super/log)-translations. The Wess-Zumino consistency conditions \cite{Wess:1971yu} are therefore obeyed. Since no holographic model for asymptotically flat spacetimes is known, we unfortunately cannot try to match the flat spacetime anomaly to a QFT model.

Even though the anomaly is zero on-shell for all metrics obeying the boundary conditions, it affects the dynamics mainly because its symplectic structure is non-zero on-shell as we now review. 

\subsubsection{A finite symplectic structure}

Now that we have found another way to regulate infinities in the action in the absence of parity conditions, we propose to fix the ambiguity \cite{Wald:1993nt,Iyer:1994ys} in the definition of the symplectic structure using the boundary terms in the action principle \eqref{action2}. Effectively, we  fix the boundary terms in the symplectic structure in such a way that the logarithmic divergences, present when parity conditions are not imposed, cancel. The final symplectic structure that we define has the form
\bea
\Omega[\delta_1 g,\delta_2 g] = \Omega_{\text{bulk}}[\delta_1 g,\delta_2 g] +  \Omega_{\text{c.t.}}[\delta_1 \s,\delta_1 k ; \delta_2 \s,  \delta_2 k]  , \label{OmegatotL}
\eea
where $\Omega_{\text{bulk}}$ is the standard bulk symplectic structure and $\Omega_{\text{c.t.}}$ is precisely the boundary symplectic structure that can be derived from the boundary action in \eqref{action2}.
Indeed, even though the logarithmic counter-term action is zero on-shell, it has a non-vanishing boundary contribution to the symplectic structure. The free actions for $\s$ and $k_{ab}$ introduced in \eqref{Ss}-\eqref{Sk} are zero on-shell but their symplectic structures  (defined with the same conventions as in the bulk) are the Klein-Gordon norm and the symplectic norm between two traceless transverse fields given by the integral of
\bea
\omega_{(\s)}[\delta_1 \s , \delta_2 \s] &=& (d^2 x)_a \sqrt{-h^{(0)}} \Big( \delta_1 \s \p^a \delta_2 \s  - (1 \leftrightarrow 2) \Big),\label{omegas} \\
\omega_{(k)}[\delta_1 k , \delta_2 k] &=& (d^2x)_a \sqrt{-h^{(0)}}  \Big( \frac{1}{4}\eps^{adc}\delta_1 B^{(1)b}_c  \delta_2 k_{db}- (1 \leftrightarrow 2) \Big) \, ,
\label{omegak}
\eea
over the sphere which is generally non-vanishing\footnote{The counter-term $\int_{\rho=\Lambda}d^3 x \sqrt{-h}(K-\hat K)$ has not the form of an off-shell action for the boundary fields and therefore it does not define a boundary symplectic structure.}. The total symplectic structure is then defined as announced in \eqref{OmegatotL} with
\bea
\Omega_{c.t.}[\delta_1 \s,\delta_1 k , \delta_2 \s,  \delta_2 k]  =  \frac{\log\Lambda}{4\pi G} \int_S \Big(
 \omega^{(\s)}[\delta_1 \s , \delta_2 \s] +  \omega^{(k)}[\delta_1 k , \delta_2 k] \Big)  \, . \label{Omegact}
\eea
The resulting prescription for fixing the boundary terms in the symplectic structure left unfixed in \cite{Lee:1990nz,Wald:1993nt,Iyer:1994ys} amounts to the prescription argued in \cite{Compere:2008us} to fix the boundary terms in the symplectic structure using the symplectic structure of the boundary terms of the action. Now, we see that this prescription is justified by the existence of a variational principle when past and future boundaries are taken into account.

\subsection{Covariant phase space charges}
\label{LagCharges}

In this section, we go back to the covariant phase space charges and look at the contributions coming from the boundary symplectic structure \eqref{Omegact}. Indeed, since the symplectic structure is finite, the conserved charges should also be finite when contributions coming from the boundary symplectic structure are taken into account. The charge one-form $\slash\hspace{-5pt} \delta Q_\xi[g]$ is now written as
\bea
\slash\hspace{-5pt} \delta Q_\xi[g] = \int_S k_\xi [\delta g ; g ] + \frac{\log\Lambda}{4\pi G} \int_S  \Big(
 \omega^{(\s)}[\delta \s , \delta_\xi \s] + \omega^{(k)}[\delta k , \delta_\xi k] \Big), \label{totkk}
\eea
evaluated on the sphere $S$ at constant time $t$ and at $\rho = \Lambda$. Here, $\delta_\xi \s$, $\delta_\xi k_{ab}$ are variations of the first order fields induced by the Lie derivative of the metric along $\xi$. 

In the previous section, we obtained the contribution from the bulk symplectic structure 
\bea\label{generalcharges2}
Q[\xi ; g] &=& \int_S  \int_{\gamma} k_\xi [\delta g ; g] \nn\\
&=& \frac{1}{16\pi G} \int_S d^2 S \, n_a \Big( \log\rho (- 2 i^a_b \hat{\xi}_{(0)}^b + 4 H \s^a - 4 \s H^a ) \nn \\ 
&& + 4 \s^a \omega - 4 \s \omega^a  - 4 \s^a H + 4\s H^a + k^{a b}H_b \nn \\
&& +2 \hat{\xi}_{(0)}^b ( -h^{(2)\: a }_b + \frac{1}{2}i^a_b+\frac{1}{2} k^{ac}k_{cb} +h_{b}^{(0)\:a} (8\s^2 +\s^c \s_c -\frac{1}{8}k_{ab}k^{ab}+k_{cd}\s^{cd} ) )  \Big) \; , \nn\\
\eea
where we observed that logarithmic translations and Lorentz charges have a divergent part.  Let us now see what happens when contributions of the boundary counter-terms are taken into account. 

 \subsubsection{Translations and supertranslations}
 
For translations and supertranslations, one checks that the boundary counter-terms do not contribute to the charges as 
\bea\label{omeja}
\int_S  \omega_{(\s)} [ \delta \s ,\delta_\xi \s ] = 0,\qquad  \int_S  \omega_{(k)} [ \delta k ,\delta_\xi k ] =0\, ,
\eea
where
\bea
 \delta_\xi \s =0, \qquad  \delta_\xi k_{ab} = 2 (\omega_{ab}+h_{ab}^{(0)}\omega ) .
\eea

For translations,  it is relatively simple to see because $ \delta_\xi \s = \delta_\xi k_{ab} = 0 $. For supertranslations, 
the first equality in \eqref{omeja} is also trivial  as these transformations do not act on $\s$. The second one can be proven after using $\delta_\xi B^{(1)}_{ab} = 0$,  since following Lemma 1 (see also \eqref{B111} and  \cite{Ashtekar:1978zz}) one can write  $B^{(1)}_{ab} \equiv - \hat \s_{ab} - h_{ab}^{(0)} \hat \s ,$ where $(\square + 3)\hat \s = 0$,  and eventually after showing that $n_a \eps^{ace}\omega_c^d \hat \s_{ed} = n_a \DD_c \Big( \eps^{ace} (\frac{1}{2} \omega^d \hat \s_{ed} - \frac{1}{2}\hat \s^d \omega_{ed} - \omega \hat \s_e)  \Big)$ is a boundary term.
  
The charges associated with translations and supertranslations are  therefore precisely the ones which can be obtained from the bulk linearized theory
\bea
Q_{(\mu)}[g ; \bar g ] = - \frac{1}{8 \pi G} \int_S d^2S E_{ab}^{(1)} n^a \DD^b \zeta_{(\mu)}\, ,\label{ch:tr}
\eea
\bea
Q_{(\omega)}[g ; \bar g ] =  \frac{1}{4\pi G}\int_S d^2 S \sqrt{-h^{(0)}} n_a \left( \s^a \omega - \s \omega^a    \right) .
\label{Qsupertransl}
\eea

Let us insist here on the fact that in the absence of parity conditions, the charges associated to supertranslations satisfying $(\Box+3)\omega=0$ are still conserved but do not vanish in general. 

 \subsubsection{Lorentz charges}

For the Lorentz charges, we saw that in the absence of parity conditions there was a logarithmically divergent part. Following our prescription, this divergence should be exactly canceled by the counter-term contributions to the total charge \eqref{totkk}. We have checked that the divergence indeed cancels after computing the following relationship between symplectic structures and Noether charges
\bea
\omega_{(\s)}( \delta \s,\mathcal L_{-\xi_{(0)}} \s) &=& \delta \Big( \sqrt{-h^{(0)}}T^{(\s)\:ab} \: \xi_{(0)\: b} (d^2x)_a \Big) - d^2S \sqrt{-h^{(0)}} n^a \DD^b (2 \xi_{(0)\: [a} \s_{b]}\delta \s)  ,\nn \\
\omega_{(k)}(  \delta k, \mathcal L_{-\xi_{(0)}} k ) &=& \delta \Big( \sqrt{-h^{(0)}}T^{(k)\:ab} \: \xi_{(0)\:b} (d^2x)_a \Big) + d^2S \sqrt{-h^{(0)}} n^a \DD^b  P_{[ab]},
\eea
where $P_{ab}=P_{[ab]}$ is an anti-symmetric tensor. Following our previous results, the Lorentz charges are thus finite and can be written in the equivalent forms
\bea
\mathcal J_{(i)}&\equiv&\frac{1}{8\pi G}\int_S d^2S  \sqrt{-h^{(0)}} (V_{ab} +2 i_{ab}) \xi^a_{\rom{rot}(i)}n^b = - \frac{1}{8\pi G}\int_S d^2S \sqrt{-h^{(0)}}   W_{ab} \xi^a_{\rom{boost}(i)}n^b ,\,\label{eq:J} \nonumber \\
\mathcal K_{(i)}&\equiv&\frac{1}{8\pi G}\int_S d^2S  \sqrt{-h^{(0)}}  (V_{ab}+2 i_{ab}) \xi^a_{\rom{boost}(i)}n^b = \frac{1}{8\pi G} \int_S d^2S  \sqrt{-h^{(0)}} W_{ab} \xi^a_{\rom{rot}(i)}n^b  \label{eq:K}. \nonumber \\
\eea
where $V_{ab}$, $W_{ab}$ are defined in \eqref{defnlV}-\eqref{defnlW}.

When parity conditions do not hold, the Lorentz charges are non-linear functionals of the asymptotic fields and therefore differ from the standard ADM and AD formulas \cite{Arnowitt:1962hi,Abbott:1981ff}. The standard ADM and AD formulas are restored when parity conditions hold.  Such asymptotic non-linearities appeared before in the context of asymptotically anti-de Sitter spacetimes. In Einstein gravity, the charges are linear functionals of field perturbations around anti-de Sitter \cite{Abbott:1981ff,Ashtekar:1984zz,Henneaux:1985tv}. However, non-linearities may appear when matter fields are present,  see e.g. \cite{Bianchi:2001de,Bianchi:2001kw,Henneaux:2006hk}.

\subsubsection{Logarithmic translations}

If we do not assume parity conditions and introduce $i_{ab}$, logarithmic translations are allowed asymptotic transformations. Remember that they modify the first order fields as
\bea
\delta_\xi \s = H ,\qquad \delta_\xi k_{ab} = 0.
\eea
Since logarithmic translations do transform  $\sigma$, the boundary terms in the symplectic structure might play a role. We find\bea
\frac{ \log\Lambda}{4\pi G} \int_{S} \omega_{(\s)}( \delta \s,\delta_H \s )  =  \frac{\log \Lambda}{8 \pi G} \delta \int_S d^2 S \sqrt{-h^{(0)}} E^{(1)}_{ab} H^a n^b \; ,
\eea
where we have used $
H \s_b - H_b \s = -\frac{1}{2} E^{(1)}_{ab} H^a -  \DD^a (H_{[a}\s_{b]})$ and discarded the total divergence term on the sphere. 
As one can see from \eqref{generalcharges2}, the bulk covariant phase space charge associated with a logarithmic translation is given by\footnote{Here,  the logarithmic translation is  defined as the sum of a logarithmic translation, as defined in chapter 2 (see also \cite{BS,B}), with a translation.}
\bea
\int_S k_\xi [\delta g ; g] = - \frac{\log \Lambda}{8 \pi G} \delta \int_S d^2 S \sqrt{-h^{(0)}} E^{(1)}_{ab}H^a n^b + \frac{1}{16 \pi G} \delta \int_S d^2 S  \sqrt{-h^{(0)}} k_{ab} n^a H^b \, .\label{infLog}
\eea
We find that the two divergent contributions are opposite of each other and exactly cancel. The remaining finite part is trivially integrable. Logarithmic translations are therefore associated with the non-trivial charges
\bea
\mathcal Q_{(H)} = \frac{1}{16 \pi G} \int_S d^2 S  \sqrt{-h^{(0)}} k_{ab} n^a H^b \; ,
\eea
which are conserved thanks to the property $\DD^a (k_{ab} H^b) = 0$.

In the restricted phase space where $k_{ab} = 0$, logarithmic translations are associated with zero charges or equivalently they are degenerate directions of the symplectic structure, in agreement with Ashtekar's result \cite{logambig}. When parity conditions are imposed, logarithmic translations are not allowed transformations and the associated charges do not exist. The presence of non-vanishing conserved charges associated with logarithmic translations is therefore a particularity of the phase space without parity conditions and with $k_{ab} \neq 0$.

\subsubsection{Some comments}

We have reviewed here that Poincar\'e charges as well as charges associated to supertranslations and logarithmic translations can be non-trivial in the generic case. To answer the question asked at the beginning of this section, this also means that logarithmic and supertranslation frames define inequivalent frames distinguished by their associated conserved charges. In this respect, the constructed phase space is bigger than the ABR phase space \cite{ABR} . 

No exact solution of vacuum Einstein's equations is known to us which breaks parity conditions. Such a solution would also possess twelve boundary Noether charges in addition to Poincar\'e, logarithmic translation and supertranslation charges. The boundary Noether charges are the Noether charges of the actions $S^{(\s)}$ and $S^{(k)}$ for the first order fields associated with the boundary Killing symmetries or equivalently with the bulk asymptotic Lorentz Killing vectors. A subclass of these solutions exists as an asymptotic series expansion at spatial infinity. Indeed, one can consistently set the logarithmic terms in the expansion \eqref{metric1} to zero and still obey Einstein's equations by fixing six linear combinations of the boundary Noether charges to zero, see section \ref{sec:LS} for details. The original Beig-Schmidt expansion \cite{BS} which uses only polynomials in $\rho$ is a consistent analytic asymptotic solution of Einstein's equations at all asymptotic orders which has six boundary Noether charges. We leave the existence, or non-existence, of a regular solution in the bulk with such charges as an open question.

It is also interesting to remark at this stage that the presence of non-vanishing charges associated with supertranslations in addition to Poincar\'e transformations is also a feature of null infinity where supertranslations along the null direction are associated with non-trivial charges as well \cite{Bondi:1962px,DrayStreubel,Wald:1999wa}. For regular asymptotic fields, one expects that supertranslation charges should be conserved at infinite past times at future null infinity or at infinite late times at past null infinity where the news tensor vanishes. Indeed, at such late or early times the expression of \cite{Bondi:1962px,DrayStreubel,Wald:1999wa} becomes conserved and proportional to the first order electric part of the Weyl tensor and matches qualitatively with our expression \eqref{Qsupertransl}. It would be interesting to make that qualitative agreement more precise by comparing the precise definition of supertranslations.

\subsection{Algebra of conserved charges}
\label{subsec:algebra}

In the last section, we obtained explicit expressions for conserved charges associated with translations, Lorentz transformations, logarithmic translations and supertranslations. We obtained that all asymptotic charges are non-trivial in general in our phase space. The set of infinitesimal diffeomorphisms form a Lie algebra defined from the commutator of generators. A natural question to ask is whether or not the algebra of translations,  logarithmic translations, Lorentz transformations and supertranslations is represented with the associated conserved charges.

General representation theorems are available \cite{Brown:1986ed,Barnich:2007bf} but one quickly realizes that they do not take into account boundary contributions to the symplectic structure. These contributions can be dealt with as follows. Every diffeomorphism in the bulk spacetime induces a specific transformation of the boundary fields through the Beig-Schmidt asymptotic expansion that identifies boundary fields from bulk fields. Therefore, the Lie algebra of infinitesimal diffeomorphisms defined from the commutator of generators also induces a Lie algebra of transformations of the boundary fields. The Poisson bracket between two charges is then defined as
\bea
\{ \mathcal Q_\xi [g ; \bar g], \mathcal Q_{\xi^\prime} [g ;\bar g] \} = - \delta_\xi \mathcal Q_{\xi^\prime} [g ;\bar g] ,\label{PB}
\eea
where the variation $\delta_\xi$ acts on the bulk fields as a Lie derivative and on the boundary fields as the transformation induced on the boundary fields from the Lie derivative of the bulk fields. It would be interesting to develop a general representation theorem which takes boundary contributions into account along the lines of \cite{Hollands:2005ya}. Here, we simply evaluate the Poisson bracket using the explicit expressions for the charges derived and taking into account the boundary field transformations.

Under an asymptotic translation $\xi = \omega(x) \p_\rho + o(\rho^{0})$ where $\DD_a \DD_b \omega +h_{ab}^{(0)}\omega = 0$, the boundary fields transform as
\bea
\delta_\omega \sigma &=& 0,\qquad \delta_\omega k_{ab} = 0,\\
\delta_\omega i_{ab} &=& 0, \qquad \delta_\omega V_{ab} = \DD_c (E^{(1)}_{ab}\omega^c ) + 2\eps_{cd (a}B^{(1)c}_{b)}\omega^d \; ,
\eea
where $E_{ab}^{(1)}$ and $B_{ab}^{(1)}$ are the first order electric and magnetic parts of the Weyl tensor while Lorentz transformations $\xi = -\xi_{(0)}$ act on the boundary fields as a Lie derivative
\bea
\delta_{-\xi_{(0)}} \sigma &=& \mathcal L_{-\xi_{(0)}}  \sigma ,\qquad \delta_{-\xi_{(0)}} k_{ab} = \mathcal L_{-\xi_{(0)}} k_{ab},\\
\delta_{-\xi_{(0)}} i_{ab} &=& \mathcal L_{-\xi_{(0)}}  i_{ab}, \qquad \delta_{-\xi_{(0)}} V_{ab} = \mathcal L_{-\xi_{(0)}} V_{ab} \,  .
\eea
Logarithmic translations\footnote{Here again, the logarithmic translation is understood as supplemented by an appropriate translation.} act as
\bea
\delta_H \sigma &=& H,\qquad \delta k_{ab} = 0,\qquad
\delta_H i_{ab}=  -\DD_c (E^{(1)}_{ab} H^c) -2 \epsilon^{mn}_{\:\:\:\:(a} B^{(1)}_{b)m} H_n \; , \\
\delta_H V_{ab}&=& -\frac{1}{2} \DD_c (k_{ab} H^c)+ 2 \DD_c (E^{(1)}_{ab} H^c) + 8 \epsilon^{mn}_{\:\:\:\:(a} B^{(1)}_{b)m} H_n \; ,
\eea
and supertranslations act as
\bea
\delta_{\omega} \sigma &=&0  ,\qquad \delta_{\omega} k_{ab} = 2 \omega_{ab}+2 h_{ab}^{(0)}\omega, \qquad \delta_{\omega} i_{ab}  = 0\; , \\
\delta_{\omega} h^{(2)}_{ab} &=& k_{c(a}\omega_{b)}^{\; c} +k_{ab}\omega +\omega^c \left( \DD_c k_{ab} - \DD_{(a}k_{b)c}\right) \nn \\
&& + \Big( \s^c \omega_{c(ab)} - \s \omega_{ab}-2 \s \omega h_{ab}^{(0)} + \omega_{(a}\s_{b)} + \s_{c(a} \omega_{b)}^c + (\s \leftrightarrow \omega) \Big) .
\eea

After an explicit evaluation, we find that all asymptotic transformations are well-represented: the Poisson bracket is anti-symmetric and is isomorphic to the semi-direct product of the Lorentz algebra with the (super)-translation and logarithmic translation algebra. In particular, the Poisson bracket between Lorentz charges and (super)-translation charges is given by
\bea
\{ \mathcal Q_{-\xi_{(0)}} , \mathcal Q_{(\omega)} \} = - \{   \mathcal Q_{(\omega)} , \mathcal Q_{-\xi_{(0)}}\} = \mathcal Q_{(\omega^\prime)} ,\qquad \omega^\prime = \mathcal L_{-\xi_{(0)}}\omega\, .
\eea
and the Poisson bracket between Lorentz charges and logarithmic translation charges is
\bea
\{ \mathcal Q_{-\xi_{(0)}} , \mathcal Q_{(H)} \} = - \{   \mathcal Q_{(H)} , \mathcal Q_{-\xi_{(0)}}\} = \mathcal Q_{(H^\prime)} ,\qquad H^\prime = \mathcal L_{-\xi_{(0)}}H\, .
\eea
Logarithmic translations and supertranslations obey the algebra
\beqs
 \{ \Q_{(\omega)} ,  \Q_{(H)} \} = - \{ \Q_{(H)} , \Q_{(\omega)} \} = \frac{1}{4\pi G} \int d^2 S \: \sqrt{-h^{(0)}} \: n_a \Big( H^a \omega - H \omega^a \Big)\; ,
\eeqs
where the right-hand side depends on the generators but does not depend on the fields. In the harmonic decomposition of $\omega$ on the sphere, the  Poisson bracket is zero for all harmonics $ l >1$ and is a Kronecker delta for the four lowest harmonics $l \leq 1$. The algebra of asymptotic conserved charges is isomorphic to the algebra of asymptotic symmetries. No non-trivial central extension of the algebra is present. Note that in order to derive these results, we made extensive use of our classification of SD tensors and their properties, as detailed in chapter 2, to simplify intermediate expressions and we discarded boundary terms. We have also used the property described in \cite{Ashtekar:1978zz,BS} (see also section \ref{sec:conservedd}) that regularity of $k_{ab}$ implies that the four conserved charges
\bea
\mathcal P_{(\mu)} = \frac{1}{8\pi G}\int_S d^2 S\, B_{ab}^{(1)} n^a \DD^b \zeta_{(\mu)},\qquad \mu=0,1,2,3,\label{noNut}
\eea
are zero. 

We obtained that all transformations: translations, logarithmic translations, Lorentz transformations and supertranslations are well-represented despite the (log/super) translation anomaly.  The fact that the Lorentz group is well-represented is not surprizing given that the cut-off needed to regularize the action, see \eqref{action2}, is invariant under asymptotic Lorentz transformations.  Now, it is also important to take into account the shifts in the action when one changes the cut-off used to regulate the action. These shifts can be analyzed as follows.

Under a change of cut-off $\Lambda$, the action will be shifted by a finite piece $S_{(0)}$ proportional to the anomaly action $ S^{(\s)} + S^{(k)}$ given in \eqref{Ss}-\eqref{Sk}. The conserved charges associated with the asymptotic Killing vector $\xi$ will then be shifted by the boundary Noether charges of the action $S^{(\s)} + S^{(k)}$ associated with the symmetry $\delta_\xi$. Using standard manipulations $\delta_\xi L = d K_\xi$, $\delta L = \frac{\delta L}{\delta \phi}\delta \phi + d \Theta [\delta \phi ]$, the boundary Noether charges are defined as $J_\xi = K_\xi - \Theta[\delta_\xi \phi]$. One then quickly sees that translations and supertranslations are associated with vanishing Noether charges $\int_S d^2 S J_\xi = 0$ while Noether charges associated with logarithmic translations are proportional to the four-momentum $\mathcal Q_{(\mu)}$ and  Noether charges associated with Lorentz transformations are given by the integral of $J_{\xi^{(0)}} = 2(T^{(\s)ab} + T^{(k) ab} )\xi^{(0)}_b (d^2 x)_a$ where $T^{(\s)ab}$ and $T^{(k)ab}$ are the stress-tensors of the actions \eqref{Ss}-\eqref{Sk}.

Therefore, under a change of regulator, the translations and supertranslation charges are invariant. Logarithmic translation charges get shifted with the four-momenta and the Lorentz charges get shifted as
\bea
\Delta \mathcal Q_{-\xi_{(0)}}[g ;\bar  g ] \sim   \int_S d^2S  \sqrt{-h^{(0)}} ( T^{(\s)}_{ab} + T^{(k)}_{ab} ) \xi^a_{(0)} n^b\, .
\eea
These shifts can be obtained similarly by varying the regulator directly into the expression for the charges associated to logarithmic translations, and rotations and boosts, before the subtraction of divergences between the bulk and the boundary. Note that these shifts agree with the analysis presented at the end of chapter 2 where it was noticed that Lorentz charges are uniquely defined for our enlarged Beig-Schmidt ansatz  up to the addition of those twelve boundary charges. Four-momenta and supertranslations are finite without needing a regulator. They are therefore manifestly unchanged by the regulator.

The situation here can be contrasted to bulk infinitesimal diffeomorphisms which induce Killing symmetries and conformal Killing symmetries of asymptotically AdS spacetimes in odd dimensions as analyzed in \cite{deHaro:2000xn,Skenderis:2000in,Hollands:2005ya,Papadimitriou:2005ii}. First, the dependence of the Lorentz charges, associated with Killing vectors, upon the choice of regulator is analogous to the shift of the stress-tensor by Weyl anomalous terms \cite{deHaro:2000xn,Skenderis:2000in}. The expression for the conserved charges \eqref{eq:K} indeed allows us to recognize $V_{ab} +2 i_{ab}$ as the second order part of the stress-tensor which generalizes the one given in \cite{Mann:2006bd,Mann:2008ay} when $k_{ab} = i_{ab} = 0$. In our case, logarithmic translations are also present and they are also shifted under a change of regulator.

In anti-de Sitter, infinitesimal diffeomorphisms associated with boundary conformal Killing vectors are well-represented even though they may act non-trivially on the action \cite{Hollands:2005ya}. Indeed, the action only varies by a c-number which depends on the boundary conditions while the dynamical phase space is preserved. The non-conservation of the associated charges is related to this c-number. In asymptotically flat spacetimes,  translations are also boundary conformal Killing vectors. Four-momenta as well as supertranslations are always exactly conserved and they do not vary under a change of regulator.

That said, the anomaly in the action can also be described as follows. When the boundary conditions do not include parity conditions, logarithmic divergent integrals in the action and in the symplectic structure should be regulated by introducing a finite cutoff which breaks asymptotic (super/log)-translation invariance. If one regulates the action by pushing the cutoff to infinity, the resulting regulated action will not be invariant under asymptotic (super/log)-translations since it would be shifted by a finite piece. Therefore, the regulated action depends on the specific (super/log)-translation frame.

Let us now try to describe how these results can be matched to the Hamiltonian formalism.

\subsection{Revisiting the Regge-Teitelboim approach}
\label{subsec:Hamiltonian}

We have seen that parity conditions on the hyperboloid are not required in order to define a consistent phase space in the hyperbolic representation of spatial infinity. Moreover, we have seen that when these conditions are relaxed, charges associated with Lorentz rotations and boosts are non-linear functionals of the first order fields and logarithmic translations and supertranslations are associated with non-trivial charges. Both these characteristics are not shared with the standard treatment of Hamiltonian charges at spatial infinity \cite{Arnowitt:1962hi,Regge:1974zd,Beig:1987aa} we reviewed in chapter 1. There, parity conditions on the sphere are imposed in order for the rotation and Lorentz boost charges to be finite. Also, charges are linear functionals of the boundary fields, logarithmic translations are not allowed transformations and parity odd supertranslations are associated with trivial Hamiltonian generators. Let us try here to resolve this tension by proposing how the results of \cite{Regge:1974zd,Beig:1987aa} can be accommodated to enlarge the phase space to fields which do not obey parity conditions. 

Once parity conditions are relaxed, the fall-off conditions are preserved by both parity even and parity odd supertranslations, and at first order in the asymptotic expansion of the fields $\g_{ij}$ and $\pi^{ij}$ by the so-called logarithmic translations which are associated to the lapse and shift functions
\bea
N^\perp &=& \log r \, K^\perp + o(r^0) ,\\
N^i &=& \log r \, K^i + o(r^0) ,
\eea
where $K^\perp$ and $K^i$ are constants. We have thus in mind that one should start by considering the phase space of metrics where the generic expansions of the fields take the form
\bea
\g_{ij} &= & \delta_{ij} + \frac{1}{r} \g_{ij}^{(1)}+ \frac{\log r}{r^2} \g_{ij}^{(ln,2)} + \frac{1}{r^2} \g_{ij}^{(2)} + o(r^{-2}) ,\label{geng1sub} \\
\pi^{ij} & = & \frac{1}{r^2} \pi^{(2)\: ij}+ \frac{\log r}{r^3} \pi^{(ln,3)\: ij}+ \frac{1}{r^3} \pi^{(3)\: ij}+ o(r^{-3})\, ,\label{geng2sub}
\eea
where, as in the covariant approach, the logarithmic branch is necessary in order to satisfy the constraints when parity conditions do not hold, as already noticed e.g. in \cite{Beig:1987aa}.

Note that we do not intend to work out in detail the Hamiltonian formalism here. Instead, our approach should be seen as a first step. Following the existence of a Lagrangian variational principle, we transpose covariant boundary conditions to the Hamiltonian formalism. By ``transpose", we mean that we consider a phase space where parity conditions imposed on the three-metric, its conjugated momenta and the supertranslations are relaxed, and so that logarithmic translations are also allowed, but where we do impose additional boundary conditions on the fields such that the angular supertranslations are fixed. These last conditions impose the, a priori, generic form of the fields present in \eqref{geng1sub}-\eqref{geng2sub} to be fixed in terms of their covariant counterparts as detailed in Appendix \ref{app:BC}. As already mentioned, it would be interesting to include mixed terms  $g_{\rho a}$ in the Beig-Schmidt expansion and allow for angular supertranslations\footnote{ In the presence of mixed terms $g_{\rho a}$, there might be a distinction between the bulk covariant phase space symplectic structure defined from the action and the one defined from the equations of motion, see \eqref{diffsymp}. One would then need to prescribe which one is the bulk symplectic structure, see \cite{Azeyanagi:2009wf} for an example where such a prescription plays an important role.}.

The canonical two-form on the canonical phase space used in the treatments of \cite{Arnowitt:1962hi,Regge:1974zd,Beig:1987aa} is the bulk canonical two-form
\bea
\Omega(\delta_1 \g,\delta_1 \pi,\delta_2 \g,\delta_2 \pi ) = \frac{1}{16 \pi G} \int_\Sigma d^3 x\Big(  \delta_1 \pi^{mn} \delta_2 \g_{mn} - \delta_2 \pi^{mn} \delta_1 \g_{mn} \Big)\; ,
\eea
defined from the bulk canonical fields $ \g_{mn}$ and $\pi^{mn}$ at the initial time surface $\Sigma$ at $t=0$. In the case of asymptotically flat spacetimes without parity conditions the bulk canonical two-form suffers from a logarithmic radial divergence. Using the boundary conditions \eqref{geng1sub}-\eqref{geng2sub}, up to first order, one can express the canonical two-form as
\bea
\Omega(\delta_1 \g,\delta_1 \pi,\delta_2 \g,\delta_2 \pi ) = (finite)  + \frac{\log \Lambda}{16 \pi G} \int_{S} d^2 S \Big( \delta_1 \pi^{(1)\: mn} \delta_2 \g^{(1)}_{mn} - \delta_2 \pi^{(1)\: mn} \delta_1 \g^{(1)}_{mn} \Big)\; , \label{can1} \nn\\
\eea
where $\Lambda$ is a large radial cut-off and $S$ is the sphere at $r = \Lambda$. Now,  exactly as in the Lagrangian formalism, we propose to modify the dynamics by adding a boundary term to the canonical form. We proceed by first writing the boundary actions \eqref{Ss}- \eqref{Sk} at $t=0$ in the 2+1 decomposition (the boundary metric becomes the real time line times the unit sphere). We then switch to the Hamiltonian formulation of the boundary action and propose to supplement the bulk canonical fields with the canonical fields of the boundary Hamiltonian. We then introduce counter-terms to the canonical form in order to minimally cancel the divergences in \eqref{can1},  following the Lagrangian prescription \eqref{Omegact}. The regulation breaks translation, supertranslation and logarithmic translation invariance. We interpret this breakdown as a consequence of the translation anomaly in the action, which is manifest only when fields have both parities.

Let us now discuss briefly the form of the Hamiltonian generators associated with asymptotic Poincar\'e transformations, logarithmic translations and supertranslations. The Hamiltonian generators contain two parts: the part coming from the bulk canonical form and the counter-term contribution that cancels the logarithmic divergences. The surface charge derived from the bulk canonical form associated with a gauge parameter $\eps^A = (\eps^\perp,\eps^m)$ is given, using the $2$-form $k_{\eps}$, by \cite{Regge:1974zd}
\bea
k_\eps^{[0m]}[\delta \g,\delta \pi ; \g,\pi] = G^{mnop}(\eps^\perp \delta \g_{op|n} - \eps^\perp_{|n} \delta \g_{op}) + 2 \eps^o \delta (\g_{on}\pi^{mn}) - \eps^m \delta\g_{no}\pi^{no} \; ,\label{RT}
\eea
where
\bea
G^{mnop} = \frac{1}{2}\sqrt{\g} \left( \g^{mo}\g^{np} +\g^{mp}\g^{no} - 2\g^{mn}\g^{op} \right)\; ,
\eea
is the inverse De Witt supermetric. Now, one can readily obtain that this expression admits at most a logarithmic divergence when one uses our boundary conditions. Indeed, the RT linear divergences trivially cancel when one uses the explicit expressions, given in Appendix \ref{app:BC}, of $\g_{ab}$ and $\pi^{ab}$ in terms of $\s$, $k_{ab}$, $h^{(2)}_{ab}$ and $i_{ab}$. The logarithmic divergence is then exactly canceled by the boundary counter-term. The resulting final expressions for the charges in Hamiltonian formalism can then be obtained by a straightforward explicit evaluation. We will not provide them here. We only note that the four-momenta are given by the usual ADM formulae, while the charges associated with rotations and boosts contain non-linear contributions in the canonical fields.

%%%%%%%%%%%%%%%%%%%%%%%%%%%%%%%%%%%%%%%%%%%%%%%%%%%%%%%%%%%%%%%%%%%

\setcounter{section}{0}

\renewcommand{\thesection}{\Roman{part}.\Alph{section}}

   \setcounter{equation}{0}

 \chapter*{Appendices Part I \markboth{Appendices Part I}{Appendices Part I}}
\addcontentsline{toc}{chapter}{Appendices Part I} 

 \hrule
\vspace{2cm}

 \setcounter{equation}{0}

 \section{Schwarzschild in Beig-Schmidt coordinates}
\label{App: BS}

In here, we would like to put the Schwarzschild solution whose metric is given by
\begin{eqnarray}\label{Schw1}
ds^2=-V(r) dt^2 + V(r)^{-1} dr^2 + r^2 (d\theta^2+\sin^2 \theta d\phi^2), \qquad
V(r)=1-\frac{2M}{r},
\end{eqnarray}
in Beig-Schmidt coordinates up to second order. For the sake of simplicity, we will set it in a gauge such that $k_{ab}=0$. The strategy to implement all this goes as follows. \\

\textbf{1st step} \\

Make a change of coordinates that brings the metric (in the usual coordinates $t,r,\theta,\phi$) to hyperbolic coordinates
\begin{eqnarray}
r &=& z \cosh \zeta \; , \nonumber \\
t &=& z \sinh \zeta \; ,
\end{eqnarray}
and expand it to second order in $z$. The metric up to second order in the radial coordinate $\rho$ is completely specified given $\sigma^{(1)}$,$\sigma^{(2)}$, $A^{(1)}_a$, $A^{(2)}_a$, $h^{(0)}_{ab}$, $h^{(1)}_{ab}$ and $h^{(2)}_{ab}$. For the Schwarzschild solution, we find
\begin{eqnarray}\label{Schw2}
ds^2&=& d\zeta^2 \biggl [ -z^2 \cosh^2 \zeta +2M z \cosh \zeta + \frac{z^4 \sinh^2 \zeta \cosh^2 \zeta}{z^2  \cosh^2 \zeta -2M z \cosh \zeta}  \biggr ] \nonumber \\
&& + 2 dz d\zeta \biggl [  -z \sinh \zeta \cosh \zeta + 2 M \sinh \zeta + \frac{z^3 \sinh \zeta \cosh^3 \zeta}{z^2  \cosh^2 \zeta -2M z \cosh \zeta} \biggr ] \nonumber \\
&& + dz^2 \biggr [ -\sinh^2 \zeta +\frac{2M}{z} \frac{\sinh^2 \zeta}{\cosh \zeta }+ \frac{z^2 \cosh^4 \zeta}{z^2  \cosh^2 \zeta -2M z \cosh \zeta}\biggr ] \nonumber \\
&& + z^2 \cosh^2 \zeta (d\theta^2 +\sin^2 \theta d\phi^2).
\end{eqnarray}
If we now expand this expression up to second order in $z$ and remember that
\begin{eqnarray}
h^{(0)}_{ab} d\phi^a d\phi^b= -d\zeta^2 +\cosh^2 \zeta (d\theta^2 + \sin^2 \theta d\phi^2),
\end{eqnarray}
we get
\begin{eqnarray}\label{Schw2approx}
ds^2&=& z^2 d\zeta^2 \biggl [  -1 + \frac{1}{z}\frac{2 M \cosh(2\zeta)}{\cosh \zeta}+\frac{4M^2 \tanh^2 \zeta}{z^2}+O(1/z^3)
\biggr ] \nonumber \\
&& + 2 z  dz d\zeta \biggl [  4 \frac{M}{z} \sinh \zeta + 4 \frac{M^2}{z^2} \tanh \zeta+O(1/z^3) \biggr ] \nonumber \\
&& + dz^2 \biggr [ 1+ \frac{2 }{z } \frac{M \cosh(2\zeta) }{\cosh \zeta}+ 4 \frac{M^2}{z^2} + O(1/z^3) \biggr ] \nonumber \\
&& + z^2 \cosh^2 \zeta (d\theta^2 +\sin^2 \theta d\phi^2),
\end{eqnarray}
and the metric, up to second order, is fully specified given
\begin{eqnarray}
\sigma^{(1)}&=&M \frac{ \cosh(2\zeta) }{\cosh \zeta} \; , \qquad (\sigma^{(1)})^2+2 \sigma^{(2)}=4M^2 \; , \nonumber \\
\sigma^{(2)}&=& -\frac{M^2}{2}(-6+2\cosh(2\zeta)+\frac{1}{\cosh^2 \zeta}) \; , \nonumber \\
A^{(1)}_{\zeta}&=& 4 M \sinh \zeta \; , \qquad A^{(2)}_{\zeta}= 4 M^2 \tanh \zeta \; , \nonumber \\
h^{(1)}_{\zeta \zeta}&=& 2 \sigma^{(1)} \; , \qquad h^{(2)}_{\zeta \zeta}= 4M^2 \tanh^2 \zeta \; .
\end{eqnarray}

\textbf{2nd step} \\

Get rid of the $A^{(1)}_{a}$ following the Beig-Schmidt algorithm by doing the generic change of coordinates
\begin{eqnarray}
z=y \: , \qquad
\phi^a=\bar{\phi}^a+\frac{1}{y} A^{(1)}_{b} h^{(0)\: ab} \; ,
\end{eqnarray}
which for the Schwarzschild solution, where only $A^{(1)}_{\zeta}$ is non-zero, is just
\begin{eqnarray}
z=y \; , \qquad  \zeta=\psi+\frac{A^{(1)}_{a} h^{(0) \: a \zeta}}{y}= \psi-\frac{ 4M \sinh \psi}{y} \; , \qquad  A^{(1)}_{\zeta}(\psi)=4M \sinh \psi.
\end{eqnarray}
Plugging this into  (\ref{Schw2})(or equivalently (\ref{Schw2approx}))  and keeping only terms up to second order, we get a metric specified by
\begin{eqnarray}\label{step2}
\sigma^{(1)}&=&M \frac{ \cosh(2\psi) }{\cosh \psi} \qquad (\sigma^{(1)})^2+2 \sigma^{(2)}=4M^2(-1+2 \frac{1}{\cosh^2 \psi})  \nonumber \\
\sigma^{(2)}&=& \frac{M^2}{4 \cosh^2 \psi}(11-4\cosh(2\psi)-\cosh (4\psi)) \nonumber \\
A^{(1)}_{\psi}&=& 0 \qquad A^{(2)}_{\psi}= -4 M^2 \tanh \psi \nonumber \\
h^{(1)}_{\psi \psi}&=& \frac{2M}{\cosh \psi} (2+3 \cosh(2\psi)) \qquad\:\: h^{(2)}_{\psi \psi}= -\frac{2M^2}{\cosh^2\psi} (3+9 \cosh(2\psi)+4 \cosh(4\psi)) \nonumber\\
h^{(1)}_{\theta \theta}&=& -8M \cosh \psi \sinh^2\psi \qquad\qquad h^{(2)}_{\theta \theta}= 16 M^2 \cosh(2\psi) \sinh^2 \psi \nonumber \\
h^{(1)}_{\phi \phi}&=& h^{(1)}_{\theta \theta} \sin^2 \theta\qquad\qquad\qquad \qquad\:\:  h^{(2)}_{\phi \phi}=  h^{(2)}_{\theta \theta} \sin^2 \theta.
\end{eqnarray}

At this point, we still need to get rid of $\sigma^{(2)}$ and $A^{(2)}$. Also, we do not have
\begin{eqnarray}
k_{ab}=0 \leftrightarrow h^{(1)}_{ab}=-2\sigma^{(1)} h^{(0)}_{ab} \: .
\end{eqnarray}
The last step takes care of all these issues.\\

\textbf{3d step} \\

The general idea is to do a supertranslation of the form
\begin{eqnarray}
y &=& \rho+\omega(\hat{\phi}^a)+ \frac{F^{(2)}(\hat{\phi}^a)}{\rho} \; , \nonumber \\
\bar{\phi}^a &=&\hat{\phi}^a+\frac{1}{\rho} h^{(0)\: ab} \omega_{,b}+\frac{G^{(2)\: a}}{\rho^2} \; ,
\end{eqnarray}
where $\bar{\phi}= \{ \psi,\theta,\phi \} $ and $\hat{\phi}=\{ \tau, \hat{\theta},\hat{\phi} \}$, and look at the more general solution $\omega$ such that
\begin{eqnarray}\label{h11}
h^{(1)}_{ab}+ 2 D_a D_b \omega + 2 \omega h^{(0)}_{ab}=-2 \sigma^{(1)} h^{(0)}_{ab}\; ,
\end{eqnarray}
where $h^{(1)}_{ab}$ was given in \eqref{step2} for the Schwarzschild solution. Then, we will fix the functions $F^{(2)}_a$ and $G^{(2)}_a$ in such a way that $\sigma^{(2)}$ and $A^{(2)}$ are set to zero.

To solve for $\omega$, we plug in \eqref{h11} a general $\omega=\omega(\tau, \bar{\theta}, \bar{\phi})$. One can see that the component $\tau \tau$ of the equation \eqref{h11}
is a differential equation for $\tau$ that reads
$\partial_{\tau}^{2} \omega -\omega +4M \cosh(\tau)=0\; $.
By writing
$\omega(\tau, \theta,\phi)=f(\tau)g(\theta,\phi) \; $,
the most general solution $f(\tau)$ of the previous differential equation is
$f(\tau)= (M+C_1+C_2)\cosh(\tau) -(2M\tau +C_1-C_2)\sinh(\tau)\;$.
If we now plug this in the other components of (\ref{h11}), we see that $\omega$ is independent of the angular coordinates and that the only equation left to satisfy is
$\sinh(\tau ) \partial_{\tau} f -\cosh(\tau) f+M(\cosh(2\tau)-2)=0\;$.
By plugging our previous solution into this last equation, we see that it is fulfilled only when
$C_1=-2M-C_2$
so that the more general non-trivial solution to the gauge fixing of supertranslations is achieved by
\begin{eqnarray}
\omega(\tau,\theta,\phi)=M c_1 \sinh(\tau) -M \cosh(\tau) -2 M \tau \sinh(\tau) \; ,
\end{eqnarray}
where we have redefined $C_1-C_2=c_1 M $.\\

Moving now to the functions appearing in the higher order terms, it is obvious after a close computer-assisted inspection that we will only need  an $F^{(2)}$ which is dependent on $\tau$ to get rid of $A^{(2)}_{\tau}$ and a function $G^{(2) \tau} (\tau)$ to get rid of $\sigma^{(2)}$, so that our general change of coordinates reads
\begin{eqnarray}
y= \rho+f(\tau)+ \frac{K(\tau)}{\rho} \; ,\qquad
\psi = \tau-\frac{\partial_{\tau} f(\tau)}{\rho} + \frac{n(\tau)}{\rho^2} \; .
\end{eqnarray}
Plugging this in the metric we see that for $\sigma^{(2)}$ to be zero, we have an algebraic equation for $K(\tau)$ which gives
\begin{eqnarray}
K(\tau)&=& \frac{M^2}{4}\biggl [ 5-(c_1-2\tau)^2+(3-(c_1-2\tau)^2) \cosh(2\tau) \nonumber \\
&& \qquad +\frac{2}{ \cosh^2(\tau)}-2(c_1-2\tau)\sinh(2\tau) \biggl ] \; .
\end{eqnarray}
Looking eventually for $A^{(2)}_{\tau}$ to be zero, we see that the equation reduces to an algebraic equation for $N(\tau)$ when using our solution for $K(\tau)$. We find
\begin{eqnarray}
N(\tau)&=&\frac{M^2}{2} \biggl [ -2(c_1 -2 \tau) -8 (c_1-2\tau) \cosh(2\tau) +(17+(c-2\tau)^2)\sinh(2\tau) \nonumber\\
&& \qquad +(\frac{1}{cosh^2(\tau)}-8) \tanh(\tau) \biggr ] \; .
\end{eqnarray}

\textbf{Conclusions}\\

At the end of this procedure, we have obtained a metric which is written in the Beig-Schmidt form and that fulfills $k_{ab}=0$. The value for $\s$ is given by
\beqs
\sigma = M \frac{ \cosh(2\psi) }{\cosh \psi} \; ,
\eeqs
while $h^{(1)}_{ab}=-2 \s h^{(0)}_{ab}$. We have also obtained the values of $h^{(2)}_{ab}$ altough we will not provide them here. Their expressions are in terms of $M^2$ quantities.
It can be checked that the value of $\s$ is precisely $\hat{\zeta}^a_{(0)}$, a solution of $(\Box+3)\s=0$ which is not one of the four solutions of $\sigma_{ab}+\s h^{(0)}_{ab}=0$. One can check that
\beqs
Q[\zeta^a_{(0)}=\partial/\partial t]=-\frac{1}{8\pi} \oint d^2 S \: E^{(1)}_{ab} \:  \zeta^a_{(0)} n^b =M\; ,
\eeqs
where $E^{(1)}_{ab}=-\DD_a \DD_b \s -\s h^{(0)}_{ab}$.

 \setcounter{equation}{0}
\section{Properties on the hyperboloid}
\label{app:properties}
In this small Appendix, we present results on the hyperboloid that are used throughout all this Part I. See also Appendix  C of \cite{Mann:2008ay} and Appendix D of \cite{Mann:2006bd}. 

The unit metric on the hyperboloid is 
\beqs
ds^2=h^{(0)}_{ab} d\phi^a d\phi^b= -d\tau^2 +\cosh^2 \tau (d\theta^2 + \sin^2 \theta d\phi^2),
\eeqs
and the covariant derivative associated to $h^{(0)}_{ab}$ is $\DD_a$.

The Riemann tensor of the metric $h^{(0)}_{ab}$ on the unit hyperboloid $\mathcal{H}$ is given by 
\beqs
\mathcal{R}^{(0)}_{abcd}= h^{(0)}_{ac} h^{(0)}_{bd} - h^{(0)}_{bc} h^{(0)}_{ad} \; .
\eeqs
The commutator of two derivatives acting on a tensor $t_{ab}$ is
\beqs
[\DD_a, \DD_b] t_c^{\:\:d}=\mathcal{R}^{(0)\:e}_{abc} t_e^{\:\:d}-\mathcal{R}^{(0)\:d}_{abe} t_e^{\:\:c} \; ,
\eeqs
which implies that if $t$, $t_a$, $t_{ab} = t_{(ab)}$ are some arbitrary fields on the hyperboloid, then
\bea
\left[ \DD_a,\square \right] t &=& -2 \DD_a t,\\
\left[ \DD_a,\DD_b \right]t_{c} &=& h^{(0)}_{ac}t_b - h^{(0)}_{bc}t_a,\\
\left[ \DD_a,\square \right]t_b &=& 2 h^{(0)}_{ab} \DD_c t^c - 4 \DD_{(a} t_{b)},\\
\left[ \DD^c,\DD_a \right] t_{cb} &=& 3 t_{ba} - h_{ab}^{(0)} t^c_{\; c},\\
\left[ \DD_a,\DD_b \right] t_{cd} &=& 2 h^{(0)}_{a(c} t_{d)b} - 2 h^{(0)}_{b(c} t_{d)a},\\
\left[ \DD_a,\square \right] t_{bc} &=& 4 h^{(0)}_{a(b}\DD^d t_{c)d} -6\DD_{(a}t_{bc)},\\
\left[ \DD^b,\square \right] t_{bc} &=& 4 \DD^d t_{cd} -2 \DD_c t .
\eea
These identities are useful for several arguments in the main text and for the proofs of the lemmae as presented in the next Appendix. Note also that upon using the equations of motion and the above identities, one can establish an infinite amount of identities for the fields $\s$ and $k_{ab}$. For example, we have 
\beqs
\sigma_c^{\:\:c}&=&-3 \s \; , \\
\sigma^c_{\;\;ca}&=&-3 \s_a \; , \\
\sigma_{ac}^{\:\:\:\:c}&=&-\s_a \; , \\
\sigma_e \sigma^{men}&=&\sigma_e \s^{mne}-\s^m \s^n +\s_e \s^e h^{(0)\: mn}\; , \\
\cdots
\eeqs

 \setcounter{equation}{0}
\section{Proofs of the Lemmas}
\label{app:proofs}
In this section, we present the  proofs of Lemmae \ref{Ash} and \ref{BS} given in section \ref{sec:firstorder}, of the Lemma  \ref{lemma8} presented in section \ref{subsec:second}, and the Lemmae  \ref{newlemma} and \ref{higheroplemma} given in \ref{subsec:lemma45}. The five lemmae have an overlapping proof.  In order to establish these lemmae we need to derive the decomposition of regular symmetric, divergence-free, and traceless (SDT) tensors on the hyperboloid. We do so in the rest of this appendix.

A general regular symmetric tensor on the hyperboloid can be expressed as a linear combination of symmetric tensors built from two-dimensional spherical harmonics. A general such tensor has the form
\bea
T_{\tau \tau} &=& f_1(\tau) Y_{lm}(\theta,\phi),\\
T_{\tau i} &=& f_2 (\tau)D^{(2)}_i Y_{lm}(\theta,\phi) + f_3(\tau) \eps_i^{\;\, j}D^{(2)}_j Y_{lm}(\theta,\phi),\\
T_{ij} &=& f_4(\tau)\left(D^{(2)}_i D^{(2)}_j +\frac{l(l+1)}{2} \eta_{ij} \right) Y_{lm}(\theta,\phi)   + f_5(\tau) \eps_{(i}^{\;\, k}D^{(2)}_{j)}D^{(2)}_k Y_{lm}(\theta,\phi) \nn  \\ & & + f_6(\tau) \eta_{ij} Y_{lm}(\theta,\phi),
\eea
where indices $i,j,k$ run over two-sphere $(\theta, \phi)$, $Y_{lm}(\theta,\phi)$ are scalar spherical harmonics on the two sphere with $l=0, 1, \ldots,$ and $m=-l, \ldots, l$, and  $D^{(2)}_i$ is the covariant derivative compatible with the round metric $\eta_{ij}$ on the two-sphere. In writing these expressions we have already made use of the identity
\be
\left( D^{(2)}_i D^{(2)}_j  + \frac{l(l+1)}{2}\eta_{ij} + \eps_{(i}^{\;\, k}\eps_{j)}^{\;\, l}D^{(2)}_k D^{(2)}_l \right) Y_{lm}(\theta,\phi) = 0,
\ee
in order to reabsorb the tensor structure $\eps_{(i}^{\;\, k}\eps_{j)}^{\;\, l}D^{(2)}_k D^{(2)}_l  Y_{lm}$ into the definition of $f_4(\tau)$. The tensor is traceless if and only if
\be
f_1(\tau) = 2 \sech^2\tau f_6(\tau).
\ee
For the case $l=0$, $m=0$, only $f_6(\tau)$ parameterizes non-zero tensors. The divergence-free condition is solved only for $f_6 \sim \sech\tau$. The general tensor then reduces to
\be
T_{ab} = \DD_a \DD_b \hat \zeta_{(0)} + h_{ab}^{(0)}\hat  \zeta_{(0)}.
\ee
When $l=1$, we have that $ \eps_{(i}^{\;\, k}D^{(2)}_{j)}D^{(2)}_k Y_{lm}(\theta,\phi) = 0$. Therefore, $f_4$ and $f_5$ do not lead to non-zero tensors. The divergence-free condition fixes
\bea
f_2(\tau) &=& -\tanh\tau f_6(\tau) - \partial_\tau f_6(\tau), \\
f_3(\tau)&=& C_1 \sech^2\tau,\\
f_6(\tau)&=& C_2 \sech^2\tau +C_3 \sech\tau \tanh\tau,
\eea
where $C_1$, $C_2$ and $C_3$ are constants. There are therefore three solutions for each value of $m = -1,0,1$, so $9$ solutions in total. Three independent solutions are the tensors admitting a scalar potentials $\hat\zeta_{(i)}$, $i=1,2,3$,
\be
T_{ab} = \DD_a \DD_b\hat  \zeta_{(i)} + h_{ab}^{(0)} \hat \zeta_{(i)}.\label{Tzetai}
\ee
These tensors are curl-free and $(\square - 3)T_{ab} = 0$. The six other tensors can be written as a linear combination of the following two sets of three tensors,
\bea
V_{(k)\tau\tau} &=& 2\sech^5\tau \zeta_{(i)},\quad V_{(k)\tau i} = \sech^3\tau \tanh \tau D^{(2)}_i \zeta_{(k)},\quad V_{(k)ij}=\eta_{ij}\sech^3\tau\zeta_{(k)}, \nn\\
W_{(k)\tau \tau} &=& 0,\quad W_{(k) \tau i} = \sech^3\tau  \eps_{i}^{\;\, j}D^{(2)}_j \zeta_{(k)},\quad W_{(k)ij} =0,
\eea
where $\zeta_{(k)} = \cosh\tau f_{(k)}$, $k = 1,2,3$. These tensors are dual to each other in the sense
\bea
\eps_a^{\;\, cd}D_c V_{d b} = -W_{ab},\qquad \eps_a^{\;\, cd}D_c W_{d b}=  V_{ab}.
\eea
Since applying two times the curl operator on a traceless, divergence-free, symmetric tensor is equivalent to applying $(\square -3)$, we deduce that both tensors obey
\bea
(\square -2)V_{ab} = 0,\qquad (\square -2)W_{ab} = 0.
\eea
These tensors also obey the orthogonality properties
\bea
\int_{S}V_{(k)ab}\xi^a_{\rom{rot}(l)} n^b d^2 S= 0, & \qquad & \int_{S} W_{(k)ab}\xi^a_{\rom{boost}(l)}n^b d^2 S= 0,\\
\int_{S} V_{(k)ab}\xi^a_{\rom{boost}(l)} n^b d^2 S = \frac{8\pi}{3} \delta_{(k)(l)}, &&
\int_{S} W_{(k)ab}\xi^a_{\rom{rot}(l)} n^b d^2 S = \frac{8\pi}{3} \delta_{(k)(l)},
\eea
where $C$ is a cut of the hyperboloid. Since $V_{(k)ab}$ and $W_{(k)ab}$ are divergence-free, these integrals are independent of the chosen cut of the hyperboloid.

For $l > 1$, we can solve the divergence-free condition in terms of $f_2$, $f_4$ and $f_5$ as
\bea
f_2(\tau) &=& -\frac{2}{l(l+1)}(\tanh \tau f_6(\tau) + \partial_\tau f_6(\tau)),\\
f_4(\tau) &=& \frac{2}{(l-1)l(l+1)(l+2)}\Big( (l+l^2+2\cosh 2\tau)f_6(\tau) \nn \\
&& +2\cosh\tau (3\sinh\tau \partial_\tau f_6(\tau)+\cosh\tau \partial_\tau \partial_\tau f_6(\tau) )\Big),\\
f_5(\tau)&=& - \frac{2}{(l-1)(l+2)}\cosh\tau (2\sinh\tau f_3(\tau)+\cosh\tau \partial_\tau f_3(\tau)).
\eea
The general tensor with harmonics $(l,m)$, $l>1$ is a linear combination of the following two tensors depending each on an arbitrary function $f(\tau)$ of $\tau$,
\bea
T_{ab}^{(I)}(f) & \equiv & T_{ab}\left(f_3(\tau) = \frac{1}{l(l+1)} f(\tau),\,f_6(\tau) = 0\right),\label{gen45}\\
T_{ab}^{(II)}(f) & \equiv & T_{ab}\left(f_3(\tau) =0,\,f_6(\tau) = \frac{1}{2} f(\tau) \right).\label{gen46}
\eea
These tensors obey the remarkable properties
\bea
\eps_a^{\;\, cd}\DD_c T_{db}^{(I)}(f) &=& T_{ab}^{(II)}(f),\label{curlTI}\\
\eps_a^{\;\, cd}\DD_c T_{db}^{(II)}(f) &=& - T_{ab}^{(I)}( \mathcal O f),\label{curlTI2}
\eea
where $ \mathcal O f$ is the following differential operator acting on $f(\tau)$,
\be
\mathcal O f \equiv (1+l(l+1)\sech^2\tau)f+2\tanh\tau \partial_\tau f+\partial_\tau \partial_\tau f\, .
\ee
We deduce also the following properties
\bea
(\square - 3)  T_{ab}^{(I)}(f) &=& - T_{ab}^{(I)}(\mathcal O f), \label{eq:box1}\\
(\square - 3)  T_{ab}^{(II)}(f) &=& - T_{ab}^{(II)}(\mathcal O f).\label{eq:box2}
\eea
Using the explicit expression for the tensors and the orthogonality of spherical harmonics, we also have
\bea
\int_ST_{ab}^{(I)}(f) \xi_{\rom{rot}(k)}^a n^b d^2 S =0, &\qquad  &\int_S T_{ab}^{(II)}(f) \xi_{\rom{rot}(k)}^a n^b d^2 S =0, \label{killJ1} \\
\int_S T_{ab}^{(I)}(f) \xi_{\rom{boost}(k)}^a n^b d^2 S =0,  &\qquad  &  \int_S T_{ab}^{(II)}(f) \xi_{\rom{boost}(k)}^a n^b d^2 S =0, \label{killJ2}\\
\int_S T_{ab}^{(I)}(f) \DD^a \zeta_{(l)} n^b d^2 S =0,  &\qquad  &  \int_S T_{ab}^{(II)}(f) \zeta_{(l)}^a n^b d^2 S =0. \label{killJ3}
\eea
The above decomposition proves lemma \ref{lemma8}. Indeed, one can isolate the $l=0,1$ harmonics and then all the higher harmonics can be regrouped in a tensor $J_{ab}$ that obeys $\int_S J_{ab} \xi_\rom{rot}^a n^b=\int_S J_{ab} \xi_\rom{boost}^a n^b=0$ as a consequence of \eqref{killJ1}-\eqref{killJ2}.

There are two special sets of two functions $f(\tau)$: the ones for which the differential operator obeys $\mathcal O f_{(0)}=0$ and the others for which $\mathcal O f_{(1)}= f_{(1)}$. The two functions obeying $\mathcal O f_{(0)}=0$ define tensors $T_{ab}^{(II)}(f_{(0)})$ such that
\bea
\DD_{[a}T_{b]c}^{(II)}(f_{(0)}) =0,\qquad (\square - 3) T_{ab}^{(II)}(f_{(0)})  = 0.
\eea
The tensor $T_{ab}^{(I)}(f_{(0)})$ is a tensor potential for $T_{ab}^{(II)}(f_{(0)})$ and is uniquely determined for the two solutions of $\mathcal O f_{(0)} =0$. From the explicit form of the tensor, we note that $T_{ab}^{(II)}(f_{(0)})$ can be written as
\bea
T_{ab}^{(II)}(f_{(0)}) = \DD_a \DD_b \Phi + h_{ab}^{(0)}\Phi, \label{scd2}
\eea
where $\Phi = \sum_{l}\sum_{m= -l}^{l}\Phi^{lm}(\tau) Y_{lm}(\theta,\phi)$ is a scalar that obeys $(\square + 3)\Phi =0$. The two independent solutions of $\mathcal O f_{(0)}=0$ correspond to the two independent solutions of the equation $(\square + 3)\Phi =0$ for fixed values of $l>1$, $-l \leq m \leq l$.

The two independent solutions for $f(\tau)$ of the differential equation $\mathcal O f_{(1)}=f_{(1)}$ can be used to define two pairs of dual tensors
\bea
W_{ab} = T_{ab}^{(I)}(f_{(1)}) , \qquad V_{ab} = T_{ab}^{(II)}(f_{(1)}),
\eea
which obey
\bea
\eps_a^{\;\, cd}\DD_c V_{db}  = -W_{ab}, \qquad  \eps_a^{\;\, cd}\DD_c W_{db}  = V_{ab} ,\\
(\square -2)V_{ab} = 0,\qquad (\square -2)W_{ab} = 0.
\eea
Given the special role of the eigenfunction of  the operator $\mathcal O$, it is natural to decompose the functions $f(\tau)$ in that basis. The equation
\be
\mathcal O f_{(n)} = (n-1)^2 f_{(n)} , \label{eqgen7}
\ee
for each positive integer $n$ is solved by associated Legendre functions of the first and second kind,
\be
f^{(1)}_{(n)} = \sech\tau P_l^n[\tanh \tau],\qquad  f^{(2)}_{(n)} = \sech\tau Q_l^n[\tanh \tau].
\ee
Lemma \ref{Ash} is then proven as follows. A symmetric traceless divergence-free tensor obeying $(\square - 3)T_{ab}=0$ can be decomposed into harmonics. The only possible tensors in harmonics $l =0,1$ have the form
\be
T_{ab} = \DD_a \DD_b \Phi + h_{ab}^{(0)} \Phi, \label{scd}
\ee
where $(\Box +3)\Phi = 0$ contains $l=0,1$ harmonics. For $l > 1$, we have seen that any tensor can be decomposed as a combination of two different tensor structures $T^{(I)}_{ab}(f^{(I)})$ and $T^{(II)}_{ab}(f^{(II)})$ depending each on one function. We then see from  \eqref{eq:box1}-\eqref{eq:box2} that such tensors obey $(\square - 3)T_{ab} =0$ if and only if $f^{(I)} = f^{(II)} = f_{(0)}$ where $f_{(0)}$ are the solutions of the differential equation $\mathcal O f_{(0)} =0$. Then, we note using \eqref{curlTI} that $T^{(I)}_{ab}(f_{(0)})$ is not curl-free and thus does not obey the preconditions of the lemma. The only remaining tensors have the form $T^{(II)}_{ab}(f_{(0)})$ and they can be written in terms of a scalar potential \eqref{scd2} as shown earlier.

The general solution of $(\Box +3)\Phi = 0$ contains $\zeta_{(i)}$, $\hat \zeta_{(i)}$, $i=0,1,2,3$ and the higher $l>1$ harmonics. For each value of $l>1$, $m$, there are two solutions for $\Phi$ that uniquely correspond to the two tensors $T^{(II)}_{ab}(f_{(0)})$. The four lower harmonics correspond to the tensor $T_{ab}$ built in \eqref{Tzetai}. The dependence in $\zeta_{(i)}$ is arbitrary since these scalars can then be added to $\Phi$ without changing $T_{ab}$. This ends the proof of lemma \ref{Ash}.

The lemma \ref{BS} is proven by noticing that by lemma \ref{Ash}, all tensors derived from a scalar using \eqref{scd} that have $l>1$ harmonics have the form $T^{(II)}_{ab}(f_{(0)})$. The tensor $T^{(I)}_{ab}(f_{(0)})$ is then the tensor potential for $T^{(II)}_{ab}(f_{(0)})$ by  \eqref{curlTI}.

Let us now prove lemma \ref{newlemma}. We consider an arbitrary SDT tensor $T_{ab}$. One can decompose it in $l=0$, $l=1$ and $l>1$ harmonics, and further the arbitrary functions $f(\tau)$ appearing in \eqref{gen45}-\eqref{gen46} can be decomposed in eigenfunctions \eqref{eqgen7} with positive integer $n$. The $l=0$, $l=1$ and $l>1$ harmonics with $n = 1$ can be written as the sum of a tensor admitting a scalar potential and the curl of an SDT tensor. From \eqref{curlTI}-\eqref{curlTI2}, the $l>1$ harmonics with $n > 1$ are explicitly the curl of an SDT tensor. Using in addition Lemma \ref{BS}, we obtain that $T_{ab}$ can be written as a sum of the curl of an SDT tensor and a sum of $\DD_a \DD_b \hat \zeta_{(i)} + h_{ab}^{(0)}\hat\zeta_{(i)}$, which proves the lemma.

Let us finally prove lemma \ref{higheroplemma}. Note that no SDT tensor obeying $(\square +n^2 - 2n -2)T_{ab} = 0$ with $n > 2$
integer can contain spherical harmonics $l =0$ or $l = 1$. This follows from the explicit form of the $l=0$ and $l=1$ SDT harmonics presented above. Therefore, any SDT tensor obeying $(\square +n^2 - 2n -2)T_{ab} = 0$ with $n > 2$ can be decomposed in the basis of tensors $T_{ab}^{(I)}(f)$ and $T_{ab}^{(II)}(f)$ \eqref{gen45}-\eqref{gen46} for $f(\tau)$ obeying the eigenvalue equation \eqref{eqgen7}. All such tensors are expanded in spherical harmonics with $l>1$.  The lemma then follows from the orthogonality of spherical harmonics \eqref{killJ1}--\eqref{killJ3}.

 \setcounter{equation}{0}
\section{Comparison of 3+1 and covariant boundary conditions}
\label{app:BC}

The hyperbolic and cylindrical representation of spatial infinity are valid in the limits $\rho \rightarrow \infty$ and $r \rightarrow \infty$, respectively. The key change of coordinates is the one mapping flat spacetime from the hyperbolic to the cylindrical representation of spatial infinity
\bea
\rho = r \sqrt{1-\frac{t^2}{r^2}},\qquad \tau = \text{arctanh}{(\frac{t}{r})}\, .
\eea
The hyperbolic and cylindrical representations coincide asymptotically in the limit where ADM time is kept finite, $t / r \rightarrow 0$ which is equivalent to $\tau \rightarrow 0$. In that case, $\rho \sim r$ asymptotically.

In order to obtain the form of the metric in $r,t$ coordinates, we expand the right-hand side of $\rho,\tau$ in powers of $t/r$ and we expand the Beig-Schmidt fields in Taylor series around $\tau = 0$,
\bea
\s(\tau,\th,\phi) &=& \s(\th,\phi) + \frac{t}{r} \s^\pi(\th,\phi) + \frac{t^2}{2 r^2} \gamma(\th,\phi) + O(r^{-3}) ,\\
k_{ab}(\tau,\th,\phi) &=& k_{ab}(\th,\phi) + \frac{t}{r} k^\pi_{ab}(\th,\phi) + \frac{t^2}{2 r^2} \gamma_{ab}(\th,\phi) + O(r^{-3}) ,\\
i_{ab} (\tau,\th,\phi) &=& i_{ab}(\th,\phi) + \frac{t}{r}  i^\pi_{ab}(\th,\phi) + O(r^{-2}),\\
h^{(2)}_{ab} (\tau,\th,\phi) &=& h^{(2)}_{ab}(\th,\phi) + \frac{t}{r}  h^{\pi,(2)}_{ab}(\th,\phi) + O(r^{-2}),
\eea
where we define $\s^\pi(\th,\phi)= \p_\tau  \s(0,\th,\phi)$, $k_{ab}^\pi(\th,\phi)= \p_\tau  k_{ab}(0,\th,\phi)$, $ \gamma(\th,\phi) = \p_\tau \p_\tau \s(0,\th,\phi)$, $ \gamma_{ab}(\th,\phi) = \p_\tau \p_\tau k_{ab}(0,\th,\phi)$, $i^\pi_{ab} =\p_\tau i_{ab}(0,\th,\phi)$, $h^{\pi,(2)}_{ab}(\th,\phi) = \p_\tau h^{(2)}_{ab}(0,\th,\phi)$. We will keep the same notation for canonical fields in Hamiltonian formalism as fields in Lagrangian formalism
\bea
\s(0,\th,\phi) = \s(\th,\phi), \qquad k_{ab} (0,\th,\phi) = k_{ab} (\th,\phi) , \nn \\
i_{ab} (0,\th,\phi) = i_{ab} (\th,\phi),\qquad h^{(2)}_{ab} (0,\th,\phi) = h^{(2)}_{ab} (\th,\phi) .
\eea
The tensors decompose into scalars, vectors and two-dimensional tensors under decomposition into temporal and spatial components. The meaning of the notation should be clear in either Hamiltonian or Lagrangian context. The fields $\gamma(\th,\phi)$ and $\gamma_{ab}(\th,\phi)$ are
 determined from the equations of motion of $\sigma$ and $k_{ab}$. After a straightforward computation, we obtain
\bea
\g_{rr} &=& 1 + \frac{2\s}{r} + \frac{\s^2 +2 t \s^\pi }{r^2} +o(r^{-2}  ) ,\nn \\
\g_{r \zeta} &=& - \frac{t}{r}k_{\tau \zeta} - t \frac{\log r (i_{\tau\zeta}) + h^{(2)}_{\tau \zeta} + t k^\pi_{\tau \zeta} }{r^2} + o(r^{-2}) \label{g2BC} ,\\
\g_{\zeta \iota} &=& r^2 g_{\zeta \iota} +(k_{\iota \zeta} - 2 \s  g_{\zeta \iota} ) r+ \log r (i_{\zeta\iota} )+ (h^{(2)}_{\zeta \iota}+ t k^\pi_{\iota \zeta} -2 t \s^\pi g_{\zeta \iota}) +o(r^0), \nn
\eea
for the canonical fields and
\bea
(det g_{\zeta \iota})^{-1/2}\pi^{rr} &=& -2 \s^\pi + \frac{1}{2}k^\pi_{\iota \zeta} g^{\iota \zeta}_{(S^2)}- D_{(S^2)}^\iota k_{\tau \iota} + \frac{\log r}{r} \Big( \frac{1}{2} g^{\zeta \iota} i^\pi_{\zeta\iota} -   D_{(S^2)}^\zeta i_{\tau \zeta}\Big) \nn \\
&&+ \frac{1}{r} \Big( \frac{1}{2} g^{\zeta \iota} h^{\pi,(2)}_{\zeta \iota} - D_{(S^2)}^\zeta h^{(2)}_{\tau \zeta}  -  2 t \gamma + 6 t \s -\frac{1}{2}t k_{\iota\zeta} g_{(S^2)}^{\iota \zeta} +\frac{1}{2} t \gamma_{\iota \zeta} g^{\iota\zeta}_{(S^2)} \nn\\
&& + 2 t k_{\tau\tau} - k_{\tau\tau}\s^\pi - t D_{(S^2)}^\iota k^\pi_\iota + (k-k \; \text{terms})\Big)  + o(r^{-1}),\nn \\
(det g_{\zeta \iota})^{-1/2}\pi^{r \iota} &=& -\frac{1}{2 r}k_\tau^\iota - \frac{\log r}{r^2} (i_{\tau}^{\iota} ) +\frac{1}{r^2} \Big(-h^{(2)\iota }_{\tau} + \frac{1}{2}i_{\tau}^{\iota} -2 t \p^\iota \s \label{g2BC2} \\
&& -\frac{t}{2}k^\pi_{\tau \zeta} g^{\zeta\iota}_{(S^2)}-\s k_{\tau \zeta} g^{\zeta\iota}_{(S^2)} - \frac{t}{2} D_{(S^2)}^\iota k_{\tau\tau} + (k-k \; \text{terms})  \Big) + o(r^{-2}) ,\nn \\
(det g_{\zeta \iota})^{-1/2}\pi^{\iota \zeta} &=& \frac{1}{r^2} \Big( - \frac{1}{2} k^{\pi, \iota \zeta} + D_{(S^2)}^{(\iota} k^{\zeta)}_\tau +g^{\iota \zeta}_{(S^2)}(\frac{1}{2}k^{\pi,\xi}_\xi - D_{(S^2)}^\xi k_{\tau \xi})  \Big) \nn\\
&&+ \frac{\log r}{r^3} \Big( - \frac{1}{2} i^{\pi, \iota \zeta} + D_{(S^2)}^{(\iota} i^{\zeta)}_\tau +g^{\iota \zeta}_{(S^2)}(\frac{1}{2}i^{\pi,\xi}_\xi - D_{(S^2)}^\xi i_{\tau \xi})  \Big) + O(r^{-3}) ,\nn
\eea
for the conjugate fields. Here, we denote by $(k-k \;\text{terms})$ terms quadratic in $k_{ab}$ which do contribute to the finite part of the conserved Lorentz charges but that we omit here for simplicity.

Let us finally discuss how the notions of parity are related between Beig-Schmidt fields and canonical fields. A field on the hyperboloid is parity-time reversal even if it is invariant under the combined transformation of inverting the hyperboloid time $\tau \rightarrow -\tau$ and doing a parity transformation $(\theta,\phi)\rightarrow (\pi-\theta,\phi+\pi)$. Fields in canonical formalism are parity-time reversal even if their components in Cartesian coordinates do not transform under three-dimensional parity and if the components of their conjugate momentum in Cartesian coordinates transform with an overall sign under parity. From the dictionary of the Beig-Schmidt asymptotic fields in 3+1 decomposition,  we see after switching from spherical to Cartesian coordinates that the even parity-time reversal conditions on $\sigma$ and $k_{ab}$ lead to parity-time reversal even first order canonical fields on the initial time slice $t=0$.

\cleardoublepage
\renewcommand{\thesection}{\arabic{chapter}.\arabic{section}}

\part{ Magnetic theory through duality}

%%%%%%%%%%%%%%%%%%%%%%%%%%%%%%%%%%%%%%%%%%%%%%%%%%%%%%%%%%%%%%%%%%%
\chapter{Electromagnetic duality for Maxwell's theory} \label{chap:emduality}

In this chapter we review the theoretical discovery of P.A.M. Dirac \cite{Dirac-1931}, in 1931, who showed that  Maxwell's equations can be made invariant under a symmetry that exchanges electric and magnetic fields at the cost of introducing sources for the magnetic field. This symmetry is known as the electromagnetic duality and the new postulated particles bear the name ``magnetic monopoles".

In section \ref{sec: EM}, we review this duality at the level of the equations of motion. In section \ref{sec: emmono}, we briefly discuss electric and magnetic charges and the Dirac string. The major importance of Dirac's work is that, even if magnetic monopoles have never been observed in Nature, the presence of at least one of them would explain the quantization of the electric charge. This is what is reviewed in section \ref{sec: Quanti}. We eventually finish by a small discussion of some other  aspects in section \ref{sec: Various}.

\setcounter{equation}{0}
\section{The electromagnetic duality}
\label{sec: EM}

In 1861, Maxwell wrote the famous equations that bear his name and that describe, in a unified way, electricity, magnetism and optics. In the vacuum, these are
\begin{eqnarray}
\vec{\nabla} . \vec{E}= 0, \qquad \qquad  \qquad  \vec{\nabla} . \vec{B}= 0 , \nonumber \\
\qquad \vec{\nabla} \times \vec{B} = \frac{\partial \vec{E}}{\partial t} , \qquad - \vec{\nabla} \times \vec{E}= \frac{\partial \vec{B}}{\partial t} .
\end{eqnarray}
It is easy to see that they are invariant under the so-called electromagnetic duality which interchanges electric and magnetic fields
\beqs\label{electromagn1}
\vec{E} \rightarrow \vec{B} , \qquad \vec{B} \rightarrow -\vec{E} .
\eeqs
As we said in the introduction of this thesis, Maxwell's equations were already fitted for special relativity. This can be seen by introducing an exact antisymmetric tensor, where the index $i$ goes from 1 to 3,
\beqs
F_{0i}= E_i , \qquad \qquad F_{ij}=\epsilon_{ijk} B^k,
\eeqs
such that Maxwell's equations can be written in a covariant form
\begin{eqnarray}
\partial_{\nu} F^{\mu \nu}=0 , \qquad
\epsilon^{\mu\nu\rho\sigma} \partial_{\nu} F_{\rho \sigma}=0 ,
\end{eqnarray}
where indices are raised and lowered with the flat metric $\eta_{\mu\nu}$ such that $F_{00}=F^{00} , F_{0i}=-F^{0i},...$. The first equation is called the equation of motion while the second is the Bianchi identity. Using differential forms, it is just
\beqs
d\star F=0, \qquad dF=0.
\eeqs
The second equation (Bianchi identity) is an identity when $F$ is obtained from a potential
\beqs
F_{\mu\nu}\equiv \partial_{\mu} A_{\nu}-\partial_{\nu} A_{\mu} ,
\eeqs
where $A_{\mu}$ is called a gauge field. Local transformations of the form
\beqs
\delta A_{\mu}=\partial_{\mu} \Lambda(x) ,
\eeqs
leave the field strength $F_{\mu\nu}$ invariant.
The electromagnetic duality \eqref{electromagn1} is rephrased as a Hodge duality on the field strength. Alternatively stated, Maxwell's equations are invariant under
\beqs\label{electromagn2}
F_{\mu\nu} \rightarrow \tilde{F}_{\mu\nu}=(\star F)_{\mu\nu} =\frac{1}{2} \epsilon_{\mu\nu\rho\sigma} F^{\rho \sigma} .
\eeqs
In the presence of  electric sources, we have
\begin{eqnarray}
\vec{\nabla} . \vec{E}= 4 \pi \rho_e , \qquad \qquad  \qquad  \vec{\nabla} . \vec{B}= 0  , \nonumber \\
\qquad \vec{\nabla} \times \vec{B} = 4 \pi \vec{j}_e + \frac{\partial \vec{E}}{\partial t}  , \qquad - \vec{\nabla} \times \vec{E}= \frac{\partial \vec{B}}{\partial t} ,
\end{eqnarray}
and the equations are not invariant anymore. To restore the symmetry, the idea of Dirac was to introduce a new density of charges and a new current in the following way
\begin{eqnarray}
\vec{\nabla} . \vec{E}= 4 \pi \rho_e , \qquad \qquad  \qquad  \vec{\nabla} . \vec{B}= 4 \pi \rho_m , \nonumber \\
\qquad \vec{\nabla} \times \vec{B} = 4 \pi \vec{j}_e + \frac{\partial \vec{E}}{\partial t} , \qquad - \vec{\nabla} \times \vec{E}= 4 \pi \vec{j}_m +\frac{\partial \vec{B}}{\partial t} ,
\end{eqnarray}
or rewritten with $F_{\mu\nu}$, $J_{e}^{\mu}=(\rho_e, j^i_e)$ and $J_{m}^{\mu}=(\rho_m, j^i_m)$
\begin{eqnarray}
\partial_{\nu} F^{\mu \nu}= 4 \pi J_{e}^{\mu} ,\nonumber \\
\frac{1}{2} \epsilon^{\mu\nu\rho\sigma} \partial_{\nu} F_{\rho \sigma}= 4 \pi J_{m}^{\mu} .
\end{eqnarray}
By introducing  a magnetic 4-current on the left hand side of the Bianchi identity,  these equations are now invariant under the duality symmetry
\begin{eqnarray}
F_{\mu\nu} \rightarrow \tilde{F}_{\mu\nu}=\frac{1}{2} \epsilon_{\mu\nu\rho\sigma} F^{\rho \sigma} , \qquad J^{\mu}_e \rightarrow J^{\mu}_m , \qquad J^{\mu}_m \rightarrow -J^{\mu}_e .
\end{eqnarray}
This duality intertwines the equation of motion with the Bianchi identity.

\setcounter{equation}{0}
\section{Electric charge versus magnetic monopole}
\label{sec: emmono}

Electromagnetic duality tells us that for every ``electric" field, there is a ``magnetic'' dual field. In the presence of sources, for every ``electric" source there is a dual ``magnetic" source. As we know, the electric charge is the source of the electric field. If electromagnetic duality is realized in Nature, Dirac postulated that there should exist a magnetic charge, source of the magnetic field. These particles are known as magnetic monopoles. A particle that contains both electric and magnetic charges is called a dyon.

Let us consider an electric source generating an electric field $\vec{E}=  Q  \frac{\vec{r}}{r^3}$ such that
\begin{eqnarray}
    A_t =\frac{Q}{r}  , \qquad \qquad \qquad F_{tr}=E_r= \frac{Q}{r^2}   .
    \end{eqnarray}
After an electromagnetic duality rotation, sending $F\rightarrow \tilde{F}$ and the source $Q\rightarrow H$, we obtain the field generated by a magnetic monopole
\beqs\label{magnmon}
    \tilde{F}_{\theta \phi}= H \: sin \theta .
\eeqs
Speaking of conserved charges, the electric charge is
\beqs
Q =-\frac{1}{4\pi} \oint F^{0i} d\Sigma_i  .
\eeqs
By analogy, one would like to associate a magnetic, topological, charge to the dual field
\beqs
H= \frac{1}{8\pi} \oint \epsilon^{ijk} \: F_{jk} \: d\Sigma_i .
\eeqs
By taking the expression \eqref{magnmon}, we can actually check that the above expression does reproduce the magnetic charge $H$.

\subsubsection{The magnetic monopole and the Dirac string}

As we have already said, in the absence of magnetic sources, the Bianchi identity $dF=0$ ensures that the field strength can be expressed as $F=dA$. Also, from Stoke's theorem, it is easy to see that if $F=dA$, we should have
\beqs
H=\oint_\Sigma F = \oint_\Sigma dA=\oint_{C} A=0,
\eeqs
where $C$ is a closed curve.

When we introduce the four-vector $J^{\mu}_m$, the magnetic charge sources the Bianchi identity and thus $F=dA$ is no longer true.  To describe the pure monopole field \eqref{magnmon}, we actually need to write
\beqs
F=dA+C, \qquad dC=J_m ,
\eeqs
where $C$ is a singular contribution as we now review.
Indeed, let us consider the gauge field
\beqs\label{aa}
\tilde{A}= - H \: (cos \theta+1) \: d\phi= - H \: \frac{1}{r(r-z)} (x dy -y dx) ,
\eeqs
which is singular along the positive $z$-axis, the so-called Dirac string singularity. For this gauge field, we see that the electric field is trivial while the magnetic field computed with (we closely follow \cite{Felsager:1981iy})
\beqs
\vec{A}=( H\frac{y}{r (r-z)}, -H \frac{x}{r(r-z)},0) ,
\eeqs
gives rise to
\beqs\label{modifmono}
\vec{B}=\vec{\nabla} \times \vec{A}= H \frac{\vec{r}}{r^3}- H \delta(x) \delta(y) \mathcal{\theta}(z) \vec{1}_{z}\; ,
\eeqs
where $\mathcal{\theta}(z)$ is the Heaviside function which is zero for $z<0$ and one for $z\geq 0 $, and $\vec{1}_{z}$ is the unit vector along the $z$-direction. Note that to obtain this last result, we used the following regularization procedure: we first set $r\rightarrow R=\sqrt{r^2+\epsilon^2}$ and obtain easily
\beqs
\vec{B}_{\text{reg}}=H \Big(\frac{\vec{r}}{R^3}- \frac{\epsilon^2 (2R-z)}{R^3 (R-z)^2} \vec{1}_{z} \Big) .
\eeqs
We then take the limit of $\epsilon\rightarrow 0$ to recover \eqref{modifmono}.

What we have described in \eqref{modifmono} is a modified magnetic field which has a singular contribution along the positive axis. This unwanted singular contribution actually sets the magnetic charge to zero.

To describe the pure monopole field \eqref{magnmon}, we actually need to write
\beqs
F=dA+C\; , \qquad C_{ij}= \epsilon_{ijz} H \delta(x) \delta(y) \mathcal{\theta}(z)\; ,
\eeqs
and we realize that $C$ is precisely a string singularity along the $z$-axis canceling the one coming from our naive guess \eqref{aa}. It is sourcing the left hand side of the Bianchi identity. One can check that under an appropriate gauge transformation, the string can be sent along the negative $z$-axis so that, for the classical theory at least, the singularity in the gauge field should be seen as an artifact as it is really the electric and magnetic fields which are the physical gauge-invariant observables.

One interesting remark, for further considerations in the next chapter, is that the integrand appearing in the surface integrals for computing the charges is the field strength $F=dA$, a gauge invariant quantity. This means that if we consider $F=dA+C$, a regular gauge transformation on $A$ does not shift the string. It can thus always be fixed so as to cancel the singular term coming from $dA$.

\setcounter{equation}{0}
\section{Quantization of the electric charge}
\label{sec: Quanti}
As we have seen in the previous section, when one wants to deal with gauge potentials, one needs to introduce the Dirac string. However, this string is not physical classically as the electric and magnetic fields, the true observables, are regular. The fact that the string should not be visible quantum mechanically led Dirac to the first explanation of the quantization of the electric charge in units of $h$.

One way to understand this result is to consider a system of an electric charge and a magnetic monopole separated by a fixed distance. This system will possess an angular momentum
\beqs
\vec{L} = \int d^3 x \:\: \vec{x} \times (\vec{E} \times \vec{B}) = \frac{QH}{4 \pi} \: \vec{n} ,
\eeqs
The fact that angular momentum is quantized in the quantum theory tells us now that the electric charge has to be quantized in units of $h$.

The importance of this result lies in the fact that the very observation of a unique magnetic monopole would explain the quantization of the electric charge.

\setcounter{equation}{0}
\section{Other comments}
\label{sec: Various}

Let us comment here on some other aspects that have not been considered in the above.

\subsubsection{Double field formalism and the invariance of the action}

Up to here, we have seen that the magnetic charge appears as a topological charge. This is due to the fact that we have described the monopole using  an ``electric" formulation as only the electric charge is considered to be dynamical. As we have pointed out in chapter 1, working in the Hamiltonian formalism, it is actually possible to introduce new gauge degrees of freedom. By doubling the number of gauge degrees of freedom, one can introduce two, dual, potentials. In this way, electric and magnetic parts can be set on an equal footing. This is known as the doubled field formalism. 

The equations of motion are invariant under the exchange of electric and magnetic fields. However, the action
\beqs
\mathcal{L}= \frac{1}{2} (E^2-B^2)\: ,
\eeqs
is obviously not invariant under such a transformation. To check that the action is invariant, it has been shown in \cite{Deser:1976iy}, using the doubled field formalism, that one should actually consider transformations of the gauge field, which represent the true dynamical variables of the theory, instead of the field strength. Although the action can be written in a manifestly invariant way, it is no longer manifestly invariant under Lorentz transformations.

\subsubsection{The strong-weak duality}

Electromagnetic duality is a strong-weak duality. Indeed, the electric charge is directly related to the coupling constant, the strength of the electromagnetic interaction. Under a duality rotation the electric charge is sent to the magnetic charge. The fact that the duality is a strong-weak duality can be understood from the quantization condition of the electric charge. Indeed, let us write it as
\beqs
QH=\frac{n}{2}\; .
\eeqs
From this, we see that if the electric theory is ``weakly" coupled, then the magnetic theory will instead be strongly coupled.

Electromagnetic duality inspired C. Montonen and D. Olive who conjectured in \cite{Montonen:1977sn} the presence of this symmetry inside non-abelian gauge theories. It was later shown by H. Osborn in \cite{Osborn:1979tq} to hold as a strong-weak duality in $\mathcal{N}=4$ super Yang-Mills.

%%%%%%%%%%%%%%%%%%%%%%%%%%%%%%%%%%%%%%%%%%%%%%%%%%%%%%%%%%%%%%%%%%%
\cleardoublepage
%%%%%%%%%%%%%%%%%%%%%%%%%%%%%%%%%%%%%%%%%%%%%%%%%%%%%%%%%%%%%%%%%%%

\chapter{Gravitational duality in linearized gravity}
\label{chap:gravduality}

In this chapter, we review how electromagnetic duality can be transposed to general relativity.
As we said at the beginning of this thesis, the main reason to believe that such a duality could be present in general relativity is the presence of the Taub-NUT solution. This solution has a mass $M$ and a parameter $N$ that play, at least in the linearized theory, the same roles as the electric and magnetic charges in electromagnetism.

Instead of providing the reader immediately with the Taub-NUT solution, we would like to take a hopefully more interesting  path. In section \ref{sec: Ehlers}, we review the fact that general relativity reduced along a Killing direction has a pair of scalars which parametrize an $SL(2,R)/SO(2)$ coset. Using this, we show that the Schwarzschild metric can be mapped under an $SO(2)$ rotation, subgroup of Ehlers's $SL(2,R)$ group, to a new solution, the Taub-NUT metric, of Einstein's equations. Gravitational duality is the name given to this $SO(2)$ rotation. It is an exact duality of the full theory in the presence of a Killing direction. We finish this section by discussing the Taub-NUT solution as a solution of general relativity.

In section \ref{sec: gravdu}, we review how linearized equations of motion, in four dimensions and in the presence of electric sources, are invariant under an $SO(2)$ duality rotation if one allows for the presence of a magnetic stress-energy tensor in complete analogy with the work of Dirac for electromagnetism. This duality is valid even in the absence of any Killing direction. The duality in the linearized theory in the absence of Killing directions has been proven to be a symmetry of the action using a double field formalism in \cite{Henneaux:2004jw}. This result was generalized in the presence of sources in \cite{Barnich:2008ts}. Note that these works only deal with the linearized theory. In \cite{Deser:2005sz}, it was proved that this duality can not be extended perturbatively to the 3-vertex in Einstein gravity using a proof similar to the one showing that electromagnetic duality of free Maxwell theory cannot be extended to Yang-Mills theory.

In section \ref{sec: charges}, we give expressions for the ten Poincar\'e charges associated to the electric stress-energy tensor and the ten dual Poincar\'e charges associated to the magnetic stress-energy tensor, generalizing in a way the Abbott-Deser construction in the presence of singularities. We show that momenta and dual momenta can be expressed as surface integrals while Lorentz and dual Lorentz charges require some gauge fixing in our ``electric" formulation.

Eventually, in section \ref{sec: gravex}, we discuss several linearized solutions such as the Schwarzschild, Kerr and Aichelburg-Sexl pp-wave and their respective  dual solutions that we refer to as the pure NUT, rotating NUT and NUT-wave. Duality considerations lead us to an exotic interpretation of the source of the Kerr metric.  The Aichelburg-Sexl shock pp-wave is obtained as an infinite boost of the Schwarzschild metric and its gravitational dual is also recovered by taking the infinite boost of the pure NUT metric.

\setcounter{equation}{0}
\section{Ehlers's symmetry and the Taub-NUT solution}
\label{sec: Ehlers}
In this section, we start by reviewing the discovery of J. Ehlers \cite{Ehlers:1957zz} who showed that Einstein's equations reduced on a circle possess an $SL(2,R)$ symmetry. We then see how the Taub-NUT solution can be obtained by an $SO(2)$ rotation of the Schwarzschild solution reduced on a circle, when the parameters $M$ and $N$ are seemingly rotated, and discuss some important aspects of this solution.

To illustrate this, let us first perform a Kaluza-Klein reduction of Einstein's equations along a timelike direction.
Start with a four dimensional metric that can be written in the form
\beqs
ds^2_4= - e^{-\phi} (dt + \mathcal{A})^2 + e^{\phi} ds^2_3.\label{ansa}
\eeqs
Plugging this ansatz for the metric into Einstein's equations, we obtain a set of three-dimensional equations: Einstein's equation for the three-dimensional metric, a Maxwell equation for the graviphoton $\mathcal{A}$ and a Klein-Gordon equation for the dilaton field $\phi$. This set of equations can be obtained from the three-dimensional Lagrangian
\beqs
\mathcal{L}_{3}=\sqrt{g_3} \Big(R_3 -\frac{1}{2} \partial_{i} \phi \: \partial^i \phi +\frac{1}{4}\: e^{-2\phi} \:  F_{ij} F^{ij}\Big) .
\eeqs
This is the standard Kaluza-Klein reduction that we will not review here, but we refer the interested reader to  C.N. Pope's lectures available on the internet for a crystal clear presentation of the subject \cite{pope}.

In three dimensions, the dual of a vector is a scalar. As we have seen for electromagnetism, duality typically exchanges equations of motion with Bianchi identities. The usual way to implement such a duality at the level of the Lagrangian is to add a term that enforces the Bianchi identity of the field to be dualized. In our case, the term is $\frac{1}{2} F_{ij} \epsilon^{ijk} \partial_{k} \chi$ and it is a boundary term when the Bianchi identity for the field strength is enforced. With this additional term, one can also integrate out $A_{\mu}$. The equation of motion for the graviphoton is the defining duality relation. Rewriting the action with the supplemented term as
\beqs
&&\mathcal{L}_{3} + \frac{1}{2} F_{ij} \epsilon^{ijk} \partial_{k} \chi= \sqrt{g_3} \Big(R_3 -\frac{1}{2} (\partial \phi)^2 -\frac{1}{2} \: e^{2\phi} \: (\partial \chi)^2  \nonumber \\
&& \qquad \qquad  \qquad +\frac{1}{4} e^{-2\phi} (F_{ij}+\frac{1}{\sqrt{g_3}}\epsilon_{ijk} e^{2\phi} \partial^k \chi) (F^{ij}+\frac{1}{\sqrt{g_3}} \epsilon^{ijl} e^{2\phi} \partial_l \chi)\Big) ,\label{lagrari}
\eeqs
and making use of the defining relation
\beqs
F_{ij}+\frac{1}{\sqrt{g_3}} \epsilon_{ijk} e^{2\phi} \partial^k \chi=0 , \label{defrell}
\eeqs
the last term in the Lagrangian \eqref{lagrari} is zero. We eventually get a theory where the dual field has now become dynamical
\beqs
\mathcal{L}_{3}=\sqrt{g_3} \Big( R_3 -\frac{1}{2} (\partial \phi)^2 -\frac{1}{2} \: e^{2\phi} \: (\partial \chi)^2 \Big) .
\eeqs
It is by now a well-known fact that the scalar sector parameterizes an $SL(2,R)/SO(2)$ coset. One can check that the Lagrangian is invariant under the non-linear transformations of the scalar fields
\beqs
e^{\phi} &\rightarrow& e^{\phi'}=(c \chi +d)^2 e^{\phi} +c^2 e^{-\phi}  , \nonumber \\
\chi e^{\phi} &\rightarrow& \chi e^{\phi'}= (a \chi +b) (c \chi +d) e^{\phi} + ac e^{-\phi} , \label{transfos}
\eeqs
where $a,b,c,d$ are the coefficients of $2\times2$ matrices such that $ad-bc=1$. In here, we will no longer discuss the $SL(2,R)$ transformations but interest ourselves in an $SO(2)$ subgroup of it
\begin{eqnarray}
 \left (
\begin{array}{cc}
a & b   \\
c & d
\end{array}
\right ) \:\:\: =\left (
\begin{array}{cc}
\cos \psi & -\sin \psi  \\
\sin \psi & \cos \psi
\end{array}
\right ) .
\end{eqnarray}
Let us now see what happens if we apply a rotation $\psi$ to the Schwarzschild metric
\beqs\label{Schwarzschildm}
ds^2_4= -\frac{(r^2-2Mr)}{r^2} dt^2 + \frac{r^2}{r^2-2Mr} dr^2 + r^2 d\Omega_2^2 .
\eeqs
One starts by performing the change of coordinates $r\rightarrow r+M$ to  cast it into the form 
\beqs
ds^2_4&=&-\frac{(r^2-M^2)}{(r+M)^2} dt^2 + \frac{(r+M)^2}{r^2-M^2} ds^2_3 , \\
ds^2_3 &=&  dr^2 + (r^2-M^2) d\Omega^2_2 \label{ds3S} ,
\eeqs
with
\beqs
\chi=0 , \qquad e^{\phi}=\frac{(r+M)^2}{r^2-M^2} =\frac{r+M}{r-M}.
\eeqs
Under a rotation of angle $\psi$, we get
\beqs
e^{\phi'}= \frac{r^2+M^2+2Mr \cos(2\psi)}{r^2-M^2} , \qquad \chi' =-\frac{2Mr \sin(2\psi)}{r^2+M^2+2Mr \cos(2\psi)}.
\eeqs
Now, one should also rotate the charges. We see that by defining
\beqs
M'= M \cos(2\psi) , \qquad N'=M \sin(2\psi) ,
\eeqs
we have
\beqs
e^{\phi'}&=& \frac{(r+M')^2+N'^2}{r^2-M'^2-N'^2} , \qquad \chi' =-\frac{2N'r}{(r+M')^2+N'^2} , \nonumber \\
ds^2_3&=&  dr^2 + (r^2-M'^2-N'^2) d\Omega^2_2 .
\eeqs
Using \eqref{defrell} and the above information, the graviphoton is
\beqs
\mathcal{A}&=& 2N' \cos \theta d\phi ,
\eeqs
Upon uplifting to four dimensions, the four-dimensional metric is
\beqs
ds^2_4 &=&- \frac{r^2-M'^2-N'^2}{(r+M')^2+N'^2} \Big(dt+2 N' \cos \theta d\phi \Big)  \nonumber \\
&&\qquad +\frac{(r+M')^2+N'^2}{r^2-M'^2-N'^2} \Big( dr^2 + (r^2-M'^2-N'^2) d\Omega^2_2 \Big) .
\eeqs
Now, by making $r\rightarrow r-M'$, we find
\beqs\label{TAUB-NUT}
ds^2& = &-\frac{\lambda}{R^2}(dt+ 2N  \cos\theta
d\phi)^2 + \frac{R^2}{\lambda} dr^2
+R^2 d\Omega^2 , \label{TNNN}
\eeqs
where $\lambda= r^2-N^2-2Mr$, $R^2=r^2+N^2$, $d\Omega^2$ is the metric on the unit two-sphere and where we have removed the primes on the charges for simplicity.

This new solution of Einstein's equations is the so-called Taub-NUT metric, name given to the solution found by A.H. Taub in \cite{Taub:1950ez} and  E. Newman, L. Tamburino and T. Unti in  \cite{Newman:1963yy}.  The Taub-NUT metric was studied by  C. Misner in  \cite{Misner:1963fr}. This metric reduces to the Schwarzschild metric  \eqref{Schwarzschildm} when $N=0$ and to the metric
\beqs
ds^2_4= -\frac{(r^2-N^2)}{(r^2+N^2)} (dt+ 2N \cos \theta d\phi) + \frac{r^2+N^2}{r^2-N^2} dr^2 +(r^2+N^2) d\Omega^2_2,
\eeqs
 when $M=0$.  In the following, we will refer to this last metric as the pure NUT metric. 
 
Let us now discuss some relevant aspects of the Taub-NUT metric.
The first important thing to notice about the Taub-NUT metric is that
\beqs
g_{t\phi} \sim 2N  \cos\theta
d\phi .
\eeqs
This resemblance with the gauge field of the magnetic monopole presented in the previous chapter (remember we had $A= H \cos \theta d\phi$) is the reason why Taub-NUT could be interpreted as a gravitational dyon, see for example \cite{ashtekar}, \cite{dowker}, \cite{dualmass}.

As we have seen for electromagnetism, the potential is singular along the $z$-axis. This singularity for the Taub-NUT metric is known as the (Dirac)-Misner string.  In  \cite{Misner:1963fr}, it was stated that the singularity has to be removed because "If one is given a manifold and on it a metric which does not at all points satisfy the necessary differentiability requirements, one simply throws away all the points of singularity". Misner showed that the singularity is a coordinate singularity (just like $r=2M$ is for the Schwarzschild black hole) and that it can be removed. Let us see how this works.

As in the electromagnetic case, the singularity can be set on the positive or negative axis by an appropriate gauge transformation. In here, it is implemented by a change of the $\phi$ coordinate. Starting from the metric \eqref{TAUB-NUT}, one can make the change of coordinate
\beqs\label{north}
t \rightarrow t^S+2N\phi  ,
\eeqs
 so that the string singularity is along the positive $z$-axis or
 \beqs\label{south}
 t \rightarrow t^N-2N\phi ,
 \eeqs
 and the string will then be along the negative $z$-axis. In the coordinate system \eqref{north}, the metric is regular around the South pole and with \eqref{south} around the North pole. For consistency, we should impose that both regular metrics are equivalent on the equator. This is just
 \beqs
 t^N=t^S+4N\phi .
 \eeqs
 Because $\phi$ is periodic with period $2\pi$, the singularity is cured if one imposes $t$ to be periodic with period $8\pi N$ as stated in  \cite{Misner:1963fr}. The drawback of identifying time is that it introduces closed timelike curves, another undesirable feature of general relativity.

 In the rest of this thesis we will not consider this identification but rather deal with the metric \eqref{TAUB-NUT}.  The main reason is that we want to deal with linearized gravity where there is an exact duality \cite{Henneaux:2004jw} that rotates the linearized Taub-NUT metric onto itself (when the mass and the NUT charge are also rotated).  Also, Taub-NUT metric is asymptotically flat (at least locally) following Regge-Teitelboim, see \cite{Bunster:2006rt}. As one can check from Part I, remark that the metric fulfills the parity conditions, even if $k_{ab}$ is now singular along the $z$-axis.  Considerations about the existence of a variational principle in the presence of NUT charge can be found in \cite{Bunster:2006rt}, \cite{Mann:2005yr} and \cite{Virmani:2011gh}.

\setcounter{equation}{0}
\section{Gravitational duality for linearized gravity}
\label{sec: gravdu}
In this section, we would like to review how gravitational duality works for linearized general relativity in the absence of Killing directions. We will re-derive the duality invariant form of the Einstein equations, cyclic and Bianchi identities, following the lines of \cite{Bunster:2006rt}.  In the rest of this thesis, we will work in vielbein formalism. This was first motivated by the study of solutions of supersymmetric theories as we detail in Part III. However, we will see that it also has its utility in the rest of this chapter.

Linearized general relativity seems to have a lot in common with electromagnetism, as they are for example both linear theories and possess both a duality symmetry. However, if one wants to generalize the electromagnetic duality to linearized gravity, there are subtleties that need to be taken care of.

The first difference with electromagnetism comes from the fact that the duality is a Hodge duality on the Riemann tensor, a tensor that has two pairs of antisymmetric indices (the Lorentz and the form indices,
respectively, in reference to the spin connection). Therefore, one is free, in the linearized theory, to pick the Lorentz indices, the form indices, or even a linear combination of these two. We argue in the following that gravitational duality is best understood when dualization is performed on Lorentz indices.

In comparison with electromagnetism, the second subtelty arises from the existence of three different ``objects" in general relativity. Indeed, we have a vielbein $e^a_{\:\:\mu}$, a connection $\omega_{\mu}^{\:\:ab}$, and a curvature $R_{ab \mu\nu}$. Because the singularity appears in the vielbein, one could reasonably wonder why the connection could not be used, instead of the Riemann tensor, to play the same role as does the field strength $F$ in electromagnetism. We show that the choice of dualization on Lorentz indices permits to lower the duality relation between Riemann tensors to a duality between spin connections. The important difference we point out is that the spin connection is a gauge-variant quantity while the field strength $F$ is not. We show however that, by duality, a gauge choice can always be made such that the dual spin connection is regular. We eventually give an expression of the spin connection in terms of the vielbein and a three index object, first introduced in \cite{Bunster:2006rt}, that contains the magnetic information of the solution. Since we linearize around flat Minkowski space in cartesian coordinates, there will be no distinction between curved and flat indices in the following.

Linearizing around flat space $g_{\mu\nu}=\eta_{\mu\nu}+h_{\mu\nu}$, the Einstein equations, cyclic and
Bianchi identities, in the absence of magnetic charges, are just
\begin{eqnarray}
&& G_{ \mu \nu}= 8 \pi G T_{\mu \nu}, \nonumber \\
&& R_{\mu[\nu\alpha\beta]}= \frac{1}{3} ( R_{ \mu \nu \alpha
\beta} + R_{ \mu \beta \nu \alpha} + R_{ \mu \alpha \beta \nu})
= 0, \nonumber \\
&&\partial_{[\alpha}\: R_{| \rho \sigma | \beta \gamma]}=
\frac{1}{3} (
\partial_{\alpha}\: R_{\rho \sigma  \beta \gamma} +
\partial_{\gamma} \: R_{ \rho \sigma  \alpha  \beta} +
\partial_{\beta} \:  R_{ \rho \sigma \gamma \alpha} ) = 0,
\end{eqnarray}
where $R_{ \rho \sigma \gamma \alpha}$ is the linearized Riemann tensor.
The Bianchi identities are solved by expressing the Riemann tensor in
terms of a spin connection. In turn, the cyclic identity is solved
when the spin connection is expressed in terms of a vielbein or, when
the local Lorentz gauge freedom is fixed, in terms of a (linearized) metric.

In the absence of sources, gravitational duality tells us that for every metric there exists a
dual metric such that their respective Riemann tensors are Hodge dual to
each other, in complete parallel with the Hodge duality in electromagnetism.
As explained above, we will prefer here a dualization
on the Lorentz indices, the first two indices in our conventions, as
is clear from the Bianchi identities above. We write the duality in the absence of sources as
\begin{eqnarray}\label{gravdual}
R_{\mu\nu\rho\sigma}\rightarrow \tilde{R}_{\mu\nu\rho\sigma}=\frac{1}{2} \:
\varepsilon_{\mu\nu\alpha\beta}R^{\alpha \beta}_{\:\:\:\:\:
\rho\sigma}, \qquad \tilde{R}_{\mu\nu\rho\sigma} \rightarrow -R_{\mu\nu\rho\sigma}=- \frac{1}{2}
\varepsilon_{\mu\nu\alpha\beta} \tilde{R}^{\alpha
\beta}_{\:\:\:\:\: \rho\sigma},
\end{eqnarray}
where $\tilde{R}_{\mu\nu\rho\sigma}$ denotes the magnetic or dual Riemann tensor. To check that linearized Einstein's equations are invariant under this duality and also to generalize the duality in the presence of electric sources, it is useful to remark that the magnetic cyclic identity in the presence of electric sources can be written as
\begin{eqnarray}
 ( \tilde{R}_{ \mu \nu \alpha \beta} + \tilde{R}_{ \mu
\beta \nu \alpha} + \tilde{R}_{ \mu \alpha \beta \nu}) &=& 3
\delta^{\rho\sigma \kappa}_{[\nu\alpha\beta]} \tilde{R}_{ \mu \rho
\sigma \kappa}
 =-\frac{1}{2} \varepsilon_{\gamma \nu
\alpha \beta} (\varepsilon^{\gamma \rho \sigma \kappa}\tilde{R}_{
\mu \rho \sigma \kappa} ) \nonumber \\
&=&-\frac{1}{2} \varepsilon_{\gamma \nu
\alpha \beta}( 2 R^{\gamma}_{\:\:\: \mu}-
\delta^{\gamma}_{\:\:\mu} R)= 8 \pi G \varepsilon_{ \nu \alpha \beta \gamma}  T^{\gamma}_{\:\:\mu}\label{rela}.
\end{eqnarray}
The electric stress-energy tensor appears at the right hand side of this last equation. Transposing Dirac's idea for electromagnetism to linearized gravity, we will add a magnetic stress-energy tensor $\Theta_{\mu\nu}$ on the right hand side of the "electric" cyclic identity. Under a gravitational duality rotation in the presence of both electric and magnetic sources, we have schematically
\begin{eqnarray}
&& R_{ \mu \nu \rho \sigma} \rightarrow  \tilde{R}_{ \mu \nu \rho
\sigma}, \qquad \tilde{R}_{ \mu \nu \rho \sigma} \rightarrow -R_{
\mu \nu \rho
\sigma},\nonumber \\
&& T_{\mu \nu} \rightarrow  \Theta_{\mu\nu}, \qquad \qquad \Theta_{\mu
\nu} \rightarrow -T_{\mu\nu}.
\end{eqnarray}
We write the full set of electric and magnetic equations respectively as
\begin{eqnarray}\label{emequations}
&& G_{ \mu \nu}= 8 \pi G T_{\mu \nu}, \nonumber \\
&& R_{ \mu \nu \alpha
\beta} + R_{ \mu \beta \nu \alpha} + R_{ \mu \alpha \beta \nu}
= - 8\pi G \varepsilon_{ \nu \alpha \beta
\gamma} \Theta^{\gamma}_{\:\:\mu}, \nonumber \\
&&\partial_{\epsilon}\: R_{\gamma \delta \alpha \beta} +
\partial_{\alpha} \: R_{ \gamma \delta  \beta \epsilon} +
\partial_{\beta} \:  R_{ \gamma \delta \epsilon \alpha} = 0, \nonumber \\
& & \nonumber \\
&& \tilde{G}_{ \mu \nu}= 8 \pi G \Theta_{\mu \nu}, \nonumber \\
&& \tilde{R}_{ \mu
\nu \alpha \beta} + \tilde{R}_{ \mu \beta \nu \alpha} +
\tilde{R}_{ \mu \alpha \beta \nu}
=  8\pi G \varepsilon_{ \nu \alpha \beta \gamma} T^{\gamma}_{\:\:\mu}, \nonumber \\
&&\partial_{\epsilon}\: \tilde R_{\gamma \delta \alpha \beta} +
\partial_{\alpha} \: \tilde R_{ \gamma \delta  \beta \epsilon} +
\partial_{\beta} \: \tilde  R_{ \gamma \delta \epsilon \alpha} = 0.
\end{eqnarray}
From these equations, the duality is manifest as soon as we write the electric and magnetic cyclic identities, by means of (\ref{rela}), as
\begin{eqnarray}
\tilde{G}_{ \mu \nu}= 8 \pi G \Theta_{\mu \nu}, \qquad
G_{ \mu \nu}= 8 \pi G T_{\mu \nu}.
\end{eqnarray}

One advantage of dualizing on Lorentz indices, as compared
to a dualization on form indices, is that we do not need to modify the Bianchi identity because
\begin{eqnarray}
 \partial_{[\alpha}\: \tilde{R}_{| \mu \nu | \beta
\gamma]}=\frac{1}{2}
\varepsilon_{\mu\nu}^{\:\:\:\:\rho\sigma}\partial_{[\alpha}\: R_{|
\rho \sigma | \beta \gamma]}.
\end{eqnarray}
Note that the vanishing of the Bianchi identity is consistent with the
cyclic identity having a non-trivial source term if and only if the
magnetic stress-energy tensor is conserved, $\partial_\mu\Theta^{\mu
  \nu}=0$, just as the ordinary stress-energy tensor. This is obviously an important property as we will construct charges from this quantity in the next section.

As already mentioned previously, the Riemann tensor can only be defined in terms of a metric
when both the cyclic and Bianchi identities have a trivial
right-hand side. To deal with the introduction of magnetic sources we
introduce, as in \cite{Bunster:2006rt}, a three-index object ${\Phi^{\mu\nu}}_{\rho}$ such that
\begin{eqnarray}
&& \partial_{\alpha} \Phi^{\alpha \beta}_{\:\:\:\:\:\gamma} = -16 \pi
G \Theta^{\beta}_{\:\:\:\gamma}, \qquad \Phi^{\alpha
\beta}_{\:\:\:\:\:\gamma}= - \Phi^{ \beta
\alpha}_{\:\:\:\:\:\gamma} ,\label{PHI} \\
&& \bar{\Phi}^{\rho\sigma}_{\:\:\:\:\:\alpha}=
\Phi^{\rho\sigma}_{\:\:\:\:\:\alpha} + \frac{1}{2} (
\delta^{\rho}_{ \:\:\: \alpha } \Phi^{\sigma} -\delta^{\sigma}_{
\:\:\: \alpha }\Phi^{\rho}), \qquad  \Phi^{\rho}= \Phi^{\rho
\alpha}_{\:\:\:\:\: \alpha}.
\end{eqnarray}
The Riemann tensor that is solution of the set of equations (\ref{emequations}) when making use of (\ref{PHI}) is
\begin{eqnarray}\label{genriemann}
R_{\alpha \beta \lambda \mu} = r_{\alpha \beta \lambda \mu}+
\frac{1}{4} \: \epsilon_{\alpha \beta \rho \sigma}
(\partial_{\lambda} \bar{\Phi}^{\rho\sigma}_{\:\:\:\:\:\mu}
-\partial_{\mu} \bar{\Phi}^{\rho\sigma}_{\:\:\:\:\:\lambda} ),
\end{eqnarray}
where $r_{\alpha \beta \lambda \mu}$ is the usual Riemann tensor verifying the usual cyclic and
Bianchi identities with no magnetic stress-energy tensor. This means that $r_{\alpha \beta
\lambda \mu}= r_{\lambda \mu \alpha \beta } $ and that it can be
derived from a potential: $r_{\alpha \beta \lambda \mu}= 2
\partial_{[\alpha} h_{\beta][ \lambda,\mu]}$.

Another advantage of the dualization on Lorentz indices comes
directly from the vanishing right hand side of the Bianchi identity. Indeed, as compared to the results presented in \cite{Bunster:2006rt}, our dualization 
gives us the right to express the linearized Riemann tensor in
terms of a spin connection by
\begin{eqnarray}\label{riemanntensor}
R_{\mu\nu\rho\sigma}=\partial_{\rho} \omega_{\mu \nu \sigma}-
\partial_{\sigma} \omega_{\mu \nu \rho}\; . \label{spinco}
\end{eqnarray}
This allows to lower
the duality relation between Riemann tensors to a duality between
spin connections. With the help of (\ref{gravdual}) and (\ref{spinco}) the gravitational duality relation becomes
\begin{eqnarray}\label{dualspinc}
\tilde{\omega}_{\mu \nu \sigma}= \frac{1}{2}
\varepsilon_{\mu\nu\alpha\beta} \: \omega^{\alpha \beta
}_{\:\:\:\:\:\sigma},
\end{eqnarray}
where this relation is true up to a gauge transformation as the spin connection is a gauge-variant object.

The linearized vielbein and the spin connection for the Riemann
tensor $r_{\mu\nu\alpha\beta}$ are defined as
\begin{eqnarray}
r_{\mu\nu\rho\sigma}&=& \partial_{\rho} \Omega_{\mu \nu \sigma}-
\partial_{\sigma} \Omega_{\mu \nu \rho}, \nonumber \\
e^\mu &=& dx^\mu+\frac{1}{2}
\eta^{\mu\nu}(h_{\nu\rho}+v_{\nu\rho})dx^\rho, \nonumber \\
\Omega_{\mu\nu} &=& \Omega_{\mu\nu\rho}e^\rho, \qquad
\Omega_{\mu\nu\rho}= \frac{1}{2}(\partial_\nu h_{\mu\rho}
-\partial_\mu h_{\nu\rho} +\partial_\rho v_{\nu\mu}),
\end{eqnarray}
where $h_{\mu\nu}=h_{\nu\mu}$ is the linearized metric and $v_{\mu\nu}=-v_{\nu\mu}$. Using this together with  relations (\ref{spinco}) and (\ref{dualspinc}), one obtains the spin connection in terms of the vielbein and the three-index object $\Phi_{\mu\nu\rho}$
\begin{eqnarray}\label{spinconnection}
\omega_{\mu\nu\rho}&=& \Omega_{\mu\nu\rho} +\frac{1}{4}
\varepsilon_{\mu\nu\gamma\delta} \bar{\Phi}^{\gamma
\delta}_{\:\:\:\:\: \rho} \nonumber \\
&=&\frac{1}{2}(\partial_\nu h_{\mu\rho} -\partial_\mu h_{\nu\rho}
+\partial_\rho v_{\nu\mu}) +\frac{1}{4}
\varepsilon_{\mu\nu\gamma\delta} \bar{\Phi}^{\gamma
\delta}_{\:\:\:\:\: \rho}. \label{omegamunurho}
\end{eqnarray}
From (\ref{dualspinc}), it is clear that there always exists a ``regular" (with respect to string singularities on the two-sphere at spatial infitinity) spin
connection  even when magnetic sources are present. From
the expression above this can be achieved for a
specific choice of $v_{\mu\nu}$ that cancels string
contributions coming from $\Phi_{\mu\nu\rho}$. One also easily sees that
\begin{eqnarray}
\tilde{\omega}_{\mu\nu\sigma}&=&\frac{1}{2}
\varepsilon_{\mu\nu\alpha\beta} \: \omega^{\alpha \beta
}_{\:\:\:\:\:\sigma} = -\frac{1}{4} [ \varepsilon_{\mu\nu\alpha
\beta} (2 \partial^{\alpha}
h^{\beta}_{\:\:\:\sigma}+\partial_{\sigma} v^{\alpha\beta})+ 2
\bar{\Phi}_{\mu \nu \sigma}].
\end{eqnarray}

\setcounter{equation}{0}
\section{Charges and dual charges}
\label{sec: charges}

Now that we have looked into more details how the duality works at the linearized level, we would like to deal with the definition of charges in the linearized theory. Since we have an electric and a magnetic stress-energy tensor, one should be able to define the usual 10 Poincar\'e charges associated to $T_{\mu\nu}$, as we presented in chapter 1, but also 10 other topological, dual, Poincar\'e charges associated to the dual stress-energy tensor $\Theta_{\mu\nu}$. This is what we explore in this section. We start by giving the generalized expressions for the ADM momenta and dual ADM momenta. We give a full treatment of the singular
string contributions, obtaining gauge-independent expressions for the surface integrals\footnote{Note that in \cite{Argurio:2008zt}, we only established them for a specific gauge choice of the vielbein.}. We eventually apply the same ideas to derive general expressions for the Lorentz charges and their duals. However, we will show that there is no possibility in this formalism to express these charges as surface integrals without partially fixing the gauge. Two copies of the Poincar\'e charges, expressed as surface integrals, have also been derived by G. Barnich and C. Troessaert in \cite{Barnich:2008ts} using a doubled (Hamiltonian) formalism. This gauge fixing is an artifact of our Lagrangian approach which is more ``electric" in spirit as the magnetic charges are topological. The charges obtained in \cite{Barnich:2008ts} should however be completely equivalent to ours.

\subsection{The momenta and dual momenta}

The generalized ADM momenta and dual ADM momenta are
\begin{eqnarray}
 P_\mu =\int T_{0\mu} d^3x  = \frac{1}{8\pi G} \int G_{0\mu}d^3x,  \label{ADMM} \\
K_\mu= \int\Theta_{0\mu} d^3x = \frac{1}{8\pi G} \int \tilde G_{0\mu}d^3x. \label{ADMM2}
\end{eqnarray}
Given the definition of the Riemann tensor in (\ref{riemanntensor}), one easily obtains
\beqs \label{g00}
G_{00} &=& \frac{1}{2} R_{ijij} = \partial_i \omega_{ijj} , \\
\label{goi}
G_{0i} & = & R_{0jij}  = \partial_i \omega_{0jj} - \partial_j \omega_{0ji}.
\eeqs
The dual Ricci tensor is
\beq
\tilde R_{\mu\rho} = \eta^{\nu\sigma} \tilde R_{\mu\nu\rho\sigma} =
\frac{1}{2} \eta^{\nu\sigma} \varepsilon_{\mu\nu\alpha\beta}
{R^{\alpha\beta}}_{\rho\sigma}.
\eeq
The dual Ricci scalar and
dual Einstein tensor are defined just as $\tilde
R=\eta^{\mu\rho}\tilde R_{\mu\rho}$ and $\tilde G_{\mu\rho} =
\tilde R_{\mu\rho} -\frac{1}{2} \eta_{\mu\rho}\tilde R$. We thus
have the following expressions
\beqs \label{tg00}
\tilde G_{00} &=& -\frac{1}{2} \varepsilon_{ijk} R_{0ijk}=\varepsilon_{ijk}\partial_i \omega_{0jk}, \\
\label{tg0j}
\tilde G_{0i} &=& \frac{1}{2} \varepsilon_{jkl} R_{klij}=
\frac{1}{2} \varepsilon_{jkl}(\partial_i \omega_{klj} - \partial_j \omega_{kli})
=\varepsilon_{jkl} \partial_l \omega_{ijk}.
\eeqs
In the last equality of (\ref{tg0j})
 we have used the identity $\partial_{[i}\omega_{jkl]}=0$.
Note also that $\tilde G_{0i}\neq \tilde G_{i0}$ for an arbitrary
(i.e. off-shell) spin connection.

Eventually, we can express the electric and magnetic Einstein tensors in the more compact form
\begin{eqnarray}\label{GGtilde}
G_{0\mu} &=&  \partial_i ({\omega^{0i}}_\mu
+\delta^0_\mu {\omega^{i\rho}}_\rho
-\delta^i_\mu {\omega^{0\rho}}_\rho ),  \nonumber \\
\tilde G_{0\mu} &=& \varepsilon^{ijk}\partial_i
\omega_{\mu jk}\; .
\end{eqnarray}
This enables us to formulate the momenta as surface integrals
\begin{eqnarray}
P_\mu & = & \frac{1}{8\pi G} \oint [ {\omega^{0l}}_\mu
+\delta^0_\mu {\omega^{l\rho}}_\rho
-\delta^l_\mu {\omega^{0\rho}}_\rho ] d \Sigma_l, \label{pmukmu1} \\
K_\mu & = & \frac{1}{8\pi G} \oint  \varepsilon^{ljk}
\omega_{\mu jk}
 d \Sigma_l . \label{pmukmu2}
\end{eqnarray}
With the help of (\ref{spinconnection}), we have
\begin{eqnarray}\label{pmukmu2}
P_0 & = & \frac{1}{16\pi G} \oint \biggl [ \partial_i h^{li}-  \partial^l {h^{i}}_{i}+ \partial_i v^{il}+\varepsilon^{ljk} \Phi_{0jk} \biggl ] d \Sigma_l, \label{p0v}\\
P_k & = & \frac{1}{16\pi G} \oint \biggl [ \partial_0 {h^{l}}_{k}- \partial^l
h_{0k} + \delta^l_k \partial^{i} h_{0i} - \delta^l_k \partial_{0}
{h^{i}}_{i}
+\partial_{k} {v_{0}}^{l}+ \delta^l_k \partial^{i} v_{i 0} \nonumber \\
&& \:\:\:\:\:\:\:\:\: \:\:\:  - \frac{1}{2}\varepsilon^{lij}[ \Phi_{ijk}+\delta_{ik} {\Phi_{j0}}^{0}+\delta_{ik }{\Phi_{jm}}^{m}]+ \frac{1}{2}\delta^{l}_{k} \varepsilon^{ijm}\Phi_{ijm} \biggl ] d \Sigma_l, \label{pkv}\\
K_0 & = & \frac{1}{16\pi G} \oint \biggl [ \varepsilon^{lij}  [ \partial_i h_{0j}+ \partial_j v_{i0} ]+{\Phi^{l0}}_{0} \biggl ]  d \Sigma_l , \\
K_k & = & \frac{1}{16\pi G} \oint  \biggl [ \varepsilon^{lij}  [
\partial_i h_{kj}+ \partial_j v_{ik} ]+{\Phi^{l0}}_{k} \biggl ]  d
\Sigma_l.
\end{eqnarray}

When there are no magnetic charges, $\Theta_{\mu\nu}$ is zero and
thus $K_{\mu}$ and $\Phi_{\mu\nu\rho}$ also by definition. Then, setting ourselves in the gauge
where $v_{\mu\nu}=0$, one easily recognizes the ADM momenta
$P_{\mu}$. Remember that in
electromagnetism the contribution of the Dirac string was
always equal to the opposite of the string contribution coming from the
regularized connection as $F$ is a gauge-independent quantity. Even if our charges are obviously gauge invariant, the important difference with electromagnetism is that here the surface integrals for calculating the charges depend on
the spin connection, a gauge-variant object.  If we want to cancel the string contributions in the expressions (\ref{pmukmu2}), we need the additional gauge freedom of the vielbein to be fixed in the right gauge. As we have seen in the previous section, this trick can always be used as the duality can always be lowered to a duality between spin connections. For each electric solution with a regular spin connection, there exists a regular spin connection for the dual magnetic solution. Note that computations can also be made using our expressions \eqref{pmukmu1}-\eqref{pmukmu2} or even the volume integrals \eqref{ADMM}-\eqref{ADMM2}.

\subsection{Electric and magnetic Lorentz charges}

In the same spirit, the general
expression for the Lorentz charges and their duals are as
follows\footnote{Note that the fixed timelike index is now upstairs,
  contrary to the definitions of the momenta. We hope that this
  (arbitrary but innocuous) switch
  in the convention will not upset the reader too much.}
\begin{eqnarray}\label{Lorentzz}
L^{\mu\nu}&=& \int   (x^{\mu} T^{0 \nu }- x^{\nu} T^{0 \mu
}) d^3 x =\frac{1}{8\pi G} \int  (x^{\mu} G^{0 \nu }- x^{\nu} G^{0
\mu }) d^3 x,
\nonumber \\
\tilde{L}^{\mu\nu}&=& \int (x^{\mu} \Theta^{0 \nu }- x^{\nu}
\Theta^{0 \mu }) d^3 x = \frac{1}{8\pi G} \int (x^{\mu}
\tilde{G}^{0 \nu }- x^{\nu} \tilde{G}^{0 \mu}) d^3 x.
\end{eqnarray}

Plugging the expression (\ref{GGtilde}) into the definition of the electric Lorentz charges
leads us to
\begin{eqnarray}\label{volumecharges}
L^{ij}&=& \frac{1}{8\pi G}\int (x^{i} G^{0 j}- x^{j} G^{0 i}) d^3 x
\nonumber \\
      &=& \frac{1}{8\pi G} \oint  \biggl [ x^{j} [\omega^{0li}-
      \delta^{li}{\omega^{0 k}}_{k}]- x^{i} [\omega^{0lj}-
      \delta^{lj}{\omega^{0 k}}_{k}] \biggl ] d\Sigma_l  +
      \frac{1}{8\pi G}\int  [\omega^{0ij}- \omega^{0ji}] \: d^3 x, \nonumber \\
L^{0i}&=&\frac{1}{8\pi G}\int (t G^{0 i}- x^{i} G^{0 0}) d^3 x
\nonumber
\\
      &=& \frac{1}{8\pi G} \oint \biggl [ -t [ \omega^{0li}-\delta^{li} {\omega^{0 k}}_{k} ]- x^{i} {\omega^{lj}}_{j}  \biggr] d\Sigma_l  +
      \frac{1}{8\pi G}\int  {\omega^{ij}}_{j} \: d^3 x.
\end{eqnarray}
We see that in the presence of non-trivial $\Phi_{\mu\nu\rho}$, we
have a priori no way to express the charges as surface integrals.
However, we know that the charges are independent of the choice of
$v_{\mu\nu}$. We can thus try to choose a gauge, an appropriate $v_{\mu\nu}$, such as
to cancel the $\Phi_{\mu\nu\rho}$ contributions present in the
volume integrals. Expanding the volume integrals in the above expressions
\begin{eqnarray}
\int 2 [\omega^{0ij}- \omega^{0ji}] \: d^3 x &=&  \int [
\partial^{i} h^{0j}  -  \partial^{j} h^{0i} +   \partial^{j} v^{i0} -
\partial^{i} v^{j0} - \varepsilon^{ijk} \Phi_{k0}^{\:\:\:\:\: 0} ]
\: d^3 x, \nonumber \\
\int 2 {\omega^{ij}}_{j} \: d^3 x &=& \int [ \partial_j
h^{ij}-\partial^{i} {h^{j}}_{j}+\partial_{j}v^{ji}+
\varepsilon^{ijk} \Phi_{0jk} ] \: d^3 x,
\end{eqnarray}
where we simplified the last equation using the relation
 $ \varepsilon^{lk[i} \bar{\Phi}_{lk}^{\:\:\:\:\:
j]}= \varepsilon^{ijk} \Phi_{k0}^{\:\:\:\:\: 0}$,
we see that we can absorb the $\Phi_{\mu\nu\rho}$ by choosing the $v_{ij}$ and the $v_{0i}$ such that
\begin{eqnarray}
\int \partial_{j}v^{ij} \: d^3 x &=& \int \varepsilon^{ijk}
\Phi_{0jk} \: d^3 x,\label{gauge1} \\
\int [\partial^{j}v^{i0} - \partial^{i}v^{j0}] \: d^3
x &=&  \int \varepsilon^{ijk} \Phi_{k0}^{\:\:\:\:\: 0} \:
d^3 x.\label{gauge2}
\end{eqnarray}
Actually, these gauge choices do not fix completely the local
Lorentz gauge, and hence $v_{\mu\nu}$. Rather, they restrict the
gauge to a choice satisfying the above integral relations. Of
course this can be done in the simplest way by choosing a
$v_{\mu\nu}$ that locally compensates the singularity contained in
$\Phi_{\mu\nu\rho}$.

Picking a gauge such that (\ref{gauge1}) and
(\ref{gauge2}) are fulfilled, we obtain
\begin{eqnarray}
L^{ij} &=& \frac{1}{8\pi G} \oint \biggl [ x^{j} [\omega^{0li}-
      \delta^{li}{\omega^{0 k}}_{k}]- x^{i} [\omega^{0lj}-
      \delta^{lj} {\omega^{0 k}}_{k}] +\frac{1}{2} [\delta^{il} h^{0j}-\delta^{jl} h^{0i}] \biggl ] d\Sigma_l, \nonumber
      \\
L^{0i} &=& \frac{1}{8\pi G} \oint \biggl [ -t [
\omega^{0li}-\delta^{li} {\omega^{0 k}}_{k} ]- x^{i}
{\omega^{lj}}_{j}  +  \frac{1}{2} [h^{il}-\delta^{il} h]\biggr]
d\Sigma_l.
\end{eqnarray}

If we  look at the dual Lorentz charges, we
have
\begin{eqnarray}\label{volumedualcharges}
\tilde{L}^{0i} &=& \frac{1}{8\pi G}\int (t \tilde {G}^{0 i}- x^{i}
\tilde{G}^{0 0}) d^3 x \nonumber
\\ &=& \frac{1}{8\pi G} \oint  -\varepsilon^{ljk}
[ t \:  {\omega^{i}}_{jk} + x^{i} \omega_{0jk}]   d\Sigma_l  +
\frac{1}{16\pi G} \int \varepsilon^{ikl}
[\omega_{0kl}-\omega_{0lk}]d^3 x\; ,
 \nonumber \\
\tilde{L}^{ij}&=& \frac{1}{8\pi G}\int (x^{i} \tilde {G}^{0 j}-
x^{j} \tilde {G}^{0 i}) d^3 x
\nonumber \\
&=&\frac{1}{8\pi G}  \oint \varepsilon^{lkm} [ x^{j}
{\omega^{i}}_{km} - x^{i} {\omega^{j}}_{km} ] d\Sigma_l +
\frac{1}{8\pi G} \int \varepsilon^{ijk} {{\omega_{k}}^{l}}_{l} d^3
x\; ,
\end{eqnarray}
where in the last equality we used $\varepsilon^{ikm}
{\omega^{j}}_{km}-
\varepsilon^{jkm}{\omega^{i}}_{km}=\varepsilon^{ijk}
{{\omega_{k}}^{l}}_{l}$.

It is amusing to observe that the pieces in $L_{\mu\nu}$ and $\tilde
L_{\mu\nu}$ that cannot be expressed as surface integrals actually enjoy
a duality relation, $\tilde L_{\mu\nu}^\mathrm{bulk}=\frac{1}{2}
\varepsilon_{\mu\nu\rho\sigma}L^{\rho\sigma}_\mathrm{bulk}$. This
surprising property cannot of course be extended to the full charges,
as is obvious from their definition in terms of the stress-energy
tensor and its dual. However, a consequence of this
observation is that with the previous choice of gauge, we can also
express the dual charges as surface integrals
\begin{eqnarray}
\tilde{L}^{0i}&=& \frac{1}{8\pi G} \oint \biggr [ -\varepsilon^{ljk}
[ t \: {\omega^{i}}_{jk} + x^{i} \omega_{0jk}]+\frac{1}{2}
\varepsilon^{ilk} h_{0k} \biggl] d\Sigma_l \; ,\nonumber \\
\tilde{L}^{ij}&=& \frac{1}{8\pi G}  \oint \biggr [\varepsilon^{lkm} [
x^{j} {\omega^{i}}_{km} - x^{i} {\omega^{j}}_{km} ] + \frac{1}{2}
\varepsilon^{ijk}[{h^{l}}_{k}-\delta^{l}_{k}h] \biggl] d\Sigma_l.
\end{eqnarray}

The expressions derived here for the electric and magnetic Lorentz charges are thus valid
in whatever gauge when expressed as volume integrals like in (\ref{volumecharges}) and (\ref{volumedualcharges}). Moreover, we have shown that there
exists a gauge choice valid for the Lorentz charges and their duals
that permits to eliminate the
$\Phi_{\mu\nu\rho}$ and express the charges in terms of surface
integrals. Note that if all $\Phi_{\mu\nu\rho}$ are zero, any gauge
is obviously fine and the charges reduce to the ADM expressions.

In the next section, we will consider several different solutions and their dual counterparts. Instead of applying  these formulas explicitly, it will prove more efficient to work out the sources of the solutions, encoded in $T_{\mu\nu}$ and $\Theta_{\mu\nu}$, and compute
the charges from their original definitions (\ref{ADMM}) and (\ref{Lorentzz}). One is ensured, following the above arguments, that the surface integrals, with a correct choice of gauge, will yield the same results.

\setcounter{equation}{0}
\section{Examples illustrating the duality}
\label{sec: gravex}
In this section, we review various linearized ``electric" solutions and their ``magnetic" counterparts. For each of them, we illustrate how the duality works and what are their associated charges.
The solutions that we will consider are the Taub-NUT metric, its infinitely boosted limit,  and its rotating version called the Kerr-NUT.

\subsection{The Taub-NUT solution}
\label{subsec: TNabc}
The metric for the Taub-NUT solution is
\beqs
ds^2& = &-\frac{\lambda}{R^2}(dt+ 2N ( k+ \cos\theta)
d\phi)^2 + \frac{R^2}{\lambda} dr^2
+R^2 d\Omega^2 , \label{TNNN}
\eeqs
where $\lambda= r^2-N^2-2Mr$, $R^2=r^2+N^2$, $d\Omega^2$ is the metric on the unit two-sphere, and $k$ is a parameter. Here, we have introduced a parameter $k$ such that for $k=1$, respectively $k=-1$, the string is along the positive, respectively negative, $z$-direction. For any other value of $k$, the metric has a singularity all along the $z$-axis. As already stated, it is easy to see that the value of $k$ can be mapped to $k+c$ if one performs the coordinate change $t\rightarrow t+ 2Nc \phi$.

Here, we would like to review the duality that maps the linearized
Schwarzschild to the linearized pure NUT solution. We show that, by gravitational
duality, the string singularity determines a magnetic stress-energy tensor and is thus
non-physical in an ``electric" theory; it is a topological charge. It is also called the magnetic mass $N$. For the sake of completeness, we eventually consider what would happen if we do not consider this singular contribution. One can easily realize that the magnetic stress-energy tensor is zero, as it is clear from its definition. The pure NUT solution is then described as a semi-infinite massless source of angular momentum $N$. This is Bonnor's interpretation \cite{bonnor69} of the pure NUT solution where the string is considered as a physical singularity in the ``electric" theory.

\subsubsection{Linearized Schwarzschild solution}

The non-trivial fluctuations of the linearized metric and spin connection for the Schwarzschild metric are
\beqs
h_{tt}&=&\frac{2M}{r}, \qquad h_{ij}=\frac{2M}{r^3}x_{i}x_{j},
\nonumber \\
\omega_{0i0}&=&\frac{1}{2} \partial_i h_{00}= -M \frac{x_i}{r^3}, \nonumber \\
\omega_{ijk}&=&\frac{1}{2} (\partial_j h_{ik}- \partial_i
h_{jk})=\frac{M}{r^3}(\delta_{jk}x_i-\delta_{ik}x_j).
\eeqs
The non-trivial components of the linearized Riemann tensor are
\begin{eqnarray}
R_{0i0j}&=& - \partial_j \omega_{0i0}=  M( - \frac{3 x_i x_j}{r^5}
+ \frac{\delta_{ij}} {r^3}+ \frac{4\pi }{3} \delta_{ij}
\delta(\textbf{x})),
\nonumber \\
R_{ijkl}&=&  \partial_k \omega_{ijl} - \partial_l \omega_{ijk}
\nonumber \\ &=&  (\frac{2M}{r^3}+\frac{8\pi M}{3}
\delta(\textbf{x})) (\delta_{ik} \delta_{jl}- \delta_{il}
\delta_{jk} )\nonumber \\ && - \frac{3M}{r^5} (\delta_{ik} \:
x_{j} \:  x_{l}- \delta_{jk} \: x_{i} \:  x_{l} -\delta_{il} \:
x_{j} \: x_{k} + \delta_{jl} \: x_{i} \:  x_{k}),
\end{eqnarray}
where we used
\begin{eqnarray}
\partial_j \frac{x_k}{r^3}&=& \frac{\delta_{jk}}{r^3}-\frac{3x_k
x_j}{r^5}+\frac{4\pi}{3} \delta_{jk} \delta(\textbf{x}).
\end{eqnarray}
We eventually see that the Ricci tensor and Ricci scalar are
\beqs
R_{00}= 4\pi M \delta(\textbf{x}), \qquad  R_{ij} =
4\pi M \delta_{ij} \delta(\textbf{x}), \qquad R =  8\pi M
\delta(\textbf{x}).
\eeqs
This also means that
\beqs
G_{00}= 8 \pi T_{00}= 8 \pi M
\delta(\textbf{x}), \qquad G_{ij}= T_{ij}=0, \qquad G_{0j}= T_{0j}=0.
\eeqs
The source for the linearized Schwarzschild solution is a point of mass $M$.

\subsubsection{The NUT solution from the dual Schwarzschild}

To obtain the ``electric" spin connection for the NUT metric, we use the duality
relation
\beqs
\omega_{\mu\nu\sigma}=-\frac{1}{2}
\varepsilon_{\mu\nu\alpha\beta}\:
{\tilde{\omega}^{\alpha\beta}}_{\:\:\:\:\:\sigma} ,
\eeqs
 where $\tilde{\omega}$ is the spin connection for the linearized
Schwarzschild after we applied the duality rotation
$\omega\rightarrow \tilde{\omega}$ and $M\rightarrow N$.  In this way we obtain a regular spin connection.
It is given by
\begin{eqnarray}
\omega_{ij0}= \varepsilon_{ijk} \tilde{\omega}_{0k0}= -N
\varepsilon_{ijk} \frac{x_k}{r^3}, \qquad
\omega_{0ij}=-\frac{1}{2} \varepsilon_{ikl} \tilde{\omega}_{klj}=
N \varepsilon_{ijk} \frac{x_k}{r^3}.
\end{eqnarray}
One can now compute the non-trivial components of the Riemann tensor
\begin{eqnarray}
R_{0i0j}&= & 0, \qquad R_{ijkl}=0, \nonumber \\
R_{ij0k} &=& N \varepsilon_{ijl} \partial_k (\frac{x^l}{r^3}) = N
\varepsilon_{ijl} (\frac{\delta_{kl}}{r^3}-\frac{3x_k x_l}{r^5}
+\frac{4\pi}{3} \delta_{kl} \delta(\textbf{x})), \nonumber \\
R_{0ijk} &=& \partial_j \omega_{0ik}- \partial_k \omega_{0ij}
\nonumber \\
         &=& -2N \varepsilon_{ijk} (\frac{1}{r^3} +
         \frac{4\pi}{3} \delta(\textbf{x})) +
         3N (\varepsilon_{ijl} \frac{x_k x_l}{r^5}-
         \varepsilon_{ikl} \frac{x_j x_l}{r^5}).
\end{eqnarray}
Einstein's equations are trivially satisfied as $R_{00}=R_{ij}=0$ and
$R_{0i}=R_{i0}=0$, meaning that $T_{\mu\nu}=0$. However, plugging the
above expressions in the cyclic identity, we obtain
\begin{eqnarray}\label{cyclicNUT}
R_{0ijk}+R_{0kij}+R_{0jki}&=&-8 \pi \varepsilon_{ijk} \Theta^{00} \nonumber \\
  &=& -2N \varepsilon_{ijk} (\frac{3}{r^3} +
         4\pi \delta(\textbf{x}))    + 6N (\varepsilon_{ijl} \frac{x_k x_l}{r^5}-
         \varepsilon_{ikl} \frac{x_j x_l}{r^5}-
         \varepsilon_{kjl} \frac{x_i x_l}{r^5}), \nonumber \\
R_{00ij}+R_{0j0i}+R_{0ij0}&=& -8\pi \varepsilon_{ijk}
{\Theta^{k}}_{0},
\nonumber \\
 R_{i0jk}+R_{ik0j}+R_{ijk0}&=&-\partial_j
 (\omega_{0ik}+\omega_{ik0})+\partial_k
 (\omega_{0ij}+\omega_{ij0})
= -8\pi \varepsilon_{jkl}{\Theta^{l}}_{i}. \nonumber \\
\end{eqnarray}
This implies that the dual solution is characterized by
\begin{eqnarray}
\Theta^{00}=N \delta(\textbf{x}), \qquad \Theta^{0k}=0, \qquad
\Theta^{li}=0.
\end{eqnarray}
Let us now remark that for a solution to describe such a magnetic particle of mass $N$, and
thus a magnetic stress-energy tensor $\Theta^{00}=
N\delta(\textbf{x})$, we need, as one can see from \eqref{PHI},
\begin{eqnarray}
{\Phi^{0z}}_{0}=-16\pi N \delta(x) \delta(y) \vartheta(z).
\end{eqnarray}
From \eqref{omegamunurho}, we see that the previous non-trivial components of the spin connection can be readily expressed as
\begin{eqnarray}
\omega_{ij0}=\frac{1}{2}(\partial_j h_{0i}-\partial_i h_{0j})+\frac{1}{4} \varepsilon_{ij0k}\: {\Phi^{0k}}_{0}, \nonumber\\
\omega_{0ij}=\frac{1}{2}(\partial_i h_{0j}+\partial_j
v_{i0})-\frac{1}{4} \varepsilon_{0ijk} \:  {\Phi^{0k}}_{0},
\end{eqnarray}
where we only assumed that the linearized vielbein is independent
on time. As we have established that the regular spin connection
is such that $\omega_{ij0}=-\omega_{0ij}$, we immediately see that
the right gauge fixing will be $h_{0i}=-v_{i0}$. The previous spin
connections are recovered with
\begin{eqnarray}
h_{0x}=v_{0x}= 2N \frac{y}{r(r-z)}, \qquad v_{0y}=h_{0y}= - 2N
\frac{x}{r(r-z)},
\end{eqnarray}
where the metric has a singularity on the positive $z$-axis, in
agreement with the form of the $\Phi_{z00}$ term. To check that
this is the right result, one can go through the same standard regularization
procedure as we used for electromagnetism ( see
e.g. \cite{Felsager:1981iy}). We set
\begin{eqnarray}
\vec{A}&=& (h_{0x}, h_{0y}, h_{0z}),\nonumber \\
\vec{B}&=& \vec{\nabla} \times \vec{A}= 2N
\frac{\vec{r}}{r^3}-8\pi N\delta(x)\delta(y)\vartheta(z) \hat{z},
\end{eqnarray}
where $\hat{z}$ is the unit vector along the $z$-axis and we obtain
\begin{eqnarray}
\partial_j h_{0i}-\partial_i h_{0j}= -
2 N \varepsilon_{ijk} \frac{x^{k}}{r^3}+ \varepsilon_{zij} 8\pi
N\delta(x)\delta(y)\vartheta(z).
\end{eqnarray}
Note that the non-trivial contribution to the
linearized metric in spherical coordinates is
\begin{eqnarray}
h_{0\phi}= - 2N(1+\cos\theta),
\end{eqnarray}
which is also the only non-trivial component for the linearized pure
NUT metric as one can directly obtain from \eqref{TNNN} with $k=1$.

Here, we interpret the singularity at $\theta=0$ as non-physical in an ``electric" way. However, it contributes to the magnetic stress-energy tensor. The solution describes thus a particle of magnetic mass $N$.

\subsubsection{The NUT solution without the string}

To recover Bonnor's interpretation, we set to zero the
${\Phi^{\mu\nu}}_{\rho}$. Then, we obviously have
$\Theta_{\mu\nu}=0$. With the previous choice of $v_{\mu\nu}$, the
non-trivial components of the spin connections are now
\begin{eqnarray}
\omega_{ij0}= -N \varepsilon_{ijk} \frac{x_k}{r^3}+
\varepsilon_{zij} 4\pi N\delta(x)\delta(y)\vartheta(z),
 \nonumber \\
\omega_{0ij}= N \varepsilon_{ijk}
\frac{x_k}{r^3}-\varepsilon_{zij} 4\pi
N\delta(x)\delta(y)\vartheta(z).
\end{eqnarray}
Note that we still have $\omega_{ij0}=-\omega_{0ij}$ so that from
(\ref{cyclicNUT}) we still have
$\Theta^{il}=\Theta^{0i}=0$. Now, we can also check that $\Theta^{00}=0$ as
it should be.

The non-trivial components for the Einstein tensor are
\begin{eqnarray}
G_{i0}=-\partial_j (\varepsilon_{zij} 4\pi N
\delta(x)\delta(y)\vartheta(z)),
\end{eqnarray}
giving non-trivial contributions to $T_{\mu\nu}$
\begin{eqnarray}
T_{x0} =  -\frac{N}{2} \delta(x)\delta'(y)\vartheta(z), \qquad
T_{y0} = \frac{N}{2}  \delta'(x)\delta(y)\vartheta(z).
\end{eqnarray}
Note that such  $T_{\mu\nu}$ is conserved.

Given this, we see that $P_\mu=0$  and $\Delta L^{xy}/\Delta z=N$ all along
the singularity. This agrees with Bonnor's interpretation of
the NUT solution as a massless source of angular momentum at the
singularity $\theta=0$.

\subsection{The Kerr and the rotating NUT}
\label{subsec: Kerrabc}
There exists in the literature a generalization of the Taub-NUT
metric with three parameters, the ADM mass $M$, the NUT charge
$N$, and a rotation parameter $a$. This solution is known as the
Kerr-NUT metric. It is a particular case of the general
Petrov type D solution found in \cite{Plebanski:1976gy}. It is given by
\begin{eqnarray}\label{MNa}
ds^2&=&-\frac{\lambda^2}{R^2} [dt- (a \sin^2\theta -2 N \cos\theta) d\phi]^2 +
\frac{\sin^2\theta}{R^2} [(r^2+a^2+N^2) d\phi - a dt]^2 \nonumber \\
&& +\frac{R^2}{\lambda^2} dr^2 +R^2 d\theta^2,
\end{eqnarray}
where $\lambda^2 = r^2-2Mr+a^2-N^2$ and  $R^2 = r^2+(N+a
\cos\theta)^2$.

If we set $a=0$ in the above solution, we recover the Taub-NUT solution \eqref{TNNN}  with $k=0$.
If we set $N=0$ in the metric (\ref{MNa}), we recover the Kerr
metric in Boyer-Lindquist coordinates
\begin{eqnarray}\label{Ma}
ds^2=-(1-\frac{2Mr}{\Sigma}) dt^2 -\frac{4M a r }{\Sigma}
\sin^2\theta dt d\phi +\frac{\Sigma}{\Delta} dr^2 +\Sigma d\theta^2
+ \frac{ B}{\Sigma} \sin^2\theta d\phi^2,
\end{eqnarray}
where $ \Delta  \equiv  \lambda^2(N=0) = r^2-2Mr+a^2$, $\Sigma
\equiv  R^2 (N=0) = r^2+ a^2 \cos^2\theta$, and $B = (r^2+a^2)^2-
\Delta a^2 \sin^2\theta$.  One can linearize this metric at first order in the
charges, meaning we only keep terms in $M$ and $Ma$ and obtain
\beqs\label{flucKerr1}
h_{00}=\frac{2M}{r}, \qquad  h_{ij}=\frac{2M}{r^3}x_{i}x_{j}, \qquad h_{0i}= \frac{2Ma}{r^3} \varepsilon_{zij} x^{j}.
\eeqs
We will review rapidly hereafter that the source for this solution
is a rotating mass $M$ with angular momentum $J_z=L^{xy}=Ma$.
A more interesting metric is the one where we set $M$ to zero  in (\ref{MNa}). This is what we refer to as the rotating NUT metric. The linearized contributions of the metric are
\begin{eqnarray}
\tilde{h}_{tx}=\frac{2Nyz}{r (x^2+y^2)}, \:\:\:\:\:\:
\tilde{h}_{ty}=\frac{-2Nxz}{r (x^2+y^2)}, \:\:\:\:\:\:
\tilde{h}_{\mu\mu}=\frac{2Naz}{r^3}.
\end{eqnarray}
We will see that this linearized metric (after we set the string along the positive $z$-axis) supplemented with the $\Phi_{\mu\nu\rho}$ contributions
\begin{eqnarray}
{\Phi^{0z}}_{0}&=&-16\pi N \delta(x) \delta(y) \vartheta(z),
\\
{\Phi^{0y}}_{x}&=& -{\Phi^{0x}}_{y}=-
{\Phi^{xy}}_{0}={\Phi^{yx}}_{0}= 8\pi N a \delta(\textbf{x}), \label{deltacontrib}
\end{eqnarray}
where $\vartheta$ is the usual Heaviside function, describes the
dual solution to the linearized Kerr. The rotating NUT solution is understood as
a point of magnetic mass $N$ and a magnetic angular momentum
$\tilde{L}^{xy}=Na$.\\

As we have seen for the Taub-NUT case, taking or not taking into account singularities contributions from $\Phi_{\mu\nu\rho}$ lead to different interpretations of the "dual" metric and thus to different interpretations of its sources. As an example, we have just discussed Bonnor's interpretation of the Taub-NUT metric. Here, we will discuss what happens if one does not include the singular delta contributions \eqref{deltacontrib} for the rotating NUT solution. In the last part, we see that by duality the Kerr metric could be given an exotic interpretation if singular contributions of this type were added.

In the following, we only present the additional information not contained in the previous Taub-NUT example as the
non-trivial contributions of the linearized Kerr-NUT metric split into
contributions that were already present in the Taub-NUT case and
additional contributions in $Ma$ or $Na$.

\subsubsection{Kerr metric}
The additional (with respect to the Schwarzschild metric) non-trivial components of the linearized metric and
linearized spin connection are
\begin{eqnarray}\label{flucKerr2}
h_{0i}&=& \frac{2Ma}{r^3} \varepsilon_{zij} x^{j}, \nonumber \\
\omega_{0ij}&=& \frac{1}{2} \partial_i h_{0j}= -Ma
\varepsilon_{zij}(\frac{1}{r^3}+\frac{4\pi}{3}
\delta(\textbf{x}))- \frac{3Ma}{r^5} \varepsilon_{zjl}x_i x^l,
\nonumber \\
\omega_{ij0}&=& \frac{1}{2} (\partial_j h_{i0}-\partial_{i}
h_{j0})= \omega_{0ji}- \omega_{0ij} \nonumber \\
&=& Ma \varepsilon_{zij}(\frac{2}{r^3}+\frac{8\pi}{3}
\delta(\textbf{x}))- \frac{3Ma x^l}{r^5}
(\varepsilon_{zil}x_j-\varepsilon_{zjl}x_i).
\end{eqnarray}
The additional non-trivial components of the linearized Riemann
tensor are
\begin{eqnarray}
R_{0ijk}&=&-Ma  \varepsilon_{zkl}(\partial_j \partial_i \partial_l
\frac{1}{r}) + Ma  \varepsilon_{zjl}(\partial_k \partial_i
\partial_l \frac{1}{r}), \nonumber \\
R_{ij0k}&=& -Ma  \varepsilon_{zjl}(\partial_k \partial_i
\partial_l \frac{1}{r}) + Ma  \varepsilon_{zil}(\partial_k
\partial_j
\partial_l \frac{1}{r}), \nonumber \\
\end{eqnarray}
where one can show that
\begin{eqnarray}
\partial_i \partial_j \partial_k \frac{1}{r}& =& -15 \frac{x_i x_j
x_k}{r^7}+ \frac{3}{r^5} (\delta_{ij} x_k +\delta_{ki} x_j +
\delta_{jk} x_i)\nonumber \\
&&- \frac{4 \pi}{5} (\delta_{ij} \partial_k \delta(\textbf{r})+
\delta_{ki} \partial_j \delta(\textbf{r}) + \delta_{jk} \partial_i
\delta(\textbf{r})).
\end{eqnarray}
Combining these results with the ones obtained for Schwarzschild, we easily find
\beqs
&& R_{j0}=R_{0j}= {R_{0ij}}^{i}=Ma
\varepsilon_{zjl}(\partial_l \Delta \frac{1}{r})=-4\pi M a
\varepsilon_{zjl}\partial_l \delta(\textbf{x}) , \nonumber \\
&& R_{00}= 4\pi M \delta(\textbf{x}), \qquad R_{ij} = 4\pi M
\delta_{ij} \delta(\textbf{x}) , \qquad  R =  4\pi M \delta(\textbf{x}).
\eeqs
Eventually, we get
\beqs
 && G_{00}= 8 \pi T_{00}= 8 \pi M
\delta(\textbf{x}), \qquad G_{ij}= T_{ij}=0, \\
&& G_{0j}= R_{0j}=8 \pi
T_{0j}=-4\pi M a \varepsilon_{zjl}\partial_l \delta(\textbf{x}).
\eeqs
This solution describes a point of electric mass $M$ with an
electric angular momentum $L^{xy}=Ma$.

\subsubsection{The rotating NUT solution from the dual Kerr}
As for the dual of linearized Schwarzschild, by duality rotation
we obtain the additional components of the spin connection of the dual Kerr metric
\begin{eqnarray}
\omega_{0i0}&=& -\frac{1}{2} \varepsilon_{ijk}
\tilde{\omega}_{jk0}= Na \frac{\delta_{zi}}{r^3} -\frac{8}{3}
\pi N a \delta_{zi} \delta(\textbf{x}) - 3 N a \frac{z x_i}{r^5}, \nonumber \\
\omega_{ijk} &=& \varepsilon_{ijl} \tilde{\omega}_{0lk} = Na (
\delta_{zi} \delta_{kj} -\delta_{zj}\delta_{ki})(-\frac{2}{r^3}+
\frac{4\pi}{3} \delta(\textbf{x})) \nonumber \\
&&\:\:\:\:\:\:\:\:\:\:\:\:\:\: +\frac{3Na }{r^5}(x_k(x_j
\delta_{zi}-x_i \delta_{zj})+ z (x_i \delta_{kj}-x_j
\delta_{ki})),
\end{eqnarray}
where we used $\varepsilon_{ijk} \varepsilon^{zjk} = 2
\delta_{i}^{z}$  and $\varepsilon_{ijk} \varepsilon^{zjl} =
\delta_{z}^{i} \delta_{l}^{k} - \delta_{z}^{k}\delta_{l}^{i}$. Following the same reasoning as before, one
can easily derive the Einstein tensor and find that this solution
corresponds to a magnetic point of mass $N$ with a magnetic
angular momentum $\tilde{L}^{xy}=Na$. This is the gravitational
dual of the Kerr solution where $\Theta_{\mu\nu}$ has a
structure equal to the stress-energy tensor for Kerr, meaning
\begin{eqnarray}
\Theta^{00}=N\delta(\textbf{x}), \qquad \Theta^{0x}= \frac{Na}{2}
\partial_{y} \delta(\textbf{x}), \qquad \Theta^{0y}= -
\frac{Na}{2} \partial_{x} \delta(\textbf{x}).
\end{eqnarray}
From this last result, we see that the non-trivial components for $\Phi_{\mu\nu\rho}$ are
\begin{eqnarray}
{\Phi^{0z}}_{0}&=&-16\pi N \delta(x) \delta(y) \vartheta(z)\; ,
\nonumber
\\
{\Phi^{0y}}_{x}&=& -{\Phi^{0x}}_{y}=-
{\Phi^{xy}}_{0}={\Phi^{yx}}_{0}= 8\pi N a \delta(\textbf{x}).
\end{eqnarray}
We also have
\begin{eqnarray}
\omega_{0i0}&=&\frac{1}{2} \partial_i h_{00} + \frac{1}{4}
\varepsilon_{0ijk} {\Phi^{jk}}_{0}= \frac{1}{2} \partial_i h_{00}+
\frac{1}{2} \delta_{iz} {\Phi^{xy}}_{0}, \nonumber \\
\omega_{ijk}&=& \frac{1}{2} (\partial_j h_{ik}- \partial_i h_{jk}+
\partial_k v_{ji}) + \frac{1}{2} \varepsilon_{ij0l}
\bar{\Phi}^{0l}_{\:\:\:\: k},
\end{eqnarray}
where for our choice of $\Phi_{\mu\nu\rho}$, we find
\begin{eqnarray}
\frac{1}{2} \varepsilon_{ij0l} \bar{\Phi}^{0l}_{\:\:\:\: k}=
\frac{1}{2} \varepsilon_{ij0l}
{\Phi^{0l}}_{k}=(\delta_{iz}\delta_{jk}-\delta_{jz}\delta_{ik}){\Phi^{0y}}_x.
\end{eqnarray}
We then easily obtain\footnote{Note that the $v_{\mu\nu}$ obtained
  here, and which lead to a regular spin connection, do not satisfy
  the gauge fixing conditions \eqref{gauge1} and \eqref{gauge2} proposed in Section \ref{sec: charges}, where the aim was rather to
  define surface integrals.}
\begin{eqnarray}
h_{00}= \frac{2Naz}{r^3}, \qquad h_{ij}= \frac{2Naz}{r^3}
\delta_{ij}, \qquad v_{ij}= \frac{2Na}{r^3}(\delta_{zi}
x_j-\delta_{zj}x_i).
\end{eqnarray}
The non-trivial components of the linearized metric in spherical
coordinates are then
\begin{eqnarray}
h_{\mu\mu}&=&\frac{2Naz}{r^3}, \qquad h_{0\phi}= 2N(1+\cos\theta).
\end{eqnarray}
These are the non-trivial components of the linearized rotating
NUT metric, where the string is along the positive $z$-axis.

\subsubsection{The rotating NUT without the delta contributions}
If we set ${\Phi^{0y}}_{x} = -{\Phi^{0x}}_{y}=-
{\Phi^{xy}}_{0}={\Phi^{yx}}_{0}= 0$, the difference with the
previous case appears for
\begin{eqnarray}
\omega_{0i0}&=& -Na\partial_i \partial_z (\frac{1}{r}), \nonumber \\
\omega_{ijk}&=& -Na [\delta_{ik} \partial_j \partial_z
(\frac{1}{r})-\delta_{jk} \partial_i \partial_z
(\frac{1}{r})+\frac{1}{2} \delta_{zi} \partial_k \partial_j
(\frac{1}{r})- \frac{1}{2} \delta_{zj} \partial_k \partial_i
(\frac{1}{r})].
\end{eqnarray}
This means that
\begin{eqnarray}
R_{00}= -4 \pi N a \delta(x) \delta(y) \delta'(z), \qquad R_{ij}=
-4 \pi N a\delta_{ij} \delta(x) \delta(y) \delta'(z),
\end{eqnarray}
and the electric Einstein tensor has now a non-trivial component
$$G_{00}=-8\pi N a
\delta(x)\delta(y)\delta'(z).$$ The solution has associated charges
$K_0=N$ and $L^{0z}=-Na$, describing
a point magnetic mass $N$ with in addition a ``boost mass" $-Na$
which can be understood as a dipole of electric masses $M$ and
$-M$ separated by a distance $\epsilon$ in the limit where
$\epsilon\rightarrow 0$ and $L_{0z}=Na=M\epsilon$ is kept
constant. Positivity of energy in general relativity tells us that
this interpretation should however be discarded.

The interested reader could eventually wonder about different
combinations of the previous considerations. One could for example
try to interpret the rotating NUT solution with only the delta
contributions and no string contribution (or respectively no
$\Phi_{\mu\nu\rho}$ contributions at all). Following our analysis
this only partially matches the proposal of J.G. Miller in
\cite{MILLER} to interpret the Kerr-NUT metric as a Kerr
black hole and an infinite source of angular momentum along the
singularity. Indeed, our calculations show that it should also be
supplemented with a magnetic angular momentum when delta
contributions are included (respectively with a dipole of electric
masses in the same limit as previously discussed).

\subsubsection{Comments about duality of the Kerr metric}

We have seen that the string singularity input from $\Phi_{\mu\nu\rho}$ in the case of the
Taub-NUT solution found its meaning in the existence of an unphysical string singularity in the
linearized metric. This is also justified by considering the Schwarzschild metric as
electric and imposing gravitational duality. From this perspective, Bonnor's proposal seems rather unphysical.

However, in the case of the Kerr-NUT solution, we have seen that some $\Phi_{\mu\nu\rho}$ terms are only singular in $r=0$. Besides duality, we do not have any a priori argument in favor of
adding these delta contributions to the rotating NUT solution. As we just described, one
could think of the linearized rotating NUT with only the string
contribution ${\Phi^{0z}}_{0}$ as another physical solution.
This interpretation is however to be rejected on physical grounds because of
the presence of a negative mass in the compound.

Let us amuse ourselves by contemplating the dual situation, i.e. the usual Kerr solution, where we insert a non-trivial
magnetic stress-energy  tensor so that the non-trivial charges become
$P_0=M$ and $\tilde{L}_{0z}=Ma$. The
sources for this solution are
\begin{equation}
T_{00}=M\delta(\textbf{x}), \qquad \Theta_{00}=Ma \delta(x)\delta(y)\delta'(z),
\end{equation}
 an electric point of mass $M$ and a di-NUT,
a dipole of NUT charges $+N$ and $-N$, separated by a distance $\epsilon$
when we take the limit $\epsilon \to 0$ and  $N \to \infty$ but with the
product $N\epsilon$  constant and equal to $\tilde L_{0z}=N\epsilon=Ma$ such that
\begin{eqnarray}
\Theta_{00}
&=& lim_{\epsilon \rightarrow 0}  [ N \delta(x)\delta(y)\delta(z+\epsilon/2)- N\delta(x)\delta(y)\delta(z-\epsilon/2)] \nonumber \\
&=& Ma  \delta(x)\delta(y)\delta'(z).
\end{eqnarray}
This situation is physical since there is no obstruction in having
negative NUT charges. Indeed, the Taub-NUT metrics with opposite signs
of $N$ are just related by a flip of the sign of the $\phi$ variable.
We should however note that this leads seemingly to a clash between the
statement of gravitational duality and positivity of the mass for the
Schwarzschild solution. In other words, according to
the above arguments the gravitational dual of a
physical situation is not necessarily physical. It would be nice to
understand this issue better, with the use for instance of positive
energy theorems.

Concerning the euclidean Kerr black hole,
this interpretation had already been noticed a long time ago in
\cite{Gibbons:1979xm}. For the Lorentzian signature, it has
recently been observed in \cite{Manko:2009xx} that the Kerr metric
could be reproduced by a non-linear superposition of two Taub-NUT
black holes of opposite NUT charges.\footnote{We would like to
thank A. Virmani and R. Emparan for pointing out this reference to
us. }
Here, we have clarified that if this is indeed true from the
perspective of the metrics, there is nevertheless a
difference depending on whether the $\delta'$ singularities find themselves
in the $T_{0i}$ components of the ordinary stress-energy tensor or in
the $\Theta_{00}$ component of the magnetic, dual, stress-energy tensor. The
difference is encoded in the tensor $\Phi_{\mu\nu\rho}$
and is reflected on which Lorentz charges are non-trivial,
the electric or the magnetic ones. We suggest to identify the Kerr metric
as a di-NUT only in the case where there is a non-trivial $\Theta_{00}$.

\subsection{Shock pp-waves}
\label{subsec: schockabc}
Pp-waves, or plane-fronted waves with parallel rays, were first introduced by Brinkmann in 1925 as metrics on Lorentzian manifolds. They are described by
\beqs
ds^2= H(u,x,y) du^2 - du \: dv +dx^2+dy^2\; ,\label{fullform}
\eeqs
where $H$ is a smooth function. Moreover, if the function $H$ is
harmonic in $x$ and $y$ then it is a solution of Einstein's equations.
Here we will consider shock pp-waves, a particular case where the function $H$
factorizes its $u$ dependence in a delta function such that
$H(u,x,y)=F(x,y) \delta(u)$ and $F(x,y)$ is a harmonic function.

To start with, we review the result of P. C. Aichelburg and R. U. Sexl in \cite{Aichelburg:1970dh} where the infinite boost of the Schwarzschild metric was considered and shown to be of the form of a shock pp-wave. It is now referred to as the Aichelburg-Sexl solution. This shock pp-wave was later re-discovered by T. Dray and G. 't Hooft in  \cite{Dray:1984ha} and understood as the gravitational radiation of a particle traveling at the velocity of light measured by an observer at rest. The method of Aichelburg and Sexl was generalized in \cite{Lousto:1988ua} and used, for example, to compute the
infinite boost of the Reissner-Nordstr\"om black hole. This more general analysis was then used in \cite{Hayashi:1994rf} for the infinite boost of  the Kerr black hole where the Aichelburg-Sexl metric is shown to be recovered in a certain limit.

Inspired by these generalized methods we describe, in a second part, the infinite boost of the pure NUT metric ($M=0$) and obtain the NUT-wave, another shock pp-wave.

In the last part, we show that the gravitational dual of the Aichelburg-Sexl pp-wave is precisely the NUT-wave, the infinitely boosted NUT wave. More generally, we establish that any pp-wave solution described by a function $F(x,y)$ of Einstein's equations possess a dual pp-wave which is characterized by the harmonic conjugate of $F(x,y)$, which we denote as $\tilde{F}(x,y)$.

\subsubsection{The Aichelburg-Sexl shock pp-wave}

In here, we reproduce the results\footnote{Note that several typos are present in the computations of this original paper.} presented in  \cite{Aichelburg:1970dh}. Let us start with the usual metric of Schwarzschild in the so-called Schwarzschild
coordinates
\beqs
ds^2=-(1-\frac{2M}{R}) dt^2 + (1-\frac{2M}{R})^{-1} dR^2 +R^2
(d\theta^2+sin^2\theta d\phi^2),
\eeqs
and make the change of variables
\beqs
R=r(1+\frac{M}{2r})^2,
\eeqs
to obtain the metric in isotropic coordinates
\beqs\label{schwiso}
ds^2=-\frac{(1-A)^2}{(1+A)^2} dt^2 + (1+A)^4 (dx^2+dy^2+dz^2),
\eeqs
where $A\equiv M/2r$ and $dx^2+dy^2+dz^2=dr^2 + r^2 (d\theta^2 + \sin^2 \theta d\phi^2)$. Let us now apply a boost along the $z$-direction
\beqs
t&=& \gamma(\tilde{t}-\beta \tilde{z}),\nonumber \\
z&=& \gamma(\tilde{z}-\beta \tilde{t}),
\eeqs
where $\gamma=1/\sqrt{1-\beta^2}$, and obtain
\beqs
ds^2&=&-\frac{(1-A)^2}{(1+A)^2}\: \gamma^2 (d\tilde{t}-\beta
\:d\tilde{z})^2 + (1+A)^4 (d\tilde{x}^2+d\tilde{y}^2+\gamma^2
(d\tilde{z}-\beta \: d\tilde{t})^2) \nonumber \\
     &=& (1+A)^4 (-d\tilde{t}^2+
d\tilde{x}^2+d\tilde{y}^2+d\tilde{z}^2)\nonumber \\
&& +\biggr [ (1+A)^4
-\frac{(1-A)^2}{(1+A)^2} \biggl ] \gamma^2 (d\tilde{t}-\beta
\:d\tilde{z})^2 , \label{schwisoboosted}
\eeqs
where $A$ is now $A=M / 2\sqrt{x^2+y^2+ \gamma^2(z-\beta \:t)^2}$ and where in the last equality we made use of
\beqs
 \gamma^2
(d\tilde{z}-\beta \:
d\tilde{t})^2=-d\tilde{t}^2+d\tilde{z}^2+\gamma^2
(d\tilde{t}-\beta \: d\tilde{z})^2.
\eeqs
We now want to consider the infinite boost limit where $\beta\rightarrow 1$, $M\rightarrow 0$ but where we keep
$M\gamma=p$, $p$ being a constant. From its definition, we see that $A$ is going to
zero, so that we can linearize the metric in $A$ and rewrite
(\ref{schwisoboosted}) as
\begin{eqnarray}\label{schwboostlin}
ds^2= (1+ 4A)(-d\tilde{t}^2+
d\tilde{x}^2+d\tilde{y}^2+d\tilde{z}^2)+ 8 \: A \: \gamma^2
(d\tilde{t}-\beta \:d\tilde{z})^2\; .
\end{eqnarray}
Immediately performing the infinite boost would mean that we only keep terms in
$\gamma^2 A$ because
\begin{eqnarray}
A\gamma^2=\frac{M\gamma}{2\sqrt{\gamma^{-2}\tilde{\rho}^2+
(\tilde{z}-\beta \: \tilde{t})^2}}\rightarrow \frac{p}{2 |
\tilde{z}- \tilde{t}|},
\end{eqnarray}
where $\tilde{\rho}^2=\tilde{x}^2+\tilde{y}^2$. The metric would be
\begin{eqnarray}\label{xdifft}
ds^2= -d\tilde{t}^2+ d\tilde{x}^2+d\tilde{y}^2+d\tilde{z}^2+ 4 \:
p \:\frac{1}{ | \tilde{z}- \tilde{t}|} (d\tilde{t}-
\:d\tilde{z})^2.
\end{eqnarray}
However, this metric is only valid for $t\neq z$. If we want to
carry the limit $ \beta\rightarrow 1$ for points where $t =z$, we
need to make the ``awkward" change of coordinates (singular in
$\tilde{t}=\tilde{z}$ when $\beta=1$ )
\begin{eqnarray}\label{T}
T(v): && z'-\beta t' = \tilde{z} - \beta \tilde{t} , \nonumber \\
      && z'+\beta t' = \tilde{z} + \beta \tilde{t}-4 p \:\: \ln \biggr [ \sqrt{(\tilde{z}-\beta \tilde{t})^2 +\gamma^{-2}}-(\tilde{z}-\tilde{t}) \biggl ].
\end{eqnarray}
To apply this change of coordinates, we first start by re-writing
(\ref{schwboostlin}) as
\begin{eqnarray}
ds^2&=& (1+4A) \biggr [
d\tilde{x}^2+d\tilde{y}^2+(d\tilde{z}-\beta
d\tilde{t})(d\tilde{z}+\beta d\tilde{t})-(1-\beta^2) d\tilde{t}^2 \biggl ] \nonumber \\
 &&+ 8 \: A \: \gamma^2 \biggr [ (d\tilde{z}-\beta
\:d\tilde{t})^2 -(1-\beta^2) d\tilde{z}^2 + (1-\beta^2)
d\tilde{t}^2 \biggl ].
\end{eqnarray}
Moreover, we see that only terms in  $A\gamma^2$ will contribute. Indeed, the contributions in $(1-\beta^2)$ will drop in the
infinite boost limit. Prior to any change of coordinates, we write the metric as
\begin{eqnarray}\label{metricc}
ds^2&=&  d\tilde{x}^2+d\tilde{y}^2+(d\tilde{z}-\beta
d\tilde{t})(d\tilde{z}+\beta d\tilde{t})+ 8 \: A \: \gamma^2
(d\tilde{z}-\beta \:d\tilde{t})^2 .
\end{eqnarray}

Now, as one can see from \eqref{T}, the change of coordinates is trivial for $d\tilde{z}-\beta
d\tilde{t}= dz'-\beta dt'$ and also
\begin{eqnarray}
d\tilde{z} + \beta d\tilde{t} = dz'+\beta dt' + 4 p \:\: \frac{ \biggr [
\frac{(\tilde{z}-\beta \tilde{t})(d\tilde{z}-\beta d\tilde{t})}{
\sqrt{(\tilde{z}-\beta \tilde{t})^2 +\gamma^{-2}}}-
(d\tilde{z}-d\tilde{t})\biggl ]}{
\sqrt{(\tilde{z}-\beta \tilde{t})^2
+\gamma^{-2}}-(\tilde{z}-\tilde{t}) } ,
\end{eqnarray}
which we could rewrite as
\begin{eqnarray}
d\tilde{z} + \beta d\tilde{t} &=&  dz'+\beta dt' + 4 p \:\:
\frac{1}{ \sqrt{(z'-\beta t')^2 +\gamma^{-2}}-(\tilde{z}- \beta
\tilde{t})+(1-\beta )\tilde{t} } \nonumber \\
&&\biggr [ \frac{(z'-\beta t')(dz'-\beta dt')}{ \sqrt{(z'-\beta
t')^2 + \gamma^{-2}}}- (d\tilde{z}- \beta d\tilde{t})
+(1-\beta)d\tilde{t}\biggl ] .
\end{eqnarray}
If we now again drop the contributions in $(1-\beta)$, we get
\begin{eqnarray}
d\tilde{z} + \beta d\tilde{t} &=&  dz'+\beta dt' - 4 p
\frac{1}{\sqrt{(z'-\beta t')^2 + \gamma^{-2}}} (dz'-\beta dt') .
\end{eqnarray}
By plugging this in (\ref{metricc}) and taking the limit
$\beta\rightarrow 1$, we get
\begin{eqnarray}\label{nearlyfinalone}
ds^2&=& -dt'^2+dx'^2+dy'^2+dz'^2+ \nonumber \\
 &&\  4 p \:\: \lim_{\beta\rightarrow 1} \biggr [  \frac{1}{\sqrt{(z'-\beta t')^2 + \gamma^{-2}
\rho'^2}}  - \frac{1}{\sqrt{(z'-\beta t')^2 + \gamma^{-2}}} \biggl
] (dt'-dz')^2 . \nonumber \\
\end{eqnarray}
To take this limit, we use the fact that
\begin{eqnarray}\label{distrib}
\lim_{\epsilon\rightarrow 0} \frac{1}{\epsilon}
f(z/\epsilon)=\delta(z) ,
\end{eqnarray}
for a function $f$ such that
\begin{eqnarray}
\int^{+\infty}_{-\infty} f(z) dz=1 .
\end{eqnarray}
In our case, if we write $Z=z'-\beta t'$ and $\gamma^{-1}=\epsilon$, we see
that
\begin{eqnarray}
g(Z/\epsilon)&=& g(\gamma (z'-\beta t'))= \frac{1}{\sqrt{(z'-\beta
t')^2 + \gamma^{-2} \rho'^2}} - \frac{1}{\sqrt{(z'-\beta t')^2 +
\gamma^{-2}}}\nonumber \\
&=& \frac{1}{\epsilon} f(Z/\epsilon) ,
\end{eqnarray}
where
\begin{eqnarray}
f(Z/ \epsilon)= \frac{1}{\sqrt{(Z/ \epsilon)^2 + \rho'^2}} -
\frac{1}{\sqrt{(Z/\epsilon)^2 + 1}} .
\end{eqnarray}
However, one can see that
\begin{eqnarray}
\int^{+\infty}_{-\infty} f(Z/ \epsilon)= -\ln(\rho'^2),
\end{eqnarray}
such that using (\ref{distrib}) the limit for the function $g(Z/\epsilon)$ is
\begin{eqnarray}
\lim_{\epsilon\rightarrow 0} g(Z/\epsilon)=
\lim_{\epsilon\rightarrow 0} \frac{1}{\epsilon}
f(Z/\epsilon)=- \ln(\rho'^2)\delta(Z) .
\end{eqnarray}
Using this result in (\ref{nearlyfinalone}), we recover the result
of Aichelburg and Sexl
\begin{eqnarray}
ds^2&=& -dt'^2+dx'^2+dy'^2+dz'^2 \nonumber \\
&& - 4 p \: \ln(x'^2+y'^2)\:
\delta(t'-z') (dt'-dz')^2\; .
\end{eqnarray}
We have thus recovered the fact that the infinitely boosted Schwarzschild metric is a shock pp-wave with function $H(u,x,y)=-4 \ln(x^2+y^2) \delta(u)$, where $F(x,y)= \ln (x^2+y^2)$ is an harmonic function as it verifies $\partial^2_x F +\partial^2_y F=0$. Note that, for $t\neq z$, this result only coincides with \eqref{xdifft} after we implement the inverse of the change of coordinates \eqref{T} such as explained in \cite{Aichelburg:1970dh}.

Another procedure to perform such limits was proposed in \cite{Lousto:1988ua} and applied to the Schwarzschild black hole in  \cite{Hayashi:1994rf}. The idea is to look at the Schwarzschild solution as a perturbation of flat space, perturbation which we linearize before taking the limit in the sense of the distribution as explained here above. We refer the reader to the original paper for this derivation. However, we will illustrate this method now by applying it to the pure NUT metric.

\subsubsection{The infinitely boosted NUT metric}

We want to perform the infinite boost of the pure NUT solution. Instead of finding an equivalent awkward change of coordinates, let us write the metric as
\begin{eqnarray}
ds^2= -d\bar{t}^2 +d\bar{x}^2+d\bar{y}^2+ d\bar{z}^2+ds_{def}^2,
\end{eqnarray}
and only consider the linearized part of the deformation which writes
\beqs\label{lindef}
ds_{def}^2=-4N \cos\bar{\theta} d\bar{t}\: d\bar{\phi}.
\eeqs
Here, for convenience, we will take the Misner string along the $x$ direction (namely interchanging $\bar{x}$ and $\bar{z}$ in \eqref{lindef}) and boost along the
$\bar{z}$ direction.
At leading order in $\gamma$, we find
\begin{eqnarray}
\bar{t}& \rightarrow & \gamma \: u  ,\nonumber \\
\bar{z}& \rightarrow & -\gamma \: u  , \nonumber \\
\bar{r}^2 & \rightarrow & \gamma^2 \: u^2+ \: (x^2+y^2)  , \nonumber \\
\cos\bar{\theta}=\frac{\bar{x}}{\bar{r}}& \rightarrow & x/ \sqrt{\gamma^2 \: u^2+(x^2+y^2)} ,\nonumber\\
d\bar{\phi} & \rightarrow & \frac{1}{\gamma} \: \frac{y du }{u^2+
\gamma^{-2} y^2} ,
\end{eqnarray}
where $\tan\bar{\phi}= \bar{y}/ \bar{z}$. Also, we defined $u=t-\beta z$ and by leading order we mean that
$\bar{z}\rightarrow \gamma (z-\beta t)= -\gamma u +\gamma
(1-\beta) (z+t)\sim -\gamma u$. Note that we can also drop in $d\bar{\phi}$  the
second term in $u \: dy$ as we will see that a contribution appears only at $u=0$,
when the infinite boost limit is considered. The deformed part of the metric becomes
\begin{eqnarray}\label{a}
ds_{def}^2&=&-4N  \frac{x}{ \sqrt{\gamma^2 \: u^2+(x^2+y^2)}}\:
\gamma du \: \: \frac{1}{\gamma} \: \frac{y du
}{u^2+\gamma^{-2}y^2} .
\end{eqnarray}
In the limit of infinite boost, we take $\gamma\rightarrow \infty
$ and $N \rightarrow 0$ while keeping $N\gamma=k$. This means we
have
\begin{eqnarray}\label{b}
ds^2_{def}= -8k \:  \lim_{\epsilon\rightarrow 0} \:  \frac{1}{\epsilon}
\:\: \frac{A \, du^2}{2\sqrt{(u/\epsilon)^2+(1+A^2)}((u/\epsilon)^2+A^2)} ,
\end{eqnarray}
where we wrote $\epsilon= \gamma^{-1}x$ and $A=y/x$. Following the same limiting procedure described for the infinitely boosted Schwarzschild metric, we find
\begin{eqnarray}
ds_{def}^2= -8k \: \arctan(1/A) \: \delta(u)\: du^2= -8k \: \arctan
(x/y) \:  \delta(u) \: du^2,
\end{eqnarray}
and the metric of the infinitely boosted pure NUT metric is just
\begin{eqnarray}\label{NUTtywave}
ds^2=-dt^2+dx^2+dy^2+dz^2 -8k \: \arctan(x/y) \: \delta(t-z) \:
(dt-dz)^2.
\end{eqnarray}
The above metric is obviously also a shock pp-wave as it has a function $H(u,x,y)$ of the required form and $\arctan(x/y)$ is an harmonic function. In the following, we refer to it as the NUT-wave.

\subsubsection{Charges of shock pp-waves}

Using previous results, one can straightforwardly consider a finite boost of the Taub-NUT metric and obtain the charges
\beqs
P_0=\gamma M , \qquad P_i=-\gamma \beta M , \qquad K_0=\gamma N , \qquad K_i=-\gamma \beta N .
\eeqs
To show this for the pure NUT metric, we can directly work with the
linearized pure NUT solution
\begin{eqnarray}
ds^2_{Lin}= - d\bar{t}^2 - 4 N \cos\bar{\theta}
d\bar{\phi} d\bar{t}+d\bar{r}^2+\bar{r}^2(d\bar{\theta}^2+\sin^2\bar{ \theta} d{\bar{\phi}}^2) ,
\end{eqnarray}
which can be written in cartesian coordinates as
\begin{eqnarray}
\label{taublin}
ds^2_{Lin}= - d\bar{t}^2 - 4N \frac{\bar{z}}{\bar{r}}\: \frac{1}{\rho^2}
(\bar{x}d\bar{y}-\bar{y}d\bar{x})d\bar{t}+d\bar{x}^2+d\bar{y}^2+d\bar{z}^2 ,
\end{eqnarray}
where $\rho^2=\bar{x}^2+\bar{y}^2$. If we now perform a boost in the $z$-direction
\begin{eqnarray}
\label{boostz}
\bar{t}&=& \gamma(t-\beta z) , \: \:\:\:\:\:
\bar{z}= \gamma(z-\beta t) ,\nonumber \\
\bar{x}&=& x , \:\:\:\:\:\:\:\:\:\:\:\:\:\:\:\:\:\:\:\:\:\: \bar{y}=y ,
\end{eqnarray}
we get
\begin{eqnarray}
\label{boostedmetric2}
ds^2&=& -dt^2 +dx^2+dy^2+ dz^2 \nonumber \\
&& -4N \frac{\gamma^2(z-\beta t)}{\bar{r}\rho^2}
(dt-\beta dz)(xdy-ydx) .
\end{eqnarray}
One should be careful while treating the coordinate $\bar{r}$ as
the large radius limit is really $ r \equiv(x^2+
y^2 +z^2)^{1/2} \rightarrow \infty$ and thus
\begin{eqnarray}
 \frac{1}{\bar{r}} &\equiv&  \biggr [ x^2+ y^2+ \gamma^2(z-\beta \: t)^2 \biggl]^{-1/2}\nonumber \\
   &=& \frac{1}{r} \biggr [(\sin^2\theta +\gamma^2 \cos^2\theta) +\gamma^2 \beta^2 (t^2/ r^2)-2\gamma^2 \beta \cos\theta
   (t/r) \biggl ]^{-1/2} \nonumber \\
 & \sim& \frac{1}{rB}
+O(1/r^2) ,
\end{eqnarray}
where we defined $B=\sqrt{sin^2\theta +\gamma^2 cos^2\theta}$. Our choice for the vielbein is
\begin{eqnarray}\label{triagviel}
e^0&=& dt- 2N \frac{\gamma^2 (z-\beta t)}{\bar{r}\rho^{2}}(ydx-xdy) ,  \nonumber \\
e^1&=&  dx , \nonumber \\
e^2&=&  dy , \nonumber \\
e^3&=& - 2N \frac{\gamma^2 \beta (z-\beta
t)}{\bar{r}\rho^{2}}  (y \: dx-
x \: dy) +  dz ,
\end{eqnarray}
where it can be checked that the spin connection is regular so that we can directly use our expressions for the charges without taking care of the singularities. Note that our
choice is precisely the triangular vielbein $e^{\bar{m}}_{\:{\bar \mu}}$ for
the linearized static pure NUT metric transformed under the boost
to $e^{m}_{\: \mu}=\Lambda^{m}_{\:\:\:\bar{n}}
\:\Lambda_{\mu}^{\:\:\: \bar{\nu}} e^{\bar{n}}_{\: \bar{\nu}}=
\delta^{m}_{\: \mu}+ \frac{1}{2} \eta^{m \nu}(h_{{\nu}{\mu}}+v_{{\nu}
{\mu}})$. Looking at (\ref{boostedmetric2}), the linear
perturbations are
\begin{eqnarray}
h_{xz}&=& -\beta h_{tx}=  -2 N  \gamma^2 \beta \frac{(z-\beta t)}{\bar{r}} \: \frac{y}{x^2+y^2} , \nonumber \\
h_{yz}&=&  -\beta h_{ty}=\:\: 2 N
\gamma^2 \beta \frac{(z-\beta t)}{\bar{r}} \:
\frac{x}{x^2+y^2} .
\end{eqnarray}
As we can directly see from (\ref{triagviel}), we have $v_{ta}= h_{ta} $ and $v_{za}= h_{za}$ for $a=x,y$. We can now easily proceed to the calculation of $K_0$
\begin{eqnarray}\label{K0}
K_{0}&=&  \frac{1}{16\pi} \oint  \epsilon^{lij}  ( \partial_i h_{0j}+ \partial_j v_{i0} )  d\hat \Sigma_l
= \frac{1}{8\pi} \oint  \epsilon^{lij}  \partial_i h_{0j}   d\hat \Sigma_l \nonumber\\
&=& \frac{N}{4\pi}\gamma^2 \oint_{S} \frac{\sin\theta}{B^3} d\theta d\phi
\nonumber \\
&=& \gamma N.
\end{eqnarray}
Note also that the time dependence in the integrand (\ref{K0}) is subleading and tends to zero when $r \rightarrow \infty$. The calculation for $K_z$ is readily the same and we find
\begin{eqnarray}\label{Kz}
K_z &=&- \frac{\beta}{8\pi} \oint  \epsilon^{lij}  \partial_i h_{0j}  d\hat \Sigma_l = -\beta K_0=-\gamma \beta N ,
\end{eqnarray}
while $K_x=K_y=0$.

With this knowledge, it is also easy to see that the charges associated to the Aichelburg-Sexl pp-wave are just $P_0=-P_z=p$ and for our NUT-wave $K_0=-K_z=k$. For the Aichelburg-Sexl pp-wave, this was done in \cite{Aichelburg:1998nm}. For the NUT-wave, one can use the symmetric vielbein
\begin{eqnarray}
e^0=  dt-\frac{F}{2} (dt-dz)\; , \qquad e^1= dx \; , \qquad
 e^2= dy\; ,  \qquad
 e^3= dz -\frac{F}{2}(dt-dz)\; ,
\end{eqnarray}
where $F$ is an harmonic function, so that  $v_{\mu\nu}=0$ and the spin connection is regular. We eventually obtain
\begin{eqnarray}
K_0 & = & \frac{1}{16\pi} \oint  \epsilon^{lij} \partial_i h_{0j}  d \Sigma_l= -\frac{k}{2\pi} \oint \frac{1}{r} \: \delta(t-z) \:   d \Sigma_r \nonumber \\
    & = & k \oint r \: \sin\theta \:  \delta(t-r\cos\theta) \:   d\theta= k \oint  \: \delta(t-r\cos\theta) \:   d(r \cos \theta) \nonumber\\
    &=& k .
\end{eqnarray}
Again, the calculation for $K_z$ is readily the same and gives $-k$.

Let us now check that the NUT-wave is the metric one obtains by acting with a gravitational duality rotation on the Aichelburg-Sexl pp-wave.

\subsubsection{Duality among shock pp-waves}

It is easy to see that the NUT-wave is the gravitational dual of the Aichelburg-Sexl pp-wave. The non-trivial fluctuations for the Aichelburg-Sexl pp-wave are
\begin{eqnarray}
h_{tt}=h_{zz}=-h_{tz}= -8\: p\: \ln(\sqrt{x^2+y^2})\; .
\end{eqnarray}

The non-trivial components of the linearized Riemann tensor defined by
$R_{\alpha \beta \gamma \delta}= 2 \partial_{[\alpha} h_{\beta
][\gamma,\delta]}$ for the Aichelburg-Sexl metric are
\begin{eqnarray}\label{Riem}
R_{tatb}= -\frac{1}{2} \: \partial_a \partial_b h_{tt} \; ,  \:\:\:\:\:
R_{tazb}= -\frac{1}{2} \: \partial_a \partial_b h_{tz} \; , \:\:\:\:\:
R_{zazb}= -\frac{1}{2} \: \partial_a \partial_b h_{zz}\; , 
\end{eqnarray}
for $a,b=x,y$. The NUT-wave has non-trivial fluctuations
\begin{eqnarray}
\tilde{h}_{tt}=\tilde{h}_{zz}=-\tilde{h}_{tz}= -8\: k\:
\arctan(x/y)\; ,
\end{eqnarray}
where $\tilde{h}_{\mu\nu}$ refers to the dual metric. The non-trivial
components of the Riemann tensor write in the same way as in  (\ref{Riem}) but with $h_{\mu\nu}$ replaced by $\tilde{h}_{\mu\nu}$.

A small computation permits to show that the duality relation is satisfied
\begin{eqnarray}\label{dudul}
\tilde{R}_{\alpha \beta \lambda \mu}=\frac{1}{2} \: \epsilon_{\alpha \beta
\gamma \delta} \: R^{\gamma \delta}_{\:\:\:\: \lambda \mu}.
\end{eqnarray}

Doing so, one checks that the gravitational dual of the Aichelburg-Sexl pp-wave
is the NUT wave. From \eqref{Riem} and \eqref{dudul}, it is easy to see that the function $\tilde{F}$ for the NUT-wave is the harmonic conjugate of the function $F$ describing the Aichelburg-Sexl pp-wave. By definition, this implies we can construct a complex variable $\zeta=y+ix$  whose logarithm is $\ln \zeta= \ln
\sqrt{x^2+y^2} + i \arctan(x/y)$ and attribute the real part
of this logarithm to the Aichelburg-Sexl metric and the imaginary part
to the dual pp-wave. This last fact can be generalized to any solution (\ref{fullform}), where $H(u,x,y)=F(x,y) \delta(u)$ and $F(x,y)$ is a harmonic function. The gravitational dual solution is characterized by $H(u,x,y)=\tilde{F}(x,y) \delta(u)$ where $\tilde{F}$ is the harmonic conjugate function of $F$ (namely ${\cal F}(\zeta)= F+i \tilde{F}$ is an holomorphic function of $\zeta$). The holomorphic nature of ${\cal F}(\zeta)$ is reminiscent of the holomorphic nature of the complex Ernst potential for BPS solutions (see for instance Section 3.4 of \cite{Englert:2007qb}).

%%%%%%%%%%%%%%%%%%%%%%%%%%%%%%%%%%%%%%%%%%%%%%%%%%%%%%%%%%%%%%%%%%%
\cleardoublepage
%%%%%%%%%%%%%%%%%%%%%%%%%%%%%%%%%%%%%%%%%%%%%%%%%%%%%%%%%%%%%%%%%%%

\part{Gravitational Duality and Supersymmetry}
%%%%%%%%%%%%%%%%%%%%%%%%%%%%%%%%%%%%%%%%%%%%%%%%%%%%%%%%%%%%%%%%%%%

\chapter{A short introduction to Supergravity} \label{chap:intropart3}

As for now, we have mainly been concerned with the Poincar\'e group which is the symmetry group of Minkowski spacetime. In the 60's, along with the understanding of the importance of internal symmetries, physicists started wondering if a larger, external, symmetry group containing the Poincar\'e group could be of any interest to describe the Nature we observe.

The first important result in this direction is the no-go theorem of S. Coleman and J. Mandula who proved in \cite{Coleman:1967ad} ``the impossibility of combining space-time and \textit{internal} symmetries in any but a trivial way". Without going into technical details, what they actually showed is that any symmetry group of the S-matrix, that would not lead to trivial physics, should be a direct product of an internal symmetry group and the Poincar\'e group, up to additional $U(1)$'s. This result describes the fact that any non-trivial mixing would bring in additional conserved quantities, and thus quantum numbers in the quantum theory, which are not observed experimentally (see also \cite{West:1990tg} for concrete examples).

No-go theorems are based on assumptions and they can a priori always be bypassed if some of these assumptions are relaxed. Independently from these considerations, Y.A. Golfand and E.S. Likhtman  achieved in \cite{Golfand:1971iw}  a non-trivial mixing of the Poincar\'e symmetry with a new type of symmetries by enlarging the concept of Lie algebra. They constructed the so-called superalgebras. The superalgebras are generalizations of the Lie algebras where one is allowed to introduce charges verifying anti-commutation relations. Following the spin-statistics theorem, these charges have half-integer spin, and are called fermionic or odd charges, as compared to integer-spin charges which are called  bosonic or even charges. This is the basic ingredient that we will need to recover the so-called supersymmetry algebra.

Actually, the work \cite{Golfand:1971iw} remained unnoticed for a long time. The real birth of supersymmetry is to be found, some years later, in the study of two-dimensional models (see \cite{Ramond:1971gb}, \cite{Neveu:1971rx}, \cite{Gervais:1971ji} ) where supersymmetry on the two-dimensional world-sheet was observed. Referring to these works, J. Wess an B. Zumino generalized the idea to four dimensions and proposed in \cite{Wess:1974tw} the first example where supersymmetry is linearly realized in four dimensions.  In a subsequent paper \cite{Wess:1973kz}, the same authors pointed out that "the supergauge transformations evade the Coleman-Mandula no-go theorem because their algebra is not a ordinary Lie algebra". See eventually \cite{Weinberg:2000cr} for more details on the birth of supersymmetry.

Nowadays, supersymmetry is the name given to a type of symmetry that relates bosons and fermions, elementary particles of different quantum nature. Under this symmetry, and because the transformation parameter has spin $1/2$, a particle of spin $s$ is mapped to its so-called superpartner particle which differs by half a unit of spin. To each known boson of integer spin, respectively known fermion of half-integer spin, there corresponds a supersymmetric partner which is a fermion, respectively a boson.  For example, the supersymmetric partner of the photon of spin 1 is a  spin 1/2 particle called the photino and to each quark of spin 1/2 there corresponds a bosonic particle of spin 0 called the s-quark.

On one side, it seems that the main reasons why physicists started considering such symmetries was that these were bringing cancellations such as in the computation of radiative corrections to the mass of the Higgs boson.  Because supercharges commute with the generator of translations, each known particle in a supermultiplet, a representation of the supersymmetry group, should have the same mass. This means that all known particles should have superpartners with the same mass. As these particles have not been observed experimentally, even if supersymmetry was part of our lives, supersymmetry must be broken. Nowadays, there exists a plethora of models to explain how supersymmetry can be broken. These models can be classified following that the breaking is explicit or spontaneous. Explicit breaking for the MSSM has been considered by adding all possible soft terms to its Lagrangian. However, a spontaneous breaking, explaining the origin of the breaking and fixing by construction the form of the soft terms, is more desirable.  Due to constraints coming from experimental facts such as flavor changing neutral currents or constraints coming from supersymmetry itself such as non-renormalization properties (see for example \cite{Seiberg:1993vc}), models to describe supersymmetry breaking are somehow constrained. For a fresh start into more experimental facts about supersymmetry, and a general introduction to supersymmetry and its breakings, see the detailed review of S. Martin \cite{Martin:1997ns}.

On another side, even if we know that supersymmetry must be broken in Nature, supersymmetry is a beautiful theoretical framework which brings in many new features. It is thus interesting to study supersymmetric theories as they are often more tractable than non-supersymmetric ones.  If gravitational duality has anything to do with the non-linear sector of Einstein's theory, we believe that the study of supersymmetric generalizations of gravity theories could help in understanding this duality.

This chapter is intended as a review of the material needed to deal with supergravity, i.e. supersymmetric extensions of gravity theories, and some of its specific solutions as we will be concerned with in the last chapter of this thesis. In section \ref{sec: supalg}, we start by reviewing  the global supersymmetry algebra. We then illustrate in section \ref{sec: Noeth} how a global symmetry is turned into a local one by means of Noether's procedure. Based on this procedure, in section \ref{sec: locgrav}, we briefly comment on how a local supersymmetric theory, where the parameter of transformation is required to be local, is a theory of gravity. In section \ref{sec: N1N2}, we provide the reader with the so-called $\mathcal{N}=1$ and $\mathcal{N}=2$ supergravities we will be interested in. We sketch how these theories are invariant under local supersymmetry transformations. We eventually recall some facts, in section \ref{sec: Boso}, about the search for bosonic solutions of these supergravities and discuss the integrability conditions, necessary conditions for the existence of Killing spinors, that lead to the BPS bound. We finish this chapter by discussing, in section \ref{sec: KSE}, two methods for solving specific Killing spinor equations.

\setcounter{equation}{0}
\section{The supersymmetry algebra}
\label{sec: supalg}
The Poincar\'e group $P$ is generated by the translations $P_{\mu}$ and the Lorentz transformations $M_{\mu\nu}$ that satisfy the relations
\beqs\label{Poinc}
&&[P_{\mu}, P_{\nu}]=0, \qquad [P_{\mu}, M_{\nu\rho}]=\eta_{\mu\nu} P_{\rho}-\eta_{\mu\rho} P_{\nu} ,\nonumber\\
&& [M_{\mu\nu} , M_{\rho \lambda}]= \eta_{\mu\lambda} M_{\nu\rho} +\eta_{\nu\rho} M_{\mu\lambda} -(\rho \leftrightarrow \lambda) ,
\eeqs
where $\eta_{\mu\nu}$ is the flat metric. As we have just discussed, the Coleman-Mandula theorem shows that there exists no extension of the Poincar\'e algebra which presents a non-trivial mixing with the Lorentz generators if we want a non-trivial S-matrix. In mathematical language, what they state is that any symmetry group made out of the Poincar\'e group  and a symmetry group $G$ with generators $T_a$ such that
\beqs
[T_a,T_b]=f_{ab}^{\:\:\:\:\:\:c} \: T_{c} , \qquad [P_{\mu},T_a]=0 ,
\eeqs
must actually be a direct product of $P$ and $G$, meaning that we should also impose
\beqs
[M_{\mu\nu}, T_a]=0.
\eeqs
In modern language, this is just the statement that only internal (global or local) symmetries can be considered under their hypothesis.  This conclusion can be bypassed if one allows the introduction of  ``odd" generators $\mathcal{Q}$ satisfying anti-commutations relations.  We will require the algebra to have a $Z_2$ graded structure
\begin{eqnarray*}
&& [even, even]=even ,\\
&& \{ odd,odd \}=even , \\
&& [even,odd]=odd .
\end{eqnarray*}
If one imposes the generalized Jacobi identities (as detailed for example in \cite{West:1990tg}), one can construct the so-called $\mathcal{N}=1 $ super-Poincar\'e algebra whose algebra is given by (\ref{Poinc}) in addition with
\beqs
&& [P_{\mu}, \mathcal{Q}_{\alpha}]=0 , \qquad [\mathcal{Q}_{\alpha}, M_{\mu\nu}]= \frac{1}{2} (\gamma_{\mu\nu})_{\alpha}^{\:\:\beta} \mathcal{Q}_{\beta} ,  \\
&& \{ \mathcal{Q}_{\alpha}, \mathcal{Q}_{\beta}\}=(\gamma^{\mu} \: C)_{\alpha \beta} \: P_{\mu}  .
\eeqs
where the last anticommutator is referred in the following as the superalgebra, the algebra of anticommuting charges.  Note that we have chosen four-component real Majorana supercharges and $C$ is the charge conjugation matrix, which we take here to be $C\equiv \gamma_0$. In our conventions, gamma matrices are also real, see Appendix A. Actually, when considering a number $\mathcal{N}$ of supercharges that we denote $\mathcal{Q}^I$, the superalgebra can be centrally extended, i.e. one can add Lorentz scalars to the superalgebras. The so-called extended superalgebras where obtained by R. Haag, J. Lopuszanski and M. Sohnius in \cite{Haag:1974qh}
\beqs
\{ \mathcal{Q}^I, \mathcal{Q}^J\}= \gamma^{\mu} \: C \: P_{\mu} \: \delta^{IJ} + C U^{IJ} + \gamma_5 \: C \: V^{IJ}\; ,
\eeqs
where $U^{IJ}=-U^{JI}$ and $V^{IJ}=-V^{JI}$ commute with all generators. Let us remark that, to be precise, in \cite{Haag:1974qh}, only point-like particles interactions were considered. With the apparition of p-branes, more  general extensions with Lorentz-tensor central charges, or brane charges, have also been considered (see for example \cite{Ferrara:1997tx}, \cite{Gauntlett:1999dt}). In these works, such extensions are permitted as the hypothesis of locality (point-like interactions) of the Coleman-Mandula theorem is relaxed.

\setcounter{equation}{0}
\section{From global to local symmetry: Noether's procedure}
\label{sec: Noeth}

In this section, we would like to review the standard procedure that is used to turn a global theory into a local one, i.e. where the symmetry parameter is made dependent on the space-time coordinates. This procedure is often shortcut by saying that it amounts to a replacement of the standard partial derivatives into covariant derivatives, where the so-called ``compensating" field has been introduced to make the action invariant under the local symmetry. In here, we would like to review the Noether procedure and re-derive this standard textbook result by applying the procedure to a simple example. This will turn out to be quite useful if one wants to understand why gravity appears when dealing with local supersymmetry, or how supersymmetric theories of gravity have been firstly constructed.

As a starting point, let us pick a generic theory whose field content is denoted by the collection of fields $\Phi_i$ and the Lagrangian is given by
\beqs
\mathcal{L}=\mathcal{L}(\Phi_i \: , \: \partial_{\mu} \Phi_i) .
\eeqs
Let us also assume that the theory is invariant under a global symmetry and denote the constant parameter of the transformation by $\Lambda$. In general, the variation of the Lagrangian is a total derivative
\beqs\label{kkuu}
\delta \mathcal{L}= \partial_{\mu} K^{\mu}\; .
\eeqs
If we want to make the symmetry local, we allow the transformation parameter to depend on the spacetime coordinates and set $\Lambda \rightarrow \Lambda(x)$. One can easily realize that the variation of the action will now be of the form
\beqs\label{delta1}
\delta \mathcal{L}= \partial_{\mu} K^{\mu} + (\partial_{\mu} \Lambda) S^{\mu}\; ,
\eeqs
where one can check that $ S^{\mu}$ is actually the Noether current $ J^{\mu}$. Indeed, a generic variation of the Lagrangian
\beqs\label{delta2}
\delta \mathcal{L}&=& \frac{\partial \mathcal{L}}{\partial \Phi_i} \delta \Phi_i + \frac{\partial \mathcal{L}}{\partial (\partial_{\mu} \Phi_i)} \delta \partial_{\mu} \Phi_i \nonumber \\
&=& \Big( \frac{\partial \mathcal{L}}{\partial \Phi_i} - \partial_{\mu} \frac{\partial \mathcal{L}}{\partial (\partial_{\mu} \Phi_i)} \Big) \delta \Phi_i + \partial_{\mu} \Big( \frac{\partial \mathcal{L}}{\partial (\partial_{\mu} \Phi_i)}  \delta \Phi_i \Big) , \nonumber \\
\eeqs
tells us that, when considering \eqref{kkuu} and \eqref{delta2}, the Noether current is given by
\beqs\label{delta3}
\Lambda J^{\mu}= \frac{\partial \mathcal{L}}{\partial (\partial_{\mu} \Phi_i)} \delta \Phi_i - K^{\mu}.
\eeqs
Comparing both expressions \eqref{delta1} and \eqref{delta2}, where \eqref{delta3} is taken into account, we find
\beqs
\Big( \frac{\partial \mathcal{L}}{\partial \Phi_i} - \partial_{\mu} \frac{\partial \mathcal{L}}{\partial (\partial_{\mu} \Phi_i)} \Big) \delta \Phi_i= \partial_{\mu} \Lambda (S^{\mu}-J^{\mu}) -\Lambda \partial_{\mu} J^{\mu} .
\eeqs
Since the left-hand side does not depend on $\partial_{\mu} \Lambda$, we must have $S^{\mu}=J^{\mu}$.

This last result is the starting point of the Noether procedure which consists, in a first step, of replacing the original Lagrangian by
\beqs\label{action1}
\mathcal{L}'=\mathcal{L}-g X_{\mu} J^{\mu},
\eeqs
where $g$ is a coupling constant and $X_{\mu}$ denotes the compensating field, and demand that
\beqs
\delta X_{\mu}=\frac{1}{g} \partial_{\mu} \Lambda .
\eeqs
We will also consider the addition of a kinetic term for the compensating field.
The usual next step is to vary the new action \eqref{action1}
\beqs
\delta \mathcal{L}'= - g X_{\mu} \delta J^{\mu},
\eeqs
and compute explicitly the variation of the Noether current. What the first implementation \eqref{action1} does is actually to make the Lagrangian invariant up to order one in the coupling constant $g$. In the second step, we will add terms in $g^2$. One can thus start an iterative procedure in powers of $g$ to make the action invariant up to all orders in $g$. 

To illustrate this procedure, let us consider the theory of a complex scalar field
\beqs
\mathcal{L}=  \partial_{\mu} \phi \:  \partial^{\mu} \bar{\phi} .
\eeqs
This theory is invariant under a global $U(1)$ symmetry because the action is invariant when the fields transform as
\beqs
\phi(x)\rightarrow e^{i\Lambda} \phi(x)\; , \qquad \bar{\phi}(x)\rightarrow e^{-i\Lambda} \bar{\phi}(x)\; ,
\eeqs
or as infinitesimal transformations
\beqs
\delta \phi =+i \Lambda \phi, \qquad  \delta \bar{\phi}= -i \Lambda \bar{\phi},
\eeqs
where $\Lambda$ is a constant parameter. If we consider the variation of this action by allowing the parameter $\Lambda$ to depend on the spacetime coordinates, we find on-shell
\beqs
\delta \mathcal{L}= (\partial_{\mu} \Lambda) (i \phi \partial^{\mu} \bar{\phi} - i \bar{\phi} \partial^{\mu} \phi) \equiv (\partial_{\mu} \Lambda) S^{\mu}.
\eeqs
One can check that $S^{\mu}$ is just the Noether current
\beqs
\Lambda J^{\mu}\equiv \frac{\partial \mathcal{L}}{\partial(\partial_{\mu} \phi)} \delta \phi+ \frac{\partial \mathcal{L}}{\partial(\partial_{\mu} \bar{\phi})} \delta \bar{\phi} .
\eeqs
Following Noether's procedure, we must introduce a gauge field $A_{\mu}$, the compensating field, by modifying the Lagrangian as
\beqs
\mathcal{L}'=\mathcal{L}-g A_{\mu} J^{\mu},
\eeqs
and impose
\beqs
\delta A_{\mu}=\frac{1}{g} \partial_{\mu} \Lambda.
\eeqs
Then, we compute the variation of $\mathcal{L}'$ and  find
\beqs
\delta \mathcal{L}' &=& -g A_{\mu} \delta J^{\mu} \nonumber \\
&=& -2 g A_{\mu} \phi \bar{\phi} \partial^{\mu} \Lambda .
\eeqs
Now, one can check that this last varied term can be exactly canceled by adding the term $g^2 A_{\mu} A^{\mu} \phi \bar{\phi}$ to the Lagrangian. We have thus completed the procedure as our theory is now invariant under the local symmetry up to any order in $g$. The final Lagrangian, where we have added a kinetic term for the gauge field, reads
\beqs
\mathcal{L}&=& F_{\mu\nu} F^{\mu\nu}+ \partial_{\mu} \phi \partial^{\mu} \bar{\phi} - g A_{\mu} J^{\mu} + g^2 A_{\mu} A^{\mu} \phi \bar{\phi} \nonumber \\
&=& F_{\mu\nu} F^{\mu\nu}+ D_{\mu} \phi D^{\mu} \bar{\phi} ,
\eeqs
where in the last equation $D_{\mu}\equiv\partial_{\mu}-i g A_{\mu}$ is the covariant derivative. Starting from a theory that is invariant under some global symmetry, one can make it invariant under a local symmetry by allowing the symmetry parameter to depend on the space-time coordinates and replacing partial derivatives by covariant derivatives. In standard language, this is the so-called minimal coupling.

One other enlightening example of this procedure is the construction of the Yang-Mills theories. To construct these theories, one starts with a collection of $N$ abelian gauge fields, i.e. each invariant under a local $U(1)$, that transforms under some global symmetry. Making this last symmetry local through the Noether procedure, one obtains a local Yang-Mills theory.

In Part II, we have been dealing with electromagnetic, respectively gravitational, duality which is a global symmetry of electromagnetism, respectively linearized gravity. One could then wonder if these duality symmetries can be gauged. It is just recently that this computation has been considered for the electromagnetic duality by M. Henneaux and C. Bunster in \cite{Bunster:2010wv} (see also the work of S. Deser for an Hamiltonian version \cite{Deser:2010it}). Following the Noether procedure, it is shown that such a gauging is impossible. In their words ``the fact of being electric or magnetic does not seem to be a space-time dependent concept". In \cite{Diffon:2011wt}, A. Diffon and H. Samtleben and M. Trigiante have considered gaugings of this type but at the cost of losing a Lagrangian description. 

The Noether procedure to gauge a global supersymmetry is the subject of the next section.

\setcounter{equation}{0}
\section{Local supersymmetry is supergravity}
\label{sec: locgrav}

The question we would like to answer here is: what happens if one considers a theory which is invariant under global supersymmetry and tries to make the spinorial parameter $\epsilon$ of the transformation dependent on the space-time coordinates ? As we said, starting from a theory invariant under global supersymmetry and following the Noether procedure, one sees that the variation of the Lagrangian, when the parameter is made local, will be of the generic form
\beqs
\delta \mathcal{L}= \partial_{\mu} K^{\mu}+ (\partial_{\mu} \bar{\epsilon}) S^{\mu},
\eeqs
where $\epsilon$ is the parameter of global supersymmetry transformations.
To restore the symmetry, one modifies the Lagrangian by adding
\beqs
- \kappa \bar{\psi}_{\mu} J^{\mu} ,
\eeqs
 where $J^{\mu}$ is the Noether current associated to the global supersymmetry invariance of the initial theory, and imposes the variation
 \beqs
 \delta \psi_{\mu} \sim \kappa^{-1} \partial_{\mu} \epsilon.
 \eeqs
For global supersymmetry, the compensating field $\psi_{\mu}$ must have a vector and a spinor index. It is a spin 3/2 field. If one goes on with the procedure by computing explicitly the variation of the Noether current $J^{\mu}$, the stress tensors associated to the matter fields of our initial theory appear. These can only be cancelled by the introduction of a new Noether coupling, the metric $g_{\mu\nu}$, which is a spin 2 field. The spin 3/2 field is the supersymmetric partner of the spin 2 field and it is in this respect that it is called the gravitino. See the excellent review  \cite{VanNieuwenhuizen:1981ae} of supergravity by P. Van Nieuwenhuizen for more details.

The first supergravity theories, or supersymmetric theories of gravity, have been constructed using the Noether procedure. After that, some theories were constructed by dimensional reduction of known supergravities. Indeed, the standard Kaluza-Klein reduction preserves supersymmetry. One should remember that Noether's procedure is an iterative procedure. Although in the simple example presented in the last section, this procedure stops at second order, it is often not the case for supergravity. This is why most supergravities constructed by the Noether procedure are only dealt with up to a certain order in the fermions.

In the next chapter, we will be dealing with the so-called $\mathcal{N}=1$ and $\mathcal{N}=2$ supergravities in four dimensions which are exactly invariant under local transformations when including quartic interactions. In the next section, instead of going through the Noether procedure to construct such theories, we will just present them and check that they are invariant under local supersymmetric variations.

\setcounter{equation}{0}
\section{The $\mathcal{N}=1$ and $\mathcal{N}=2$ supergravities}
\label{sec: N1N2}

General relativity is a theory for a spin 2 field, known as the graviton. In its simplest supersymmetric version, we will couple it to a spin 3/2 particle known as the gravitino. Those fields form the $\mathcal{N}=1$ supergravity multiplet. The (3/2,\:2)-theory is known as the $\mathcal{N}=1$ supergravity. The action is invariant under local supersymmetry where the parameter of transformation is allowed to depend on the space-time coordinates. This $\mathcal{N}=1$ supergravity was constructed first in  \cite{Freedman:1976xh} (see also \cite{Deser:1976eh}) by applying Noether's procedure (and a lot of intuition) to the sum of the free actions of Einstein-Hilbert and Rarita-Schwinger, describing respectively the spin 2 and spin 3/2 free fields. In this way, an interacting theory invariant under local supersymmetry was found. As we explain below, it can be written as \cite{Romans:1991nq} \footnote{ We have the same conventions as in \cite{Romans:1991nq}. We would like to warn the reader that supergravity is a jungle of conventions and one should always pay great attention when using conventions of others.}
\beqs\label{sugra15}
e^{-1} \mathcal{L}= -\frac{1}{4} R +\frac{1}{2} \bar{\psi}_\mu \gamma^{\mu\nu\rho} \hat{D}_{\nu} \psi_{\rho},
\eeqs
where $\hat{D}_{\mu}\equiv\partial_{\mu} +\frac{1}{4} \hat{\omega}_{\mu}^{\:\:ab} \gamma_{ab}$ and where the spin connection is fixed by its own equation of motion
\beqs
\hat{\omega}_{\mu ab}=\Omega_{\mu ab}-\Omega_{\mu ba} -\Omega_{ab \mu} , \\
\Omega_{\mu \nu}^{\:\:\:\:\:a}=\partial_{[\mu} e_{\nu]}^{\:\:a}-\frac{1}{2} \bar{\psi}_{\mu} \gamma^a \psi_{\nu} .
\eeqs
Note that the ``hat" for spin connection or covariant derivative is used in the presence of fermions.
It is invariant under the supersymmetry transformations
\beqs\label{sugravar}
\delta e_{\mu}^{\:\:a}=\bar{\epsilon} \gamma^a \psi_{\mu} , \qquad \delta \psi_{\mu}= \hat{D}_{\mu} \epsilon .
\eeqs
We also have $\delta e_a^{\:\:\mu}=-\bar{\epsilon} \gamma^\mu \psi_{a}$.
In the original work \cite{Freedman:1976xh}, the independent fields are the vielbein and the gravitino. The disadvantage of their method is that the computations are quite involved: ``A term in fifth power of the gravitino field has been shown to vanish by a computer calculation". Their method is referred to as the second order formalism because the Einstein-Hilbert Lagrangian depends on the metric, actually the vielbein, and provide second order equations of motion. In \cite{Deser:1976eh}, they considered the spin connection as an independent field. This is a first order formalism. For general relativity, the first order formalism is known as the Palatini formalism. Instead of considering the Einstein-Hilbert Lagrangian, one considers the action
\beqs
\mathcal{L}_P= -\frac{1}{4}\: |e|\:  e_a^{\:\:\mu} e_{b}^{\:\:\nu} R_{\mu\nu}^{\:\:\:\:\:\:ab} [\omega], \qquad R_{\mu\nu}^{\:\:\:\:\:\:ab}= 2\:  \partial_{[\mu} \:  \omega_{\nu]}^{\:\:\:\:ab} +2 \: \omega_{[\mu}^{\:\:\:\:ac} \: \omega_{\nu]c}^{\:\:\:\:\:\:b} ,
\eeqs
which depends on both independent quantities $e$ and $\omega$.  By varying this action, one finds
\beqs\label{deltaP}
\delta \mathcal{L}_P=-\frac{1}{2} |e| (R_{\mu}^{\:\:a} -\frac{1}{2} e_{\mu}^{\:\:a} R)\:  \delta e_{a}^{\:\:\mu}-\frac{3}{2} |e| (D_{\mu} e_{\nu}^{\:\:a}) \: e_{[a}^{\:\:\mu}e_{b}^{\:\:\nu}e_{c]}^{\:\:\sigma} \: \delta \omega_{\s}^{\:\:bc},
\eeqs
upon using $\delta R_{\mu\nu}^{\:\:\:\:\:\:ab}=2 D_{[\mu} \omega_{\nu]}^{\:\:\:ab}$. The equations of motion are thus
\beqs
&&R_{\mu}^{\:\:a}[\omega] -\frac{1}{2} e_{\mu}^{\:\:a} R=0 ,\\
&&D_{[\mu} e_{\nu]}^{\:\:a}=0 \rightarrow \omega=\omega[e].
\eeqs
This is a set of two first order equations. Upon implementing the defining equation for the spin connection into the first equation, we go back to the second order Einstein equations. First and second order formalisms are thus equivalent on-shell. This first order formalism, where the vielbein and the spin connections are taken as independent fields, was used in \cite{Deser:1976eh} to construct the $\mathcal{N}=1$ supergravity. One disadvantage of this method is that it requires the appropriate variation of the spin connection, as it is considered to be an independent field, under local supersymmetry.

In  \eqref{sugra15}, we have provided the reader with an action that combines benefits from both methods and simplifies a lot the verification of its invariance under local supersymmetry. This is referred as the 1.5 order formalism. We will start by considering that the vielbein and spin connections are independent fields just as in the first order formalism. However, while varying the action, we implement the fact that the spin connection is not completely an independent field but can be solved in terms of the vielbein and the gravitino. Indeed, one can check that the variation of the spin connection in the action
\beqs\label{sugra16}
e^{-1} \mathcal{L}= -\frac{1}{4} R +\frac{1}{2} \bar{\psi}_\mu \gamma^{\mu\nu\rho} D_{\nu} \psi_{\rho}=-\frac{1}{4} R +\frac{1}{2} \epsilon^{\mu\nu\rho\sigma} \bar{\psi}_\mu \gamma_\nu \gamma_5 D_{\rho} \psi_{\sigma} ,
\eeqs
 yields its defining equation in terms of the vielbein and the gravitino
\beqs
\frac{\delta \mathcal{L}}{\delta \omega}=0 \qquad \rightarrow \qquad D_{[\mu} e_{\nu]}^{\:\:a}=\frac{1}{2} \bar{\psi}_{\mu} \gamma^a \psi_{\nu} \qquad \rightarrow \qquad \omega=\omega[e,\psi] \; .
\eeqs
Pay attention to the fact that the action \eqref{sugra16} has a non-hatted covariant derivative.

Following the 1.5 order formalism, let us check that we have an action invariant under the local supersymmetry transformations \eqref{sugravar}. As we look at the variation of our supergravity action and implement $\omega=\omega[e,\psi]$, we see that
\beqs
\delta \mathcal{L}&=& \frac{\delta L}{\delta e}\Big|_{\omega=\hat{\omega}[e,\psi]} + \frac{\delta L}{\delta \psi}\Big|_{\omega=\hat{\omega}[e,\psi]} +  \frac{\delta L}{\delta \omega}\Big|_{\omega=\hat{\omega}[e,\psi]} \Big(  \frac{\delta \hat{\omega}}{\delta e} \delta e+ \frac{\delta \hat{\omega}}{\delta \psi} \delta \psi\Big) \nonumber \\
&=& \frac{\delta L}{\delta e}\Big|_{\omega=\hat{\omega}[e,\psi]} + \frac{\delta L}{\delta \psi}\Big|_{\omega=\hat{\omega}[e,\psi]}.
\eeqs
This method is thus much easier as one does not need to provide the variation of the spin connection, such as in the first order formalism, or vary it when checking invariance under supersymmetry as in the second order formalism, only variations with respect to the vielbein and the gravitino are needed. 

 Variation of the Einstein-Hilbert term, with respect to the vielbein, was already provided in  \eqref{deltaP} and is just
\beqs
\delta \mathcal{L}_{EH}=-\frac{1}{2} |e| (R_{\mu}^{\:\:a} -\frac{1}{2} e_{\mu}^{\:\:a} R)\:  \delta_{\epsilon} e_{a}^{\:\:\mu}= \frac{1}{2} |e| (R_{\mu}^{\:\:a} -\frac{1}{2} e_{\mu}^{\:\:a} R)\: \bar{\epsilon} \gamma^\mu \psi_{a}\; .
\eeqs
The variation of the Rarita-Schwinger term is a bit more involved but we find, up to quartic order in fermions, that it is precisely of the form
\beqs
\delta \mathcal{L}_{RS}= -\delta \mathcal{L}_{EH}\; .
\eeqs
One can also rapidly obtain the quartic order terms and chek, using a Fierz-identity, that they cancel. This concludes our verification.

Although we will quickly consider the $\mathcal{N}=1$ theory to study supersymmetric pp-waves, we will mostly focus on $\mathcal{N}=2$ supergravity. From representation theory, one sees that the content of the gravity multiplet is a metric $g_{\mu\nu}$, a pair of real gravitini and a Maxwell field. It can also be understood as the coupling  of the $\mathcal{N}=1$ gravitino multiplet, composed of a gauge field and a gravitino, to the $\mathcal{N}=1$ supergravity just described. One can rapidly imagine that the construction of this supergravity in \cite{Ferrara:1976fu} through Noether's procedure has been a real nightmare.

The pure $\mathcal{N}=2$ supergravity lagrangian is given in 1.5 formalism by \cite{Romans:1991nq}
\beqs
e^{-1} \mathcal{L}= -\frac{1}{4} R +\frac{1}{4} F_{\mu\nu} F^{\mu\nu} +\frac{1}{2} \bar{\psi}_\mu \gamma^{\mu\nu\rho} \hat{\nabla}_{\nu} \psi_{\rho}  +\frac{i}{8} (F+\hat{F})^{\mu\nu} \bar{\psi}_{\sigma} \gamma_{[\mu} \gamma^{\s \kappa} \gamma_{\nu]} \psi_\kappa,
\eeqs
where $\psi_{\mu}=(1/\sqrt{2})(\psi^1_{\mu}+i\psi_{\mu}^2)$ is a complex gravitino, $\hat{\nabla}_\mu\equiv\hat{D}+\frac{i}{4} \hat{F}_{ab} \gamma^{ab} \gamma_{\mu}$ is called the super-covariant
derivative,  $F_{\mu\nu}\equiv2 \: \partial_{[\mu} A_{\nu]}$ and $\hat{F}_{\mu\nu}\equiv F_{\mu\nu}-\text{Im}(\bar{\psi}_{\mu} \psi_{\nu})$. The spin connection is defined just as for the $\mathcal{N}=1$ supergravity but with $\Omega_{\mu \nu}^{\:\:\:\:\:a}=\partial_{[\mu} e_{\nu]}^{\:\:a}-\frac{1}{2} \text{Re} (\bar{\psi}_{\mu} \gamma^a \psi_{\nu})$. The lagrangian is invariant under
\beqs
\delta e_{\mu}^{\:\:a}=\text{Re} (\bar{\epsilon} \gamma^a \psi_{\mu}), \qquad \delta \psi_{\mu}=\hat{\n}_{\mu} \epsilon, \qquad \delta A_{\mu}=\text{Im} (\bar{\epsilon} \psi_{\mu} )\; ,
\eeqs
where now $\epsilon$ is a complex spinor, 
as one can check following an equivalent procedure as the one described for the $\mathcal{N}=1$ case.

\setcounter{equation}{0}
\section{Bosonic solutions of supergravities}
\label{sec: Boso}

When looking at supersymmetric solutions of supergravity theories, one is often only interested in bosonic solutions where all fermions have been set to zero. In this case, variations of the bosons are trivial. However, for consistency, we should impose that variations of the fermions do not introduce bosons. Schematically, we need to set
\beqs\label{delferm}
\delta (\text{fermions})=0 .
\eeqs

In this thesis, we will not discuss the classification of the solutions but pay attention to particular solutions. Let us consider only the $\mathcal{N}=2$ supergravity. The results for $\mathcal{N}=1$ are obtained by setting the Maxwell field, and one gravitino, to zero. In the $\mathcal{N}=2$ case, the requirement \eqref{delferm} is just
\beqs\label{deltapsi}
\delta \psi_{\mu}= \Big( \partial_{\mu} +\frac{1}{4} \omega_{\mu}^{\:\:ab} \gamma_{ab} +\frac{i}{4} F_{ab} \gamma^{ab} \gamma_{\mu} \Big) \epsilon=0.
\eeqs
This equation is known as the Killing spinor equation. When we say that a bosonic solution (of the equations of motion of the supergravity theory) is supersymmetric, or more exactly preserves some supersymmetry, we mean that one can find non-trivial solutions to the Killing spinor equations. Remark that if one starts by finding a configuration of bosonic fields that solves the Killing spinor equation such that non-trivial Killing spinors exist, one still needs to check that it is a solution of the equations of motion. Killing spinor identities, developed in \cite{Kallosh:1993wx}, have however been useful to show that most of the equations of motion will be trivially fulfilled in this case, see also \cite{Bellorin:2005hy}. This, and the use of G-structures, has enabled the classification of all supersymmetric solutions of various supergravities (see for example the illuminating work of J. Gauntlett, J. Gutowski, C. Hull, S. Pakis and H. Reall in \cite{Gauntlett:2002nw} where they deal with the minimal supergravity in five-dimensions).

Coming back to the Killing spinor equations, consistency conditions for the existence of such Killing spinors is of the general form
 \beqs\label{consistency}
[\nabla_{\mu}, \nabla_{\nu}]\epsilon=0\; ,
\eeqs
where $\nabla_{\mu}\equiv \hat{\nabla}_{\mu}|_{\psi=0}$.
These are also often referred as integrability conditions. From \eqref{deltapsi}, one easily sees, using  $[D_{\mu},D_{\nu}]\epsilon= ( \frac{1}{4} R_{\mu\nu}^{\:\:\:\:ab} \gamma_{ab} ) \epsilon$, that
\beqs
[\hat{\nabla}_{\mu}, \hat{\nabla}_{\nu}] \epsilon= \Big( \frac{1}{4} R_{\mu\nu}^{\:\:\:\:ab} \gamma_{ab}+\frac{i}{2} \gamma^{ab} \gamma_{[\nu} D_{\mu]} F_{ab} \Big) \epsilon=0.
\eeqs

 For all solutions we will be concerned with, one can check that \eqref{consistency} is of the generic form
\begin{eqnarray}\label{integrability}
[\nabla_{\mu},\nabla_{\nu}]\epsilon=X_{\mu\nu} \: \Theta \:  \epsilon=0\; .
\end{eqnarray}
This equation is an algebraic equation and has non-trivial solutions if and only if the operator $\Theta$
acting on the supersymmetry parameter $\epsilon$ has vanishing eigenvalues, i.e.
\begin{eqnarray}
\text{det} \: \Theta=0 .
\end{eqnarray}
This last equation involves a relation among the sources of the solutions ( such as the mass, the NUT charge and the electromagnetic charges as we describe after for the charged Taub-NUT metric). Actually, upon implementing this relation, we see that $\Theta$ becomes a projector. The trace of this projector gives us the number of supersymmetries that are preserved by the solution. The consistency conditions for Minkowski space are trivial such that all supersymmetries are preserved. Minkowski spacetime is said to be maximally supersymmetric.

\setcounter{equation}{0}
\section{Solving Killing spinors equations}
\label{sec: KSE}

Even if it is already non-trivial to find supersymmetric solutions, it is an even more difficult task to actually solve the Killing spinor equations to obtain the expressions for the Killing spinors. A method to solve these equations in a specific set-up was presented in \cite{Romans:1991nq}.  We will start by reviewing this algorithm and we will then present a more generic, although more intuitive, alternative procedure to compute the Killing spinors we will be interested in.
In the next chapter we will apply both methods to compute the Killing spinors of the Reissner-Nordstrom and Taub-NUT solutions with electric charge $Q$ and magnetic charge $H$ . Although the final expressions obtained with each method seem rather different, we provide at the end of this section a small generic argument to check that they are actually equivalent.

\subsection{Romans' algorithm}
\label{subsec:Romans}

As in \cite{Romans:1991nq}, we will be interested in solving a system composed of the integrability condition \eqref{integrability}, an algebraic equation on the Killing spinors, and the Killing spinor equation for $r$ obtained after all $t,\theta,\phi$ dependence has been worked out. The system we want to solve is thus of the form
\begin{eqnarray}
\Pi \epsilon(r)= \frac{1}{2}\Big(1+ x(r) \Gamma_1+y(r) \Gamma_2\Big) \epsilon(r)=0 \: ,\label{system1} \\
\partial_r \epsilon(r)=\Big(a(r)+b(r)\Gamma_1+c(r)\Gamma_2\Big)\epsilon(r)\:, \label{system2}
\end{eqnarray}
where $\Gamma_1$ and $\Gamma_2$ are such that
\begin{eqnarray}
(\Gamma_1)^2=(\Gamma_2)^2=1 \:,  \qquad \Gamma_1 \Gamma_2=-\Gamma_2 \Gamma_1 \:,
\end{eqnarray}
where $\Pi$ is a projector $\Pi^2=\Pi$ which imposes that $x^2+y^2=1$, and where we also assume that $y\neq0$. This last condition permits us to set, without loss of generality, $c=0$ as we can always use (\ref{system1}) into (\ref{system2}).
There is eventually a last consistency condition that needs to be imposed on the parameters. This arises when one takes the derivative of the projection equation
\begin{eqnarray}
\Pi \epsilon=0 \rightarrow \partial_r(\Pi \epsilon)= (\partial_r \Pi) \epsilon + \Pi \partial_r \epsilon=0.
\end{eqnarray}
It is now rather easy to see, using $\Gamma_2 \epsilon=-(1/y)(1+x\Gamma_1)\epsilon$ and $y'/y=-x x'/y^2$, that the last equation is just
\begin{eqnarray}
\frac{(x' +2 b y^2)}{2 y^2} \Gamma_1 (1+x \Gamma_1) \epsilon= -\frac{(x' +2 b y^2)}{2 y} \Gamma_1 \Gamma_2 \epsilon=0 \: ,
\end{eqnarray}
which is satisfied (when $y\neq 0$) if and only if
\begin{eqnarray}
x'+2by^2=0.
\end{eqnarray}
The more general solution can be written as
\begin{eqnarray}
\epsilon(r)=\frac{1}{2}[A(r)+B(r)\Gamma_1+C(r)\Gamma_2+D(r)\Gamma_{12}] \epsilon_0.
\end{eqnarray}
For simplicity, we will moreover assume that $A=-B$ and $C=D$. Plugging this in the projection equation (\ref{system1}), we see that a non-trivial solution requires
\begin{eqnarray}\label{projectoreq}
C=-A \frac{(1-x)}{y},
\end{eqnarray}
while the Killing spinor equation requires
\begin{eqnarray}
A'=(a-b)A \: , \qquad
C'=(a+b)C.
\end{eqnarray}
Summing those last two equations such as to get rid of $b$ and using (\ref{projectoreq}), we obtain a differential equation for $A$
\begin{eqnarray}
A'=[a+ \frac{x'}{2(1-x)}+\frac{y'}{2y}]A.
\end{eqnarray}
One can easily check that the solution is
\begin{eqnarray}
A(r)=p(x,y) e^{w} ,   \qquad w=\int^r a(r') dr'  ,  \qquad p(x,y)=\sqrt{\frac{y}{1-x}}=\sqrt{\frac{1+x}{y}},
\end{eqnarray}
and then we also get
\begin{eqnarray}
C(r)= q(x,y) e^{w} , \qquad q(x,y)=-\sqrt{\frac{1-x}{y}} .
\end{eqnarray}
We eventually recover Romans' result
\begin{eqnarray}
\epsilon(r)=(A(r)+C(r)\Gamma_2)P(-\Gamma_1)\epsilon_0 ,
\end{eqnarray}
where $P(\Gamma_1)$ is the projector
\begin{eqnarray}
P(\Gamma_1)=\frac{1}{2}(1+\Gamma_1) .
\end{eqnarray}

\subsection{Alternative method}

Let us consider once more the system \eqref{system1}-\eqref{system2} and write the projection equation in the generic form
\beqs
\Pi \epsilon= \frac{1}{2} (1+ Y) \epsilon=0\; ,
\eeqs
where $\Pi^2=\Pi$ implies that $Y^2=1$. One could try to shortcut Romans's procedure by writing the solution directly as
\beqs
\epsilon(r)= f(r)\frac{1}{2} (1-Y) \epsilon_0.
\eeqs
This is indeed a solution of the projection equation $\Pi \epsilon\sim (1-Y^2)\epsilon=0$. We can then solve for the function $f(r)$ by directly plugging it into the differential equation. We will see that this method is much more efficient to obtain the Killing spinors of the solutions we will consider in the next chapter.

As we will check on specific examples, one may be tormented by the fact that this method will generically provide  different expressions for the Killing spinors than obtained with Romans' method. However, one can see that the respective projections on $\epsilon_0$ actually project on the same space. Let us see how this works.

Let us denote the two Killing spinors derived using each method as
\begin{eqnarray}
\epsilon_1 (r) =f(r) \Pi_1 \epsilon_0 , \qquad  \epsilon_2 (r) =f(r) \Pi_2 \epsilon_0 ,
\end{eqnarray}
where $\Pi_1$ and $\Pi_2$ are projectors. We will say that the apparently different projections are equivalent if we have
\begin{eqnarray}\label{pp1}
\Pi_1 \Pi_2=\Pi_2=\Pi_2^2 ,  \qquad \Pi_2 \Pi_1=\Pi_1=\Pi_1^2 ,
\end{eqnarray}
and $\text{Tr}(\Pi_1)=\text{Tr}(\Pi_2)$. Indeed, introducing $\Pi_3=\Pi_1-\Pi_2$, one can rewrite \eqref{pp1} as
\begin{eqnarray}\label{p3p21}
\Pi_3 \Pi_2=0 , \qquad \Pi_3 \Pi_1=0 .
\end{eqnarray}
Substracting these last equations implies also that
\begin{eqnarray}
\Pi_3^2=0   \rightarrow \Pi_1=\Pi_2+ X ,
\end{eqnarray}
where $X^2=0$ and  $\text{Tr}(X)=0$. We thus see that $X \epsilon$ does not bring in additional information. Both projections are thus equivalent.

%%%%%%%%%%%%%%%%%%%%%%%%%%%%%%%%%%%%%%%%%%%%%%%%%%%%%%%%%%%%%%%%%%%

\chapter{Gravitational duality in $\mathcal{N}=1$ and $\mathcal{N}=2$ supergravity} \label{chap:supergravduality}

In this chapter, we want to discuss the supersymmetric properties of the bosonic solutions, and their gravitational duals, discussed in Part II. Our main concern will be to understand how the supersymmetry algebra copes with the charges generated under gravitational duality. In particular, in the context of $D=4$, ${\cal N}=2$ supergravity we review how the BPS
equation is generalized in presence of NUT charge to \cite{Kallosh:1994ba}
\begin{equation}
\sqrt{M^2+N^2}=|Z|, \nonumber
\end{equation}
and in turn we show how the superalgebra itself takes into
account the possibility of turning on a NUT charge, or more generally a dual
momentum. We eventually see how these conclusions apply to the $\mathcal{N}=1$ superalgebra by studying supersymmetric pp-waves.

We start by reviewing, in section \ref{sec:TNN2}, the charged Taub-NUT solution of the coupled Einstein-Maxwell equations and its BPS bound in the $\mathcal{N}=2$ supergravity theory. We apply the methods described in the previous chapter to find the Killing spinors of the Reissner-Nordstrom solution, helping us in deriving those of the charged Taub-NUT solution. In section \ref{sec:Asproj}, we comment on the form of the asymptotic projection found for the charged Taub-NUT solution and how this is related to the left hand side of the superalgebra. In section \ref{sec:supalg}, we show that the complexified Witten-Nester form is actually holding all the information that we would require on the right hand side of the superalgebra.  In section \ref{sec:Modifsupalg}, we justify the presence of this extra term, containing the dual momenta information, as a topological extension of the algebra of bosonic supercharges. We see that supercharges undergo a transformation under gravitational duality which tells us to consider a generalized superalgebra that includes the dual momenta.  In section \ref{sec:ppw}, we show that the same phenomena arises when considering the NUT-wave in $\mathcal{N}=1$ supergravity. We relegate properties of gamma matrices to Appendix \ref{App: Gamma}.

\setcounter{equation}{0}
\section{The charged Taub-NUT solution of $\mathcal{N}=2$ supergravity}
\label{sec:TNN2}

As we explained in section \ref{sec: Boso}, when searching for supersymmetric solutions of supergravities, one is usually focusing on purely bosonic supergravity solutions, i.e. where all fermionic
fields have been set to zero. We will thus consider the purely bosonic part of the ${\cal N}=2$ supergravity Lagrangian, which is just the
Einstein-Maxwell Lagrangian
\begin{equation}
e^{-1} \: \mathcal{L}= -\frac{1}{4} R + \frac{1}{4}
F_{\mu \nu} F^{\mu\nu} , \label{action}
\end{equation}
and impose the Killing spinor equation
\begin{equation}\label{kilspineq}
\delta{\psi}_{\mu}= \nabla_{\mu}\epsilon= D_{\mu}
\epsilon + \frac{i}{4} F_{ab} \gamma^{ab}\: \gamma_\mu \: \epsilon
=0.
\end{equation}

All supersymmetric solutions of the $\mathcal{N}=2$ supergravity have been classified by P. Tod in \cite{Tod:1983pm}.

A particular solution to the equations of motion derived
from the action (\ref{action}) is the charged Taub-NUT black hole solution carrying,
besides the mass $M$, a NUT charge $N$, and both electric $Q$  and magnetic $H$ Maxwell charges
\begin{eqnarray}
ds^2& = &-\frac{\lambda}{R^2}(dt+2N\cos\theta
d\phi)^2 + \frac{R^2}{\lambda} dr^2
+ R^2 \:d\Omega^2 , \label{tnmetric}
\end{eqnarray}
\begin{equation}\label{tngauge}
A_{t}=\frac{Qr+NH}{R^2}, \:\:\:\:\:\:\:
A_{\phi}=\frac{-H(r^2-N^2)+2NQr   }{R^2}\cos\theta,
\end{equation}
where $d\Omega^2$ is the metric on the unit two-sphere and where
\beqs
\lambda\equiv r^2-N^2-2Mr+Z^2\: , \qquad
R^2\equiv r^2+N^2 \: , \qquad
Z^2\equiv Q^2+H^2 \: .
\eeqs
Note that in the following, we will also describe the Reissner-Nordstrom solution with electric and magnetic charges. It is obtained from the above solution by setting $N=0$. We get
\begin{eqnarray}
ds^2& = &-(1-\frac{2M}{r}+\frac{Z^2}{r^2})dt^2 + (1-\frac{2M}{r}+\frac{Z^2}{r^2})^{-1} dr^2
+r^2 d\Omega^2 , \label{rnmetric}
\end{eqnarray}
\begin{equation}\label{rngauge}
A_{t}=\frac{Q}{r}, \:\:\:\:\:\:\:
A_{\phi}=-H \cos\theta .
\end{equation}

As we are interested in the supersymmetry properties of the charged Taub-NUT solution, we will need to solve for the Killing spinor equations. For this reason, let us introduce here the vielbein
\begin{eqnarray}
e^0&=& \frac{\sqrt{\lambda}}{R}(dt+2N\cos\theta d\phi), \:\:
\:\:\:\:\:\: e^1= \frac{R}{\sqrt{\lambda}}dr,
\nonumber \\
e^2&=& Rd\theta,
\:\:\:\:\:\:\:\:\:\:\:\:\:\:\:\:\:\:\:\:\:\:\:\:\:\:\:\:\:\:\:\:\:\:\:\:\:
e^3=R\sin\theta  d\phi. \label{vielbein}
\end{eqnarray}
With this, one can also compute the non-trivial
components of the spin connection
\begin{eqnarray*}
\omega_{t}^{\:\:01}&=&
\frac{\lambda'}{2R^2}-\frac{\lambda}{R^3}R' ,\:\:\:\:\:\:\:\:\:\:\:\:\:\:\:\:\:\:\:\:\:\:\:\:\:
\omega_{\theta}^{\:\:12}=-\frac{\sqrt{\lambda}}{R}R'  , \\
 \omega_{\phi}^{\:\:13}&=& -\frac{\sqrt{\lambda}}{R}R'\sin\theta ,
\:\:\:\:\:\:\:\:\:\:\:\:\:\:\:\:\:\:\:\:\:\:\:\:\:
 \omega_{\phi}^{\:\:23}=- \cos\theta (1+\frac{2\lambda N^2}{R^4}) ,
 \\
 \omega_{\phi}^{\:\:02}&=&
 -\frac{\sqrt{\lambda}}{R^2}N\sin\theta ,  \:\:\:\:\:\:\:\:\:\:\:\:\:\:\:\:\:\:\:\:\:\:\:\:\:
  \omega_{\theta}^{\:\:03}=\frac{\sqrt{\lambda}}{R^2}N , \\
   \omega_{t}^{\:\:23}&=&-\frac{\lambda}{R^4}N , \:\:\:\:\:\:\:\:\:\:\:\:\:\:\:\:\:\:\:\:\:\:\:\:\:\:\:\:\:\:\:\:\:\:\:
\omega_{\phi}^{\:\:01}=2N\cos\theta
(\frac{\lambda'}{2R^2}-\frac{\lambda}{R^3}R').
\end{eqnarray*}
The non-zero components of $F_{ab}$ are
\begin{eqnarray*}
F_{01}=\frac{1}{R^4}(Q(r^2-N^2)+2HNr) = -\frac{Q}{R^2} +2r \frac{Q r +NH}{R^4} ,
\nonumber \\
F_{23}=\frac{1}{R^4}(H(r^2-N^2)-2QNr) = \frac{H}{R^2} - 2N \frac{Q r +NH}{R^4} ,
\end{eqnarray*}
so that
\begin{eqnarray}
F_{ab} \gamma^{ab} & = & -2 F_{01} \gamma_{01} +2 F_{23} \gamma_{23} 
\nonumber\\
&= & -\frac{2}{R^4} \gamma_{01} (r+\gamma_5 N)^2 (Q - \gamma_5 H).
\end{eqnarray}

The expressions for $\omega_\mu^{\:\: ab}\gamma_{ab}$ are
\beqs
\omega_t^{ab}\gamma_{ab} &=& \frac{2}{R^4} \gamma_{01}
\left[ (r-M)R^2 - \lambda (r+\gamma_5 N) \right], \\
\omega_r^{ab}\gamma_{ab} &=& 0 , \\
\omega_\theta^{ab}\gamma_{ab} &=& -2 \frac{\sqrt{\lambda}}{R^2}
\gamma_{12} (r+\gamma_5 N), \\
\omega_\phi^{ab}\gamma_{ab} &=& -2 \frac{\sqrt{\lambda}}{R^2}
\sin\theta \gamma_{13} (r+\gamma_5 N) -2\cos\theta \gamma_{23}\nonumber \\
& &
+4N \cos\theta \frac{1}{R^4} \gamma_{01}\left[ (r-M)R^2 - \lambda (r
+\gamma_5 N)\right].
\eeqs
We also have
\beq
\gamma_t=\frac{\sqrt{\lambda}}{R}\gamma_0, \quad
\gamma_r= \frac{R}{\sqrt{\lambda}}\gamma_1, \quad
\gamma_\theta= R\gamma_2, \quad
\gamma_\phi= R\sin\theta \gamma_3 +2N \frac{\sqrt{\lambda}}{R}\cos\theta
\gamma_0.
\eeq
Plugging all this into the supersymmetric variation of the gravitino, we find
\beqs
\delta \psi_t & = & \partial_t \epsilon + \frac{1}{2R^4} \gamma_{01}
\left\{(r-M)R^2 - \lambda (r+\gamma_5 N) - i(r+\gamma_5 N)^2 (Q-\gamma_5 H)
\frac{\sqrt{\lambda}}{R}\gamma_0\right\}\epsilon ,\nonumber \\
\delta \psi_r & = & \partial_r \epsilon - i \frac{1}{2R^4} \gamma_{01}
(r+\gamma_5 N)^2 (Q-\gamma_5 H) \frac{R}{\sqrt{\lambda}}\gamma_1 \epsilon,
\nonumber\\
\delta \psi_\theta & = & \partial_\theta \epsilon + \frac{1}{2R^4}
\left\{-\sqrt{\lambda} R^2 \gamma_{12}(r+\gamma_5 N) \nonumber
 - i\gamma_{01}(r+\gamma_5 N)^2 (Q-\gamma_5 H) R\gamma_2\right\}\epsilon ,\\
\delta \psi_\phi & = & \partial_\phi \epsilon + \frac{1}{2R^4}
\left\{-\sqrt{\lambda} R^2 \sin\theta \gamma_{13} (r+\gamma_5 N)
-R^4 \cos\theta \gamma_{23} \right. \nonumber\\ & & \qquad
+2N \cos\theta \gamma_{01}\left[ (r-M)R^2 - \lambda (r+\gamma_5 N)\right]
\nonumber \\ & & \qquad \left.
- i\gamma_{01}(r+\gamma_5 N)^2 (Q-\gamma_5 H)\left(R\sin\theta \gamma_3
+2N \frac{\sqrt{\lambda}}{R}\cos\theta \gamma_0\right)\right\}\epsilon.
\eeqs

As we said in the previous chapter, necessary conditions for the existence of Killing spinors will generally introduce relations among the constants. The BPS bound for the charged Taub-NUT solution was recovered using integrability conditions in \cite{Kallosh:1994ba,AlonsoAlberca:2000cs}. For simplicity, we introduce the following expressions
\beqs
r\pm \gamma_5 N & = & R e^{\pm \beta(r) \gamma_5}, \nonumber \\
M\pm \gamma_5 N & = & U e^{\pm \alpha_m \gamma_5}, \nonumber \\
Q\pm \gamma_5 H & = & Z e^{\pm \alpha_q \gamma_5} ,\label{angles}
\eeqs
where $U=\sqrt{M^2+N^2}$, and as discussed in section \ref{sec: Boso}, we check that integrability conditions impose the algebraic equation
\beqs\label{thetaa}
 \Theta \epsilon= \frac{1}{2} \Big(1 + \frac{iZ R}{\sqrt{\lambda } } (\frac{e^{\beta \gamma_5}}{R}-\frac{U e^{\alpha_m \gamma_5}}{Z^2}) e^{-\alpha_q \gamma_5} \gamma_0 \Big) \epsilon=0 .
\eeqs
As an example, for the one willing to explicitly check this, we find
\beqs
[\nabla_{r}, \nabla_{\theta}]\epsilon= X_{r\theta} \Theta \epsilon ,  \qquad X_{r\theta}=\frac{iZ}{R^2} e^{(3\beta-\alpha_q)\gamma_5} .
\eeqs
The equation \eqref{thetaa} has non-trivial solutions if the determinant of $\Theta$ is zero. Imposing this, we recover the BPS condition
\beqs\label{bpsbound}
M^2+N^2=Q^2+H^2 .
\eeqs
Plugging this last relation back into (\ref{thetaa}), we immediately get
\beqs
\Theta \epsilon= \frac{1}{2} \Big(1 -i  e^{(\beta+\alpha_m-\alpha_q) \gamma_5} \gamma_0 \Big) \epsilon=0\; .
\eeqs
We can then easily check that $\Theta$ is a projector, as $\Theta^2=\Theta$, with $\tr \Theta=2$. As the projection will halve the number of independent Killing spinors, we say that we have an half-BPS solution.
In the case we set $N=0$, and thus also $\beta=\alpha_m=0$, this simplifies to
\beqs
M^2=Z^2 , \qquad \Theta \epsilon=\frac{1}{2} \Big(1 -i  e^{-\alpha_q  \gamma_5} \gamma_0 \Big) \epsilon=0,
\eeqs
and we thus obviously also describe half-BPS states. In our paper \cite{Argurio:2008zt}, a more intuitive derivation of the same condition \eqref{bpsbound} was provided. We do not wish to reproduce it here and refer the reader to this paper for more details.

Given the above BPS bound \eqref{bpsbound}, the SUSY variations can be rewritten as
\beqs
\delta \psi_t&=& \partial_t \epsilon + \frac{r-M}{2R^3}Z \gamma_{01}
e^{(\beta-\alpha_m)\gamma_5}\left\{1-ie^{(\beta+\alpha_m-\alpha_q)\gamma_5}
\gamma_0\right\}\epsilon, \\
\delta \psi_r &=& \partial_r \epsilon - \frac{Z}{2R(r-M)} i
e^{(2\beta-\alpha_q)\gamma_5}\gamma_0\epsilon, \\
\delta \psi_\theta &=& \partial_\theta \epsilon -\frac{1}{2} \gamma_{12}
\epsilon + \frac{Z}{2R}\gamma_{12} e^{(\beta-\alpha_m)\gamma_5}
\left\{1-ie^{(\beta+\alpha_m-\alpha_q)\gamma_5}\gamma_0\right\}\epsilon, \\
\delta \psi_\phi &=& \partial_\phi \epsilon
-\frac{1}{2} (\sin\theta \gamma_{13}
+  \cos\theta \gamma_{23})\epsilon + \\
& & + \left[ \frac{Z}{2R} \sin\theta \gamma_{13} + \frac{NZ(r-M)}{R^3}
\cos\theta \gamma_{01}\right] e^{(\beta-\alpha_m)\gamma_5} \nonumber
\left\{1-ie^{(\beta+\alpha_m-\alpha_q)\gamma_5}\gamma_0\right\}\epsilon.
\eeqs

Upon imposing the projection equation, we immediately see that the set of equations reduces to
\beqs
\delta \psi_t&=& \partial_t \epsilon , \\
\delta \psi_r &=& \partial_r \epsilon - \frac{Z}{2R(r-M)} i
e^{(2\beta-\alpha_q)\gamma_5}\gamma_0\epsilon, \\
\delta \psi_\theta &=& \partial_\theta \epsilon -\frac{1}{2} \gamma_{12}
\epsilon , \\
\delta \psi_\phi &=& \partial_\phi \epsilon
-\frac{1}{2} (\sin\theta \gamma_{13}
+  \cos\theta \gamma_{23})\epsilon ,
\eeqs
and we can solve the Killing spinor equations for the $t, \theta$ and $\phi$ dependence by writing the solution in the form
\begin{eqnarray}
\epsilon(t,r,\theta,\phi)=
e^{\frac{1}{2}\gamma_{12}\theta}\:
e^{\frac{1}{2}\gamma_{23}\phi} \epsilon(r), \label{sphkill}
\end{eqnarray}
where $\epsilon(r)$ is a solution of the system
\beqs\label{systema}
&&\Theta \epsilon= \frac{1}{2} \left\{1-ie^{(\beta+\alpha_m-\alpha_q)\gamma_5}\gamma_0\right\}\epsilon=0 \: ,
\label{proj} \\
&&\partial_r \epsilon = \frac{Z}{2R(r-M)} i
e^{(2\beta-\alpha_q)\gamma_5}\gamma_0\epsilon \label{systeq}.
\eeqs

The above projection equation (\ref{proj}) will be essentially enough for the rest
of the discussion on the relation between the Killing spinor and the
supersymmetry algebra. However, for the sake of completeness, and in order
to show that a solution indeed exists, we would like to provide the reader with the complete expressions of the Killing spinors (see also \cite{Argurio:2008zt}).

The system \eqref{proj}-\eqref{systeq} is exactly of the form presented in section \ref{sec: KSE}.
We will start by applying Romans's algorithm and our alternative algorithm to the Reissner-Nordstrom black hole with electric charge and with electric and magnetic charges. These results are, for example, presented in \cite{Romans:1991nq}. We start by reviewing this firstly because it is easier to illustrate both methods, and secondly because it strongly suggests how one should deal, as we will do just after, with the charged Taub-NUT black hole.

\subsubsection{RN with only electric charge}

Here, the BPS condition is $M=Q$, and also $\beta=\alpha_m=\alpha_q=0$. The projection simplifies to
\begin{eqnarray}\label{projoo}
\Pi \epsilon=\frac{1}{2}(1-i\gamma_0)\epsilon=0 ,
\end{eqnarray}
Following Romans's procedure, we need $y\neq 0$ and we thus set $x=0, y=1$ and $\Gamma_2=-i\gamma_0$.
The Killing spinor equation \eqref{systeq} simplifies to
\begin{eqnarray}
\partial_r \epsilon= \frac{M}{2r(r-M)}  i\gamma_0 \epsilon= \frac{M}{2r(r-M)}  \epsilon .
\end{eqnarray}
Following Romans's algorithm, we have
\begin{eqnarray}
a=\frac{M}{2r(r-M)} , \qquad b=0,
\end{eqnarray}
and the integrability condition is trivially satisfied as $x=b=0$. We obtain
\begin{eqnarray}
p=1, \qquad q=-1, \qquad e^{w}=\sqrt{1-\frac{M}{r}}.
\end{eqnarray}
The general solution is
\begin{eqnarray}
\epsilon(r)=\sqrt{1-\frac{M}{r}}(1+i \gamma_0) \epsilon_0\; .
\end{eqnarray}

Note that we will obtain exactly the same result if we apply our alternative method. Indeed, from \eqref{projoo}, we start by stating that
\beqs
\epsilon(r)=f(r)(1+i \gamma_0) \epsilon_0\; .
\eeqs
The differential equation for $f(r)$ gives precisely the same result as above.

\subsubsection{RN with electric and magnetic charges}

In this case, the projection equation is
\begin{eqnarray}
\Theta \epsilon&=&\frac{1}{2} (1-i e^{-\alpha_q \gamma_5} \gamma_0) \epsilon = \frac{1}{2}(1+ \frac{Q}{M}(-i\gamma_0) +\frac{H}{M} (i\gamma_{123})) \epsilon=0\; ,
\end{eqnarray}
and the Killing spinor equation is the same as in the previous case
\begin{eqnarray}
\partial_r \epsilon&=&-\frac{i}{4} (1-\frac{M}{r})^{-1} (\frac{2}{r^2}(-Q \gamma_{01}+H \gamma_{23})) \gamma_1 \epsilon = \frac{M}{2r(r-M)} \epsilon\; ,
\end{eqnarray}
upon using the projection equation. Given this, we have
\begin{eqnarray}
a= \frac{M}{2r(r-M)} \: , \qquad b=0 \: , \nonumber \\
x=H/M \: , \qquad y=Q/M  \: , \nonumber \\
\Gamma_1=i\gamma_{123}  \: , \qquad \Gamma_2=-i\gamma_0 \: ,
\end{eqnarray}
and the solution is
\begin{eqnarray}\label{RN2a}
\epsilon (r)=\sqrt{1-\frac{M}{r}}(\sqrt{\frac{M+H}{Q}}+i\gamma_0 \sqrt{\frac{M-H}{Q}})\frac{1}{2}(1-i\gamma_{123}) \epsilon_0\; .
\end{eqnarray}

In here, our alternative method is actually equivalent to an approach presented in  \cite{Romans:1991nq}  which relies on the use of electromagnetic duality. It is implemented by setting
\begin{eqnarray}
Q=M \cos \alpha \: , \qquad H=M \sin \alpha \: ,
\end{eqnarray}
and realizing that
\begin{eqnarray}
\nabla(\alpha)=exp(-\frac{1}{2}\gamma_5 \alpha) \nabla(0) exp(+\frac{1}{2}\gamma_5 \alpha) \: .
\end{eqnarray}
Given the Killing spinors for the purely electric RN case ($\alpha=0$), we immediately find
\begin{eqnarray}
\epsilon(r) &=& \sqrt{1-\frac{M}{r}} \frac{1}{2}(1+i\gamma_0(\cos \alpha+\gamma_5 \sin \alpha)) \epsilon_0 \nonumber \\
             &=&  \sqrt{1-\frac{M}{r}} \frac{1}{2} (1+ i e^{-\alpha_q \gamma_5} \gamma_0 ) \epsilon_0 \: . \label{RN2}
\end{eqnarray}
This last result is obviously what one obtains using our alternative method.
Although the projections, let us call them $\Pi_1$ and $\Pi_2$, appearing in \eqref{RN2a} and \eqref{RN2} seem rather different, one can check that they are equivalent following the argument presented in section \ref{sec: KSE}. Indeed, we have $\Pi_1 \Pi_2=\Pi_1$ and $\Pi_2 \Pi_1=\Pi_2$, but also  Tr$(\Pi_1)=$Tr$(\Pi_2)=2$.

\subsubsection{TN with electric  and magnetic charges}
Here, if one wants to follow Romans's algorithm, the system to solve is
\beqs
&&\Theta \epsilon= \frac{1}{2} \left\{1-ie^{(\beta+\alpha_m-\alpha_q)\gamma_5}\gamma_0\right\}\epsilon=0 \: ,  \\
&&\partial_r \epsilon = \frac{Z}{2R(r-M)} i
e^{(2\beta-\alpha_q)\gamma_5}\gamma_0\epsilon ,
\eeqs
meaning that we set
\begin{eqnarray}
a&=& \frac{(Qr^2+2NHr-N^2Q)}{2y R^3(r-M)} \; , \qquad b=-\frac{N}{2R^2 y} \; , \nonumber \\
x&=&\frac{(MH-NQ)r-N(MQ+NH)}{R Z^2} \; , \qquad y=\frac{(MQ+NH)r+N(MH-NQ)}{R Z^2} \; , \nonumber \\
\Gamma_1&=&i\gamma_{123} \; , \qquad \Gamma_2=-i\gamma_0 \; ,
\end{eqnarray}
and that
\begin{eqnarray}
exp[w]=exp[\int a dr]= K \sqrt{\frac{Z^2 y (r-M)}{R}}  ,
\end{eqnarray}
where $K$ is an integration constant  that we fix such that in the limit $N\rightarrow 0$ we recover the result for the charged RN. By setting $K=1/\sqrt{MQ}$, we have
\begin{eqnarray}
exp[w](N=0)=\sqrt{1-\frac{M}{r}}\; ,
\end{eqnarray}
where $y(N=0)=Q/M$, $Z(N=0)=M$. The final result takes the rather complicated form
\begin{eqnarray}
\epsilon(r)=\sqrt{\frac{r-M}{R}}(\sqrt{1+x}+i\gamma_0 \sqrt{1-x})(1-i\gamma_{123})\tilde{\epsilon}_0\; .
\end{eqnarray}

Now, one can try to solve it using our alternative method. However, it is a bit more involved as the projection is now an $r$-dependent projection. Let us first rewrite this projection into a more enlightening form
\beqs
\frac{1}{2} \Big( 1-ie^{(\beta+\alpha_m-\alpha_q)\gamma_5}\gamma_0 \Big) \epsilon= \frac{1}{2} e^{(\frac{\beta}{2}-\alpha_m)\gamma_5} \Big( e^{\alpha_m \gamma_5}- i e^{\alpha_q \gamma_5} \gamma_0 \Big)  e^{-\frac{\beta}{2} \gamma_5} \epsilon\; ,
\eeqs
where we made use of $e^{2\alpha_m \gamma_5}= e^{2\alpha_q \gamma_5}$ which is valid when  $M^2+N^2=Z^2$ and also $\gamma_5 \gamma_0=-\gamma_0 \gamma_5$.
Given all this, it is now easy to realize that the solution will be
\begin{eqnarray}
\epsilon(r)&=& \sqrt{\frac{r-M}{R}} \frac{1}{2} \Big( 1+i e^{(\beta+\alpha_m-\alpha_q)\gamma_5}\gamma_0\Big) e^{(\frac{\beta}{2}+\alpha_m)\gamma_5} \epsilon_0 \label{solTN1}\\
                  &=& \sqrt{\frac{r-M}{R}} e^{\frac{\beta}{2} \gamma_5} \frac{1}{2} (e^{\alpha_m \gamma_5}+i e^{-\alpha_q \gamma_5}\gamma_0)  \epsilon_0 \; . \label{solTN2}
\end{eqnarray}
Indeed, it is easy to see that it reduces to the result found for the RN solution \eqref{RN2} when $N=0$ and that it trivially fulfills the projection equation. Eventually, one can check that it also fulfills the Killing spinor equation
\begin{eqnarray}
\partial_r \epsilon= \frac{Z}{2R(r-M)} i e^{(2\beta-\alpha_q)\gamma_5}\gamma_0 \epsilon .
\end{eqnarray}
Indeed, by plugging our guess for $\epsilon$, we rewrite it as
\begin{eqnarray}
\partial_r \epsilon= \frac{Z}{2R(r-M)}  e^{(\beta-\alpha_m)\gamma_5}  \epsilon .
\end{eqnarray}
Now, we are just left with checking that
\begin{eqnarray}
\partial_r (\sqrt{\frac{r-M}{R}} e^{\frac{\beta}{2} \gamma_5})=\frac{Z}{2R(r-M)} e^{(\beta-\alpha_m) \gamma_5} (\sqrt{\frac{r-M}{R}} e^{\frac{\beta}{2} \gamma_5}) .
\end{eqnarray}
This is trivial after computing
\begin{eqnarray}
\partial_r  \sqrt{\frac{r-M}{R}} &=& \frac{Mr+N^2}{2(r-M)R^2}  \sqrt{\frac{r-M}{R}} \; , \nonumber \\
\partial_r e^{\frac{\beta}{2}\gamma_5} &=& -\frac{\gamma_5 N}{2R^2} e^{\frac{\beta}{2}\gamma_5}\; ,
\end{eqnarray}
and also
\begin{eqnarray}
\frac{Mr+N^2}{2(r-M)R^2}-\frac{\gamma_5 N}{2R^2} &=& \frac{(M-\gamma_5 N) (r+\gamma_5 N)}{2(r-M)R^2}\nonumber \\
&=&  \frac{Z}{2R(r-M)}  e^{(\beta-\alpha_m)\gamma_5}\; .
\end{eqnarray}

We have thus shown that, provided the BPS bound $M^2+N^2=Z^2$ is satisfied, the metric has a Killing spinor, which actually depends on two complex numbers. The metric thus preserves
half of the 8 supersymmetries. This is in agreement with the results  presented in
\cite{Tod:1983pm} (see also \cite{Kallosh:1994ba,AlonsoAlberca:2000cs}).

As a last word, we could worry about the issue whether the Killing spinor
is globally defined. Indeed, as we discussed in Part II, the metric has a coordinate singularity
along the $z$ axis. However, we said that one can remove the
singularity along half of the axis by a coordinate transformation.
Essentially, one obtains two completely regular patches
on the upper and lower hemispheres, where the metric is the same
as (\ref{tnmetric}), but with $\cos \theta$ replaced by $\cos\theta \pm 1$.
It amounts to shift the time coordinate $t$ by $\pm 2N\phi$. Since the
Killing spinor is $t$-independent, we can already see that it will be
the same on the two patches. This can be verified by re-deriving its expression
as above with the regular metric in each patch. As expected one
finds the same result as above.

\setcounter{equation}{0}
\section{Asymptotic projection and the superalgebra}
\label{sec:Asproj}
In this section, we analyze in more details the solution for the Killing spinor found in the
previous section. In particular, we consider the projection that defines the Killing spinor
and take its limit of large radius, where the metric is asymptotically flat. The projection can be
recast in a form which is similar to the right hand side of the ${\cal N}=2$ supersymmetry
algebra. However, we point out that the term containing the NUT charge has the wrong hermiticity condition
and thus does not seem to fit in any of the central (or else) extensions of the most general
${\cal N}=2$ supersymmetry algebra.

The projection defining the four independent real components of the Killing
spinor is given by
\beq
\left\{1-ie^{(\beta(r)+\alpha_m-\alpha_q)\gamma_5}\gamma_0\right\}\epsilon=0.
\label{projr}
\eeq
We have emphasized that it is $r$-dependent. There are two
observations one can make about this dependence.
Recalling that $\tan \beta(r) = N/r$ and that $\tan \alpha_m =N/M$, we
see that when the NUT charge is absent, both $\beta=0$ and $\alpha_m=0$.
The projector becomes $r$-independent. However, even when $N\neq 0$,
in the limit of large radius, $r\to \infty$, we observe that $\beta \to 0$
and the $r$-dependence also disappears. We are left with a constant asymptotic
projector which depends on all of the four charges (where it is of course
understood that they satisfy the BPS bound (\ref{bpsbound})).

Let us rewrite the projector in a more readable form. By setting $\beta=0$
and multiplying by $e^{-\alpha_m \gamma_5}$, we obtain
\beq
\left\{ M-\gamma_5 N -i (Q -\gamma_5 H )\gamma_0\right\}\epsilon=0.
\label{projasymp}
\eeq

Let us now try to motivate the fact that the NUT charge should find its place in the supersymmetry algebra.
Remember that the most general ${\cal N}=2$ superalgebra including the scalar central
charges (see e.g. \cite{West:1990tg}) was already presented in section \ref{sec: supalg} for Majorana supercharges $\Q^I$ (with $I=1,2$ in our present case). It is
\beq
\{\Q^I, \Q^J\} = \gamma^\mu C P_\mu \delta^{IJ}
+ C U^{IJ} + \gamma_5 C V^{IJ},
\eeq
where both $U^{IJ}=-U^{JI}\equiv U \varepsilon^{IJ}$ and
$V^{IJ}=-V^{JI}\equiv V \varepsilon^{IJ}$, and $C\equiv \gamma_0$ is the charge
conjugation matrix. In our
conventions, Majorana spinors are real and we can define
a single complex Dirac supercharge
\beq
\Q = \frac{1}{\sqrt{2}} \left( \Q^1 + i \Q^2 \right).\label{dirac}
\eeq
The only non trivial relation of the superalgebra becomes
\beq
\{\Q, \Q^\star\} = \gamma^\mu C P_\mu -i (U+ \gamma_5 V) C.
\label{superalg}
\eeq

When there is a multiplet of BPS saturated states,
some combinations of the supercharges have to be represented trivially,
i.e. they have to
vanish. This translates into the statement that the matrix $\{\Q^I, \Q^J\}$,
or equivalently $\{\Q, \Q^\star\}$, is not of maximal rank.
This means also that the right hand side of (\ref{superalg})
must have vanishing eigenvalues.
In the present case, for a massive state at rest, we identify $P_0\equiv M$.
Further, if we set $U\equiv Q$ and $V\equiv H$, we see that
we have preserved supersymmetries if the equation
\beq
\left\{ M -i (Q -\gamma_5 H )\gamma_0\right\}\epsilon=0 ,
\label{projRN}
\eeq
has solutions (note that we have multiplied the expression
in (\ref{superalg}) by $\gamma^0$ on the left and $C$ on the right).

We recognize the equation (\ref{projasymp}) for $N=0$. So, we see that for
a Reissner-Nordstr\"om black hole with electric and magnetic charges, the projection on the Killing spinor
in the extremal case maps directly to the right hand side of the
${\cal N}=2$ superalgebra. Actually, we could have
guessed the superalgebra (\ref{superalg}) from the expression for the
projector (\ref{projRN}). It is
thus tempting to do this for the case where $N\neq 0$. From (\ref{projasymp}),
we see that $N$ must belong to a ``charge'' which carries a Lorentz index.
The most straightforward guess is that $N\equiv K_0$ of a vectorial
charge $K_\mu$ which enters the superalgebra as
\beq
\{\Q, \Q^\star\} \stackrel{?}{=} \gamma^\mu C P_\mu +\gamma_5 \gamma^\mu C K_\mu
-i (U + \gamma_5 V) C.
\label{superalgnut}
\eeq
We see that the NUT charge $N$ seems to belong to an extension of the
superalgebra which is not central in the sense that it is not a Lorentz
scalar. Such extensions have been studied \cite{vanHolten:1982mx},
and the most general
${\cal N}=2$ superalgebra taking them into account has been written
\cite{Ferrara:1997tx,Gauntlett:1999dt}.
It is however straightforward to see that our term with $K_\mu$ is not
part of any extension considered so far. The reason why Eq.(\ref{superalgnut}) is wrong  is extremely simple:
it violates hermiticity. Indeed, we have that
$(\gamma_5 \gamma^\mu C)^\dagger = - \gamma_5 \gamma^\mu C$, while
any term on the right hand side must be hermitian since
$\{\Q, \Q^\star\}^\dagger= \{\Q, \Q^\star\}$. Alternatively, it is stated in \cite{West:1990tg}, that if one tries in general to include such a term, the Jacobi identity of the schematic form
\beqs
[\{Q^I,Q^J\},J]+[\{Q^I,J\},Q^J]+[\{Q^J,J\},Q^I]=0 ,
\eeqs
can never be satisfied unless the additional term mixes non-trivially with the Poincar\'e generators. This is however excluded by the Coleman-Mandula no-go theorem.

Before seeking a way to solve this puzzle, we will provide the reader with a more stronger procedure to understand how the left hand side of the superalgebra is related to the projection acting on the spinors. We will see in the following section that the dual momenta $K_\mu$ arise naturally through a generalization of this construction.

\setcounter{equation}{0}

\section{The Witten-Nester form}
\label{sec:supalg}

A general theorem proving positivity of energy in general relativity was first proposed by R. Schoen and S.T. Yau in \cite{Schoen:1979zz}. A simplified proof was given by E. Witten in \cite{Witten:1981mf} where spinorial techniques are used. This last construction was even more simplified by J. Nester in   \cite{Nester:1982tr}  where a covariant tensor is introduced, known as the Witten-Nester two-form. Positivity of the energy in  supergravity was proved in \cite{Deser:1977hu} (see also \cite{Gibbons:1982fy}) by showing that the Hamiltonian is the square of the spinor supercharges just as for globally supersymmetric theories. 

In here, we are not interested in positivity energy theorems and refer the reader to these papers for more information. However, 
for our purposes, it will be interesting to review the fact that these constructions for general relativity have, surprisingly, a deep connection with supergravity constructions as was pointed out in  \cite{Witten:1981mf} and made precise by C. M. Hull in \cite{Hull:1983ap}. Especially, we want to review the fact that the Witten-Nester two-form can be seen as describing the right hand side of the superalgebra. This was explicitly shown   for $\mathcal{N}=1$ supergravity in \cite{Hull:1983ap} and $\mathcal{N}=2$ in \cite{Gibbons:1982fy}. At first sight, it is not so astonishing that such a connection exists. Indeed, the positivity energy theorems have connected this two-form to the usual ADM definitions of energy and momenta, while we have reviewed that $P_{\mu}$ appears in the superalgebra.

Let us begin by showing how the Nester form \cite{Nester:1982tr} is related to the variation of the supercharge, expressed as a surface integral (see  \cite{Teitelboim:1977hc,Gibbons:1982fy,Hull:1983ap}), which is just the algebra of charges. The bosonic algebra is then linked to the superalgebra. This construction enables us to see how the Witten-Nester two-form provides an expression of the bosonic charges, appearing on the right hand side of the superalgebra, in terms of surface integrals at infinity and coincides with the usual ADM expressions. Actually, we show that complexifying the Witten-Nester form, the ADM momenta appear together with the dual, magnetic, ADM momenta and are equivalent to the expressions already derived in Part II. This complexification is triggered by considering a ``topological", symmetric, contribution to the algebra of bosonic supercharges. We will show that these contributions are the ones describing the dual charges.  Evaluated on the charged NUT black hole, the right hand side of the modified superalgebra reduces exactly to the
asymptotic expression contained in the definition of the Killing spinor, discussed in the previous section. We follow closely \cite{Hull:1983ap}.

Using the Noether method one computes the  generator of supertranslations\footnote{Here, supertranslations have obviously nothing to do with the supertranslations discussed in Part I.}. It can be written as a volume integral, which in turn can be expressed as a surface integral
\beqs
\tilde \Q[\epsilon, \bar \epsilon] &=&\frac{i}{2\pi} \int \varepsilon^{\mu\nu\rho\sigma}
\bar\epsilon  \gamma_5 \gamma_\nu \hat{\nabla}_{\rho} \psi_\sigma d\Sigma_\mu
+c.c. \nonumber \\
&= & -\frac{i}{4\pi} \oint  \varepsilon^{\mu\nu\rho\sigma}
\bar\epsilon  \gamma_5 \gamma_\rho \psi_\sigma d\Sigma_{\mu\nu} + c.c.,
\label{surface}
\eeqs
where $\hat{\nabla}_{\rho}$ is the supercovariant derivative acting on a
spin-3/2 field, $c.c.$ denotes complex conjugate, $\bar\epsilon= \epsilon^\dagger C \equiv \epsilon^\dagger \gamma_0$
and we take the convention $\varepsilon_{0123}=-\varepsilon^{0123}=1$. Although we are being quick here, note that we are just stating that the supercharge is obtained by taking the volume integral of the linearized Rarita-Schwinger field equations contracted with a Killing spinor (see \cite{Hull:1983ap}), in complete analogy with the Poincar\'e charges obtained in chapter 1 by contracting the linearized Einstein equations with a Killing vector, ``\`a la Abbott-Deser".

The charge $\tilde \Q[\epsilon, \bar \epsilon]$
is bosonic and transforms the supergravity fields according
to a supertranslation. When acting for instance on the
bosonic fields, which are real, it generates a variation which is
also real. We recall that  in the present ${\cal N}=2$ case,  the gravitino
$\psi_\mu$ is Dirac and hence complex. In terms of the fermionic Dirac
supercharges defined in (\ref{dirac}), and because $(\bar \epsilon \Q)^\star = -\bar \Q \epsilon$, we have
\beq
\tilde \Q[\epsilon, \bar \epsilon] = i(\bar \epsilon \Q + \bar \Q \epsilon) . \label{complex}
\eeq

It follows from the theory of surface charges (see for instance
\cite{Barnich:2001jy,Barnich:2007bf}) that the variation
of the supercharge should define its bracket in the usual way
\beq
\delta_{\epsilon_1,\bar \epsilon_1}\tilde  \Q[\epsilon_2, \bar \epsilon_2] = i \left[
\tilde \Q[\epsilon_1, \bar \epsilon_1],\tilde \Q[\epsilon_2, \bar \epsilon_2] \right]  \label{algebra0}.
\eeq
In terms of the fermionic supercharges (\ref{dirac}), using  the expression
(\ref{complex}), we would then obtain
\beq i \left[
\tilde \Q[\epsilon_1, \bar \epsilon_1],\tilde \Q[\epsilon_2, \bar \epsilon_2]
\right]  =
i \bar \epsilon_2 \{ \Q , \Q^\star\} C \epsilon_1 -
i \bar \epsilon_1 \{ \Q , \Q^\star\} C \epsilon_2.\label{total}
\eeq

However we will see that our analysis will force us to consider a possible ``topological extension"
\beq
\delta_{\epsilon_1,\bar \epsilon_1} \tilde \Q[\epsilon_2, \bar \epsilon_2] = i \left[
\tilde \Q[\epsilon_1, \bar \epsilon_1],\tilde \Q[\epsilon_2, \bar \epsilon_2] \right] + T. \label{algebra}
\eeq
The crux of the matter is that
$\delta_{\epsilon_1,\bar \epsilon_1} \tilde \Q[\epsilon_2, \bar \epsilon_2]$
is {\em not} antisymmetric in the exchange of $\epsilon_1$ and $\epsilon_2$,
as we now show.

Using (\ref{surface}) one finds for the bracket term and the ``topological"
term the following expressions
$$
i \left[
\tilde \Q[\epsilon_1, \bar \epsilon_1],\tilde \Q[\epsilon_2, \bar \epsilon_2] \right]
=\frac{1}{2} (\delta_{\epsilon_1,\bar \epsilon_1}\tilde  \Q[\epsilon_2, \bar
\epsilon_2] -\delta_{\epsilon_2,\bar \epsilon_2} \tilde \Q[\epsilon_1, \bar
\epsilon_1]) \qquad \qquad\qquad \qquad \mbox{} 
$$
\beq
=-\frac{i}{4\pi} \oint  \varepsilon^{\mu\nu\rho\sigma}
\bar\epsilon_2  \gamma_5 \gamma_\rho \nabla_\sigma \epsilon_1   \;
d\Sigma_{\mu\nu} +\frac{i}{4\pi} \oint  \varepsilon^{\mu\nu\rho\sigma}
\nabla_\rho \bar\epsilon_1  \gamma_5 \gamma_\sigma  \epsilon_2   \;
d\Sigma_{\mu\nu}- (1 \leftrightarrow 2),\label{variaanti}
\eeq
and
\beqs
T &\equiv & \frac{1}{2} (\delta_{\epsilon_1,\bar \epsilon_1}\tilde  \Q[\epsilon_2, \bar \epsilon_2] +\delta_{\epsilon_2,\bar \epsilon_2}\tilde  \Q[\epsilon_1, \bar \epsilon_1]) \nonumber\\ &=&\frac{i}{4\pi} \oint  \varepsilon^{\mu\nu\rho\sigma}
\nabla_\rho( \bar\epsilon_1  \gamma_5 \gamma_\sigma  \epsilon_2 +\bar\epsilon_2  \gamma_5 \gamma_\sigma  \epsilon_1)\; d\Sigma_{\mu\nu} . \label{variasym}
\eeqs
Note that obviously (\ref{variaanti}) is identically zero when $\epsilon_1=\epsilon_2$ but $T$ is non-zero.

We now focus on the following expression which is the ``building
block" of the expressions appearing in (\ref{variaanti})-(\ref{variasym})
\beq
\hat E^{\mu\nu}\equiv \frac{1}{4\pi}  \varepsilon^{\mu\nu\rho\sigma}
\bar\epsilon  \gamma_5 \gamma_\rho \nabla_\sigma \epsilon
\label{nestercom} .
\eeq

This is precisely the two-form presented by Nester \cite{Nester:1982tr}
and generalized by Gibbons and Hull \cite{Gibbons:1982fy},
albeit in its complex version\footnote{ In references \cite{Nester:1982tr} and
  \cite{Gibbons:1982fy}, they indeed considered  $\hat E^{\mu\nu}+(\hat
  E^{\mu\nu})^*$.} (recall that $\epsilon$ is Dirac in our
set up). One can see that the (antisymmetric) bracket term (\ref{variaanti})
and the (symmetric) topological term (\ref{variasym}) map respectively to
the real and imaginary parts of the Nester form (\ref{nestercom}).

We are now going to use the expression (\ref{nestercom})
to obtain a linear combination
of purely bosonic surface integrals, which correspond to space-time momenta
and Maxwell charges.
In order to proceed, we
linearize gravity around Minkowski spacetime, in cartesian coordinates.
As we have already seen, we consider space-time endowed with a NUT charge
as asymptotically flat, at least as far as spacelike surface integrals
are concerned \cite{Bunster:2006rt}.

First of all, following \cite{Gibbons:1982fy}, we rewrite the
complex  Nester form as
\beq
 \hat E^{\mu\nu}= E^{\mu\nu} +
H^{\mu\nu},
\eeq
where
\beq
E^{\mu\nu}= \frac{1}{4\pi}
\varepsilon^{\mu\nu\rho\sigma} \bar\epsilon \gamma_5 \gamma_\rho
D_\sigma \epsilon, \qquad H^{\mu\nu} = \frac{i}{16\pi}
\varepsilon^{\mu\nu\rho\sigma} F_{ab} \bar\epsilon \gamma_5
\gamma_\rho \gamma^{ab} \gamma_\sigma \epsilon.
\eeq
One can readily check that $H^{\mu\nu}$ is actually real, hence any
surprise will necessarily come from the purely gravitational term
$E^{\mu\nu}$.

In the following, we will express everything in terms of the linearized
spin connection $\omega_{\mu\nu\rho}$.
Hence the covariant derivative on a spinor becomes
(note that we no longer distinguish between flat and
curved indices, since they are the same at first order)
\beq
D_\mu \epsilon = \partial_\mu \epsilon
+\frac{1}{4} \omega_{\nu\rho\mu} \gamma^{\nu\rho} \epsilon.
\eeq
We now plug back this expression in $E^{\mu\nu}$. Note that in the
surface integral, the piece proportional to $\partial_\mu \epsilon$
will drop out as explained in detail in \cite{Witten:1981mf,Hull:1983ap}.
The spinors will henceforth be identified with the constant value that
they take asymptotically.\footnote{Indeed, we can actually take the spinors to
be the Killing spinors of flat space in cartesian coordinates.}
Hence we restrict to
\beq
E^{\mu\nu}=\frac{1}{16\pi}  \varepsilon^{\mu\nu\rho\sigma}
\omega_{\alpha \beta \sigma}
\bar\epsilon  \gamma_5 \gamma_\rho \gamma^{\alpha\beta}\epsilon.
\eeq
Using the relation (see appendix \ref{App: Gamma})
\beq
\gamma_\rho \gamma_{\lambda\tau} =
\eta_{\rho\lambda} \gamma_\tau -\eta_{\rho\tau}\gamma_\lambda
-\varepsilon_{\rho \lambda\tau \xi}\gamma^\xi \gamma_5 ,
\eeq
we obtain
\beqs
E^{\mu\nu}= \frac{1}{8\pi} \bar\epsilon \gamma^\lambda \epsilon
\left( {\omega^{\mu\nu}}_\lambda
+\delta^\mu_\lambda {\omega^{\nu\rho}}_\rho
-\delta^\nu_\lambda {\omega^{\mu\rho}}_\rho \right)
+ \frac{1}{8\pi}  \bar\epsilon \gamma^\lambda\gamma_5 \epsilon\:
\varepsilon^{\mu\nu\rho\sigma} \omega_{\lambda\rho\sigma}.
\eeqs
Note that the first term
above is real while the second is imaginary.

Integrating the above 2-form at spatial infinity, we select the
$E^{0i}$ component, with $i=1,2,3$. We can then reexpress the integral
in terms of purely bosonic surface integrals as
\beq
\oint E^{0i} d\hat \Sigma_i =
\bar\epsilon \gamma^\lambda P_\lambda \epsilon
+  \bar\epsilon \gamma_5\gamma^\lambda K_\lambda \epsilon ,
\eeq
where we recover the following expressions for the ADM momenta and the
dual magnetic momenta, already obtained in Part II,
\beqs
P_\lambda & = & \frac{1}{8\pi} \oint ( {\omega^{0i}}_\lambda
+\delta^0_\lambda {\omega^{i\rho}}_\rho
-\delta^i_\lambda {\omega^{0\rho}}_\rho ) d\hat \Sigma_i, \label{admp}\\
K_\lambda & = & \frac{1}{8\pi} \oint
\varepsilon^{ijk} \omega_{\lambda jk}
 d\hat \Sigma_i.\label{admk}
\eeqs
At last, we can also address the second term of the generalized
Nester form, which is treated as in \cite{Gibbons:1982fy}. By writing
\beq
H^{\mu\nu} = \frac{i}{32\pi} \varepsilon^{\mu\nu\rho\sigma} F_{\lambda\tau}
\bar\epsilon  \gamma_5 (\gamma_\rho \gamma^{\lambda\tau} \gamma_\sigma
-\gamma_\sigma \gamma^{\lambda\tau} \gamma_\rho)\epsilon,
\eeq
and using (see appendix \ref{App: Gamma})
\beq
\gamma_\rho \gamma_{\lambda\tau} \gamma_\sigma
-\gamma_\sigma \gamma_{\lambda\tau} \gamma_\rho
= 2\varepsilon_{\rho \lambda\tau \sigma}
\gamma_5 + 2 ( \eta_{\rho\lambda} \eta_{\tau \sigma}
-\eta_{\rho\tau} \eta_{\lambda \sigma} ),
\eeq
we obtain
\beq
H^{\mu\nu} =\frac{i}{4\pi} F^{\mu\nu} \bar\epsilon\epsilon
+\frac{i}{8\pi} \varepsilon^{\mu\nu\rho\sigma} F_{\rho\sigma}
\bar\epsilon\gamma_5 \epsilon.
\eeq
The surface integral then becomes
\beq
\oint H^{0i} d\hat \Sigma_i = -i \bar\epsilon U \epsilon
-i \bar\epsilon\gamma_5 V \epsilon,
\eeq
with the central charges defined by
\beqs
U & = & -\frac{1}{4\pi} \oint F^{0i} d\hat \Sigma_i,\\
V & = & \frac{1}{8\pi} \oint \varepsilon^{ijk} F_{jk} d\hat
\Sigma_i.\label{magncharge} \eeqs
It can be checked that $U=Q$ and $V=H$ on the electromagnetically charged RN or TN solutions.

Summing up all the terms, we have
\beq
\oint \hat E^{0i} d\hat \Sigma_i
=\bar\epsilon \gamma^\lambda P_\lambda \epsilon
+  \bar\epsilon\gamma_5 \gamma^\lambda K_\lambda \epsilon
-i \bar\epsilon U \epsilon
-i \bar\epsilon\gamma_5 V \epsilon \label{finalex}.
\eeq
It is clear that the above expression cannot be equated to
$\bar \epsilon \{ \Q, \Q^\star\} C \epsilon$, which would then result
in the ``wrong'' superalgebra (\ref{superalgnut}).
But now we see that the obstruction to do so is precisely the presence
of the topological term $T$ in (\ref{algebra}).

Using the definitions of $T$ (\ref{variasym}) and of  the complex Nester form
(\ref{nestercom}) we see that
\beq
T(\epsilon, \bar{\epsilon}) =-i \oint (\hat E -\hat E^*)
\label{relnecharge}.
\eeq
Using then the result (\ref{finalex}) we finally indeed find
\beq
T(\epsilon, \bar{\epsilon})= -2 i \bar\epsilon\gamma_5 \gamma^\lambda K_\lambda \epsilon.
\label{topo}
\eeq

To sum up, we see that a refined analysis of the Nester form in its complex version permits to
 recover precisely the additional term which was guessed from the
asymptotic projection acting on the Killing spinor.
In this context, we see that this additional  term  is actually violating the
 relation (\ref{algebra0}) and corresponds to a ``topological" term leading to
 the bosonic algebra (\ref{algebra}). It would be interesting, but beyond
 the scope of this thesis,
to understand better under the lines of \cite{Barnich:2007bf}, the
 appearance of such  topological terms.

\setcounter{equation}{0}
\section{A modified superalgebra}
\label{sec:Modifsupalg}

As we have already discussed in Part II, Taub-NUT spaces are notoriously problematic for the time
identifications that they imply \cite{Mueller:1985ij},
and for the presence of the Misner strings \cite{Misner:1963fr},
which are gauge-variant singularities. It has been suggested that these
pathologies are enough to conclude that such spacetimes are not globally
supersymmetric \cite{Ortin:2006xg},
even though they have locally (and globally as well)
Killing spinors. However from the point of view of the surface integrals that
define both the bosonic
and the fermionic charges of the superalgebra, the spacetime with NUT charge
is asymptotically flat according to the simplest definition
\cite{Bunster:2006rt}. If we were to assume
that the presence of Killing spinors implies that the spacetime is
supersymmetric, we would be faced with the challenge of including the NUT
charge in
the superalgebra. The (asymptotic) projection acting on the Killing spinor
must be the same as the projection acting on the supercharges which are
represented trivially on a BPS multiplet. However, as we have shown, the NUT
charge enters in a term which cannot be part of the superalgebra because of
its wrong hermiticity. Below, we suggest a tentative path to
trivialize this problem.

A logical possibility is to write the corrected variation of the supercharge
(\ref{algebra}) in a different form, by introducing a new supercharge
$\tilde \Q'$
\beq
\delta_{\epsilon_1,\bar \epsilon_1} \tilde \Q[\epsilon_2, \bar \epsilon_2] = i \left[
\tilde \Q'[\epsilon_1, \bar \epsilon_1],\tilde \Q[\epsilon_2, \bar \epsilon_2] \right]. \label{algebranew}
\eeq
The above expression is not antisymmetric under the exchange of
$\epsilon_1$ and $\epsilon_2$, which is another way of encoding the presence
of the (symmetric) topological term.
In terms of the fermionic supercharges $\Q$ and $\Q'$, (\ref{algebranew})
reads
\beq
\delta_{\epsilon_1,\bar \epsilon_1} \tilde \Q[\epsilon_2, \bar \epsilon_2] =
i \bar \epsilon_2 \{ \Q , {\Q'}^\star\} C \epsilon_1 -
i \bar \epsilon_1 \{ \Q' , \Q^\star\} C \epsilon_2,\label{totalnew}
\eeq
where we have supposed that $\{\Q,\Q'\}=0$.
Then, equating the above to the expression obtained through the
Nester form, we get
\beq
\{\Q, {\Q'}^\star\} = \gamma^\mu C P_\mu +\gamma_5 \gamma^\mu C K_\mu
-i (U + \gamma_5 V) C.
\label{superalgnutnew}
\eeq
Now the l.h.s. is no longer hermitian, so there are no obstructions
to having the antihermitian term containing $K_\mu$ in the r.h.s.
The question is of course what is $\Q'$. It must be related to $\Q$
otherwise we would be doubling the number of supercharges.
We now show that it is related to $\Q$ through an ``axial" phase shift.

Let us rewrite for definiteness the relation (\ref{superalgnutnew}) on
our particular static massive, charged states with NUT charge
\beq
\{\Q ,  {\Q'}^\star \} = M+\gamma_5 N -i (Q +\gamma_5 H )\gamma_0.
\label{proposal}
\eeq
Using the angles defined in \eqref{angles}, it can be rewritten
as
\beq
\{\Q , {\Q'}^\star\} = \sqrt{M^2+N^2} e^{\alpha_m \gamma_5}
-i Z e^{\alpha_q \gamma_5} \gamma_0.
\eeq
If the charge
$\Q'$ is related to  $\Q$ by a simple
phase rotation
\beq
{\Q'}^\star = \Q^\star e^{\alpha_m \gamma_5},\label{phase}
\eeq
then eq.~(\ref{proposal}) takes a more standard, hermitian form
\beq
 \{\Q , \Q^\star \} = M' -i (Q' +\gamma_5 H' )\gamma_0,
\label{new}
\eeq
with
\beq
M'=
\sqrt{M^2+N^2},\qquad  Q'= \frac{QM-HN}{\sqrt{M^2+N^2} }, \qquad
H'=  \frac{HM+QN}{\sqrt{M^2+N^2} }.
\eeq
Hence, through a non-linear redefinition of the charges, we obtain the
relation (\ref{new}) that in the new variables defines an hermitian
superalgebra. Actually, the new variable $M'$ is precisely the result of a
gravitational duality rotation that eliminates the NUT charge
\beq
\left( \begin{array}{cc} \cos\alpha_m & \sin \alpha_m \\
-\sin \alpha_m &  \cos\alpha_m  \end{array}\right)
\left( \begin{array}{c} M \\ N  \end{array}\right) =
\left( \begin{array}{c} M' \\ 0  \end{array}\right).
\eeq
Note that also $Q'$ and $H'$ are obtained from $Q$ and $H$ through an
electromagnetic duality rotation of the same angle.

The phase rotation (\ref{phase}) depends on dynamical quantities,
such as $N$ and $M$. The latter however commute with
the supercharges for consistency of the superalgebra, hence for instance
we are assured that $\{\Q , \Q' \} = 0$. Moreover, one could wonder what
modified supersymmetry variation is induced by $\Q'$. This clearly deserves
to be investigated, though for consistency we anticipate that we should not
find any modification in the transformation
laws of the elementary fields.

In a more
general case where both ordinary and NUT momenta $P_i$ and $K_i$ are non zero
the situation is a bit subtler.
Indeed, focusing only on the ``gravitational'' part, we would have
\beq
\{\Q , {\Q'}^\star\} = P_0 +\gamma_5 K_0 +
(P_i + \gamma_5 K_i ) \gamma^i\gamma_0.
\eeq
After a rotation similar to (\ref{phase}) we would get
\beq
 \{\Q , \Q^\star \} = \sqrt{P_0^2+K_0^2} + \frac{1}{\sqrt{P_0^2+K_0^2}}
\left[ P_i P_0 + K_i K_0 + \gamma_5 ( K_i P_0 - P_i K_0) \right]
\gamma^i\gamma_0.\label{offending}
\eeq
We thus still have an offending anti-hermitian term, which is however
proportional to $K_i P_0 - P_i K_0$ and is thus not present when $K_\mu$
is parallel to $P_\mu$. Now, under a general gravitational duality rotation
\cite{Bunster:2006rt}
we have that
\beq
\left( \begin{array}{cc} \cos\alpha & \sin \alpha \\
-\sin \alpha &  \cos\alpha  \end{array}\right)
\left( \begin{array}{c} P_\mu \\ K_\mu  \end{array}\right) =
\left( \begin{array}{c} P_\mu' \\ K'_\mu  \end{array}\right),
\label{gravrot}
\eeq
and a NUT 4-momentum $K_\mu$ can be completely eliminated only if it is
parallel to $P_\mu$. We thus seem to be able to make sense out of a
superalgebra in the presence of NUT charges only when the latter can be
eliminated by a gravitational duality rotation.

When this is not possible, we do not seem to be able to define a superalgebra.
Note that we are not aware of solutions with non-aligned $K_\mu$ and
$P_\mu$ charges. Actually, it can be shown on simple examples that the
r.h.s. of (\ref{superalgnutnew}) does not have vanishing eigenvalues when
$K_\mu$ and $P_\mu$ are non parallel.

In the case $K_\mu = \lambda P_\mu$, we have $\lambda=N/M =
\tan \alpha_m $ and performing the rotation (\ref{gravrot}) with
$\alpha=\alpha_m$, the relation (\ref{offending}) becomes the usual
superalgebra
\beq
 \{\Q , \Q^\star \} =\gamma^\mu C P_\mu' .
\eeq

Note that $K_\mu$ is always parallel to $P_\mu$
if the spatial components $K_i$ and $P_i$ are
obtained by boosting a static object with $K_0$ and $P_0$ charges.
Indeed, remember that we have seen in Part II that boosting a pure Taub-NUT solution,
one indeed obtains a solution with $K_i\neq 0$. Notice that this solution will however not be supersymmetric. Actually, in the infinite
boost limit, one recovers the magnetic dual of the usual pp-wave,
which is moreover half-BPS. This latter fact lends support to
the presence of the dual magnetic momenta even in the ${\cal N}=1$
superalgebra, along the same lines as above. We will discuss that in the last section.

We could thus sum up in the following way the answer to the question that
motivated this work, namely how does the NUT charge enter in the supersymmetry
algebra. When $K_\mu$ is parallel to $P_\mu$, which seems to be the only
situation where we have Killing spinors, we can eliminate $K_\mu$ by a gravitational duality rotation
(\ref{gravrot}). The superalgebra then incorporates
the NUT charges through the (duality invariant) combination $P_\mu'$.
Alternatively, we can define a generalization of the superalgebra
(\ref{superalgnutnew}) where the NUT charges appear on the r.h.s. but where
we have to define a new supercharge through the axial phase rotation
(\ref{phase}). It is this latter generalized superalgebra that can be directly
related to the complex Nester form. Nevertheless,
both alternatives give the same BPS bound and projection
on the supercharges, and are hence compatible with the projection on the
Killing spinor.  In conclusion, this is evidence that backgrounds which are
obtained through gravitational duality rotations from ordinary BPS solutions, such as
Reissner-Nordstr\"om black holes, are indeed supersymmetric.

In the following section, we review how this argument can be understood in $\mathcal{N}=1$ supergravity.

\setcounter{equation}{0}
\section{Supersymmetric pp-waves}
\label{sec:ppw}

Here, we want to review the fact that the shock pp-wave
is a supersymmetric solution of $\mathcal{N}=1$
supergravity\footnote{Note that all supersymmetric solutions of
$\mathcal{N}=1$ supergravity were classified in \cite{Tod:1983pm}.}. As the
BPS bound is $P_0=-P_3$ for the Aichelburg-Sexl metric, we want
to establish that the BPS bound is $K_0=-K_3$ for our dual
pp-wave. From Part II, we see that the charges for the
dual pp-wave verify this BPS bound.

As explained in Part II, we will work with a regular spin connection. Let us start with a pp-wave of the form
\begin{eqnarray}
ds^2&=& -dt^2+dx^2+dy^2+dz^2+ F (dt-dz)^2
\nonumber \\
 &=& -du(dv-F du)+ dx^2 + dy^2 ,
\end{eqnarray}
where $F=F(x,y)$ and where we defined light-cone coordinates
$u=t-z$ and
$v=t+z$. Note that we dropped again the delta function for simplicity. An obvious vielbein choice in light-cone coordinates is
\begin{eqnarray}
e^-&=& du , \:\:\:\:\:\:\:\:\: e^+= dv-F du , \nonumber \\
e^1&=& dx , \:\:\:\:\:\:\:\:\: e^2= dy ,
\end{eqnarray}
and the metric is $ds^2=\eta_{ab} e^a \: e^b$ where
the non-vanishing components are $\eta_{11}=\eta_{22}=1$,
$\eta_{+-}=\eta_{-+}=-1/2$. Going back to cartesian coordinates, we obtain the symmetric
vielbein
\begin{eqnarray}
e^0&=& \frac{1}{2}(e^+ + e^-)= dt-\frac{F}{2} (dt-dz) ,
 \nonumber \\
e^1&=& dx , \nonumber \\
 e^2&=& dy ,  \nonumber \\
 e^3&=& \frac{1}{2}(e^+ - e^-)=dz -\frac{F}{2}(dt-dz) ,
\end{eqnarray}
where symmetricity is understood by the fact that
$v_{\mu\nu}=-v_{\nu \mu}=0$. The non-trivial components of the spin
connection are
\begin{equation}
\omega_{0a}=-\omega_{3a}= \frac{1}{2} \partial_a F(x,y)(dt-dz) ,
\end{equation}
where $F(x,y)=-4 p \ln(x^2+y^2)$ for the Aichelburg-Sexl pp-wave and $F(x,y)=-8\: k \:  \arctan(x/y)$ for the dual pp-wave. Even
if in the case of our dual pp-wave the metric has a string
singularity, one can see that the spin connection is ``regular" in
the $x-y$ plane.

It can be easily seen that the pp-wave solution is an half-BPS
solution of $\mathcal{N}=1$ supergravity when looking at the
Killing spinor equation
\begin{eqnarray}
\delta \psi_{\mu}= \biggr [ \partial_{\mu}+\frac{1}{4}
\omega_{\mu}^{mn}\: \gamma_{mn} \biggl ] \epsilon=0 .
\end{eqnarray}
Indeed, we obtain
\begin{eqnarray}
\delta \psi_{t}&=& \biggr [ \partial_{t}- \frac{1}{4}
\partial_a F(x,y) (\gamma_{0}+\gamma_{3})\gamma_a \biggl ] \epsilon=0 \; ,\nonumber \\
\delta \psi_{x}&=& \partial_{x} \epsilon =0 \; ,\nonumber \\
\delta \psi_{y}&=& \partial_{y} \epsilon =0 \; ,\nonumber \\
\delta \psi_{z}&=& \biggr [ \partial_{z}+\frac{1}{4}
\partial_a F(x,y) (\gamma_{0}+\gamma_{3})\gamma_a \biggl ] \epsilon=0\; .
\end{eqnarray}
As the second and third equations show that $\epsilon$ does not
depend on $x$ and $y$, then the first and fourth equations imply
the projection $(\gamma_0+\gamma_3)\epsilon=0$, which is also what one obtains when using the integrability conditions.
This determines that the solution preserves half of the supersymmetries and has a constant
Killing spinor. This projection corresponds to the BPS bound
$P_0=-P_3$ for the Aichelburg pp-wave and $K_0=-K_3$ for our dual pp-wave.

Our computation shows that the infinite boost of Taub-NUT, a shock
pp-wave, is also a half-supersymmetric solution of
$\mathcal{N}=1$ supergravity. This provides more evidence
that the NUT charge should be included in the $\mathcal{N}=1$
supersymmetry algebra in the same lines as described in the previous section
\begin{equation}
\{\Q, {\Q'}\} = \gamma^\mu C P_\mu +\gamma_5 \gamma^\mu C K_\mu,
\label{superalgnutnew2}
\end{equation}
where ${\Q'}$ is related to $\Q$ by a phase ${\Q'} = \Q \, e^{\alpha \gamma_5}$ with $\tan \alpha= K_0/P_0$.
Indeed, the ``modified" superalgebra (\ref{superalgnutnew2}) is consistent with the projection and the BPS bound just derived.

%%%%%%%%%%%%%%%%%%%%%%%%%%%%%%%%%%%%%%%%%%%%%%%%%%%%%%%%%%%%%%%%%%%
%%%%%%%%%%%%%%%%%%%%%%%%%%%%%%%%%%%%%%%%%%%%%%%%%%%%%%%%%%%%%%%%%%%%%%%%%%%%
%%%%%%%%%%%%%%%%%%%%%%%%%%               Appendices                     %%%%%%%%%%%%%%%%%%%%%%%%%
%%%%%%%%%%%%%%%%%%%%%%%%%%%%%%%%%%%%%%%%%%%%%%%%%%%%%%%%%%%%%%%%%%%%%%%%%%%%
%\appendix
\setcounter{section}{0}

\renewcommand{\thesection}{\Roman{part}.\Alph{section}}
\renewcommand{\thesubsection}{\thesection .\Roman{subsection}}

   \setcounter{equation}{0}

  \chapter*{Appendices Part III  \markboth{Appendices Part III}{Appendices Part III}}
\addcontentsline{toc}{chapter}{Appendices Part III} 

 \hrule
\vspace{2cm}

 \setcounter{equation}{0}

 \section{Gamma matrices}
\label{App: Gamma}

We will consider 4 real 4x4 matrices $\gamma_a$, called gamma matrices, that satisfy the Clifford algebra
\beqs
\{\gamma_{a}, \gamma_{b} \}=\gamma_a \gamma_b+\gamma_b \gamma_a  = 2  \eta_{ab} \: I.
\eeqs
For definiteness, we list below a choice of real gamma matrices:
\begin{eqnarray}
\gamma_0 &=& \left (
\begin{array}{cccc}
0 & 0 & 0  & -1 \\
0 & 0 & 1  &  0 \\
0 &-1 & 0  &  0 \\
1 & 0 & 0  &  0
\end{array}
\right ) , \:\:\: \gamma_1= \left (
\begin{array}{cccc}
1 & 0 & 0  & 0 \\
0 & -1 & 0  &  0 \\
0 & 0 & 1  &  0 \\
0 & 0 & 0  &  -1
\end{array}
\right ) , \nonumber \\
 \gamma_2 &=& \left (
\begin{array}{cccc}
0 & 0 & 0  & 1 \\
0 & 0 & -1  &  0 \\
0 & -1 & 0  &  0 \\
1 & 0 & 0  &  0
\end{array}
\right ) , \:\:\: \gamma_3= \left (
\begin{array}{cccc}
0 & -1 & 0  & 0 \\
-1 & 0 & 0  &  0 \\
0 & 0 & 0  &  -1 \\
0 & 0 & -1  &  0
\end{array}
\right ) .
\end{eqnarray}
We also have
\beqs
\gamma^a=\eta^{ab} \gamma_a \qquad \rightarrow \qquad \gamma^{0}=-\gamma_0 , \qquad \gamma^i=\gamma_i
\eeqs
Let us now define the parity matrix
\beqs
\gamma_5=\gamma_0 \gamma_1 \gamma_2 \gamma_3 , \qquad (\gamma_5)^2=-1 ,
\eeqs
which is real and antisymmetric. The conjugation matrix used in the main text is defined as $C \equiv \gamma_0$.
Let us eventually define
\beqs
\gamma_{a_1...a_n}=\gamma_{[a_1}...\gamma_{a_n]} ,
\eeqs
which implies as one can check
\beqs
\gamma_{ab}&=& \frac{1}{2} (\gamma_a \gamma_b - \gamma_b \gamma_a)=\gamma_a \gamma_b-\eta_{ab} , \\
\gamma_{abc}&=& \gamma_a \gamma_b \gamma_c -\eta_{ab} \gamma_c +\eta_{ac}\gamma_b - \eta_{bc} \gamma_a  \\
             &=& \gamma_a \gamma_{bc}-\eta_{ab} \gamma_c +\eta_{ac}\gamma_b , \label{gabc2}\\
\gamma_{abcd}&=&\gamma_{a} \gamma_b \gamma_c \gamma_d +\gamma_a \gamma_c \eta_{bd}-\gamma_a \gamma_b \eta_{cd}-\gamma_a \gamma_d \eta_{bc}\nonumber \\
&& \:\:\: -\gamma_b \gamma_c \eta_{ad}-\gamma_c \gamma_d \eta_{ab}+\gamma_b \gamma_d \eta_{ac}-\eta_{ac} \eta_{bd}+\eta_{ab} \eta_{cd}+\eta_{ad} \eta_{bc}\\
&=& \frac{1}{2} (\gamma_{a} \gamma_{bc} \gamma_d - \gamma_d \gamma_{bc} \gamma_a)-\eta_{ab} \eta_{cd} +\eta_{ac}\eta_{bd} , \label{gabcd2}\\
...\nonumber
\eeqs
One can also verify the following identities
\beqs
&& \gamma_a \gamma^a= 4 \: I , \qquad \gamma_b \gamma^a \gamma^b=-2 \gamma^a , \qquad \gamma_{c}\gamma^a \gamma^b \gamma^c= 4 \eta^{ab}\: I , \\
&& \gamma_{d} \gamma^a \gamma^b \gamma^c \gamma^{d}=2 \gamma^a \gamma^b \gamma^c - 4 \gamma^a \eta^{bc} + 4 \gamma^b \eta^{ac} - 4 \gamma^c \eta^{ab} .
\eeqs
From the above definitions and identities, one can check that $\gamma^{abc}=-\gamma_d \gamma^{abcd}$. This also implies that
\beqs
\gamma^{abc}=\epsilon^{abcd}\gamma_d \gamma_5 ,
\eeqs
where $\epsilon_{0123}=+1$.
Indeed, $\gamma_5$ can be written as
\beqs\label{g5re}
\gamma_5&=&\gamma_0 \gamma_1 \gamma_2 \gamma_3= \delta^{abcd}_{0123} \gamma_a \gamma_b \gamma_c \gamma_d \nonumber\\
&=& \frac{1}{4!} \epsilon^{abcd} \gamma_a \gamma_b \gamma_c \gamma_d ,
\eeqs
and one readily sees that
\beqs
\epsilon^{abcd} \gamma_d \gamma_5= \epsilon^{abcd} \gamma_d \frac{1}{4!} \epsilon_{efgh} \gamma^e \gamma^f \gamma^g \gamma^h=-\gamma_d \gamma^{abcd} .
\eeqs
Another useful relation implied by the previous one is
\beqs
\gamma_d \gamma_5 \gamma_{ab}&=&\gamma_{d} \gamma_{ab} \gamma_5 \nonumber \\
&=& \gamma_{dab} \gamma_5 + 2 \eta_{d[a}\gamma_{b]}\gamma_5 \nonumber \\
&=& -\epsilon_{dabc} \gamma^c+ 2 \eta_{d[a}\gamma_{b]}\gamma_5 ,
\eeqs
where in the second equality we have used \eqref{gabc2}.

A last useful relation that can be obtained from \eqref{gabc2} is 
\beqs
\gamma_{a} \gamma_{bc} \gamma_d - \gamma_d \gamma_{bc} \gamma_a= 2 \epsilon_{abcd} \gamma_5+ 2(\eta_{ab} \eta_{cd} -\eta_{ac}\eta_{bd}), \label{gabcd2}\\
\eeqs
which can be checked by replacing $\gamma_5$ with \eqref{g5re}.

%%%%%%%%%%%%%%%%%%%%%%%%%%%%%%%%%%%%%%%%%%%%%%%%%%%%%%%%%%%%%%%%%%%%%%%%%%%%
%%%%%%%%%%%%%%%%%%%%%%%%%%               Bibliographie                    %%%%%%%%%%%%%%%%%%%%%%%
%%%%%%%%%%%%%%%%%%%%%%%%%%%%%%%%%%%%%%%%%%%%%%%%%%%%%%%%%%%%%%%%%%%%%%%%%%%%
\bibliographystyle{jhep}%{utphysmod2}
\addcontentsline{toc}{chapter}{Bibliography}

%\bibliography{bibliographie} \markboth{Bibliography}{Bibliography}

%% This BibTeX bibliography file was created using BibDesk.
%% http://bibdesk.sourceforge.net/

%% Created for Francois Dehouck at 2011-12-15 16:43:23 +0100 

%% Saved with string encoding Unicode (UTF-8) 
\providecommand{\href}[2]{#2}\begingroup\raggedright

  %%%%%%%%%%%%%%%%%%%%%%%%%%%%%%%%%%%%%%%%%%%%%%%%%%%%%%%%%%%%%%%%%%%%%%%%%%%

\end{document}